\documentclass[11pt]{article}
\usepackage{amsfonts, amssymb, a4, setspace, lscape,color}
\usepackage{amsmath}
\usepackage{bbm}
\usepackage{newinutile}
\usepackage{graphicx,calc}
\usepackage{cite}

\usepackage{mathrsfs}

\newcommand{\rr}[1]{{\textrm{#1}}}

\newcommand{\bb}[1]{{\mathbb{#1}}}

\begin{document}

\title{Heterogeneous perturbation of fluid density and solid elastic strain 
in consolidating porous media
}

\author{P.\ Artale Harris}
\affiliation{Dipartimento di Scienze di Base e Applicate per l'Ingegneria, 
              Sapienza Universit\`a di Roma, 
              via A.\ Scarpa 16, I--00161, Roma, Italy.}
\email{pietro.artale.h@gmail.com}
\author{E.N.M.\ Cirillo}
\affiliation{Dipartimento di Scienze di Base e Applicate per l'Ingegneria, 
              Sapienza Universit\`a di Roma, 
              via A.\ Scarpa 16, I--00161, Roma, Italy.}
\email{emilio.cirillo@uniroma1.it}
\author{G.\ Sciarra}
\affiliation{Dipartimento di Ingegneria Chimica Materiali Ambiente,
             Sapienza Universit\`a di Roma, 
             via Eudossiana 18, 00184 Roma, Italy}
\email{giulio.sciarra@uniroma1.it}

\maketitle

\begin{abstract}
The occurrence of heterogeneous perturbations of fluid mass density 
and solid elastic strain of a porous continuum, as a consequence of its undrained 
response is a very important topic in 
theoretical and applied poromechanics. 
The classical Mandel--Cryer effect
provides an explanation of fluid overpressure 
in the central region of a porous sample,
immediately after the application of the loading. 
However this effect fades away when the fluid
leaks out of the porous network.
Here this problem is studied within the framework of a 
second gradient theory and a thorough description 
of the static and the dynamics of the phenomenon is given. 
\end{abstract}

\section{Introduction}
\label{s:introduzione}
When a mechanical pressure is exherted on the solid skeleton 
of a porous medium 
and its
elastic strain is a consequence of the variation of the fluid mass content 
inside the pores, 
several interesting phenomena can occur which accompany shrinkage or swelling of the solid skeleton.
In this paper attention is focused on the occurrence of heterogeneous perturbations of the fluid mass density 
and the skeleton elastic strains, as a consequence of the undrained 
response of the porous medium. The classical Mandel--Cryer effect, see \cite{Mandel} and \cite{Cryer63},
provides an explanation, within multidimensional consolidation, of fluid overpressure, in the central region of the sample,
immediately after the application of the loading. However, in that case this is a definitely non--permanent effect 
which fades away when the fluid leaks 
out the boundary and the pore pressure reverses and dissipates. 
The physical background of the Mandel--Cryer effect 
is that the generation of fluid over-pressure due to loading is immediate, but the dissipation 
due to the fluid flow is retarded by the permeability and the distance to the drainage boundary.
On the other hand several authors discussed the onset of strain localization during globally undrained triaxial
tests, in particular for loose granular materials, 
see \textit{e.g.}, \cite{Mooney}, \cite{Mokni1998}, or \cite{Sulem2006}.
In this case local fluid exchange is allowed, even in presence of localized strain, inside the specimen until, at high level of confinement, 
the pore pressure generation inside the band leads locally to fluidization of the crushed 
material, which results into the formation of connected channels in the heart of the band.
Similar confinement effects have also been recorded in a fluidized column test , see \cite{Nichols94}, 
where a fluid is forced to flow through a saturated sample from the bottom. By tuning the velocity of the fluid, 
the drag force acting on the solid grains, possibly causing unbalance of gravity force, is controlled
\cite{Vardoulakis04_1,Vardoulakis04_2}. These experimental results demonstrate that 
under consolidation loading, and because of porosity change, the fluid can migrate through the pores and 
eventually remain segregated, possibly enhancing localized overpressurization and fluidization of the soil, 
see e.g. \cite{Kolymbas94,Nichols94}. 

In the papers \cite{CIS2011,CIS2009,CIS2010} and \cite{CIS2013} we have attacked this problem from the 
point of view of bifurcation theory and we have shown that it is possible to describe interesting phenomena 
(still in the range of non--linear elasticity) taking place when the confining pressure exerted on the solid 
exceeds a suitable limiting value. The idea is to get a formulation capable for describing 
the onset of a fluid--rich and a fluid--poor phase, eventually coexisting inside the porous skeleton
at equilibrium. Introducing a non--local energy contribution, which penalizes gradients 
of strain and fluid mass density, a smooth transition between phases of the porous medium,
associated with different fluid content, has been modeled, so accounting for the arising of the 
above mentioned heterogeneous elastic strains. 
Undrained conditions are therefore locally achieved where fluid segregation is attained, even if
a standard Darcean dissipative process, associated to the fluid flowing out of the
drainage boundary, occurs.

Assuming the potential energy to be quadratic in the first derivatives of the strain 
and of the fluid mass density variation, the evolution is described by 
a Cahn--Hilliard--like equation provided 
that the dissipative forces are proportional to the seepage
velocity, say the velocity of the fluid with respect to the solid.
This means that the above mentioned assumption of Darcean flow
still remains valid. 

Within this modeling framework, a generalized consolidation problem for a 
one--dimensional 
porous continuum is analyzed so extending the classical results due to Terzaghi
and those ones, relative to a gradient model, previously obtained by one of the authors, and co--workers \cite{SIIM},
in which only the fluid--poor phase was admissible at equilibrium. The equation governing 
the behavior of the fluid constituent is of higher (fourth) order with respect to the Laplace equation 
which classically prescribes the behavior of the pore--water pressure. Following previous results 
reported in \cite{CIS2013}, different boundary conditions can be considered, 
in particular essential or natural boundary conditions relative to the velocity of the fluid relative to the solid the fluid chemical potential
as well as to the fluid mass density or its spatial gradient; where the dependence of the boundary value problem on
higher order derivatives has been taken into account. Here we shall address the two cases in which zero chemical potential, see 
section \ref{s:zch}, or zero fluid velocity, say impermeability of the porous skeleton, see section \ref{s:osi}, 
have been assumed on the whole or part of the boundary, together with 
essential boundary conditions on the strain of the solid and the density of the fluid.
As already mentioned the interest will be in the occurrence of heterogeneous elastic strains 
of the solid skeleton and variations of the fluid density; the confining pressure is therefore chosen
so as to guarantee the coexistence of phases and, consequently, the onset and the propagation,
up to its stationary placement, of the interface 
between them, see \cite{CIS2012}.

\section{The model}
\label{s:model}
We introduce the one dimensional poromechanical model
\cite{CIS2013}
whose geometrically linearized version will be studied 
in the following sections. Kinematics will be briefly resumed starting from the
general statement of the model \cite{Coussy} together with some
particular issue introduced in \cite{SIIM}. 
The equations governing 
the behavior of the porous system will then be deduced prescribing 
the conservative part of
the constitutive law through a suitable potential energy density 
$\Phi$ and the dissipative 
contributions through purely Darcy terms. 

\subsection{Poromechanics setup}
\label{s:setup}
Let $B_\rr{s}:=[\ell_1,\ell_2]\subset\bb{R}$, with 
$\ell_1,\ell_2\in\bb{R}$, 
and $B_\rr{f}:=\bb{R}$ be the \textit{reference}
configurations
for the solid and fluid components~\cite{Coussy}. 
The \textit{solid placement} 
$\chi_\rr{s}:B_\rr{s}\times\bb{R}\to\bb{R}$ is a $C^2$ function such that 
the map $\chi_\rr{s}(\cdot,t)$, 
associating to each $X_\rr{s}\in B_\rr{s}$
the position occupied at time $t$ by the particle labeled 
by $X_\rr{s}$ in the reference configuration $B_\rr{s}$,
is a $C^2$--diffeomorphism.
The \textit{fluid placement} map 
$\chi_\rr{f}:B_\rr{f}\times\bb{R}\to\bb{R}$
is defined analogously.
The \textit{current configuration} $B_t:=\chi_\rr{s}(B_\rr{s},t)$ at time 
$t$ is the set of positions 
of the superposed solid and fluid particles.

Consider the $C^2$ function
$\phi:B_\rr{s}\times\bb{R}\to B_\rr{f}$ 
such that 
$\phi(X_\rr{s},t)$ is 
the fluid particle that at time $t$ occupies the 
same position of the solid particle $X_\rr{s}$; 
assume, also, that $\phi(\cdot,t)$ is a $C^2$--diffeomorphism 
mapping univocally a solid particle 
into a fluid one.
The three fields\footnote{In the sequel we shall often use the inverse 
functions of the
space sections of the field $\chi_\rr{f}$, $\chi_\rr{s}$, and $\phi$. 
We shall misuse the notation and let 
$\phi^{-1}(\cdot,t)$ be the inverse of the map 
$X_\rr{s}\to\phi(X_\rr{s},t)$ at a given time $t$.
Similarly we shall also consider 
$\chi_\rr{s}^{-1}(\cdot,t)$ and 
$\chi_\rr{f}^{-1}(\cdot,t)$.}
$\chi_\rr{s}$, $\chi_\rr{f}$, and $\phi$ are 
not at all independent; indeed, by definition, we immediately have that 
$\chi_\rr{f}(\phi(X_\rr{s},t),t)=\chi_\rr{s}(X_\rr{s},t)$ 
for any $X_\rr{s}\in B_\rr{s}$ and $t\in\bb{R}$.

The Lagrangian velocities are two maps associating with each time and 
each point in the solid and fluid reference space the velocities of the 
corresponding solid and fluid particles at the 
specified time. 
More precisely,
the \textit{Lagrangian velocities} are the two maps
$u_\alpha:B_\alpha\times\bb{R}\to\bb{R}$ 
defined by setting
\begin{equation}
\label{vel-lagr}
u_\alpha(X_\alpha,t):=\frac{\partial\chi_\alpha}{\partial t}(X_\alpha,t)
\end{equation}
for any $X_\alpha\in B_\alpha$, where $\alpha=\rr{s},\rr{f}$.
We also consider the \textit{Eulerian velocities} 
$v_\alpha:B_t\times\bb{R}\to\bb{R}$ associating
with each point $x\in B_t$ and for each time $t\in\bb{R}$ the velocities
of the solid and fluid particle occupying the place $x$ at time $t$;
more precisely we set 
$v_\alpha(x,t):=u_\alpha(\chi^{-1}_\alpha(x,t),t)$.

In studying the dynamics of the porous system one can arbitrarily 
choose two among the three fields 
$\chi_\rr{s}$, $\chi_\rr{f}$, and $\phi$. 
Since the reference configuration $B_\rr{s}$ of the solid component 
is know a priori, 
a good choice appears to be that of expressing all the dynamical observables 
in terms of the fields $\chi_\rr{s}$ and $\phi$ which are defined 
on $B_\rr{s}$. Consider the Lagrangian velocity $u_\rr{f}$ of the 
fluid component which is defined on $B_\rr{f}\times\bb{R}$; we prove 
that for any $X_\rr{s}\in B_\rr{s}$ and $t\in\bb{R}$
\begin{equation}
\label{vf}
u_\rr{f}(\phi(X_\rr{s},t),t)
=
u_\rr{s}(X_\rr{s},t)
-
\frac{\chi_\rr{s}'(X_\rr{s},t)}
     {\phi'(X_\rr{s},t)}\, \dot\phi(X_\rr{s},t)
\end{equation}
which means that the velocity of the fluid relative to the solid
$v(X_\rr{s},t):=u_\rr{f}(\phi(X_\rr{s},t),t)-u_\rr{s}(X_\rr{s},t)$, in the following seepage velocity,
is proportional to the opposite of $\dot \phi(X_\rr{s},t)$.
The dot and the prime denote, here and in the sequel, 
the partial derivative with respect to time and to the 
space variable $X_\rr{s}$ respectively. 

\subsection{Equations of motion}
\label{s:variational}
In \cite{CIS2013} we have derived 
the equations of motion by using a
variational approach much similar to that developed in~\cite{SIIM}.
We shall recall, here, the main steps in the computation.

It is natural to assume that, if the system is acted upon only by 
conservative forces, its dynamics is described by a 
\textit{Lagrangian density} 
$\mathscr{L}$,
relative to the solid reference configuration space volume,
depending on the space variable $X_\rr{s}$ and on time 
through (in principle)
$\dot\chi_\rr{s}$, $\dot\phi$, $\chi''_\rr{s}$, $\phi''$,
$\chi'_\rr{s}$, $\phi'$, $\chi_\rr{s}$, and $\phi$.
The Lagrangian density is
equal to the 
\textit{kinetic energy density} minus
the \textit{overall potential energy density} accounting for both 
the internal and the external conservative forces.
The equations of motion for the two fields $\chi_\rr{s}$ and $\phi$ can be 
derived assuming that the possible motions of the system in an 
interval of time $(t_1,t_2)\subset\bb{R}$
are those such that the fields $\chi_\rr{s}$ and $\phi$ are 
extremals
for the \textit{action functional}
\begin{equation}
\label{azione}
 A(\chi_\rr{s},\phi)
 :=
 \int_{t_1}^{t_2}\rr{d}t
 \int_{B_\rr{s}}\rr{d}X_\rr{s}
 \,
 \mathscr{L}(
            \dot\chi_\rr{s}(X_\rr{s},t),
            \dots,
            \phi(X_\rr{s},t))
\end{equation}
in correspondence of the independent variations 
of the two fields
$\chi_\rr{s}$ and $\phi$ on $B_\rr{s}\times(t_1,t_2)$.
In other words any possible motion of the system in the 
considered interval is a solution of the Euler--Lagrange equations
associated to the variational principle $\delta A=0$.

A different variational principle is needed if the 
fluid component of the system is acted upon by dissipative
forces; the virtual work made by these forces
must be taken into account.
Consider the independent variations 
$\delta\chi_\rr{s}$ 
and $\delta\phi$ of the two fields $\chi_\rr{s}$ and $\phi$ and
denote by $\delta W$ the corresponding elementary 
\textit{virtual work}
made by the dissipative forces acting on the fluid component.
The possible motions of the system, see for instance 
\cite[Chapter~5]{Blanchard},
in an interval of time $(t_1,t_2)\subset\bb{R}$
are those such that the fields $\chi_\rr{s}$ and $\phi$ satisfies
the variational principle
\begin{equation}
\label{vardiss}
\delta A
=
-\int_{t_1}^{t_2}\delta W\,\rr{d}t
\end{equation}
namely, the variation of the the action integral in 
correspondence of a possible motion is equal to the 
integral over time of minus the virtual work of the dissipative 
forces corresponding to the considered variation of the fields. 

The way in which dissipation has to be 
introduced in saturated porous media 
models is still under debate, see for instance \cite{Nield}.
In particular according to the effectiveness of the hypothesis 
of separation of scales, between the local and macroscopic level,
Darcy's or Stokes' effects are accounted for. We refer 
the interested reader to \cite{CIS2013} for a complete discussion of this 
problem.

In this paper we shall 
consider the so--called Darcy effect, i.e., the dissipation 
due to forces proportional to the velocity of the 
fluid component measured with respect to the solid.
By following phenomenological arguments \cite{Bear} or
by developing suitable homogenization schemes \cite{Ene&SanchezP75}
it is seen that 
the fluid flow is, in this case, controlled by the permeability of
the porous material.
A natural expression for the virtual work of the dissipation forces 
acting on the fluid component and taking into account the Darcy 
effect is 
\begin{equation}
\label{lv01}
 \delta W:=
 -
 \int_{B_t}
 D[v_\rr{f}(x,t)-v_\rr{s}(x,t)]
 [\delta\chi_\rr{f}(\chi^{-1}_\rr{f}(x,t),t)
 -
 \delta\chi_\rr{s}(\chi^{-1}_\rr{s}(x,t),t)]
\,\rr{d}x
\end{equation}
where $D>0$ is a constant proportional to the inverse of permeability and
$\delta\chi_\rr{f}$ is the variation of the field $\chi_\rr{f}$
induced by the independent variations $\delta\chi_\rr{s}$ and 
$\delta\phi$.
The quantity 
$-D[v_\rr{f}-v_\rr{s}]$ is the 
dissipative force density
(relative to the current configuration space volume)
which depends on the kinematic fields only through the 
velocity of the fluid component relative to the solid. 
Apparently equation (\ref{lv01}) for the virtual work of Darcy dissipative
forces is consistent with the classical expression of Darcy dissipation 
associated with the viscous flow through a porous continuum, see \cite{Coussy}.

Following the recipe described above we have to express the
virtual work in terms of the field $\chi_\rr{s}$ and $\phi$.
In \cite{CIS2013} it has been proven that 
\begin{equation}
\label{lv18}
\delta W
=
-\int_{B_\rr{s}}R\delta\phi
\;\rr{d}X_\rr{s}
\;\;\;\textrm{ with }\;\;\;
R:=D\frac{1}{(\phi')^2}\dot\phi(\chi'_\rr{s})^3
\end{equation}

In order to write explicitly the variation of the action we 
specify, now, the form of the Lagrangian density.
In the sequel we shall not consider the inertial 
effects, so that, the Lagrangian density will be the opposite of 
the potential energy 
density associated to both the internal and external 
conservative forces. 
As it has been shown in~\cite{SIIM} (see equation~(18) therein)
it is reasonable to assume that the potential energy density 
depends on the space and time variable only via two 
physically relevant functions: 
the strain of the solid and a properly normalized fluid mass 
density~\cite{CIS2011,CIS2009,CIS2010,SIIM}.

More precisely, 
consider the Jacobian $\chi'_\rr{s}(X_\rr{s},t)$ of the solid placement map,
which  measures the ratio between current and 
reference volumes of the solid component, and let 
\begin{equation}
\label{deformazione}
\varepsilon(X_\rr{s},t):=[(\chi'_\rr{s}(X_\rr{s},t))^2-1]/2
\end{equation}
be the \textit{strain field}.
Let $\varrho_{0,\rr{f}}:B_\rr{f}\to\bb{R}$ 
be the fluid reference \textit{density}; we define the 
\textit{fluid mass density} field
\begin{equation}
\label{densita}
m_\rr{f}(X_\rr{s},t)
 :=\varrho_{0,\rr{f}}(\phi(X_\rr{s},t))
   \phi'(X_\rr{s},t)
\end{equation}
Assuming that the mass is conserved, it is proven~\cite{SIIM}
that the field $m_\rr{f}$ can be interpreted as 
the fluid mass density measured with respect to the 
solid reference volume.

We assume the total potential energy density $\Phi$ to depend 
on the fields 
$m_\rr{f}$ and $\varepsilon$ and on their space
derivative
$m'_\rr{f}$ and $\varepsilon'$.
Since 
$m_\rr{f}=\varrho_{0,\rr{f}}(\phi)\phi'$,
$m'_\rr{f}=\varrho'_{0,\rr{f}}(\phi)(\phi')^2+\varrho_{0,\rr{f}}(\phi)\phi''$,
$\varepsilon=((\chi'_\rr{s})^2-1)/2$,
and $\varepsilon'=\chi'_\rr{s}\chi''_\rr{s}$, we have that 
the Lagrangian density $\mathscr{L}=-\Phi$ depends on the space and time 
variables through the fields 
$\phi$, $\phi'$, $\phi''$, $\chi'_\rr{s}$, and 
$\chi''_\rr{s}$.

By a standard variational computation, see \cite{CIS2013} for 
details, we finally get the equations of motion
\begin{equation}
\label{em04}
 \Big[\chi'_\rr{s}
 \Big(\frac{\partial\Phi}{\partial\varepsilon}
      -
      \Big(
      \frac{\partial\Phi}{\partial\varepsilon'}
      \Big)'
 \Big)
 \Big]'
 =0
\;\textrm{ and }\;
 \varrho_{0,\rr{f}}(\phi)
 \Big[\frac{\partial\Phi}{\partial m_\rr{f}}
      -
      \Big(
      \frac{\partial\Phi}{\partial m'_\rr{f}}
      \Big)'
 \Big]'
 =
 R
\end{equation}
and the boundary conditions
\begin{equation}
\label{em04bc}
\begin{array}{l}
{\displaystyle
 \Big\{
 \frac{\partial\Phi}{\partial m'_\rr{f}}
 \varrho_{0,\rr{f}}(\phi)
 \delta\phi'
 +
 \Big[
 \Big(\frac{\partial\Phi}{\partial m_\rr{f}}
      -
      \Big(
      \frac{\partial\Phi}{\partial m'_\rr{f}}
      \Big)'
 \Big)
 \varrho_{0,\rr{f}}(\phi)
 +
 \frac{\partial\Phi}{\partial m'_\rr{f}}
 \varrho'_{0,\rr{f}}(\phi)\phi'
 \Big]
 \delta\phi
}
\vphantom{\bigg\{_\{}
\\
{\displaystyle
 \phantom{aaaaaaaaaaaaaaaaaaaaa}
 + 
 \frac{\partial\Phi}{\partial\varepsilon'}
 \chi'_\rr{s}
 \delta\chi'_\rr{s}
 +
 \Big(\frac{\partial\Phi}{\partial\varepsilon}
      -
 \Big(
      \frac{\partial\Phi}{\partial\varepsilon'}
 \Big)'
 \Big)
 \chi'_\rr{s}
 \delta\chi_\rr{s}
 \Big\}_{\ell_1}^{\ell_2}
 =0
}
\end{array}
\end{equation}
We remark that, though they are partially written in terms of 
the fields $m_\rr{f}$ 
and $\varepsilon$, the equations (\ref{em04}) are evolution 
equations for the two fields $\chi_\rr{s}$ and $\phi$. Only under a
brute approximation, see the geometrical linearization discussed in 
Section~\ref{s:lin},
they will reduce to a set of evolutionary equations 
for the fields $m_\rr{f}$ and $\varepsilon$. 

\subsection{Equations of motion under geometrical linearization}
\label{s:lin}
In this section we rewrite the equations of motion of the 
poromechanical system under the so called 
\textit{geometrical linearization} assumption, namely, 
in the case 
when only small elastic strains are present in the system.
From now on we shall assume that $\varrho_{0,\textrm{f}}$ is constant. 
We first introduce the \textit{displacement fields} 
$u(X_\rr{s},t)$ and $w(X_\rr{s},t)$ by setting
\begin{equation}
\label{gd00}
\chi_\rr{s}(X_\rr{s},t)=X_\rr{s}+u(X_\rr{s},t)
\;\textrm{ and }\;
\phi(X_\rr{s},t)=X_\rr{s}+w(X_\rr{s},t)
\end{equation}
for any $X_\rr{s}\in B_\rr{s}$ and $t\in\bb{R}$. We then 
assume that $u$ and $w$ are small, together with 
their space and time derivatives, and write the equations of motion 
up to the first order in $u$, $w$, and derivatives.
By using 
(\ref{deformazione}), (\ref{densita}), and recalling the definition 
of $R$ in (\ref{lv18}),
we get   
\begin{equation}
\label{gd01}
m_\rr{f}=\varrho_{0,\rr{f}}(1+w'),\;
m:=m_\rr{f}-\varrho_{0,\rr{f}}=\varrho_{0,\rr{f}}w',\;
\varepsilon\approx u',\;
\textrm{ and }
R\approx D\dot{w},
\end{equation}
where $\approx$ means that all the terms of order larger than one 
have been neglected.

We have introduced above the field $m$. In the following we shall imagine 
$\Phi$ as a function of $m$ and $m'$ and the equations of motion and 
the boundary conditions will be written in terms of this field. 
We get the equations of motion 
\begin{equation}
\label{gd02}
 \Big[
 (1+\varepsilon) 
 \Big(
 \frac{\partial\Phi}{\partial\varepsilon}
 -
 \Big(
 \frac{\partial\Phi}{\partial\varepsilon'}
 \Big)'
 \Big)
 \Big]'
 =0
\;\;\textrm{ and }\;\;
 \varrho_{0,\rr{f}}
 \Big[
 \frac{\partial\Phi}{\partial m}
      -
      \Big(
      \frac{\partial\Phi}{\partial m'}
      \Big)'
 \Big]'
 =
 D\dot{w}
\end{equation}
and the associated boundary conditions
\begin{equation}
\label{gd02bc}
 \Big\{
 \frac{\partial\Phi}{\partial\varepsilon'}
 \delta\varepsilon
 +
 \frac{\partial\Phi}{\partial m'}
 \delta m
 +
 \Big[
 \Big(\frac{\partial\Phi}{\partial m}
      -
      \Big(
      \frac{\partial\Phi}{\partial m'}
      \Big)'
 \Big)
 \varrho_{0,\rr{f}}
 \Big]
 \delta w
 +
 \Big(\frac{\partial\Phi}{\partial\varepsilon}
      -
 \Big(
      \frac{\partial\Phi}{\partial\varepsilon'}
 \Big)'
 \Big)
 \delta u
 \Big\}_{\ell_1}^{\ell_2}
 \!\!\!\!
 =0
\end{equation}

Recalling that in our approximation $m=\varrho_{0,\textrm{f}}w'$, see 
the second among equations (\ref{gd01}), we have that 
the equations (\ref{gd02}) are evolution equations for the 
fields $m$ and $\varepsilon$.
Moreover, 
as remarked in \cite{CIS2013},
provided the potential energy density 
$\Phi$ depends on the space derivatives $m'$ and $\varepsilon'$ of the 
fields $m$ and $\varepsilon$ as a quadratic form with constant coefficients, 
it follows that the second between the equations of motion (\ref{gd02}) 
becomes a Cahn--Hilliard--like equation for the field $m$ 
with driving field still depending 
parametrically on $\varepsilon$.

Note, also, that 
the equations (\ref{gd02bc}) are an extension to the case of 
second gradient elasticity of classical natural and essential 
boundary conditions \cite{ISV}.
By choosing in two different ways the boundary condition we shall 
obtain two different physically relevent boundary value problems. 

\subsection{The zero chemical potential problem}
\label{s:zch}
A set of boundary conditions implying that (\ref{gd02bc}) 
are satisfied is
\begin{equation}
\label{gd03bc}
\Big(
 \frac{\partial\Phi}{\partial\varepsilon'}
 \delta\varepsilon
 +
 \frac{\partial\Phi}{\partial m'}
 \delta m
\Big)_{\ell_1,\ell_2}
 \!\!\!\!
=
 \Big[\frac{\partial\Phi}{\partial m}
      -
      \Big(
      \frac{\partial\Phi}{\partial m'}
      \Big)'
 \Big]_{\ell_1,\ell_2}
 \!\!\!\!
=
 \Big[\frac{\partial\Phi}{\partial\varepsilon}
      -
 \Big(
      \frac{\partial\Phi}{\partial\varepsilon'}
 \Big)'
 \Big]_{\ell_1,\ell_2}
 \!\!\!\!
=0
\end{equation}
where the notation above means that the functions in brackets are 
evaluated both in $\ell_1$ and $\ell_2$.
With this choice it is possible to fix the boundary conditions 
directly on fields $m$ and $\varepsilon$ (and derivatives).

The first equation (\ref{gd03bc}) is the additional 
boundary condition due to the presence of the gradient terms in the 
potential energy density $\Phi$. This equation specifies essential boundary 
conditions on the derivatives of the displacement fields or natural 
boundary conditions on the so called double forces, see \cite{germain73}.
The generalized essential boundary conditions can be read as a 
prescription on the derivative of the independent fields $\chi_\rr{s}$ 
and $\phi$, see equations (\ref{deformazione}) and (\ref{densita}); whilst 
the extended natural boundary conditions prescribe, on one hand, 
the additional forces which the solid continuum is able to balance at
the boundary and, on the other, the wetting properties of the fluid
which fills the pores \cite{Seppecher89}.

The second and the third equations (\ref{gd03bc}) provide 
natural boundary conditions prescribing 
the chemical potential of the fluid
and the traction exerted on the overall porous solid, 
respectively.
These conditions extend the classical ones (which can be easily found in 
the literature) for a porous solid suffering internal stresses due to 
applied tractions and for a saturating fluid with fixed 
chemical potential; see \cite{baek} for more details. 
In this case we use the words generalized traction and chemical potential 
because of the additional contribution to stress 
($\partial \Phi/\partial\varepsilon$) 
and chemical potential ($\partial\Phi/\partial m$) provided by the spatial 
derivatives of the corresponding 
hyperstress fields, say $(\partial \Phi/\partial\varepsilon')'$
and $(\partial \Phi/\partial m')'$.

Finally, we note that, by exploiting the third boundary 
condition (\ref{gd03bc}) in $\ell_1$, we get the PDE problem
\begin{equation}
\label{problema-d}
\left\{
\begin{array}{l}
{\displaystyle
 \frac{\partial\Phi}{\partial\varepsilon}
 -
 \Big(
 \frac{\partial\Phi}{\partial\varepsilon'}
 \Big)'
 =0
\;\;\textrm{ and }\;\;
 \varrho_{0,\rr{f}}^2
 \Big[
      \frac{\partial\Phi}{\partial m}
      -
      \Big(
      \frac{\partial\Phi}{\partial m'}
      \Big)'
 \Big]''
 =
 D\dot{m}
\vphantom{\bigg\{_\big\{}
}
\\
{\displaystyle
\Big(
 \frac{\partial\Phi}{\partial\varepsilon'}
 \delta\varepsilon
 +
 \frac{\partial\Phi}{\partial m'}
 \delta m
\Big)_{\ell_1,\ell_2}
=
 \Big[\frac{\partial\Phi}{\partial m}
      -
      \Big(
      \frac{\partial\Phi}{\partial m'}
      \Big)'
 \Big]_{\ell_1,\ell_2}
=0
}
\\
\end{array}
\right.
\end{equation}
that will be called the \textit{zero chemical potential} problem.

\subsection{The one--side impermeable problem}
\label{s:osi}
In the zero chemical potential problem we can imagine 
that the porous one-dimensional system is plunge into 
a fluid reservoir and that at the two boudaries 
the fluid can get in and out from the sample freely. 
A rather different very interesting situation in 
applications, see the discussion provided in the 
Introduction, is the one in which one of the two 
boudaries is not permeable to the fluid. 

This physical situation can be mathematically implemented as follows.
Consider the third term of boundary condition (\ref{gd02bc}), which is zero when
zero chemical potential is prescribed at the boundary, see equation (\ref{problema-d}).
According with classical arguments of variational calculus it vanishes
also when the variation of the fluid displacement, $\delta w$, vanishes, which 
corresponds to account for essential rather than natural boundary conditions.
From the physical point of view, assuming $\delta w=0$ implies the fluid particle, occuping
the same current place as the solid particle which remains at the boundary, 
to be always the same during evolution. In other words assuming that, at any time
$t$, $\delta w$ equals zero, implies the seepage velocity, 
say $v=u_{\rr f}-u_{\rr s}=-\dot w$, to identically vanish and therefore the boundary to be
impermeable to the fluid flow. Considering the bulk equation (\ref{gd02})$_2$ 
this boundary condition can therefore be rephrased as follows:
\begin{equation}
\label{vel000}
v=-\frac{\varrho_{0,\rr{f}}}{D}\Big(
     \frac{\partial\Phi}{\partial m}
     -
     \Big(\frac{\partial\Phi}{\partial m'}\Big)'
\Big)'=0
\end{equation}

Thus, it follows that a set of boundary conditions 
implying that (\ref{gd02bc}) are satisfied is
\begin{equation}
\label{vel010}
\begin{array}{rcl}
{\displaystyle
 \Big(
  \frac{\partial\Phi}{\partial\varepsilon'}
  \delta\varepsilon
  +
  \frac{\partial\Phi}{\partial m'}
  \delta m
 \Big)_{\ell_1}
  \!\!\!\!
}
&\!\!=&\!\!
{\displaystyle
 \Big[\Big(\frac{\partial\Phi}{\partial m}
      -
      \Big(
      \frac{\partial\Phi}{\partial m'}
      \Big)'
      \Big)'
 \Big]_{\ell_2}
 \!\!\!\!
 \vphantom{\bigg\{_\big\}}
}\\
&\!\!=&\!\!
{\displaystyle
 \Big[\frac{\partial\Phi}{\partial m}
      -
      \Big(
      \frac{\partial\Phi}{\partial m'}
      \Big)'
 \Big]_{\ell_1}
 \!\!\!\!
=
 \Big[\frac{\partial\Phi}{\partial\varepsilon}
      -
 \Big(
      \frac{\partial\Phi}{\partial\varepsilon'}
 \Big)'
 \Big]_{\ell_1,\ell_2}
 \!\!\!\!
=0
}\\
\end{array}
\end{equation}

Finally, we note that, by exploiting the third boundary 
condition (\ref{gd03bc}) in $\ell_1$, we get the PDE problem
\begin{equation}
\label{problema-d-osi}
\left\{
\begin{array}{l}
{\displaystyle
 \frac{\partial\Phi}{\partial\varepsilon}
 -
 \Big(
 \frac{\partial\Phi}{\partial\varepsilon'}
 \Big)'
 =0
\;\;\textrm{ and }\;\;
 \varrho_{0,\rr{f}}^2
 \Big[
      \frac{\partial\Phi}{\partial m}
      -
      \Big(
      \frac{\partial\Phi}{\partial m'}
      \Big)'
 \Big]''
 =
 D\dot{m}
\vphantom{\bigg\{_\big\{}
}
\\
{\displaystyle
\Big(
 \frac{\partial\Phi}{\partial\varepsilon'}
 \delta\varepsilon
 +
 \frac{\partial\Phi}{\partial m'}
 \delta m
\Big)_{\ell_1,\ell_2}
\!\!
\!\!
\!
=
 \Big[\Big(\frac{\partial\Phi}{\partial m}
      -
      \Big(
      \frac{\partial\Phi}{\partial m'}
      \Big)'
      \Big)'
 \Big]_{\ell_2}
\!\!
\!\!
\!
=
 \Big[\frac{\partial\Phi}{\partial m}
      -
      \Big(
      \frac{\partial\Phi}{\partial m'}
      \Big)'
 \Big]_{\ell_1}
\!\!
\!\!
\!
=0
}
\\
\end{array}
\right.
\end{equation}
that will be called the \textit{one--side impermeable} problem.

\subsection{The stationary problem}
\label{s:stazionario}
In the above section we have introduced two different physically 
interesting problems for the porous medium we are studying by 
specifying two different sets of boundary conditions. 
The stationary solution of those two problems are the solution of 
the two differential equations
\begin{equation}
\label{stazionario000}
 \frac{\partial\Phi}{\partial\varepsilon}
 -
 \Big(
 \frac{\partial\Phi}{\partial\varepsilon'}
 \Big)'
 =0
\;\;\textrm{ and }\;\;
 \Big[
      \frac{\partial\Phi}{\partial m}
      -
      \Big(
      \frac{\partial\Phi}{\partial m'}
      \Big)'
 \Big]''
 =
0
\end{equation}
with the boundary conditions provided respectiveley in (\ref{problema-d}) 
and (\ref{problema-d-osi}) for the zero chemical potential and 
the one--side impermeable problems.

It is easy to show that in both cases the boundary conditions 
imply that, see also \cite[equation~(40) and related discussion]{CIS2013},
$\partial\Phi/\partial m-(\partial\Phi/\partial m')'=0$, so that 
for both the problems the stationary solutions are the 
solutions of the PDE problem
\begin{equation}
\label{stazionario020} 
\left\{
\begin{array}{l}
{\displaystyle
 \frac{\partial\Phi}{\partial\varepsilon}
 -
 \Big(
 \frac{\partial\Phi}{\partial\varepsilon'}
 \Big)'
 =0
\;\;\textrm{ and }\;\;
 \Big[
      \frac{\partial\Phi}{\partial m}
      -
      \Big(
      \frac{\partial\Phi}{\partial m'}
      \Big)'
 \Big]''
 =
0
\vphantom{\bigg\{_\big\{}
}
\\
{\displaystyle
\Big(
 \frac{\partial\Phi}{\partial\varepsilon'}
 \delta\varepsilon
 +
 \frac{\partial\Phi}{\partial m'}
 \delta m
\Big)_{\ell_1,\ell_2}
=0
}
\\
\end{array}
\right.
\end{equation}

\subsection{Dissipative character of the evolution}
\label{s:diss}
Besides the stationary solutions, another common features of the 
zero chemical potential and the one--side impermeable problems 
is the dissipative character of the dynamics. 

Indeed, 
the dissipative character of the physical problem we are studying 
reflects in the existence of the not increasing energy functional 
\begin{equation}
\label{liapunov}
F(\chi_\rr{s},\phi)
:=
\int_{\ell_1}^{\ell_2}
\Phi(m',\varepsilon',m,\varepsilon)\,\rr{d}X_\rr{s}
\end{equation}
for both the systems
(\ref{problema-d}) and (\ref{problema-d-osi}).
To prove this we compute the time derivative  of the functional $F$ 
evaluated on the fields 
$\chi_\rr{s}(X_\rr{s},t)$ and $\phi(X_\rr{s},t)$.
We first get 
\begin{displaymath}
\frac{\rr{d}F}{\rr{d}t}
=
\int_{\ell_1}^{\ell_2}
\Big[
     \frac{\partial\Phi}{\partial\varepsilon'}\dot{\varepsilon}'
     +
     \frac{\partial\Phi}{\partial m'}\dot{m}'
     +\frac{\partial\Phi}{\partial \varepsilon}\dot{\varepsilon}
     +\frac{\partial\Phi}{\partial m}\dot{m}
\Big]
\rr{d}X_\rr{s}
\end{displaymath}
and, by integrating by parts, we obtain 
\begin{equation}
\label{derF}
 \frac{\rr{d}F}{\rr{d}t}
=
\Big[
     \frac{\partial\Phi}{\partial\varepsilon'}\dot{\varepsilon}
     +
     \frac{\partial\Phi}{\partial m'}\dot{m}
\Big]_{\ell_1}^{\ell_2}
 +\int_{\ell_1}^{\ell_2}
 \Big[
     \Big(\frac{\partial\Phi}{\partial m}
      -\Big(\frac{\partial\Phi}{\partial m'}\Big)'\Big)
      \dot{m}
     +
     \Big(\frac{\partial\Phi}{\partial \varepsilon}
      -\Big(\frac{\partial\Phi}{\partial \varepsilon'}\Big)'\Big)
      \dot{\varepsilon}
 \Big]
 \rr{d}X_\rr{s}
\end{equation}

Assume, now, that the fields $m$ and $\varepsilon$ are solution 
either the problem (\ref{problema-d}) or (\ref{problema-d-osi}).
Then, 
\begin{displaymath}
 \frac{\rr{d}F}{\rr{d}t}
=
 \frac{\varrho_{0,\rr{f}}^2}{D}
 \int_{\ell_1}^{\ell_2}
     \Big(\frac{\partial\Phi}{\partial m}
      -\Big(\frac{\partial\Phi}{\partial m'}\Big)'\Big)
 \,
     \Big(\frac{\partial\Phi}{\partial m}
      -\Big(\frac{\partial\Phi}{\partial m'}\Big)'\Big)''
 \,
 \rr{d}X_\rr{s}
\end{displaymath}
Finally, by integrating by parts and by using the boundary conditions, 
in both cases we find 
\begin{displaymath}
 \frac{\rr{d}F}{\rr{d}t}
=
 -\frac{\varrho_{0,\rr{f}}^2}{D}
 \int_{\ell_1}^{\ell_2}
     \Big[
     \Big(\frac{\partial\Phi}{\partial m}
      -\Big(\frac{\partial\Phi}{\partial m'}\Big)'\Big)'
     \Big]^2
 \rr{d}X_\rr{s}
\end{displaymath}
which proves that the functional $F$ in not increasing along 
the motions of the Cahn--Hilliard--like systems
(\ref{problema-d}) and (\ref{problema-d-osi}).

\subsection{Summary of the section}
\label{s:sum}
In conclusion, in this section, we have written the equations of motions 
(\ref{gd02}) and the associated boundary conditions (\ref{gd02bc}) 
under the geometrical linearization assumption.
We have then introduced suitable boundary conditions 
to model mathematically the zero chemical potential and 
one-side impermeability problems. 
We have thus obtained the two PDE problems
(\ref{problema-d}) and (\ref{problema-d-osi}). 
Moreover, 
we have remarked that the stationary solutions 
of those problems are given by the same system of equations 
(\ref{stazionario020}). 
Finally, we have proven that the energy functional (\ref{liapunov}) 
does not increase 
along the solutions of both the two systems 
(\ref{problema-d}) and (\ref{problema-d-osi}), which suggests 
that in both cases the motions will tend asymptotically to the 
stationary profiles. 

\section{A special model}
\label{s:risultati}
We apply, now, the theory developed above to the special model 
introduced in~\cite{CIS2011,CIS2009,CIS2010} and whose stationary 
behavior has been widely discussed in those papers.

We first specialize the model we are studying by choosing the 
second gradient part of the dimensionless potential energy, that is we assume 
\begin{equation}
\label{sec010}
\Phi(m',\varepsilon',m,\varepsilon)
 :=
 \frac{1}{2}[k_1(\varepsilon')^2+2k_2\varepsilon' m'+k_3(m')^2]
 +
 \Psi(m,\varepsilon)
\end{equation}
with $k_1,k_3>0$, $k_2\in\bb{R}$ such that $k_1k_3-k_2^2\ge0$.
These parameters provide energy penalties for 
the formation of interfaces; they have the physical dimensions of
squared lengths and, according with the above mentioned 
conditions, provide a well--grounded identification of the intrinsic
characteristic lengths of the one--dimensional porous continuum.

Moreover, 
we consider the following expression for the total potential 
energy density in the perspective of describing the transition between
a fluid--poor and a fluid--rich phase 
\begin{equation}
\label{sec015}
\Psi(m,\varepsilon)
 :=
 \frac{\alpha}{12}m^2(3m^2\!-8b\varepsilon m+6b^2\varepsilon^2)
 +
 \Psi_\rr{B}(m,\varepsilon)
\end{equation}
where 
\begin{equation}
\label{sec020}
\Psi_\rr{B}(m,\varepsilon):=
 p\varepsilon+\frac{1}{2}\varepsilon^2+\frac{1}{2}a(m-b\varepsilon)^2
\end{equation}
is the Biot potential energy density~\cite{biot01},
$a>0$ is the ratio between the fluid and the solid rigidity, 
$b>0$ is a coupling between the fluid and the solid component, 
$p>0$ is the external pressure,
and
$\alpha>0$ is a material parameter responsible for the showing 
up of an additional equilibrium.

In the papers~\cite{CIS2011,CIS2009,CIS2010} we have studied 
the stationary problem (\ref{stazionario020}) 
corresponding to the potential energies (\ref{sec010}) and (\ref{sec015}). 
In this section we recall briefly 
those results and 
extend the paremetrical study in \cite{CIS2013}
with respect to the second gradient coefficients $k_1$, $k_2$, and $k_3$. 

First of all we write explicitly the stationary problem 
corresponding to the potential energies (\ref{sec010}) and (\ref{sec015}). 
By (\ref{stazionario020}) we get 
\begin{equation}
\label{problema-staz00}
\left\{
\begin{array}{l}
-(2/3)\alpha b m^3+\alpha b^2m^2\varepsilon+p+\varepsilon-ab(m-b\varepsilon)
-k_1\varepsilon''-k_2m''=0
\\
 \alpha m^3-2\alpha b m^2 \varepsilon+\alpha b^2m\varepsilon^2
 +a(m-b\varepsilon)
 -k_2\varepsilon''-k_3m'' 
=0
\\
{\displaystyle
\Big(
 (k_1\varepsilon'+k_2m')
 \delta\varepsilon
 +
 (k_2\varepsilon'+k_3m')
 \delta m
\Big)_{\ell_1,\ell_2}
=0
}
\\
\end{array}
\right.
\end{equation}
where the last line is the boundary condition.

\subsection{Phases: constant stationary solutions}
\label{s:fasi}
In~\cite{CIS2011,CIS2009}
we have studied the constant solutions of (\ref{problema-staz00}) 
which are called 
\textit{phases} of the system. We have 
proven that there exists a pressure $p_\rr{c}$, called 
\textit{critical pressure}, such that for any $p\in[0,p_\rr{c})$ there 
exists a single phase 
$(m_\rr{s}(p),\varepsilon_\rr{s}(p))$, 
called 
the \textit{standard phase}, which is very similar to the usual 
solution of the Biot model. For $p> p_\rr{c}$ a second phase 
$(m_\rr{f}(p),\varepsilon_\rr{f}(p))$, richer in fluid with respect to the 
standard phase and hence called \textit{fluid--rich} phase, appears.
We underline that from now on $m_\rr{f}(p)$ indicates the increment
of the fluid mass density $m$ evaluated in the fluid--rich phase
of the porous system, parametrized by the consolidating pressure.

We have shown that the standard phase 
$(m_\rr{s}(p),\varepsilon_\rr{s}(p))$ 
is the solution of the two equations 
$m=b\varepsilon$ and 
$p=f_1(\varepsilon)$, for any $p>0$, where 
$f_1(\varepsilon):=-\varepsilon-\alpha b^4\varepsilon^3/3$.
On the other hand 
the fluid--rich phase 
$(m_\rr{f}(p),\varepsilon_\rr{f}(p))$ 
is the solution, with the smallest value of 
$\varepsilon$ (recall $\varepsilon\in(-1/2,0)$, so that the smallest 
value has indeed largest modulus),
of the two equations 
$m=m_+(\varepsilon)$ and 
$p=f_+(\varepsilon)$, where 
\begin{displaymath}
m_+(\varepsilon)=
\frac{b}{2}
\Big[
     \varepsilon+\sqrt{\varepsilon^2-\frac{4a}{\alpha b^2}}
\Big]
\;\;\textrm{ and }\;\;
f_+(\varepsilon)
\!:=\!
-\varepsilon+ab[m_+(\varepsilon)-b\varepsilon]
  -\alpha b^2\varepsilon m_+^2(\varepsilon)
  \vphantom{\bigg\{}
  +\frac{2}{3}\alpha bm_+^3(\varepsilon)
\end{displaymath}
For $\varepsilon\le-2/(b\sqrt{\alpha/a})$ 
the function $f_+(\varepsilon)$ is positive, diverging to $+\infty$ 
for $\varepsilon\to-\infty$, and has a minimum at $\varepsilon_\rr{c}$ 
such that $f_+(\varepsilon_\rr{c})=p_\rr{c}$; this explains why the fluid--rich
phase is seen only for $p>p_\rr{c}$.

Moreover it has been shown that for any $p>0$ the point 
$(m_\rr{s}(p),\varepsilon_\rr{s}(p))$ 
is a minimum of the two variable potential 
energy $\Psi(m,\varepsilon)$ with $p$ fixed, while
$(m_\rr{f}(p),\varepsilon_\rr{f}(p))$ 
is a minimum for $p>p_\rr{c}$ and 
a saddle point for $p=p_\rr{c}$.
For more details we refer to \cite{CIS2011}.

\subsection{Profiles: not constant stationary solutions}
\label{s:profili}
In \cite{CIS2010} 
it has been proven that there exists a unique value $p_\rr{co}$ 
of the pressure, called \textit{coexistence pressure}, 
such that the potential energy of the two phases is equal.
More precisely, 
it has been proven that the equation 
$\Psi(m_\rr{s}(p),\varepsilon_\rr{s}(p))
 =\Psi(m_\rr{f}(p),\varepsilon_\rr{f}(p))$ has the single 
solution $p_\rr{co}$. 

The behavior of the system at the coexistence pressure 
is particularly interesting; from now on we shall always 
consider $p=p_\rr{co}$ and, for this reason, we shall drop $p$ 
from the notation. 
When the external pressure is equal to $p_\rr{co}$,
none of the two above phases is 
favored and we ask if profiles connecting one phase to the other exist.
More precisely, in \cite{CIS2010} we have addressed the 
problem of the existence of a \textit{connection} between the two phases, 
that is, a solution of the stationary problem (\ref{problema-staz00}) 
on $\bb{R}$ tending to the standard phase 
for $X_\rr{s}\to-\infty$ 
and to the fluid--rich one for $X_\rr{s}\to+\infty$.
Using results in \cite{AF}  
we have proven that such a connection does exist provided
$k_1k_3-k_2^2>0$.

We have also shown that, for $k_1k_3-k_2^2=0$ 
(\textit{degenerate} case), the 
problem of finding a solution of the stationary problem 
can be reduced to the computation of a definite integral via a 
rephrasing of the problem as a one dimensional conservative mechanical 
system.
We do not repeat here the discussion, but we refer the interested 
reader to \cite[Section~4.2]{CIS2013} and references therein.

In this section we just want to check that the behavior 
of the system out of the degenerate case does not change too much 
with respect to the degenerate case itself. 
In the not degenerate case, i.e., 
$k_1k_3-k_2^2>0$, 
in \cite{CIS2010} we could prove the existence of a connection, 
but we were not 
able to find it explicitly.
Studying in more detail the behavior of the solutions of the 
system (\ref{problema-staz00}) is a difficult task since, in this 
case, the problem cannot be rewritten as a one dimensional 
conservative mechanical system.
We shall then study the problem numerically and solve the 
system (\ref{problema-staz00}) by means of the finite difference 
method powered by the Newton--Rapson algorithm. The substitution 
rule we use are standard, see, for instance 
\cite[Appendix~B]{CIS2013}.

\begin{figure}[t]
\begin{picture}(200,300)(-20,0)
\put(0,200)
{
\resizebox{5cm}{!}{\rotatebox{0}{\includegraphics{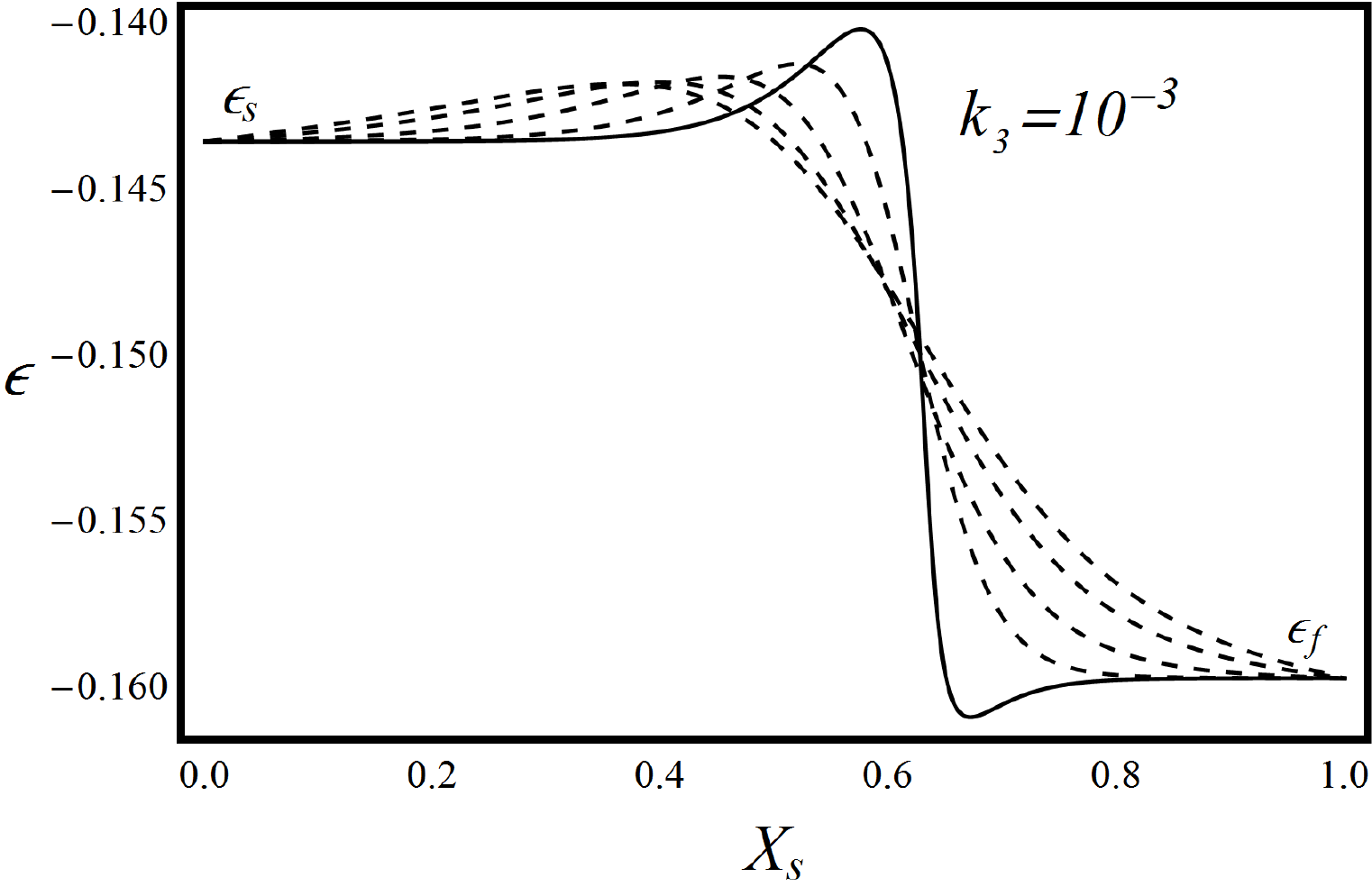}
}} 
}
\put(180,200)
{
\resizebox{5cm}{!}{\rotatebox{0}{\includegraphics{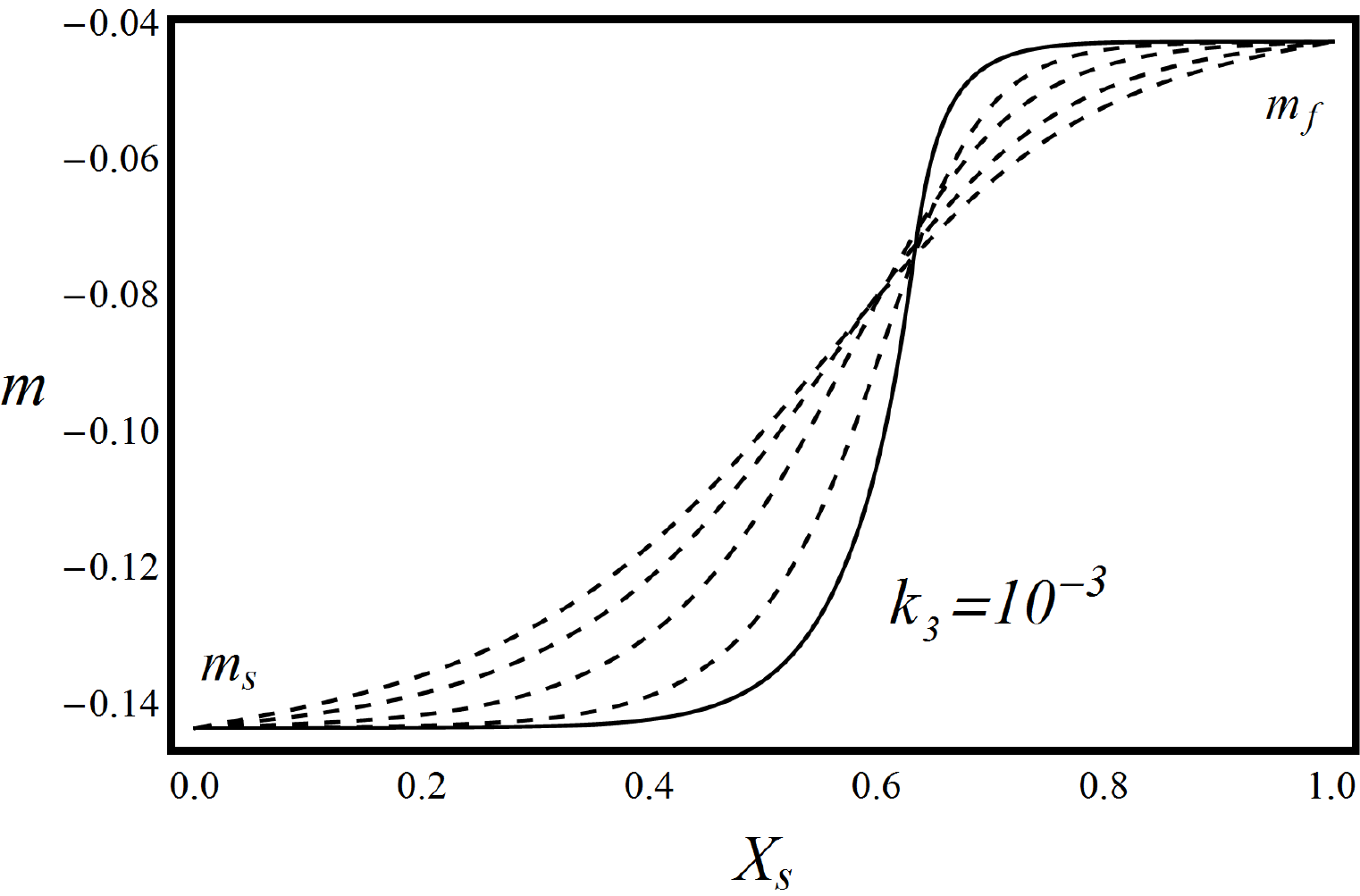}}} 
}
\put(0,100)
{
\resizebox{5cm}{!}{\rotatebox{0}{\includegraphics{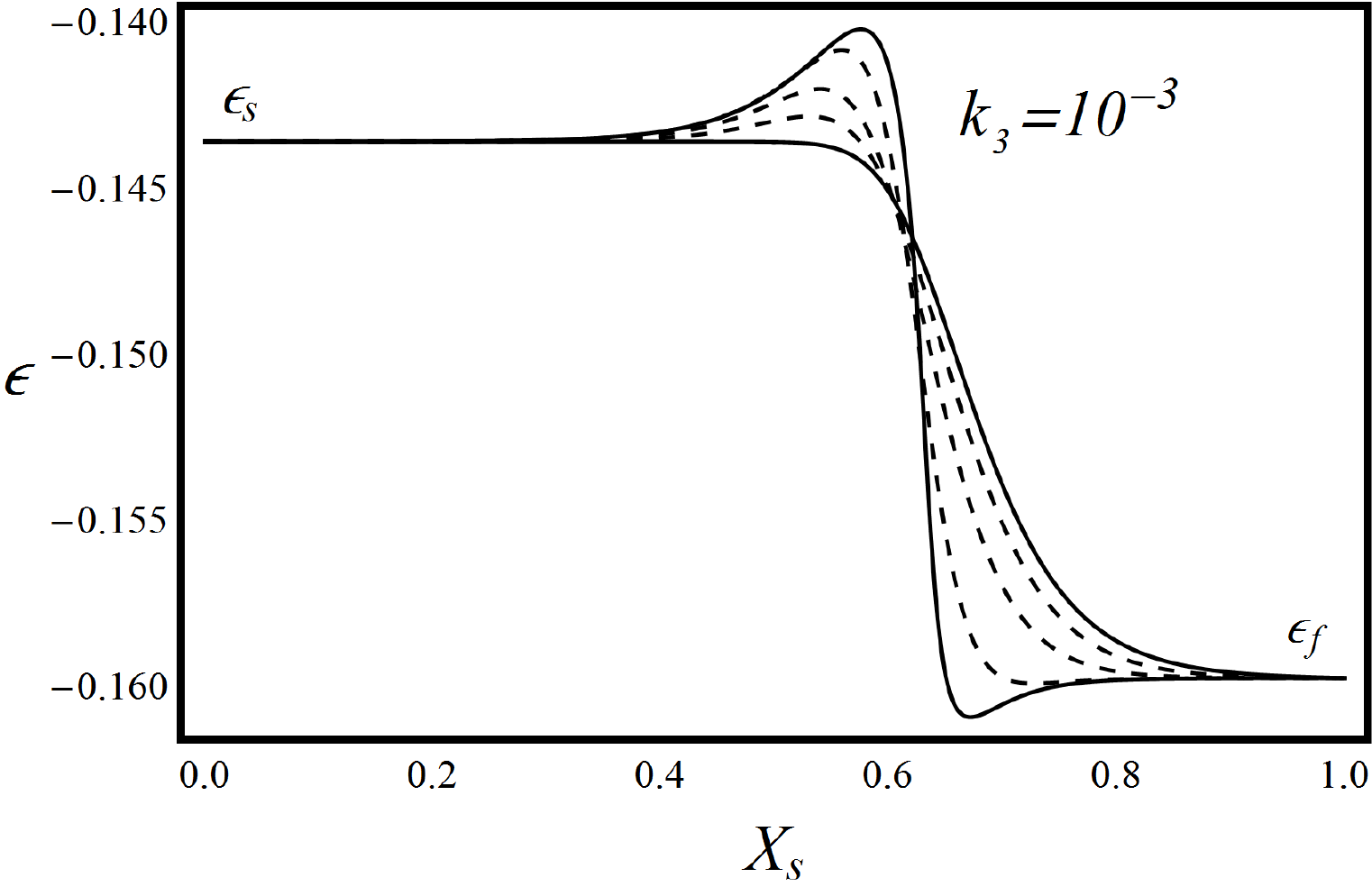}
}} 
}
\put(180,100)
{
\resizebox{5cm}{!}{\rotatebox{0}{\includegraphics{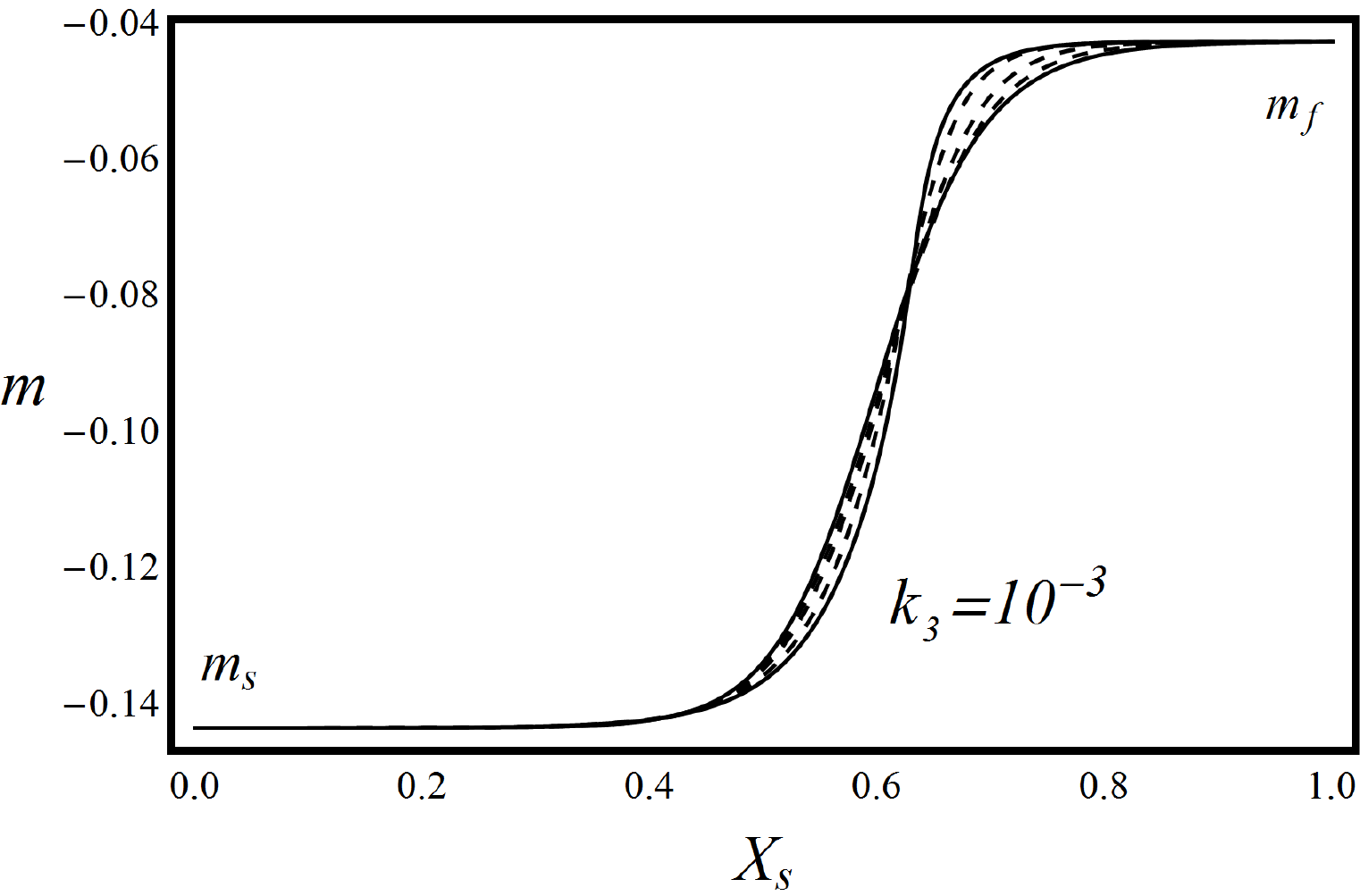}}} 
}
\put(0,0)
{
\resizebox{5cm}{!}{\rotatebox{0}{\includegraphics{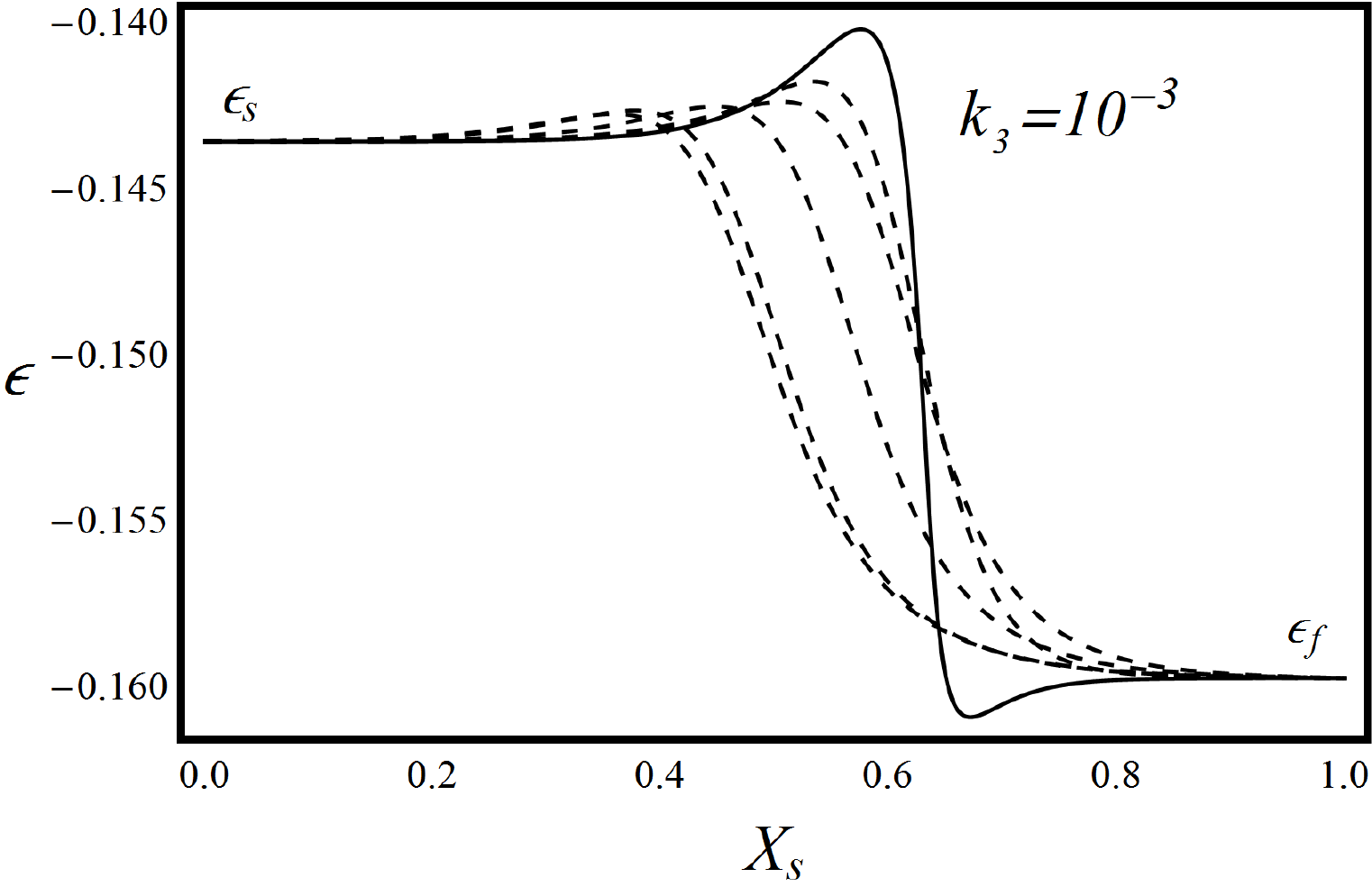}
}} 
}
\put(180,0)
{
\resizebox{5cm}{!}{\rotatebox{0}{\includegraphics{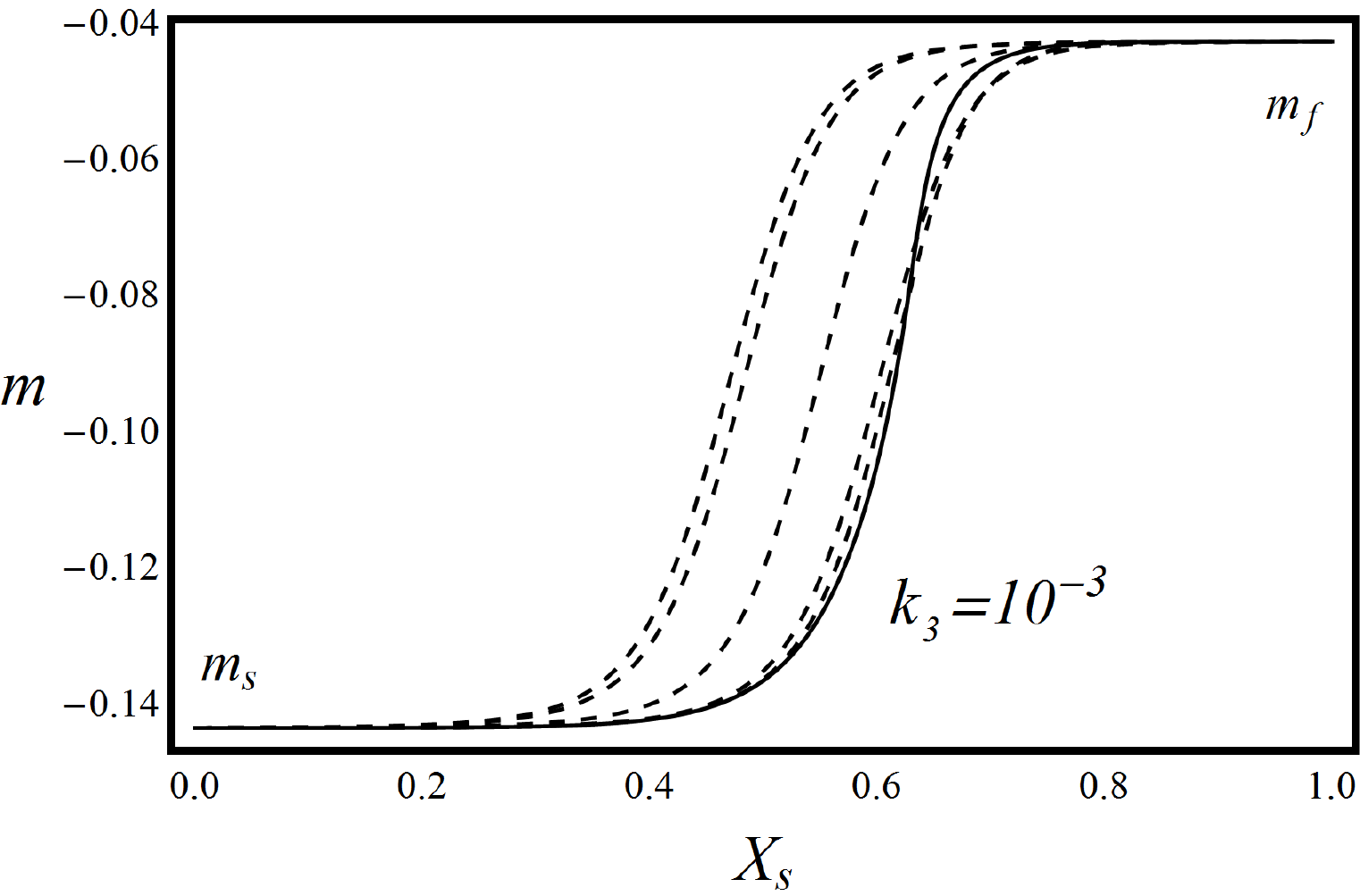}}} 
}
\end{picture}  
\caption{Soutions ($\varepsilon(X_s)$ on the left and $m(X_s)$ on 
the right) of the stationary problem (\ref{problema-staz00}) with the 
Dirichlet boundary conditions 
$m(0)=m_s$, $\varepsilon(0)=\varepsilon_s$, $m(1)=m_f$, and 
$\varepsilon(1)=\varepsilon_f$ on the finite interval $[0,1]$, at 
the coexistence pressure for 
$a=0.5,\,b=1$, and $\alpha=100$.
The second gradient parameters are as follows.
Bottom pictures:
$k_2=k_3=10^{-3},\,k_1=10^{-3}$ (solid line), and 
$k_1=10^{-2},\,0.4\times 10^{-2},\,0.7\times 10^{-2},\,0.8\times 10^{-2},
\,0.9\times 10^{-2}$ (dotted lines).
Middle pictures:
$k_1=k_3=10^{-3},\,k_2=\pm 10^{-3}$ (solid lines), 
and $k_2=-0.4\times10^-3,\,0.2\times10^-3,\,0.8\times10^-3$ (dotted lines).
Top pictures:
$k_1=k_2=10^{-3},\,k_3=10^{-3}$ (solid line), and $k_3=10^{-2},\,0.2\times 10^{-2},\,0.4\times 10^{-2},\,0.7\times 10^{-2},\,10^{-2}$ (dotted lines).}
\label{variak1}
\end{figure}

Our results are depicted in the figure \ref{variak1}. 
Morally, in the three pictures, we first consider the case 
$k_1=k_2=k_3=10^{-3}$ and then vary one of the three 
parameters at time. Our results show that the 
general behavior of the connections does not change very much and 
that only minor details change. 

\section{The not stationary problem}
\label{s:dar-results}
In this section we discuss the not stationary solution of 
the problems (\ref{problema-d}) and (\ref{problema-d-osi}). 
As stated in the Introduction our aim is that 
of finding out the effect of the impermeable boundary 
to the evolution fo the profiles. 

The equations of motion describing the evolution of the system 
in the two cases read
\begin{equation}
\label{num05}
\left\{
\begin{array}{l}
-(2/3)\alpha b m^3+\alpha b^2m^2\varepsilon+p+\varepsilon-ab(m-b\varepsilon)
-k_1\varepsilon''-k_2m''=0
\\
 \varrho^2_{0,\rr{f}}
 ( 
  \alpha m^3-2\alpha b m^2 \varepsilon+\alpha b^2m\varepsilon^2
  +a(m-b\varepsilon)
  -k_2\varepsilon''-k_3m'' 
 )''
 =D\dot{m}
\\
\end{array}
\right.
\end{equation}
with the boundary conditions
\begin{equation}
\label{num01}
\left\{
\begin{array}{l}
((k_2m'+k_1\varepsilon')\delta\varepsilon
+
(k_3m'+k_2\varepsilon')\delta m
)_{\ell_1,\ell_2}
=0
\\ \\
( 
 \alpha m^3-2\alpha b m^2 \varepsilon+\alpha b^2m\varepsilon^2
 +a(m-b\varepsilon)
 -k_2\varepsilon''-k_3m'' 
)_{\ell_1,\ell_2}
=0
\end{array}
\right.
\end{equation}
for the zero chemical potential problem (\ref{problema-d}) 
and 
\begin{equation}
\label{num02}
\left\{
\begin{array}{l}
((k_2m'+k_1\varepsilon')\delta\varepsilon
+
(k_3m'+k_2\varepsilon')\delta m
)_{\ell_1,\ell_2}
=0
\\ \\
( 
 3\alpha m^2m'
 -4\alpha b mm' \varepsilon
 -2\alpha b m^2 \varepsilon'
 +\alpha b^2m'\varepsilon^2
\\
\phantom{(3\alpha m^2m'}
 +2\alpha b^2m\varepsilon\varepsilon'
 +a(m'-b\varepsilon')
 -k_2\varepsilon'''-k_3m''' 
)_{\ell_2}
=0
\\ \\
( 
 \alpha m^3-2\alpha b m^2 \varepsilon+\alpha b^2m\varepsilon^2
 +a(m-b\varepsilon)
 -k_2\varepsilon''-k_3m'' 
)_{\ell_1}
=0
\end{array}
\right.
\end{equation}
for the one--side impermeable problem 
(\ref{problema-d-osi}).

We have studied the above PDE problem with the same numerical 
approach used for the stationary problem.

The dissipative evolution of the elastic strain $\varepsilon$ and 
the fluid mass density variations $m$, relative to the zero chemical 
potential boundary condition and the one--side impermeability condition, 
are discussed for different initial conditions, say: $(i)$ \textit{linear} 
initial condition connecting the two phases, 
$(ii)$ \textit{fluid--poor phase} initial condition, 
$(iii)$ \textit{fluid--rich phase} intial condition.
We remind that in the case of one side impermeability condition the 
chemical potential is still required to vanish at $X_{\rr s}=0$.
In figures \ref{rettaduepuntivecchio}--\ref{tappofluid} the 
corresponding profiles of the elastic strain, the fluid mass density
variation and the seepage velocity (defined as in \eqref{vel000})
are reported at different times.

\subsection{Linear intial condition}
\label{s:retta}
\begin{figure}[t]
 \begin{minipage}[b!h!]{3cm}
   \includegraphics[width=4cm]{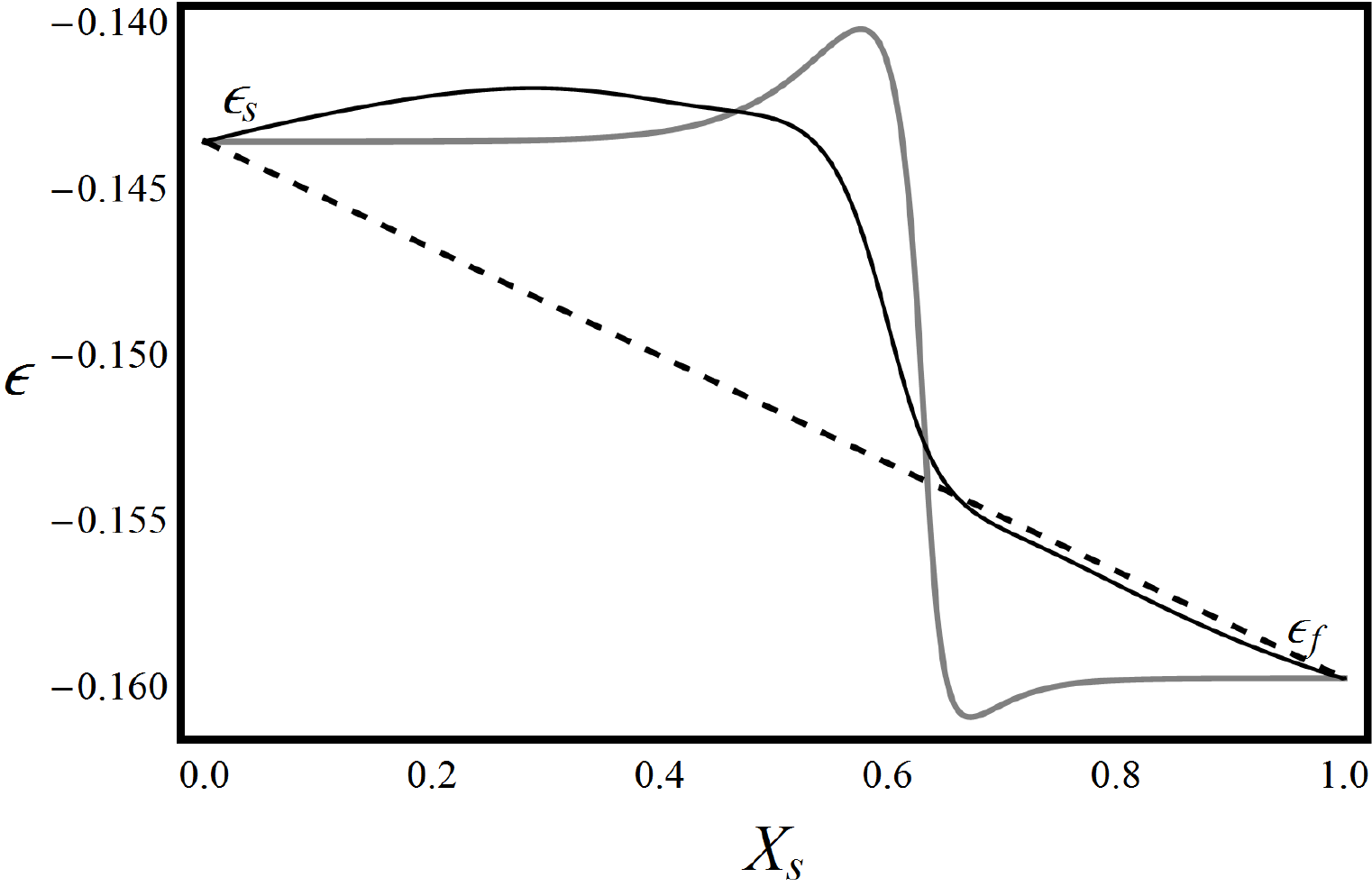}
 \end{minipage}
 \hspace{11mm}  
 \begin{minipage}[b!h!]{3cm}
  \includegraphics[width=4cm]{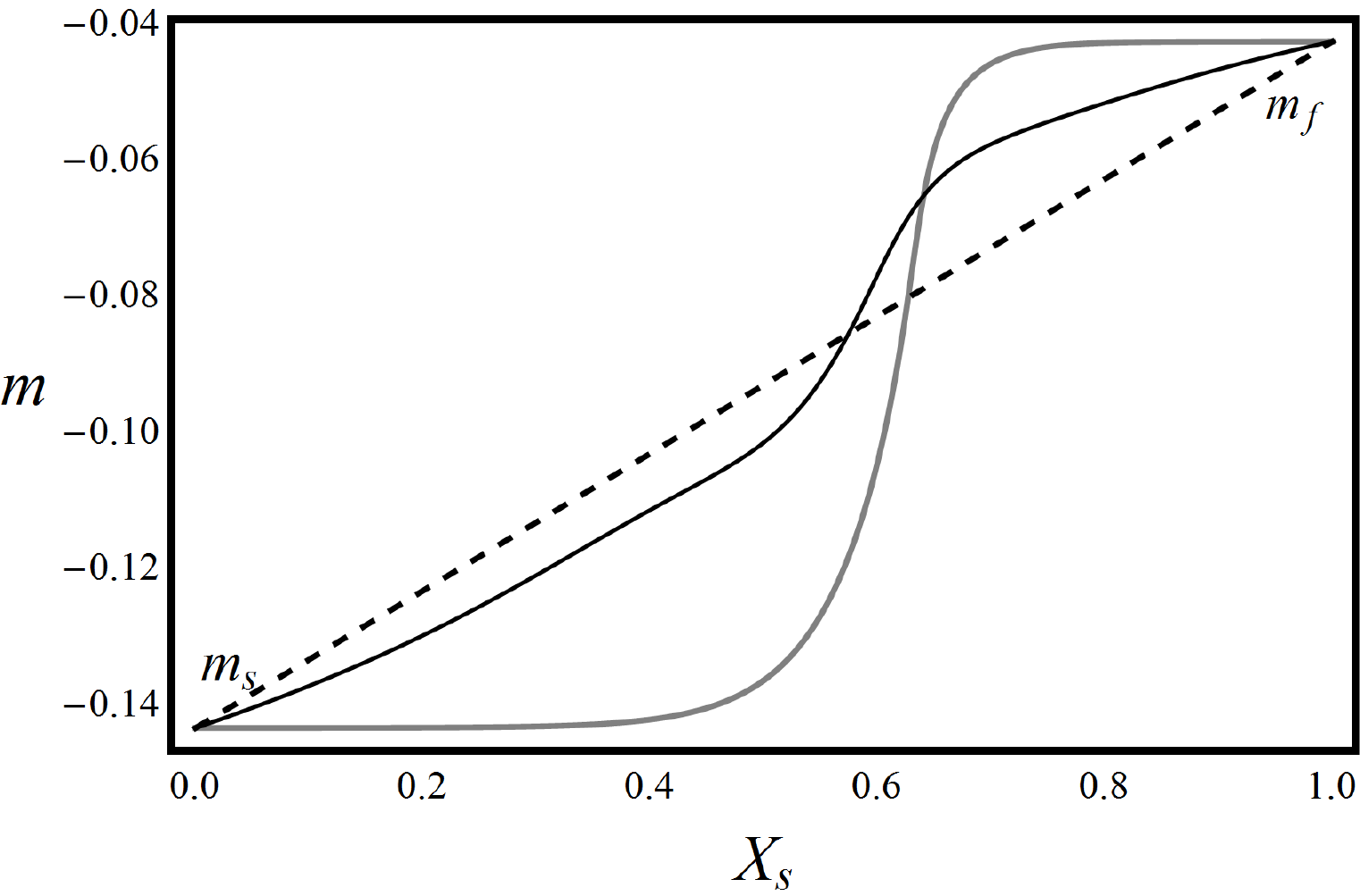}
   \end{minipage}
  \hspace{11mm} 
\begin{minipage}[b!h!]{3cm}
  \centering
   \includegraphics[width=4cm]{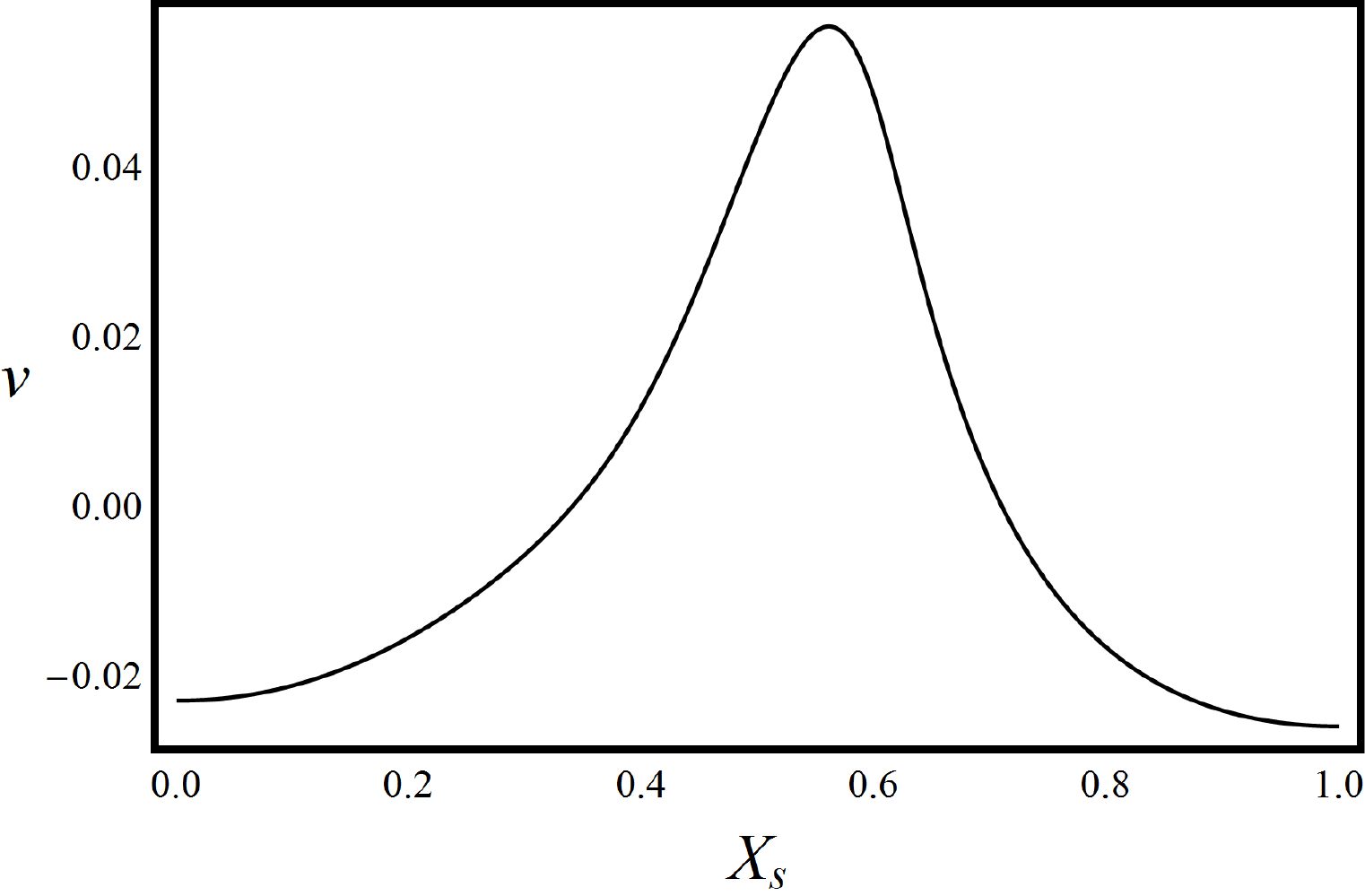}
 \end{minipage}
 \vspace{2mm}
\centering
 \begin{minipage}[b!h!]{3cm}
\centering   
   \includegraphics[width=4cm]{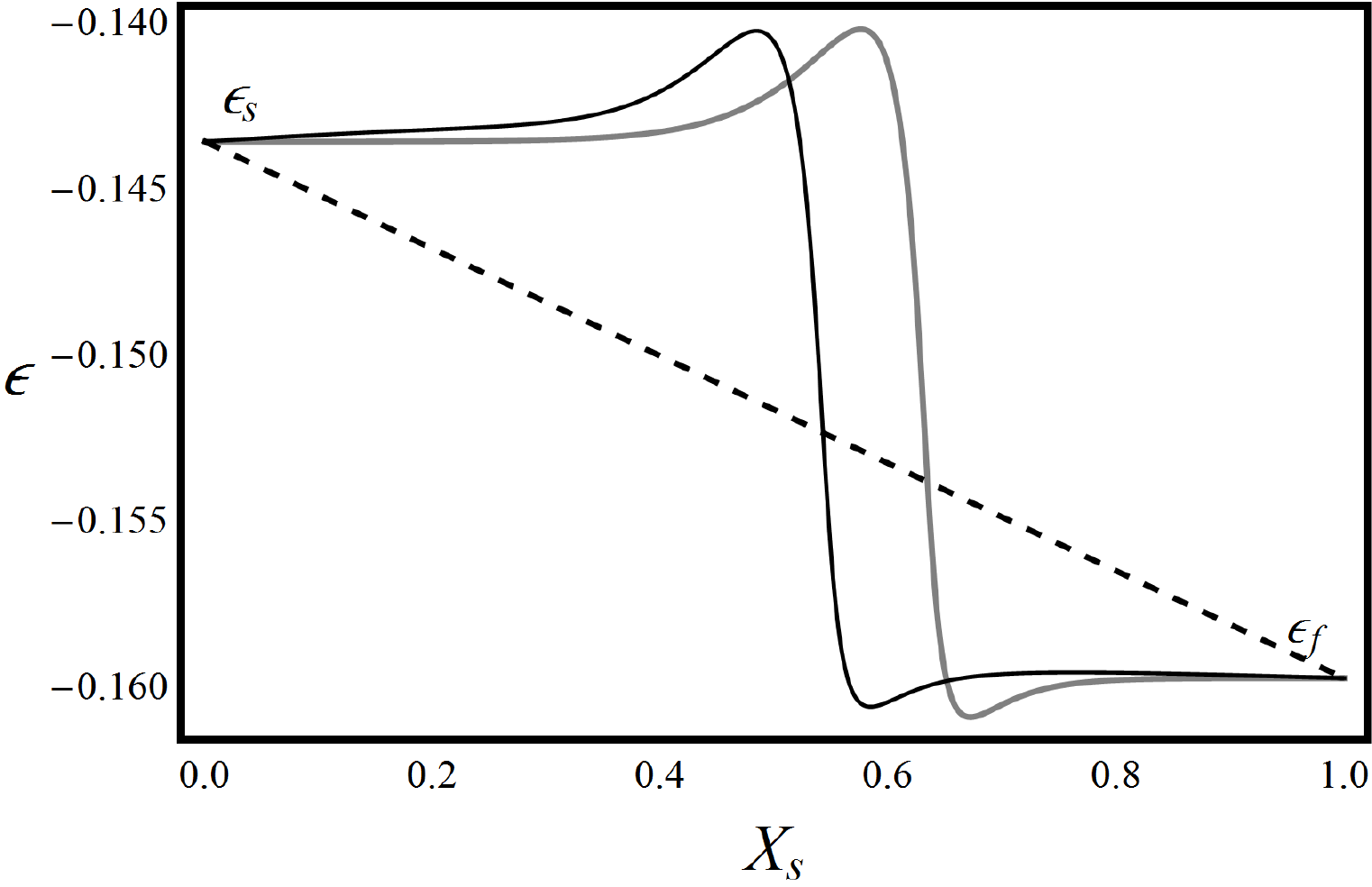}
 \end{minipage}
 \hspace{11mm}  
 \begin{minipage}[b!h!]{3cm}
  \includegraphics[width=4cm]{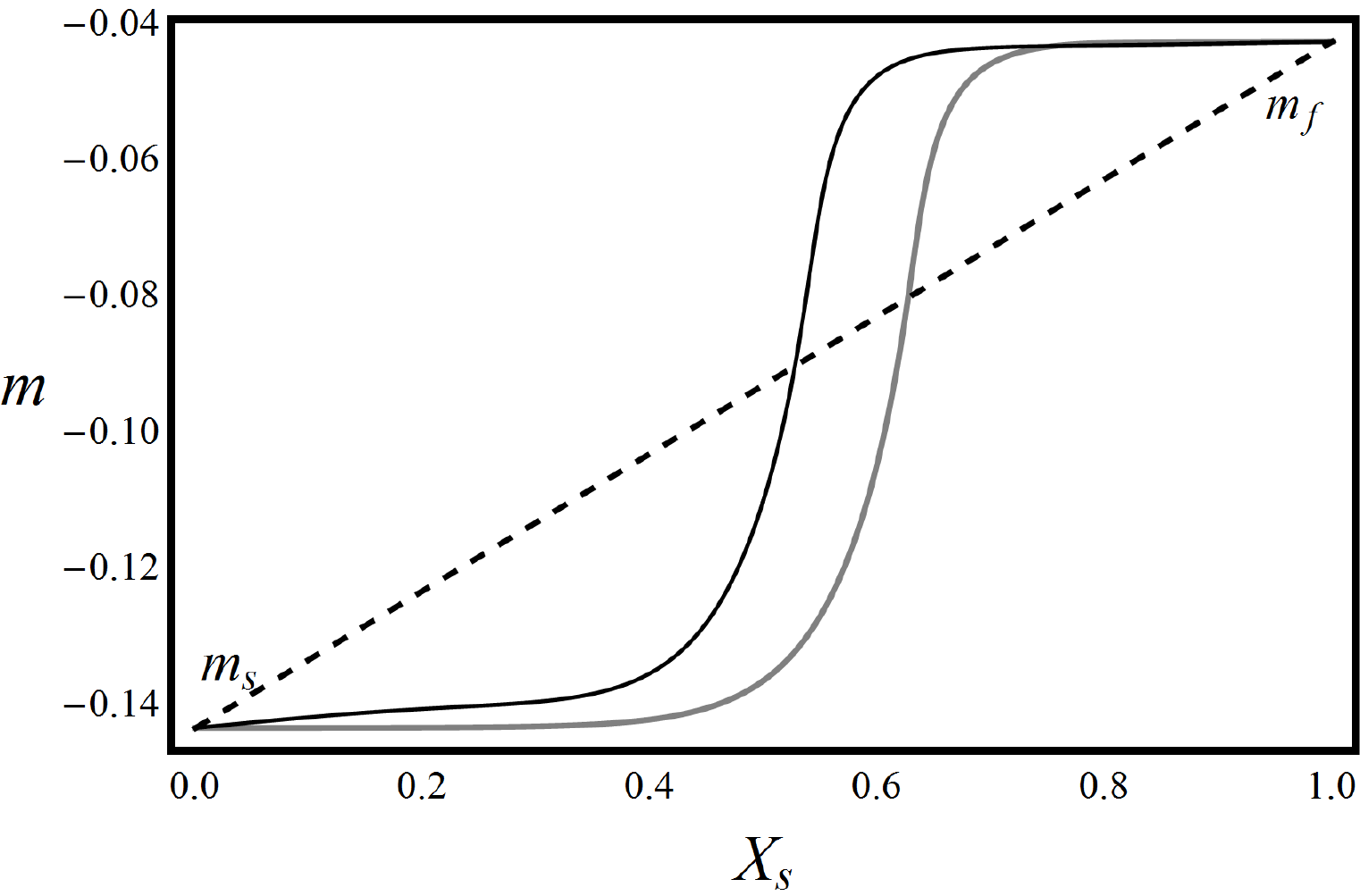}
   \end{minipage}
  \hspace{11mm} 
\begin{minipage}[b!h!]{3cm}
  \centering
   \includegraphics[width=4cm]{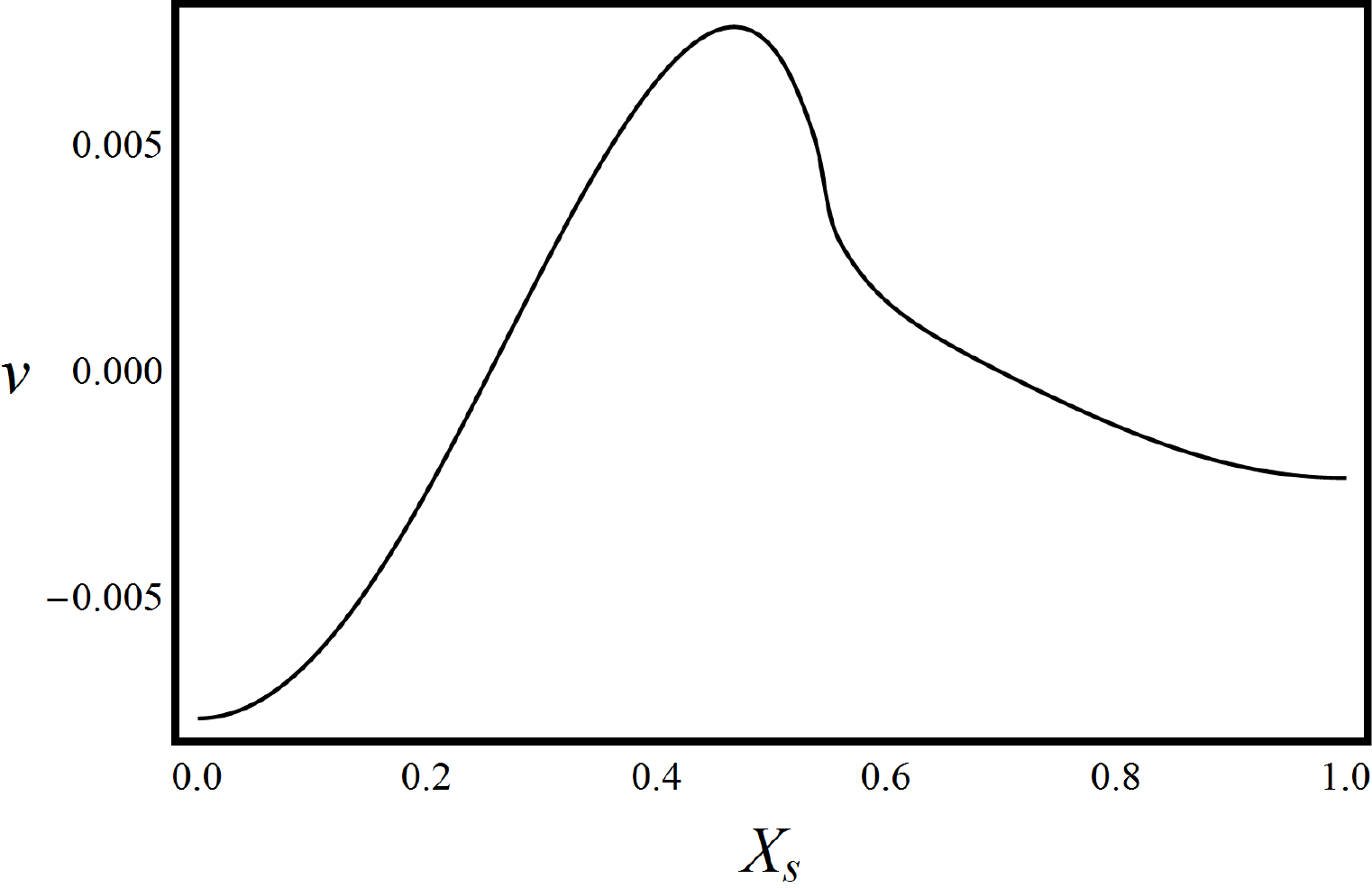}
 \end{minipage}
 \vspace{2mm}
 \centering
 \begin{minipage}[b!h!]{3cm}
\centering   
   \includegraphics[width=4cm]{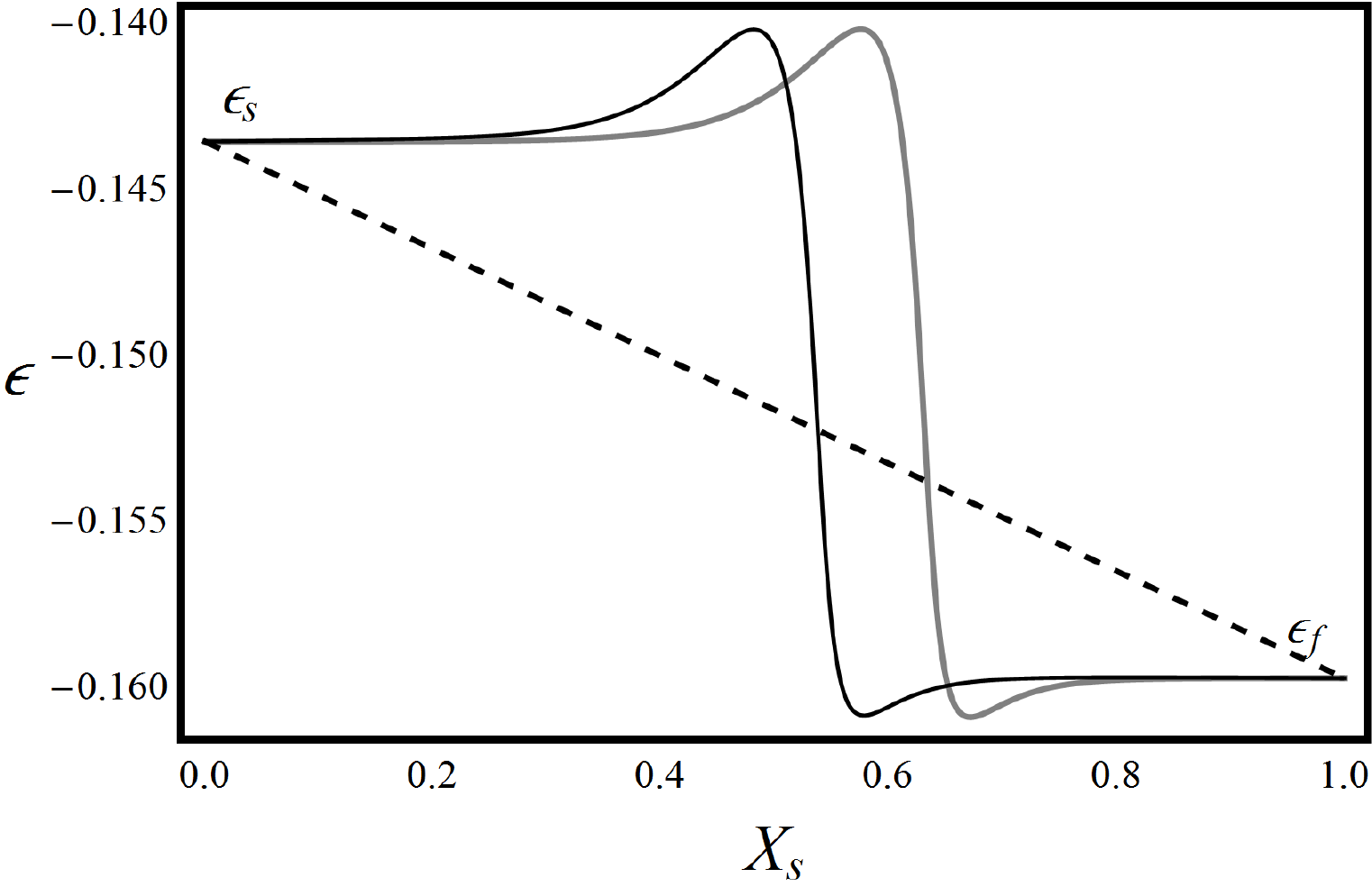}
 \end{minipage}
 \hspace{11mm}  
 \begin{minipage}[b!h!]{3cm}
  \includegraphics[width=4cm]{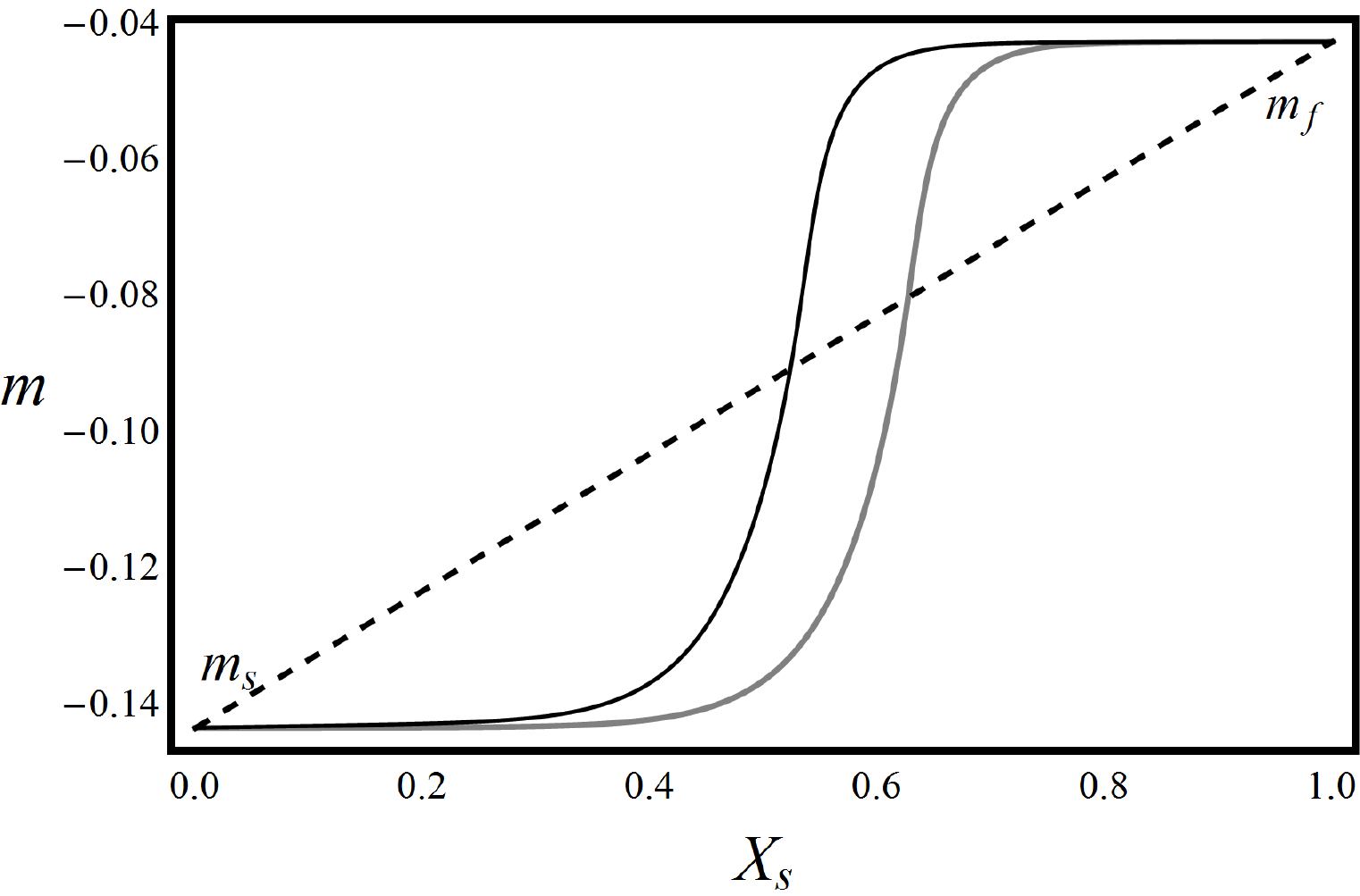}
   \end{minipage}
  \hspace{11mm} 
\begin{minipage}[b!h!]{3cm}  
  \centering
   \includegraphics[width=4cm]{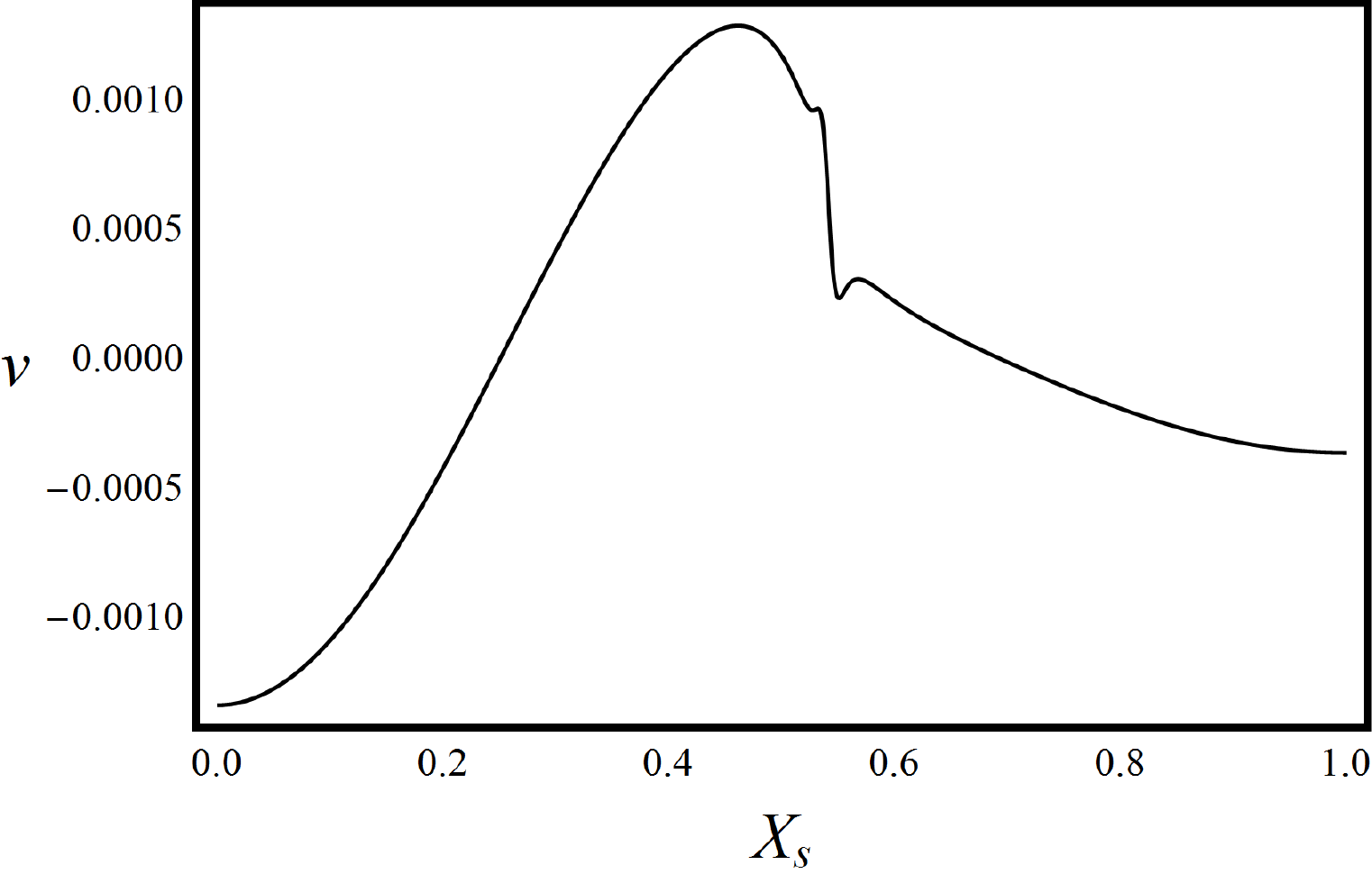}
 \end{minipage}
  \vspace{2mm}
 \centering
 \begin{minipage}[b!h!]{3cm}
\centering   
   \includegraphics[width=4cm]{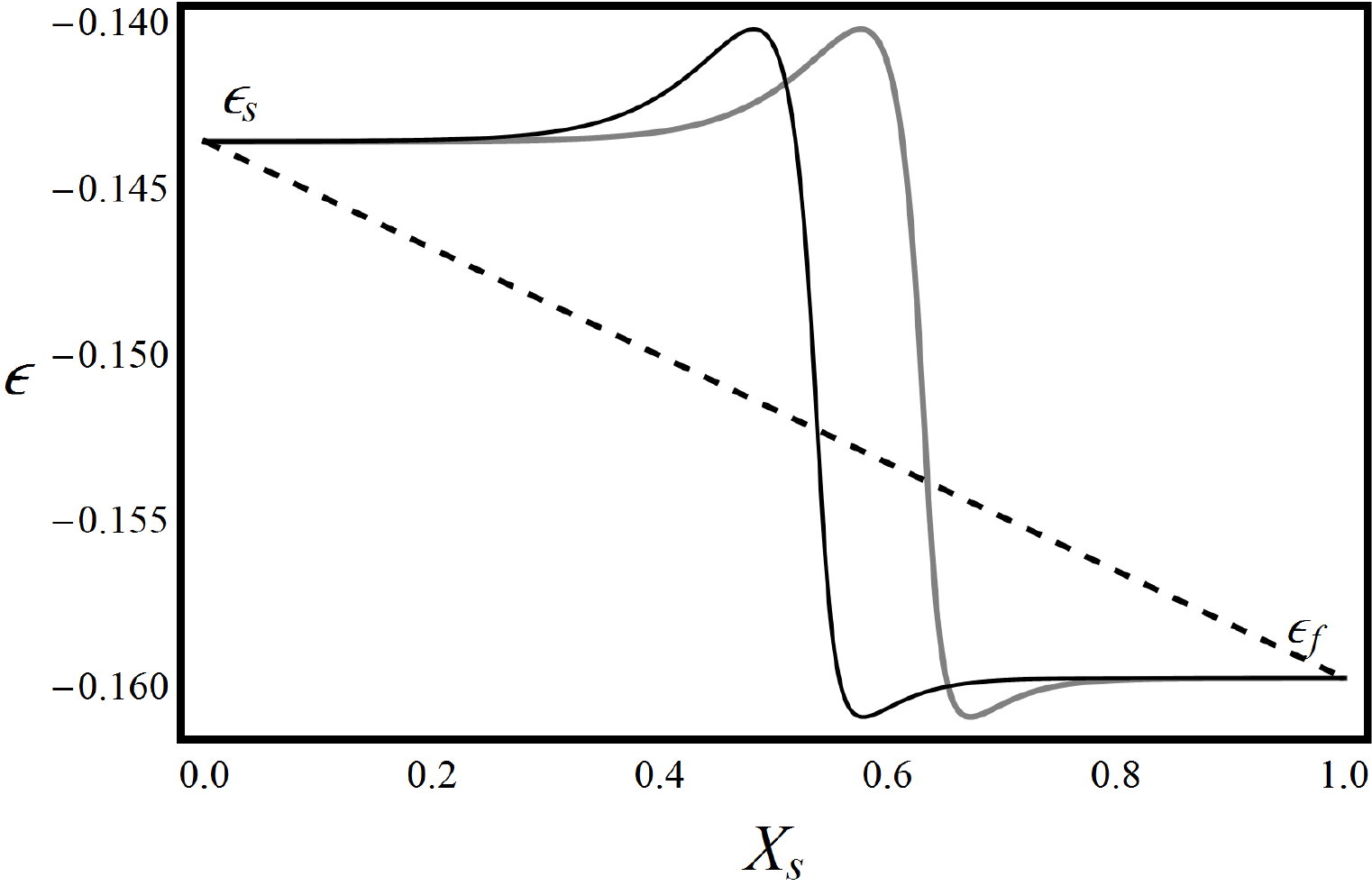}
 \end{minipage}
 \hspace{11mm}  
 \begin{minipage}[b!h!]{3cm}
  \includegraphics[width=4cm]{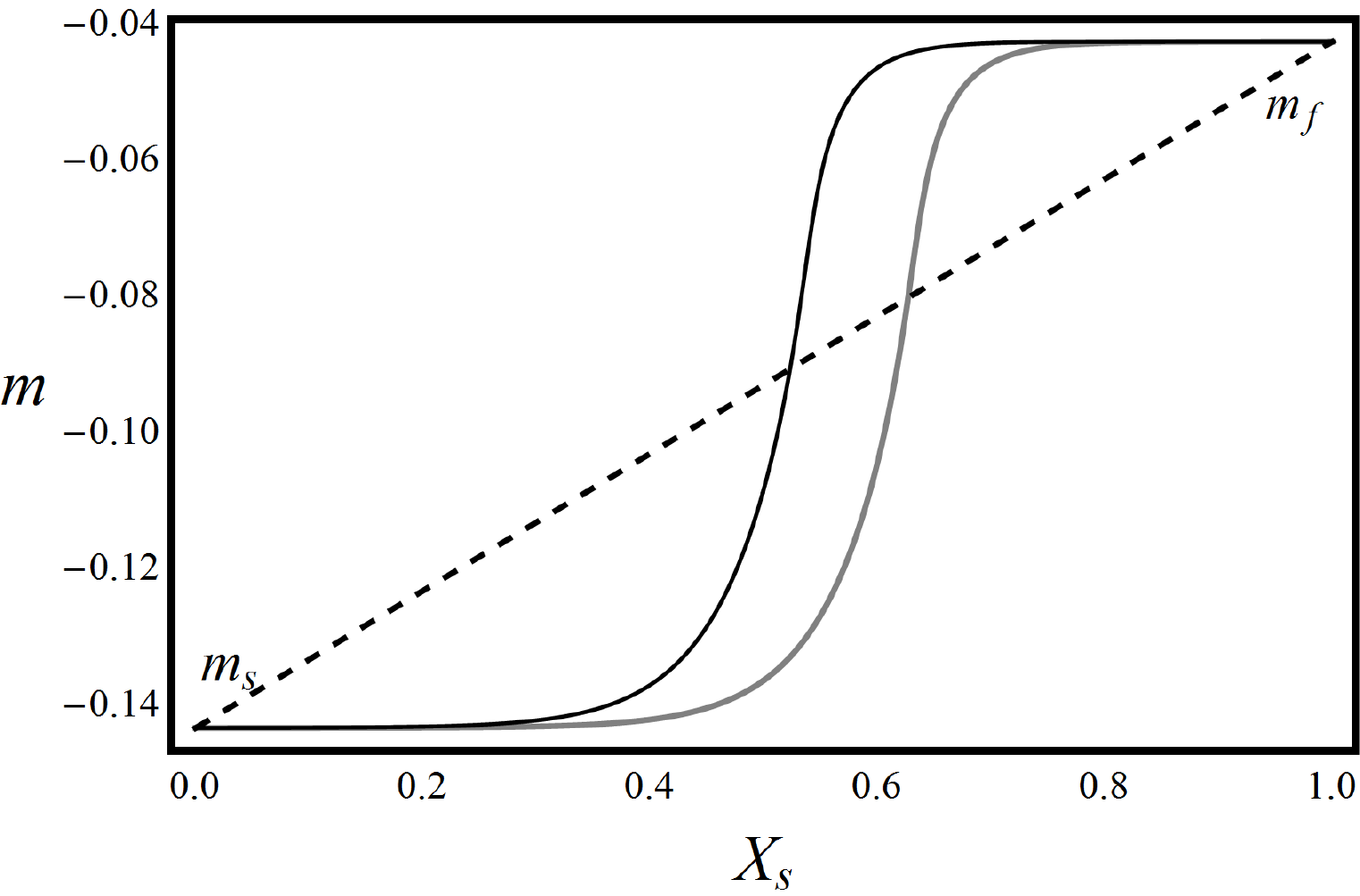}
   \end{minipage}
  \hspace{11mm} 
\begin{minipage}[b!h!]{3cm}
  \centering
   \includegraphics[width=4cm]{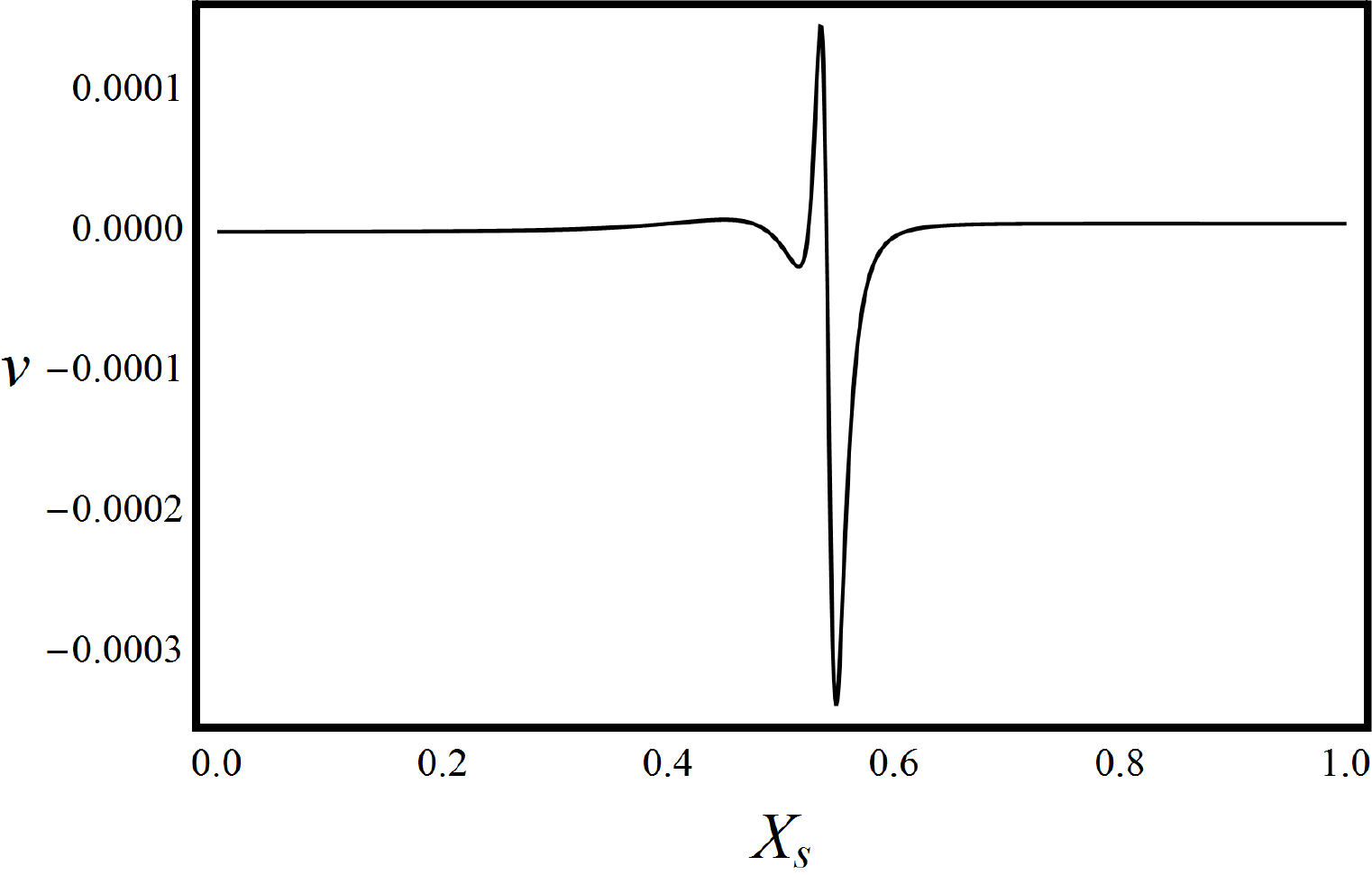}
 \end{minipage}
  \vspace{2mm}
 \centering
 \begin{minipage}[b!h!]{3cm}
\centering   
   \includegraphics[width=4cm]{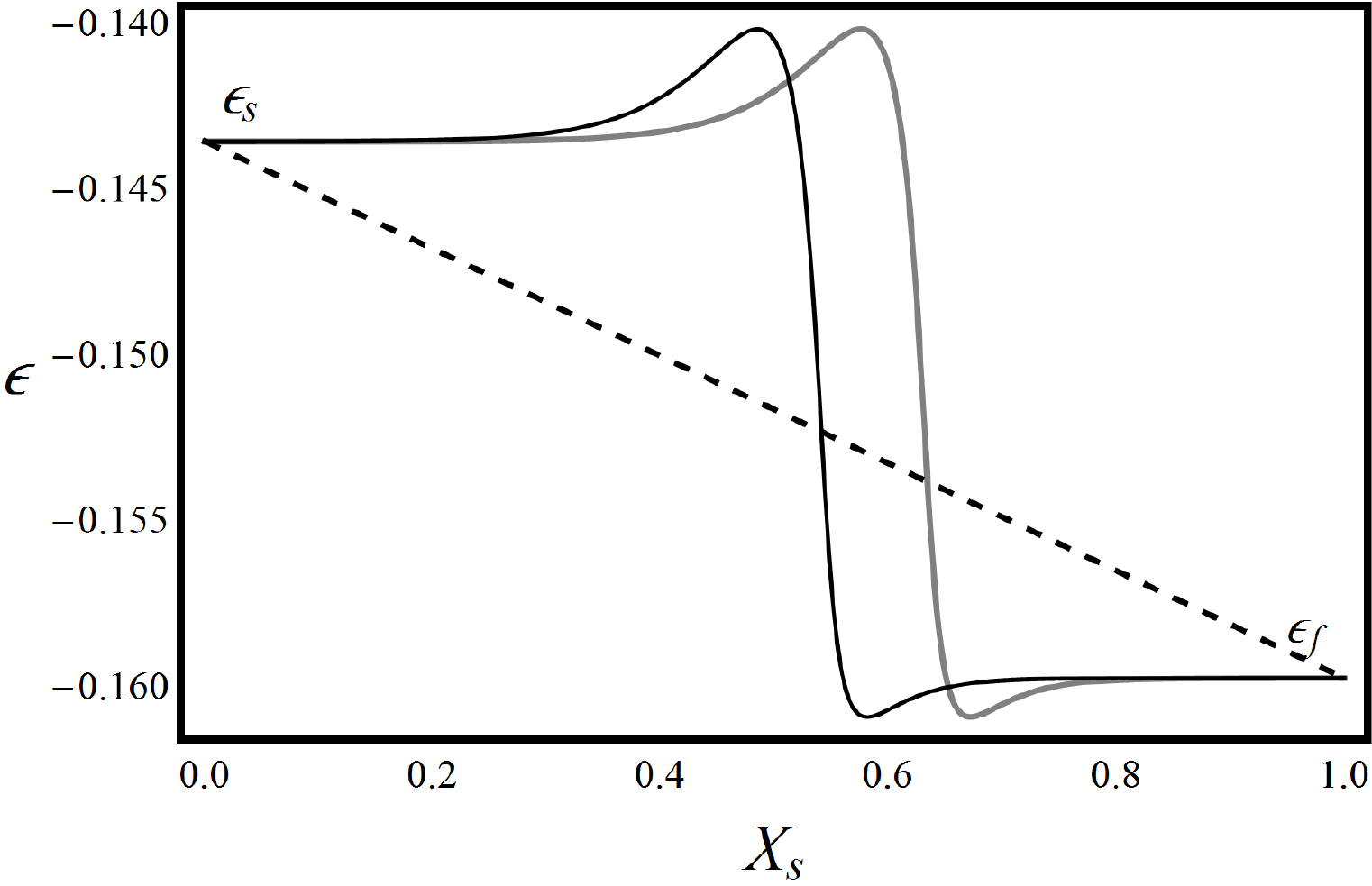}
 \end{minipage}
 \hspace{11mm}  
 \begin{minipage}[b!h!]{3cm}
  \includegraphics[width=4cm]{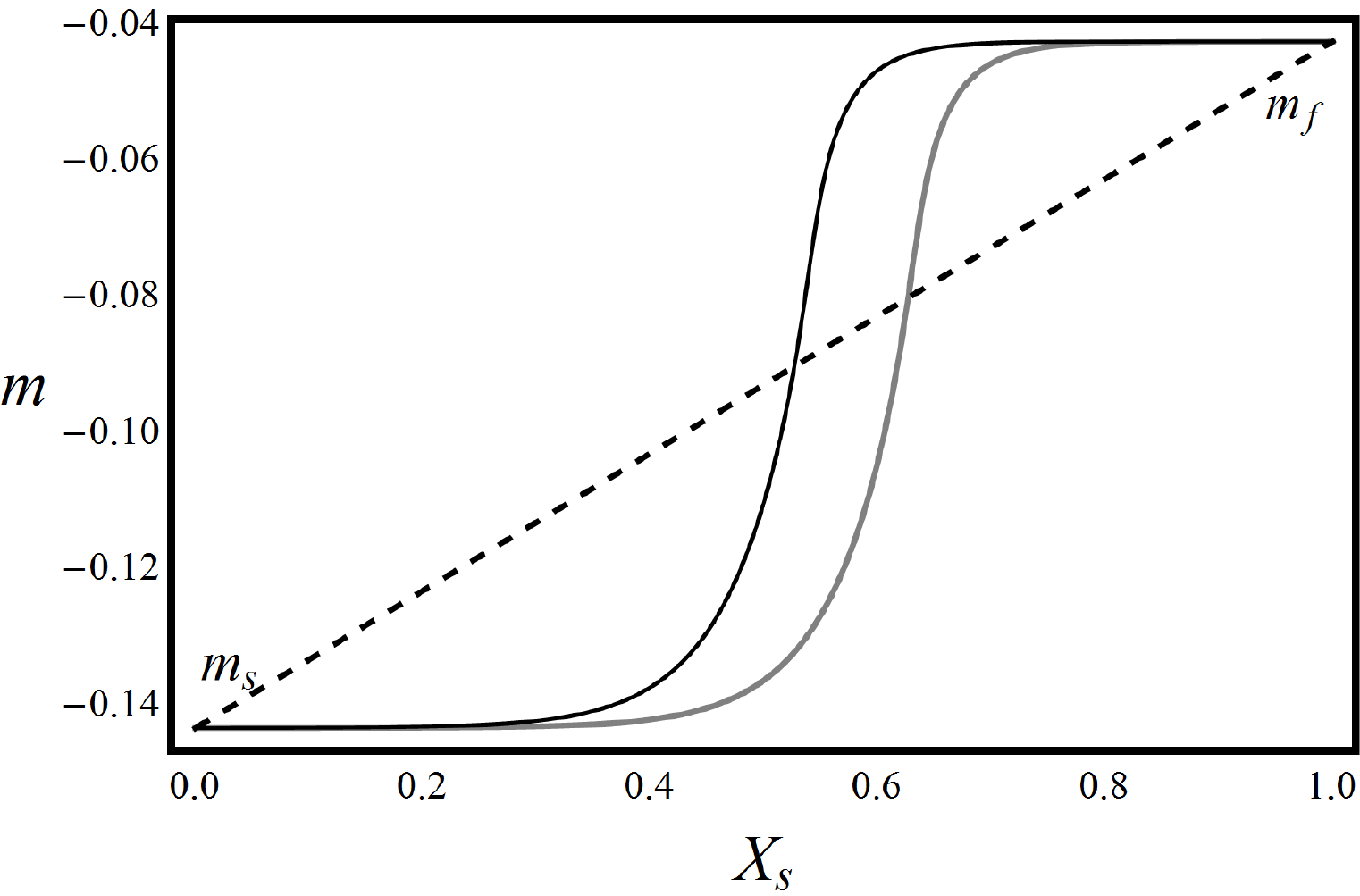}
   \end{minipage}
  \hspace{11mm} 
\begin{minipage}[b!h!]{3cm}
  \centering
   \includegraphics[width=4cm]{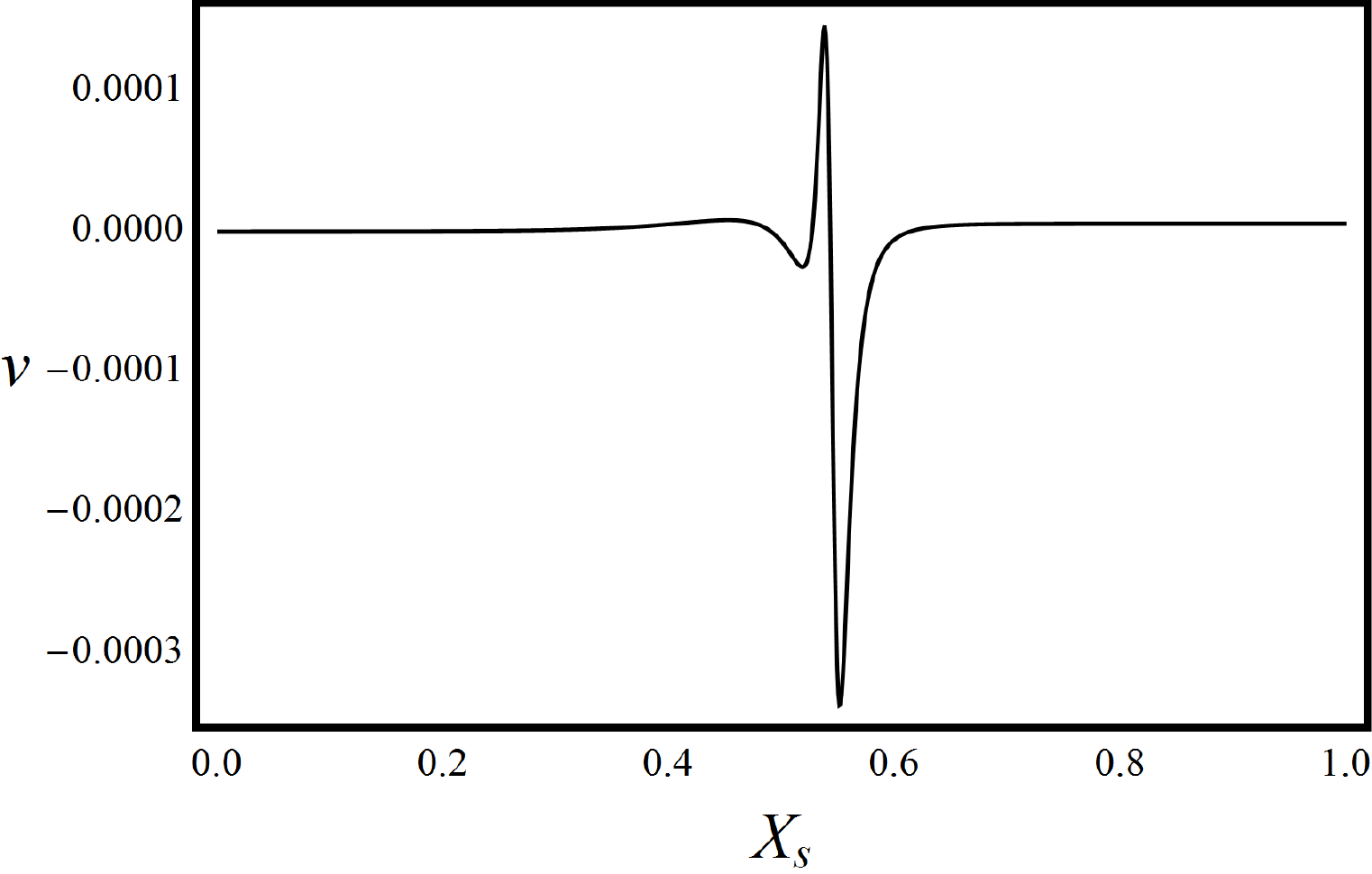}
 \end{minipage}
  \vspace{2mm}
 \centering
 \begin{minipage}[b!h!]{3cm}
\centering   
   \includegraphics[width=4cm]{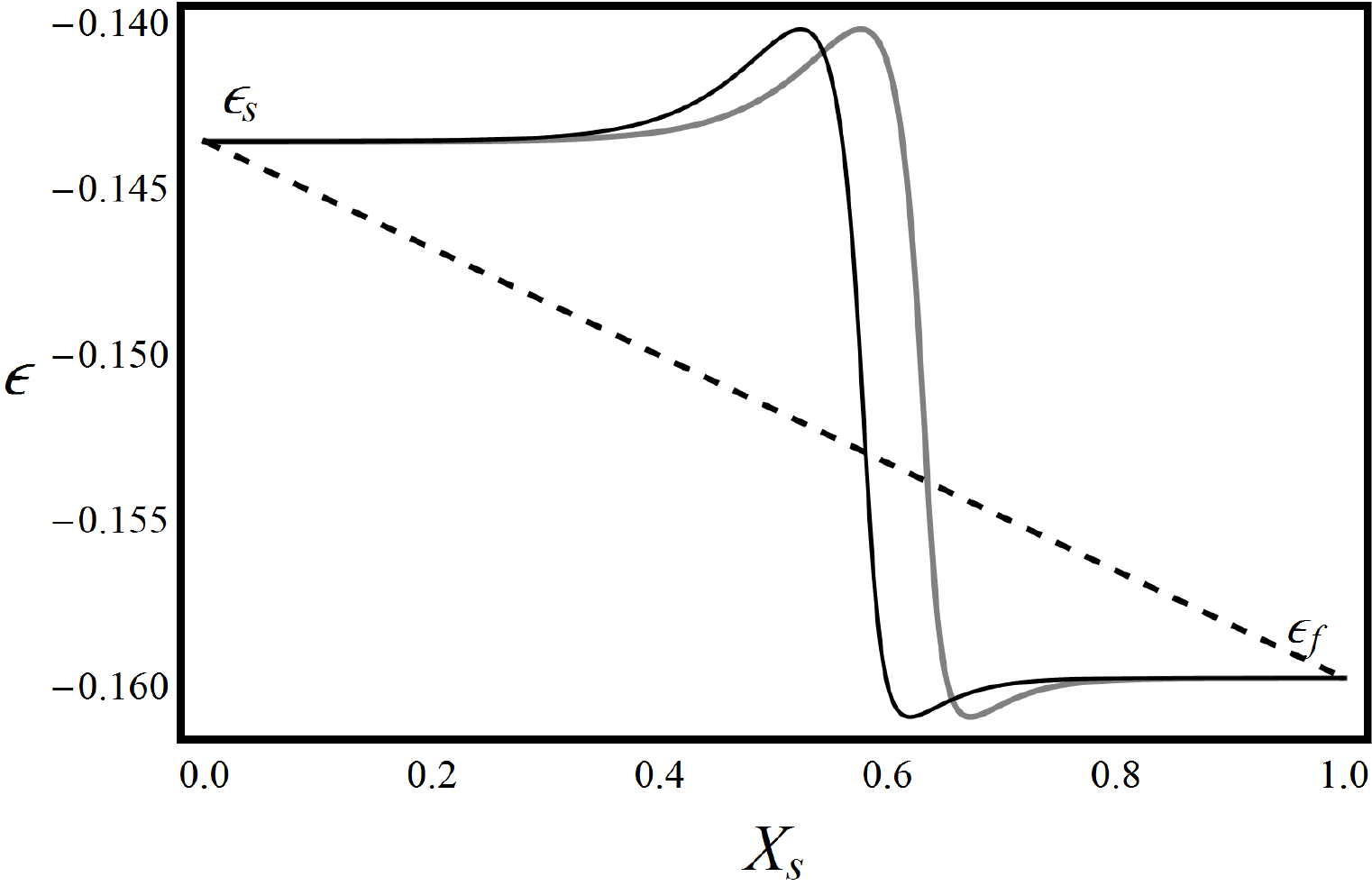}
 \end{minipage}
 \hspace{11mm}  
 \begin{minipage}[b!h!]{3cm}
  \includegraphics[width=4cm]{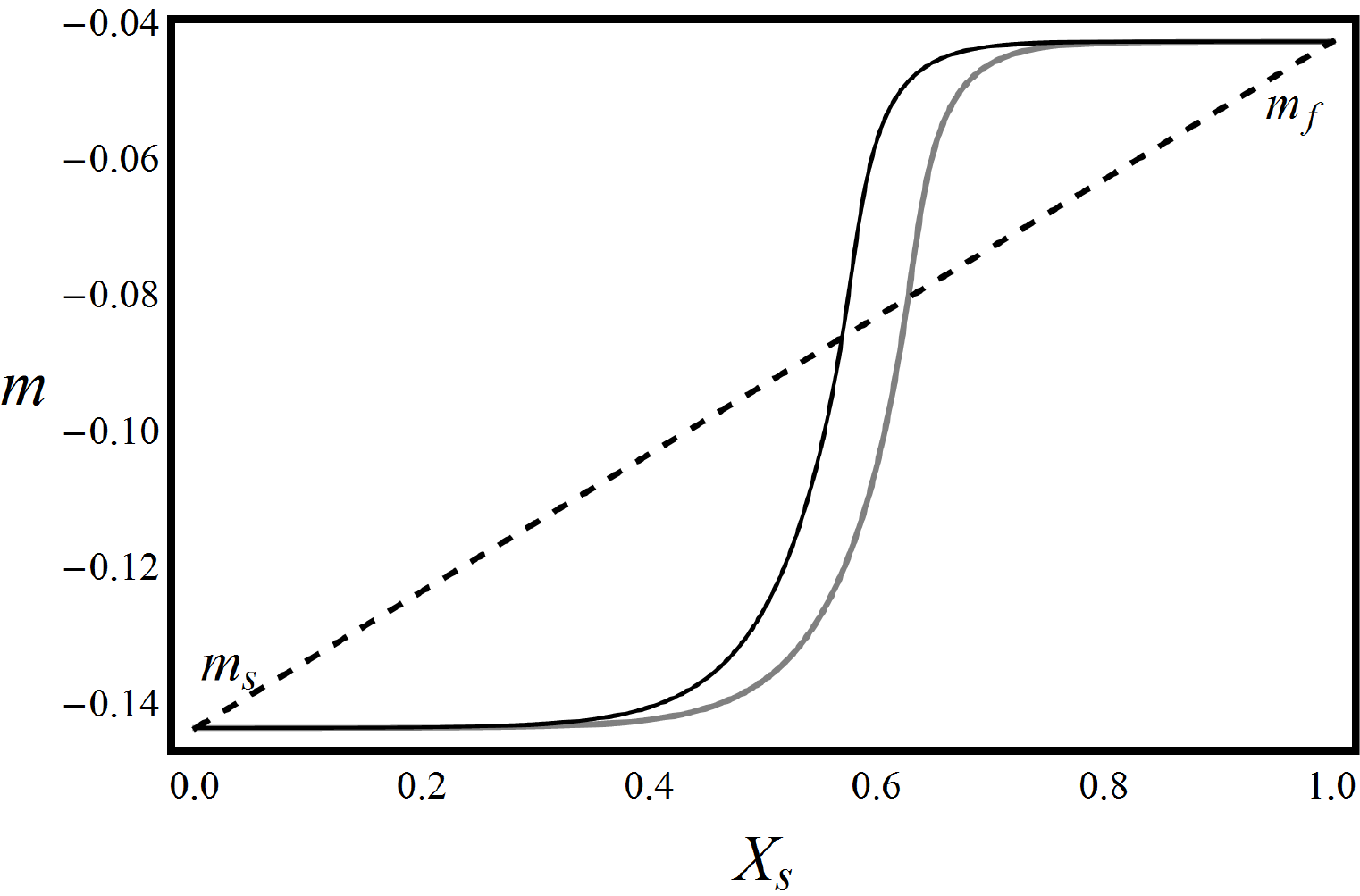}
   \end{minipage}
  \hspace{11mm} 
\begin{minipage}[b!h!]{3cm}
  \centering
   \includegraphics[width=4cm]{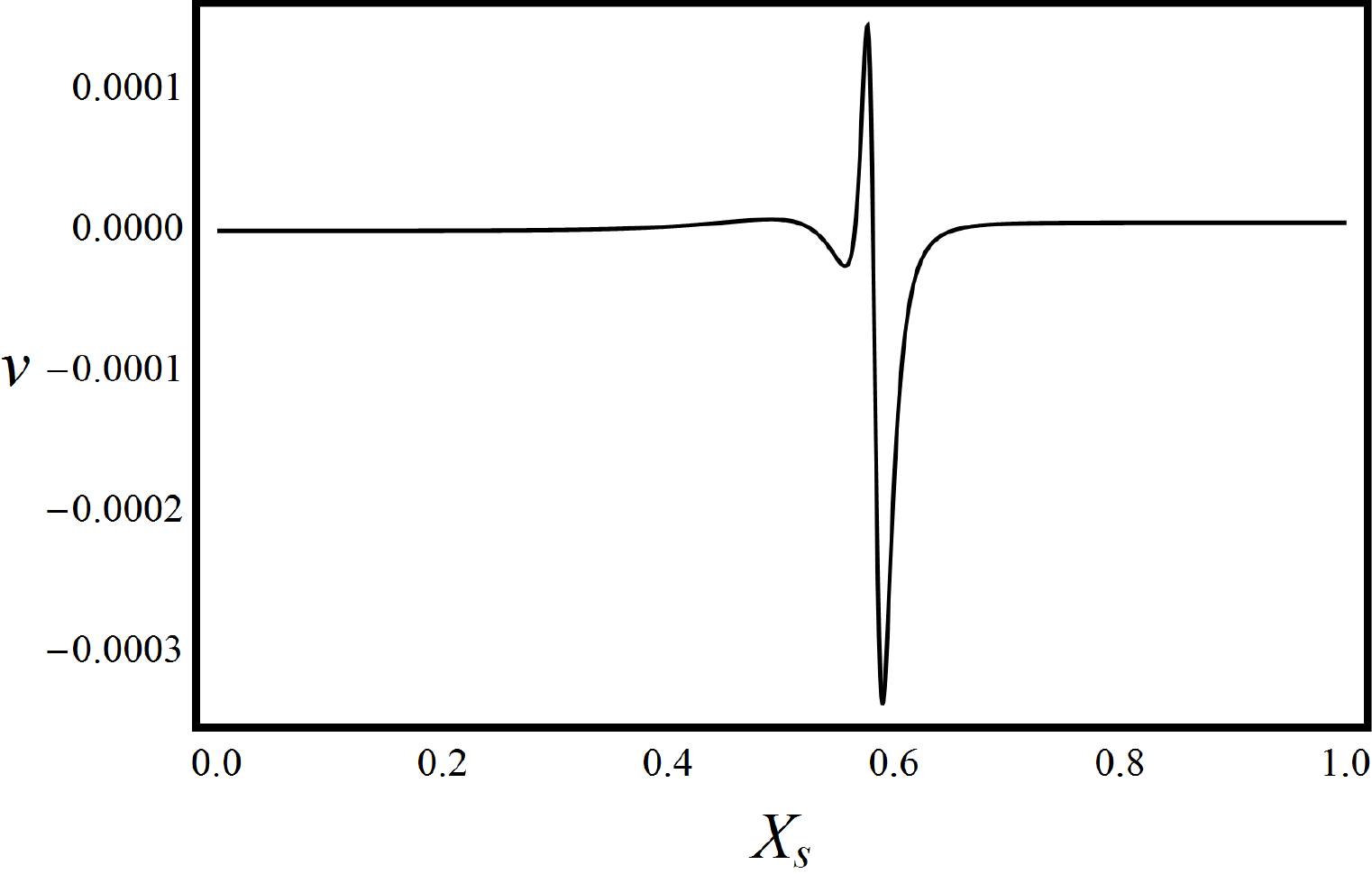}
 \end{minipage}
 \centering
\caption{Profiles (black lines) $\varepsilon(X_s,t)$, $m(X_s)$ and $v(X_s)$ for the zero chemical potential problem obtained by solving 
the non-stationary system with a linear initial state (dashed lines). We used Dirichelet boundary conditions $m(0)=m_s,\,\varepsilon(0)=\varepsilon_s,\,m(1)=m_f$, 
$\varepsilon(1)=\varepsilon_f$ on the finite interval $[0,1]$, at the coexistence pressure for $a=0.5,\,b=1,\,\alpha=100,\,k_1=k_2=k_3=10^{-3}$. 
Profiles at times $t=0.05,\,t=0.2,\,t=0.3,\,t=0.75,\,t=4,\,t=40$ in lexicographic order. The gray lines represent stationary profiles.}
\label{rettaduepuntivecchio}
\end{figure}

\begin{figure}[h!]
\centering
 \begin{minipage}[b!h!]{3cm}
\centering   
   \includegraphics[width=4cm]{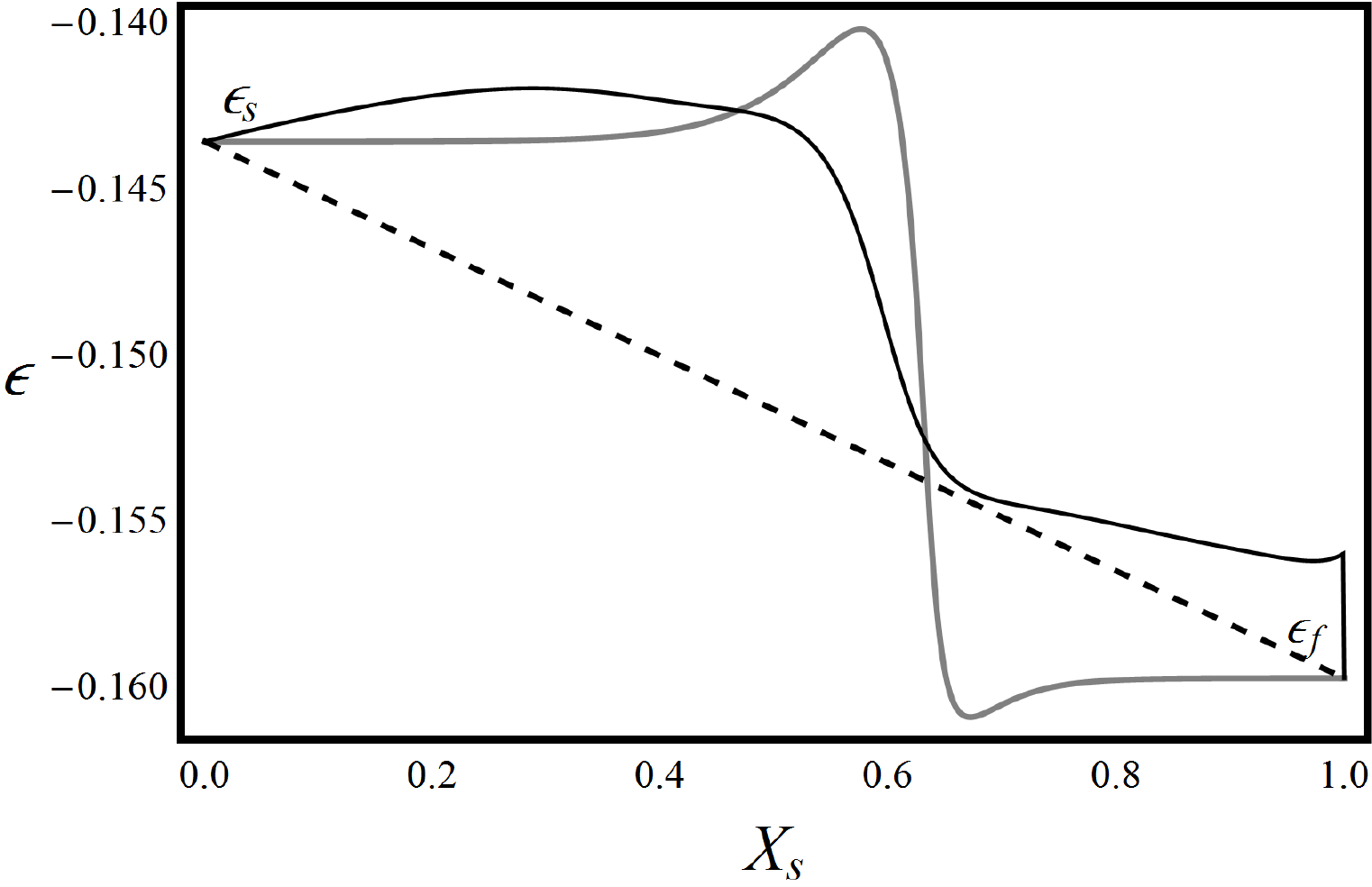}
 \end{minipage}
 \hspace{11mm}  
 \begin{minipage}[b!h!]{3cm}
  \includegraphics[width=4cm]{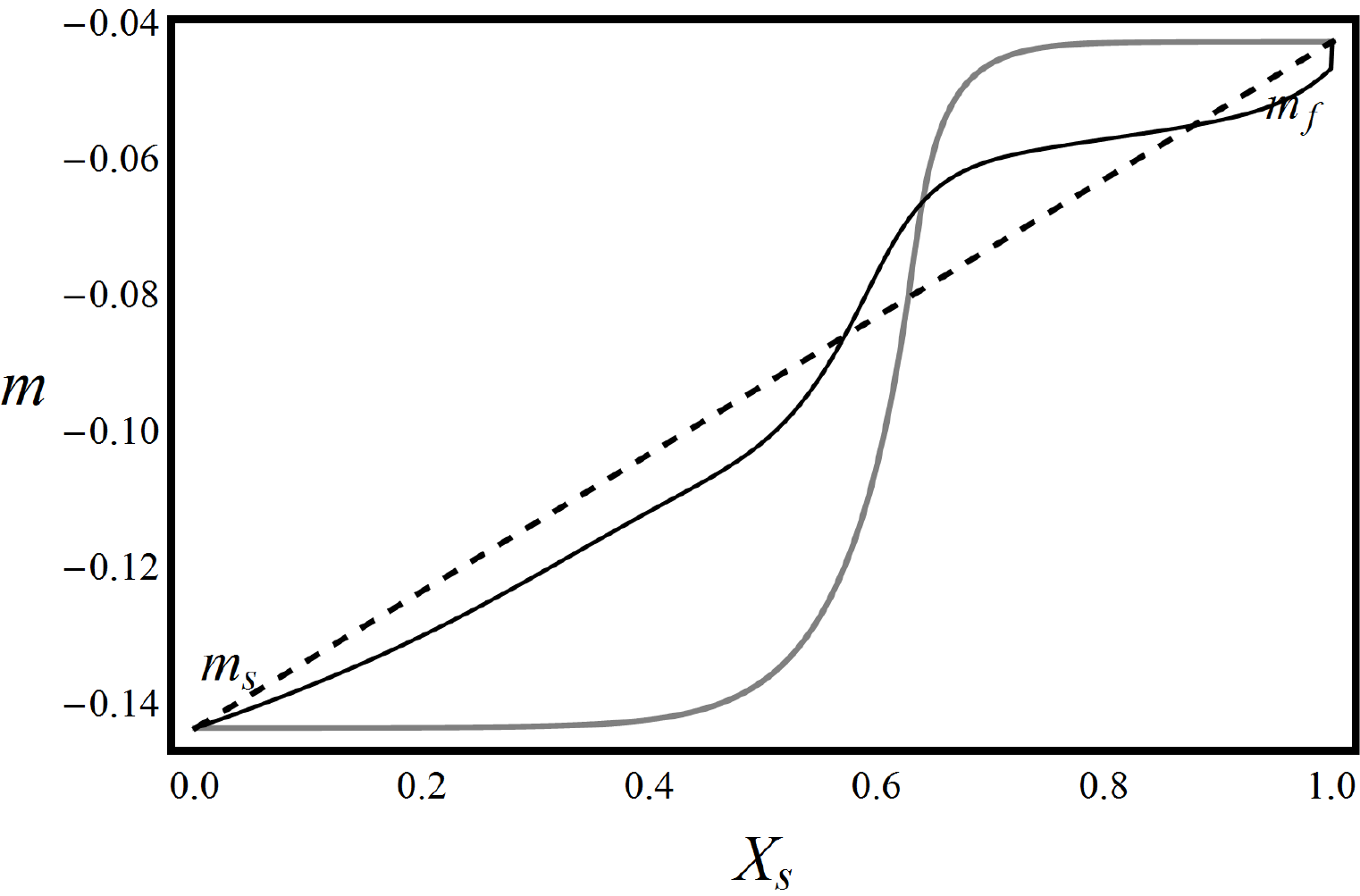}
   \end{minipage}
  \hspace{11mm} 
\begin{minipage}[b!h!]{3cm}
  \centering
   \includegraphics[width=4cm]{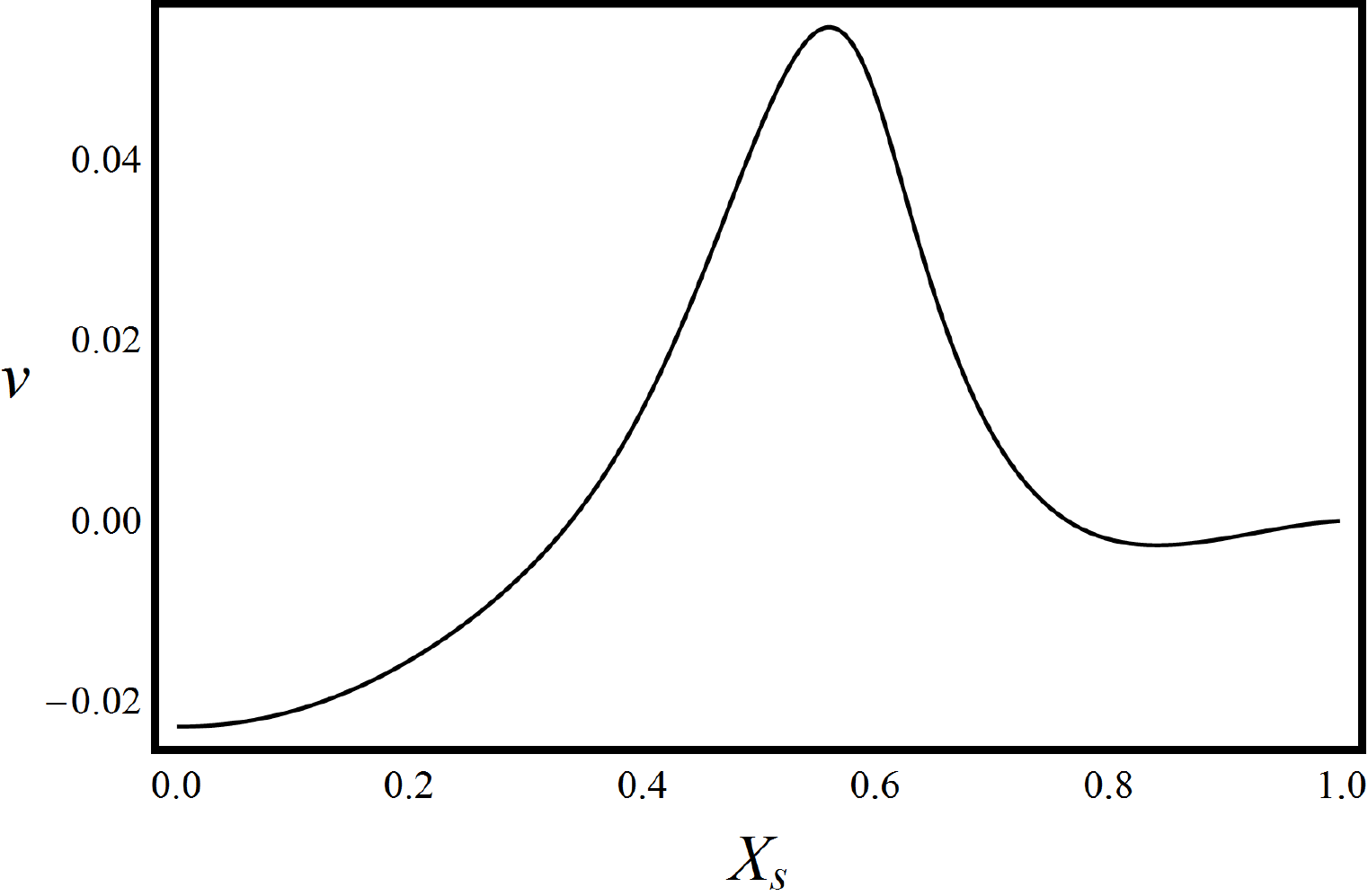}
 \end{minipage}
 \vspace{2mm}
\centering
 \begin{minipage}[b!h!]{3cm}
\centering   
   \includegraphics[width=4cm]{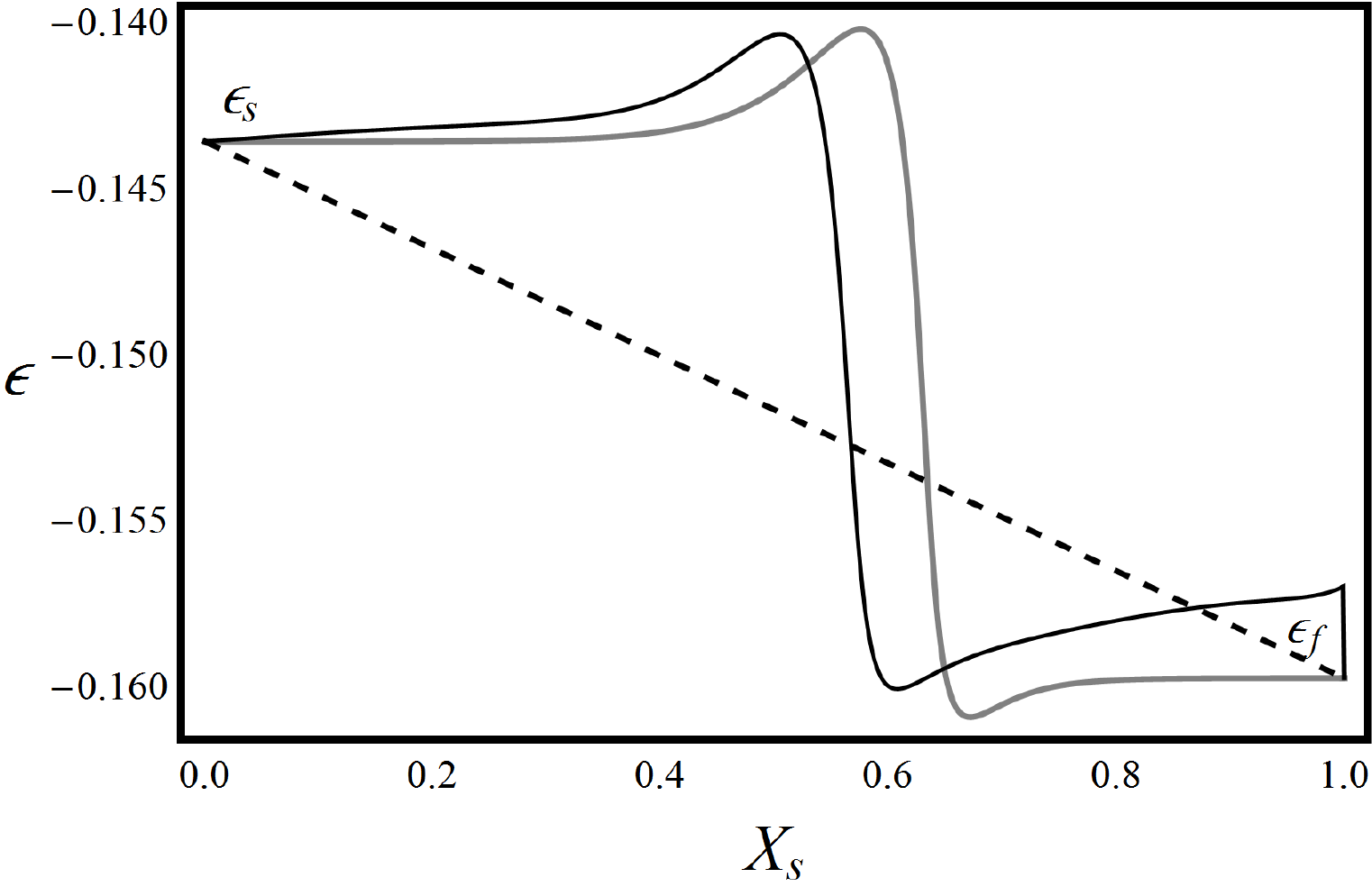}
 \end{minipage}
 \hspace{11mm}  
 \begin{minipage}[b!h!]{3cm}
  \includegraphics[width=4cm]{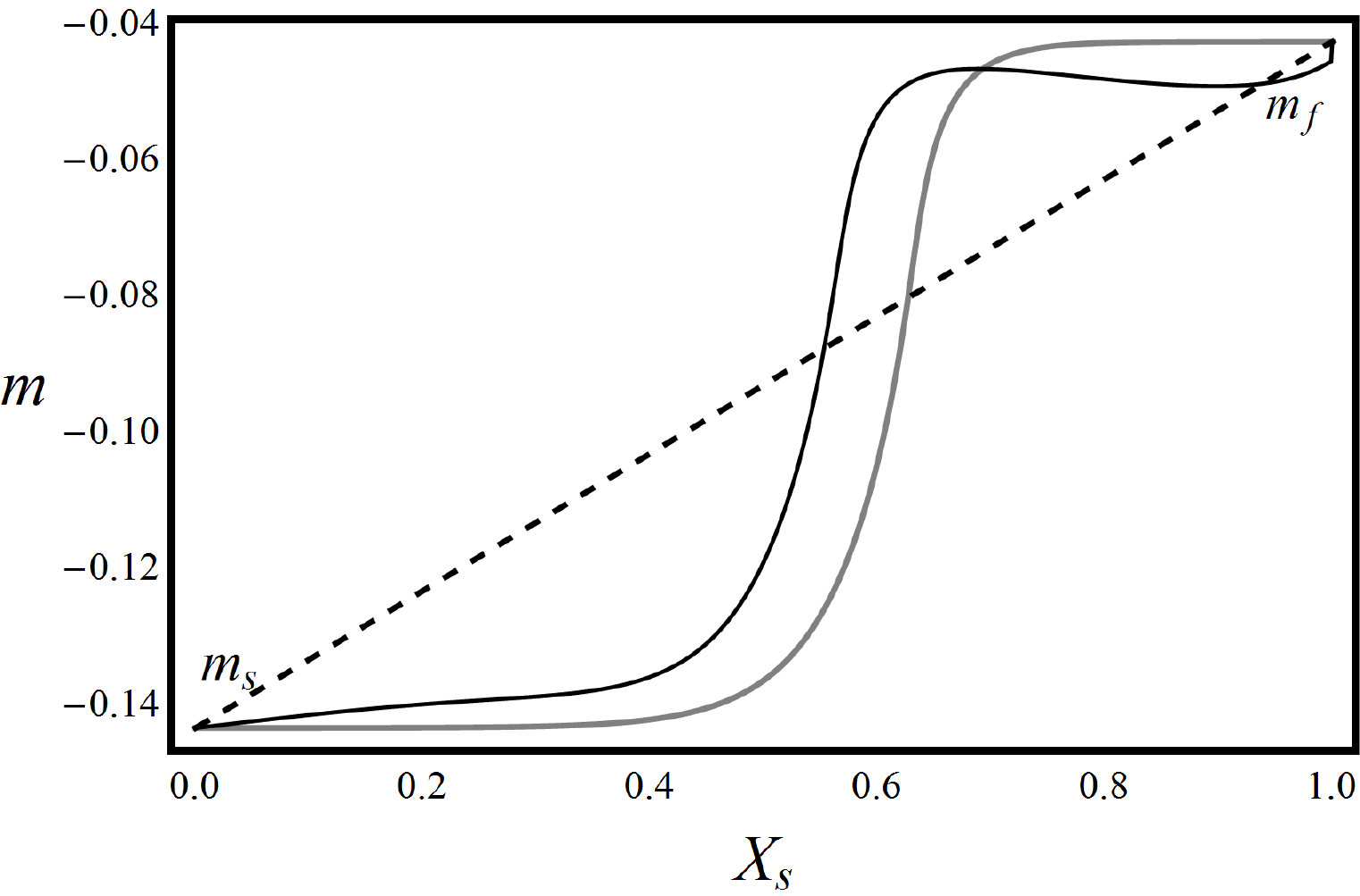}
   \end{minipage}
  \hspace{11mm} 
\begin{minipage}[b!h!]{3cm}
  \centering
   \includegraphics[width=4cm]{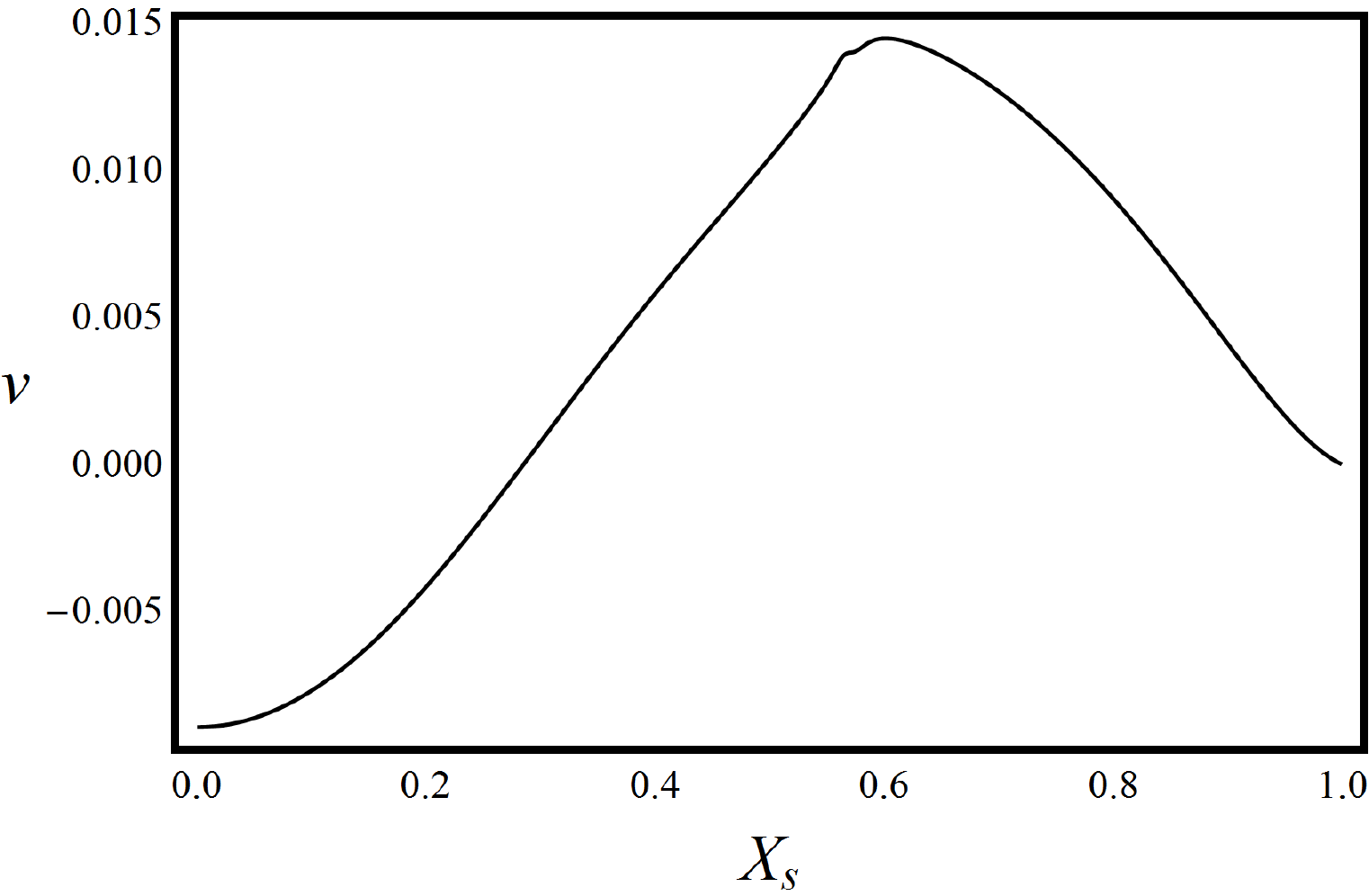}
 \end{minipage}
 \vspace{2mm}
 \centering
 \begin{minipage}[b!h!]{3cm}
\centering   
   \includegraphics[width=4cm]{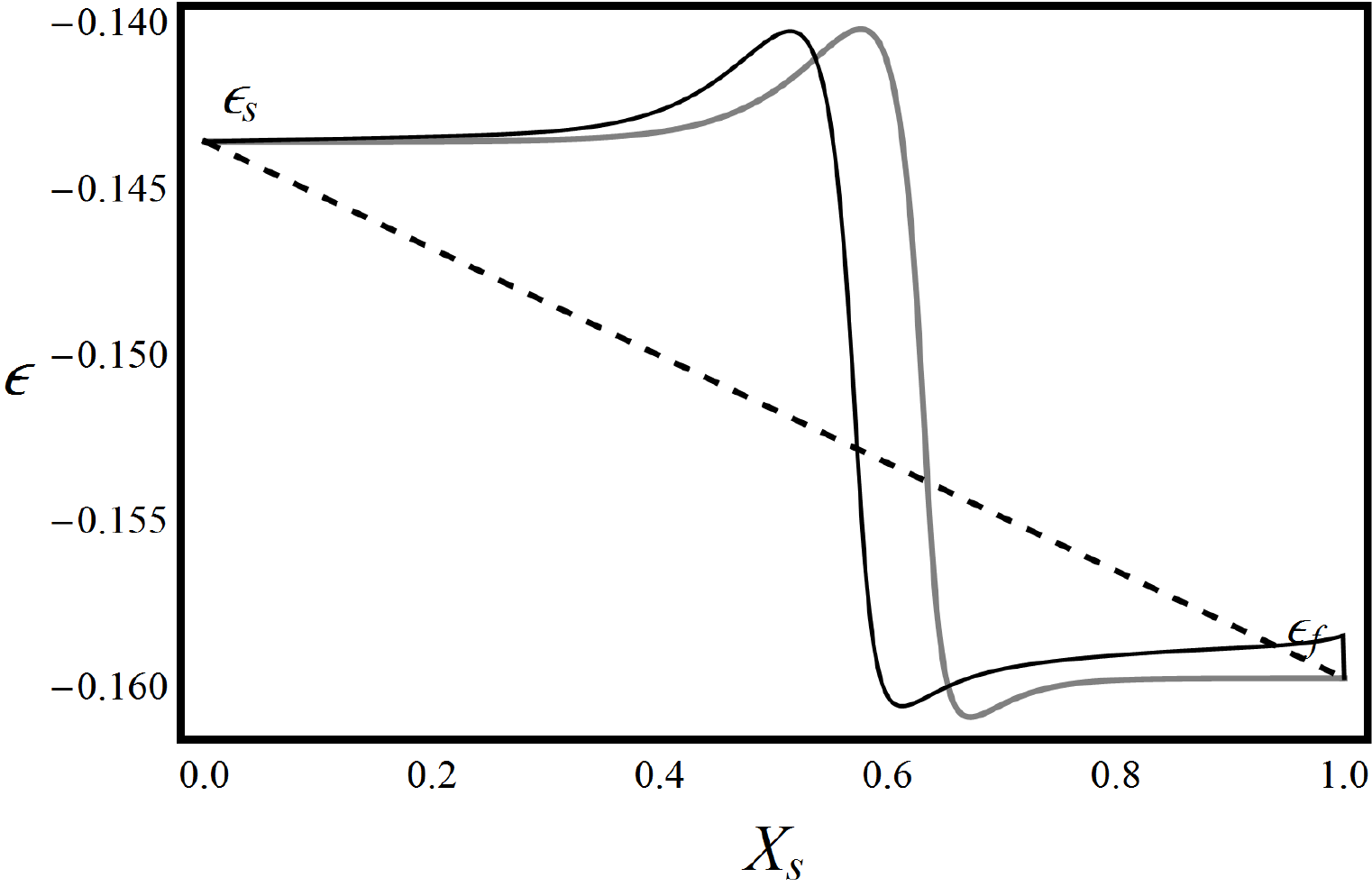}
 \end{minipage}
 \hspace{11mm}  
 \begin{minipage}[b!h!]{3cm}
  \includegraphics[width=4cm]{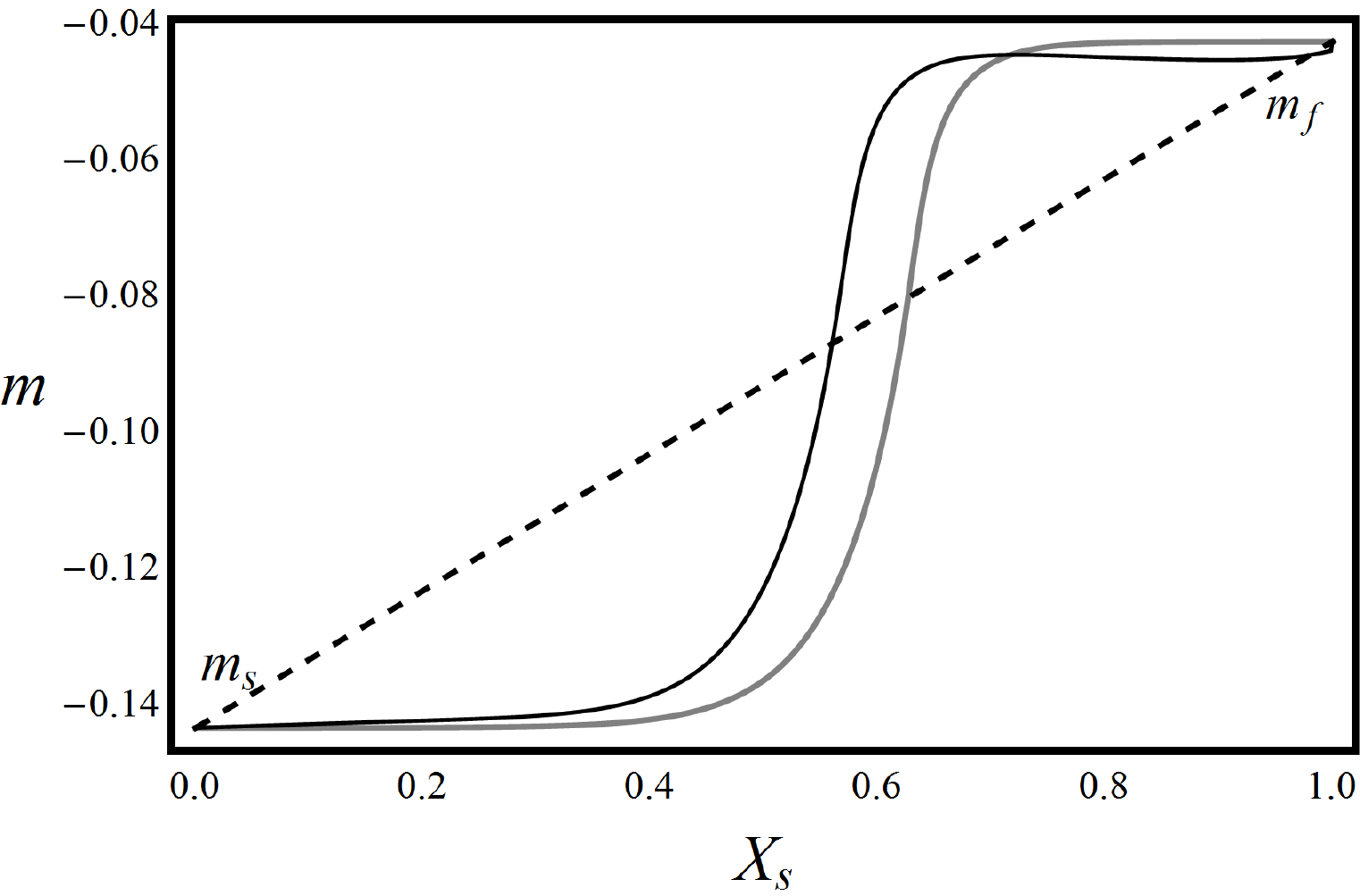}
   \end{minipage}
  \hspace{11mm} 
\begin{minipage}[b!h!]{3cm}  
  \centering
   \includegraphics[width=4cm]{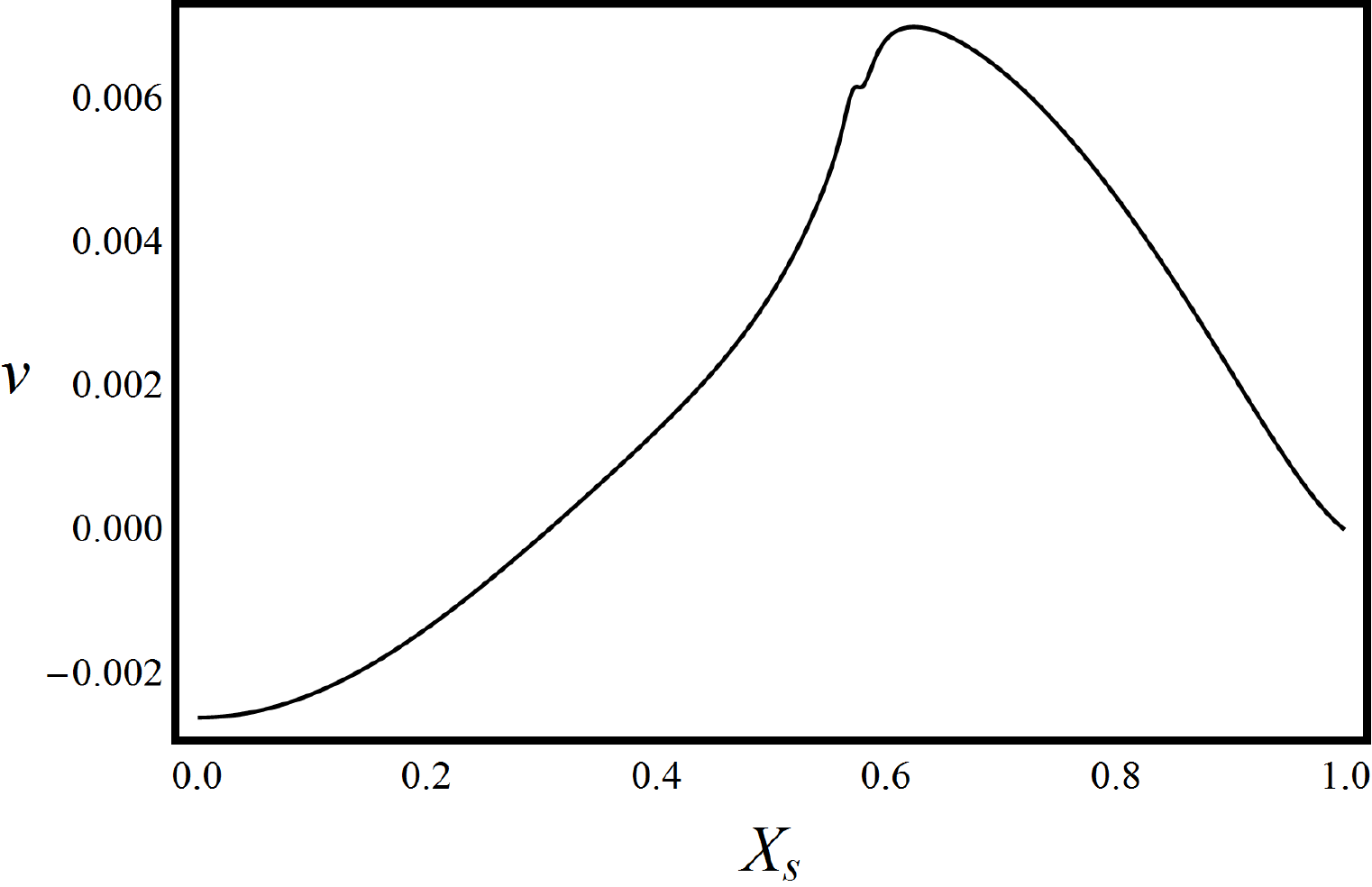}
 \end{minipage}
  \vspace{2mm}
 \centering
 \begin{minipage}[b!h!]{3cm}
\centering   
   \includegraphics[width=4cm]{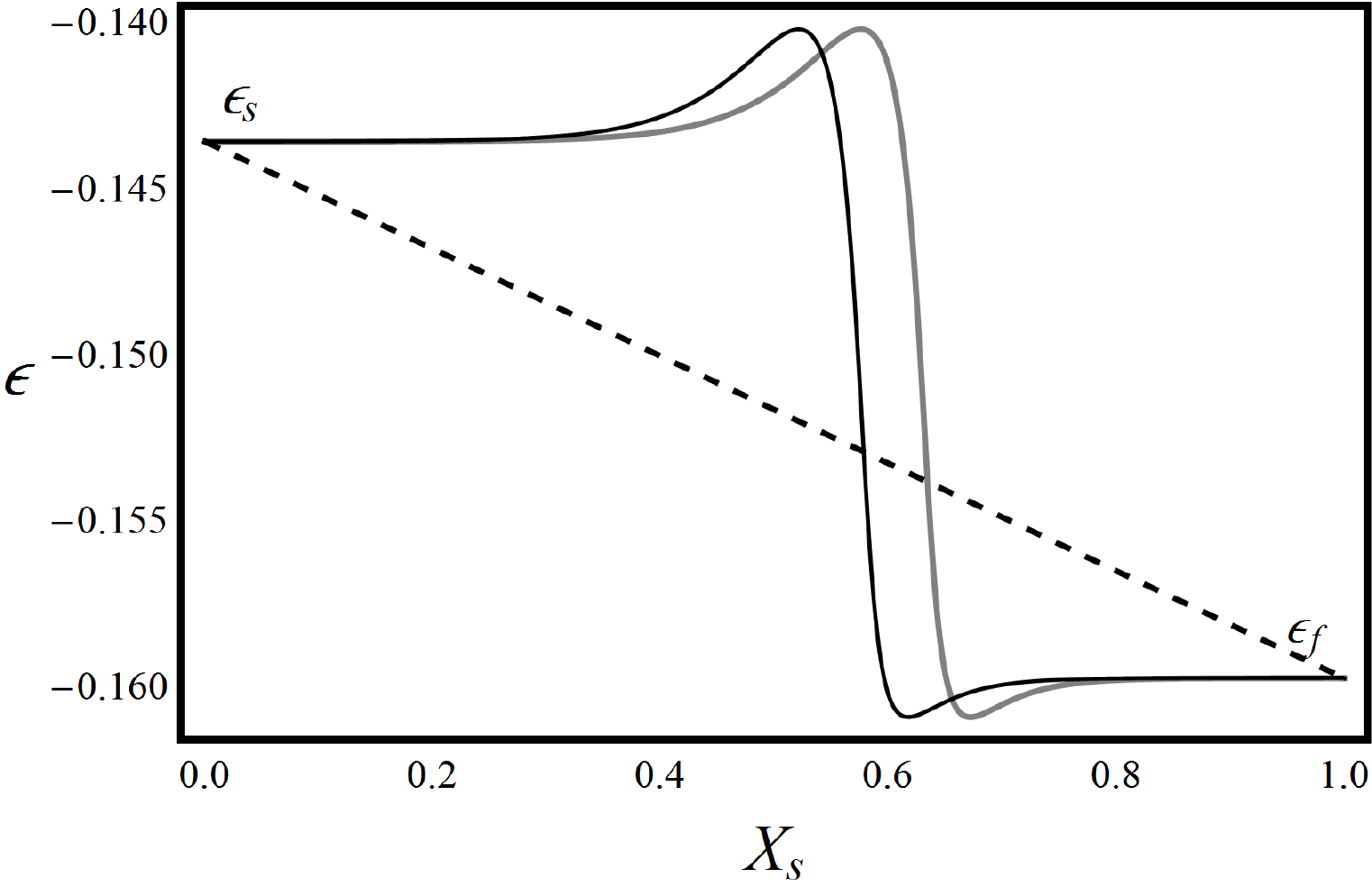}
 \end{minipage}
 \hspace{11mm}  
 \begin{minipage}[b!h!]{3cm}
  \includegraphics[width=4cm]{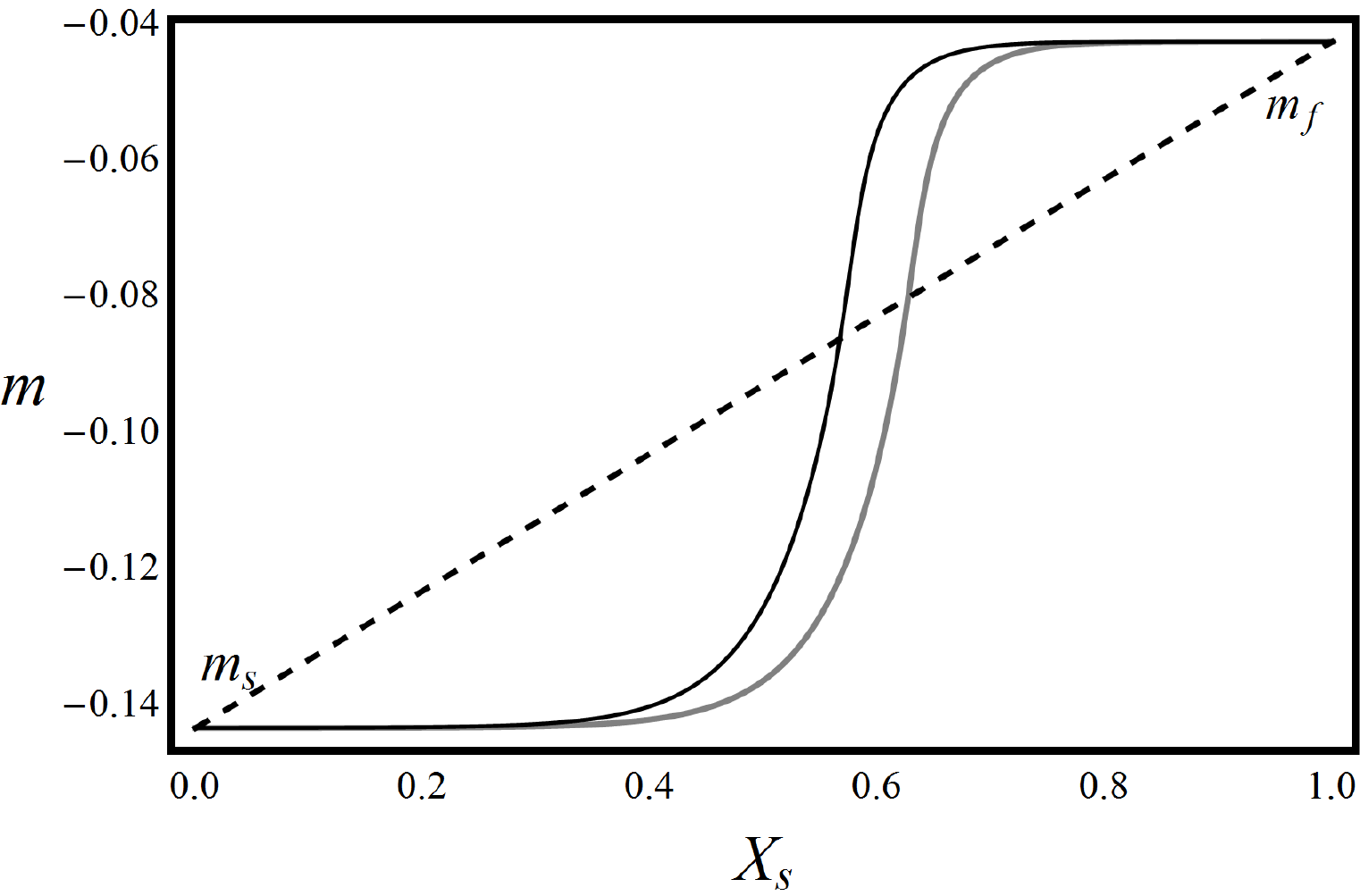}
   \end{minipage}
  \hspace{11mm} 
\begin{minipage}[b!h!]{3cm}
  \centering
   \includegraphics[width=4cm]{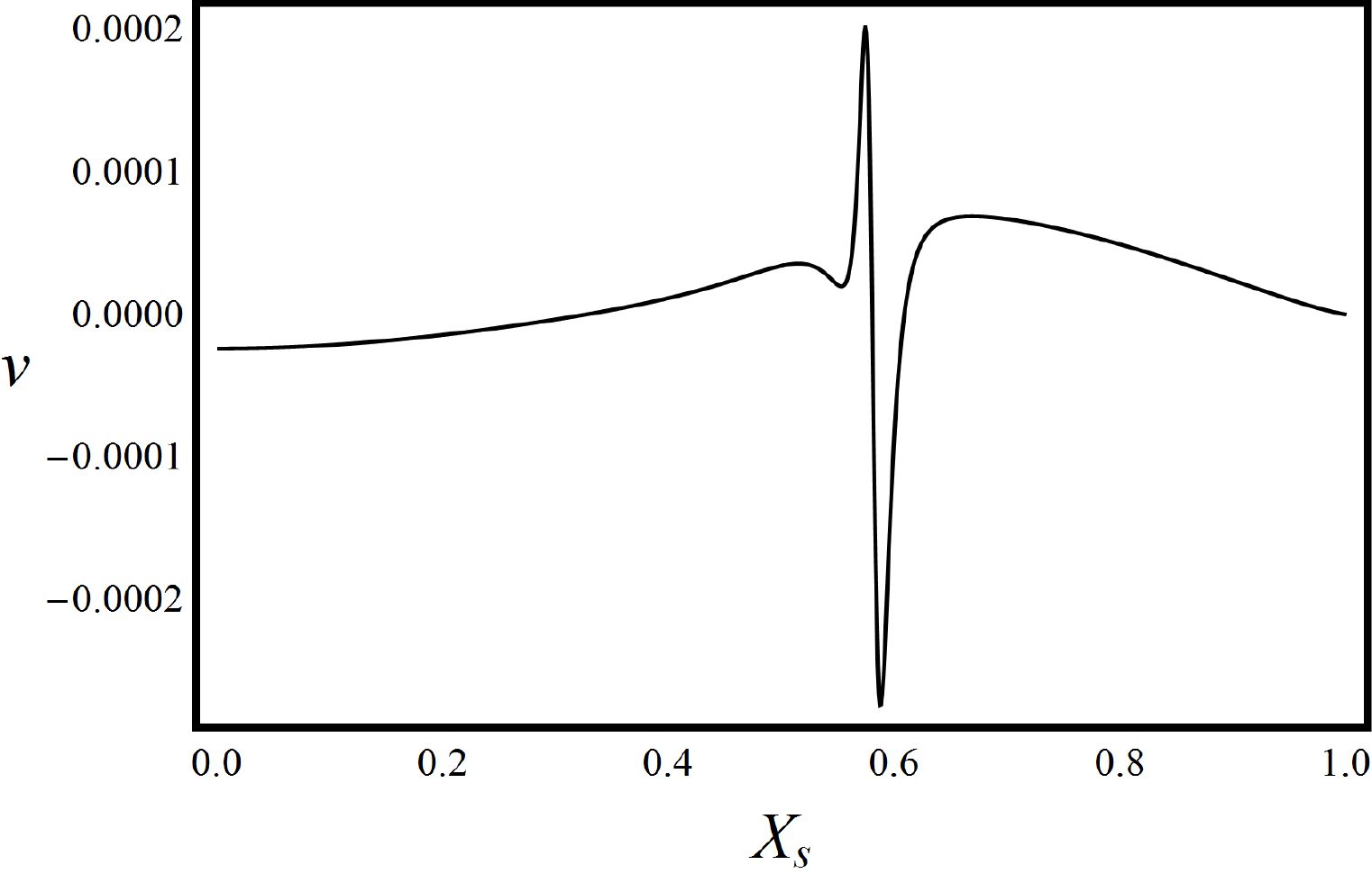}
 \end{minipage}
 \vspace{2mm}
 \centering
 \begin{minipage}[b!h!]{3cm}
\centering   
   \includegraphics[width=4cm]{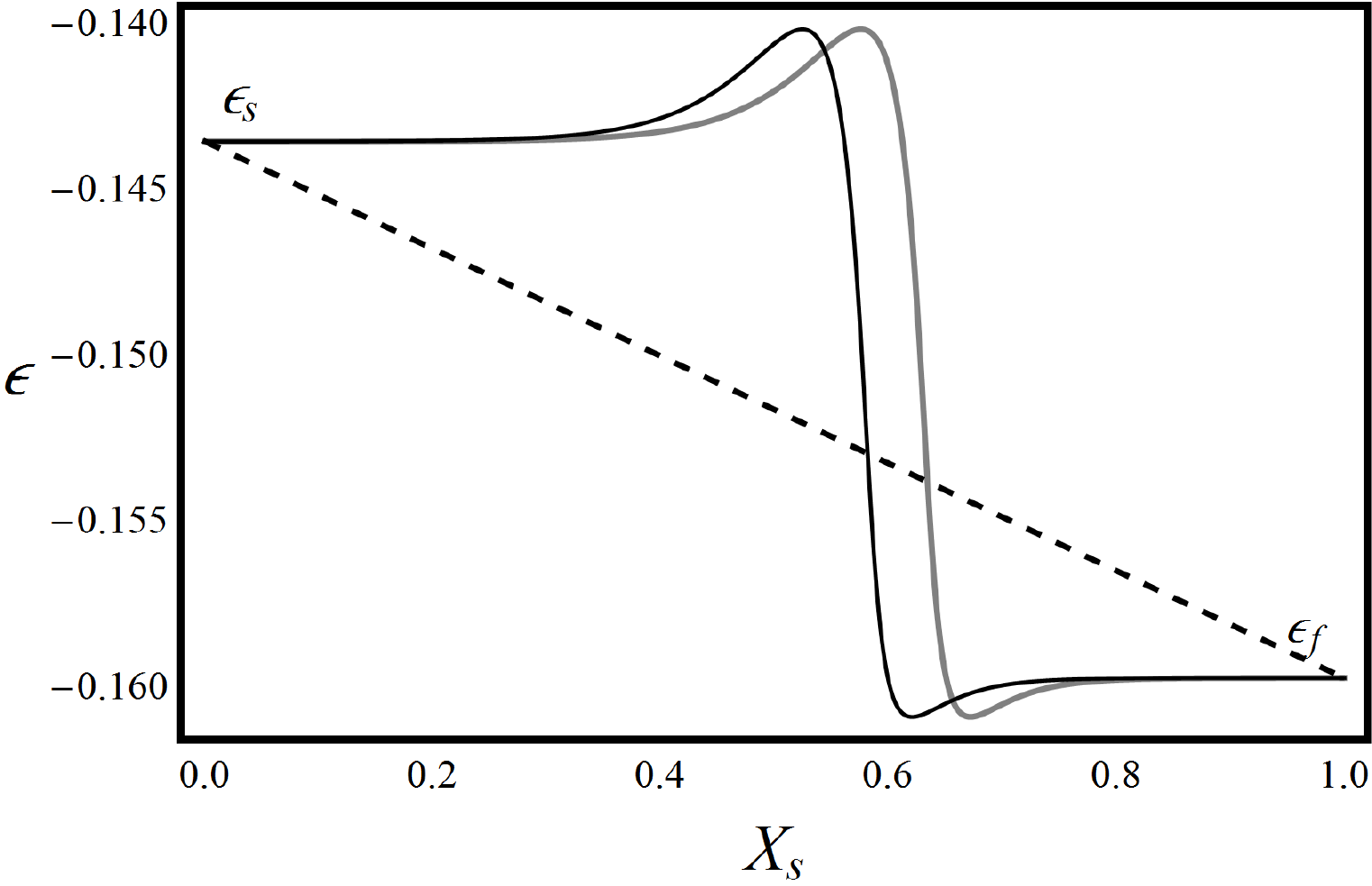}
 \end{minipage}
 \hspace{11mm}  
 \begin{minipage}[b!h!]{3cm}
  \includegraphics[width=4cm]{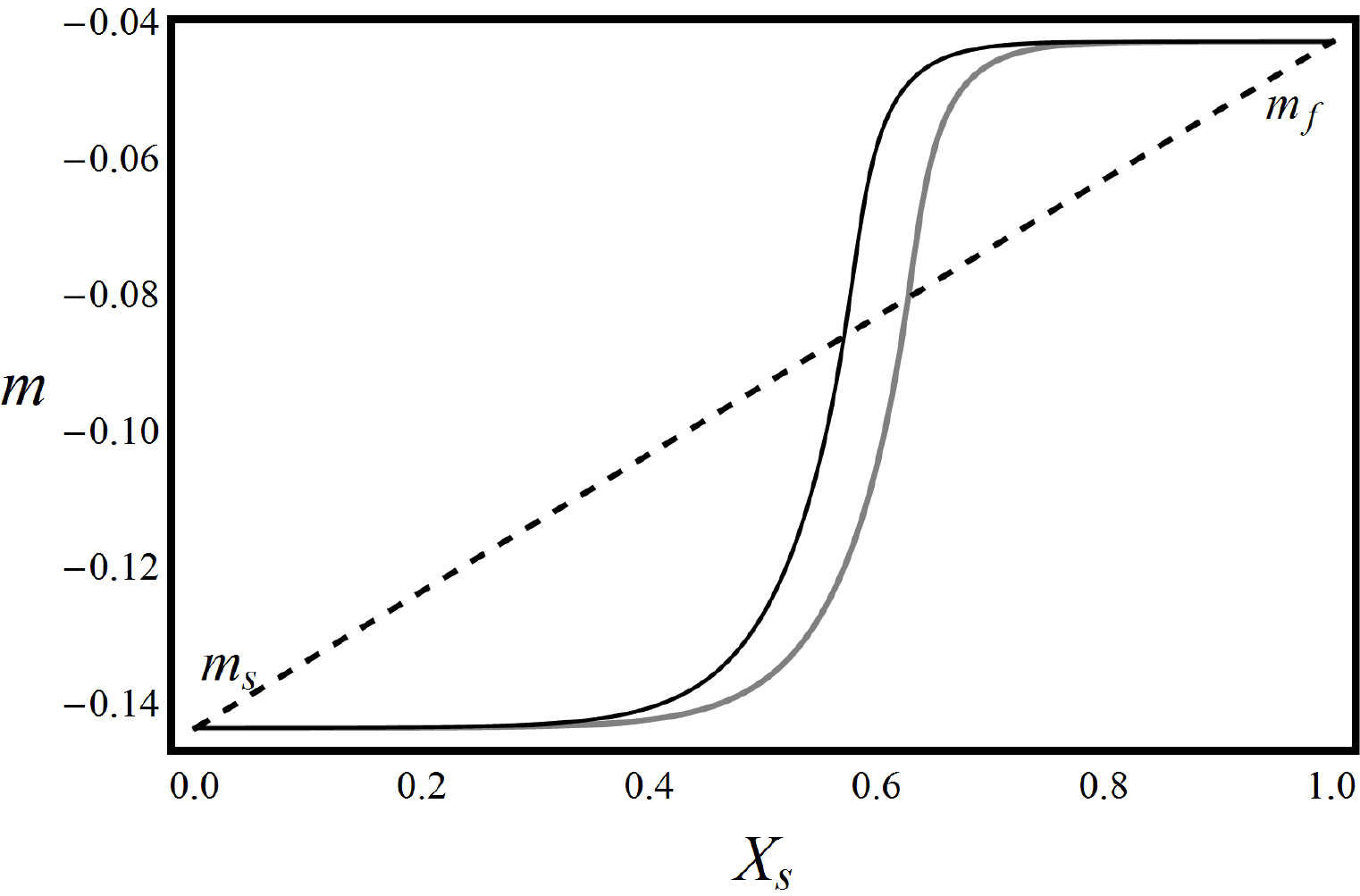}
   \end{minipage}
  \hspace{11mm} 
\begin{minipage}[b!h!]{3cm}
  \centering
   \includegraphics[width=4cm]{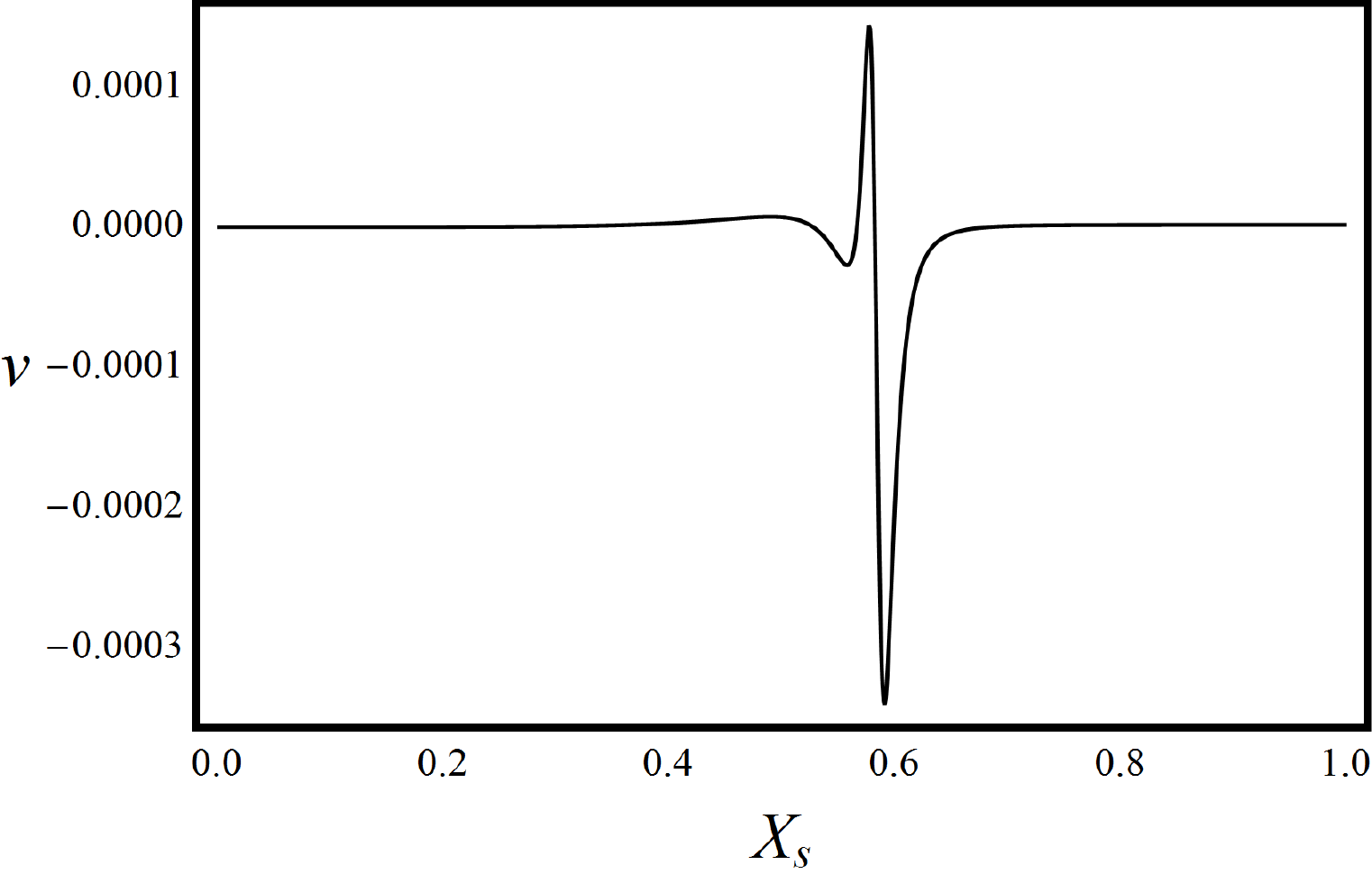}
 \end{minipage}
 \vspace{2mm}
 \centering
  \begin{minipage}[b!h!]{3cm}
\centering   
   \includegraphics[width=4cm]{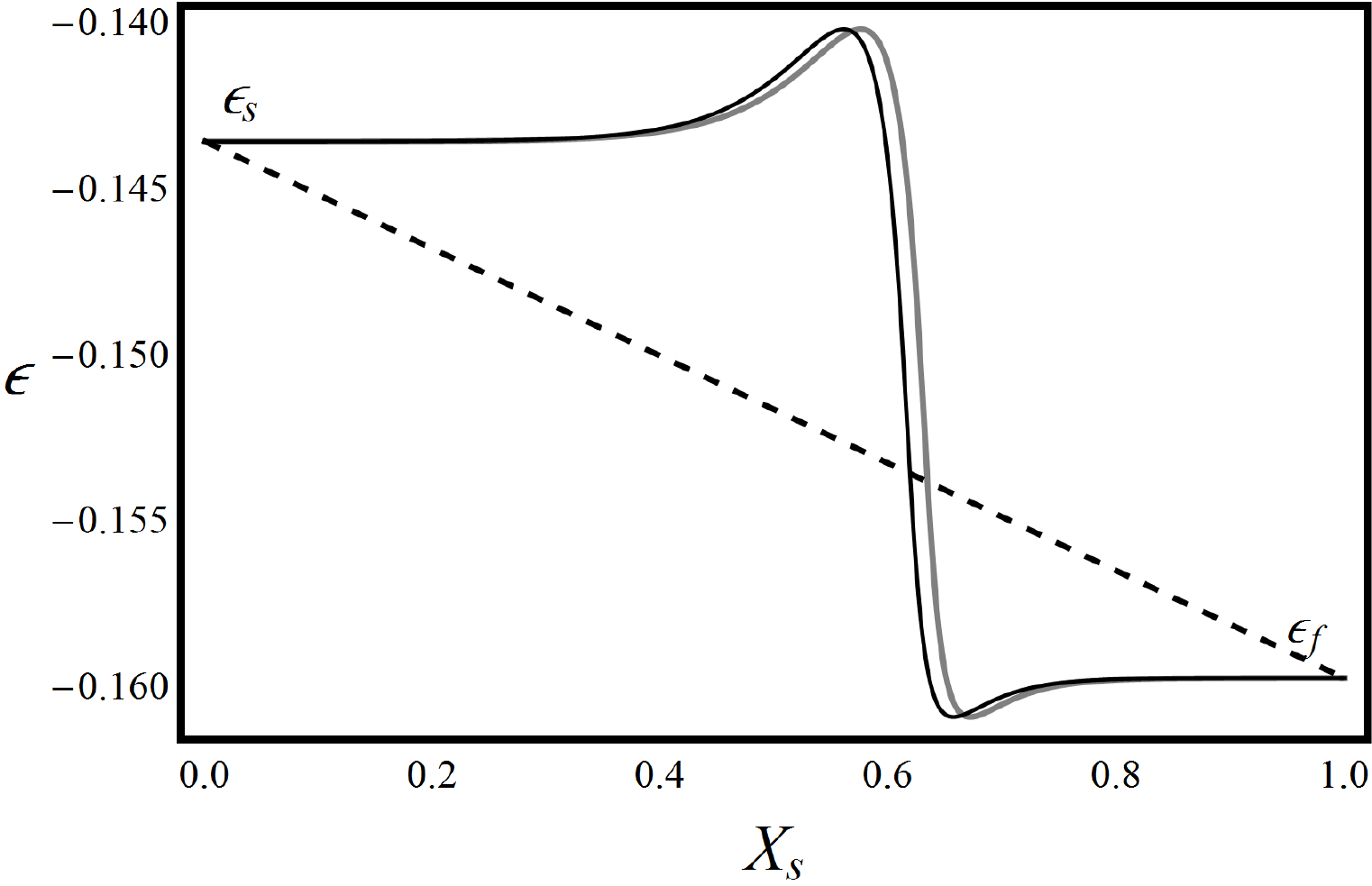}
 \end{minipage}
 \hspace{11mm}  
 \begin{minipage}[b!h!]{3cm}
  \includegraphics[width=4cm]{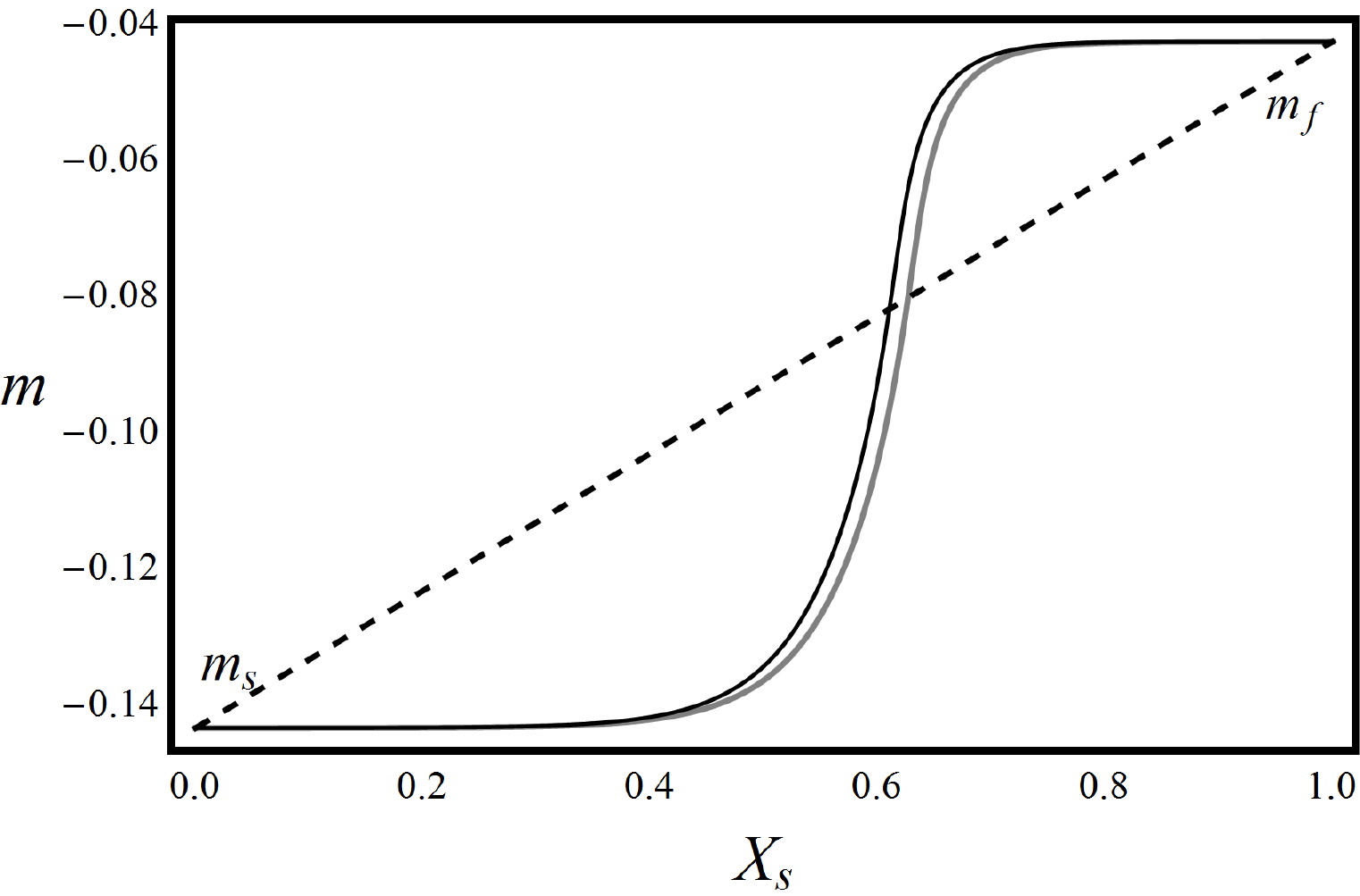}
   \end{minipage}
  \hspace{11mm} 
\begin{minipage}[b!h!]{3cm}
  \centering
   \includegraphics[width=4cm]{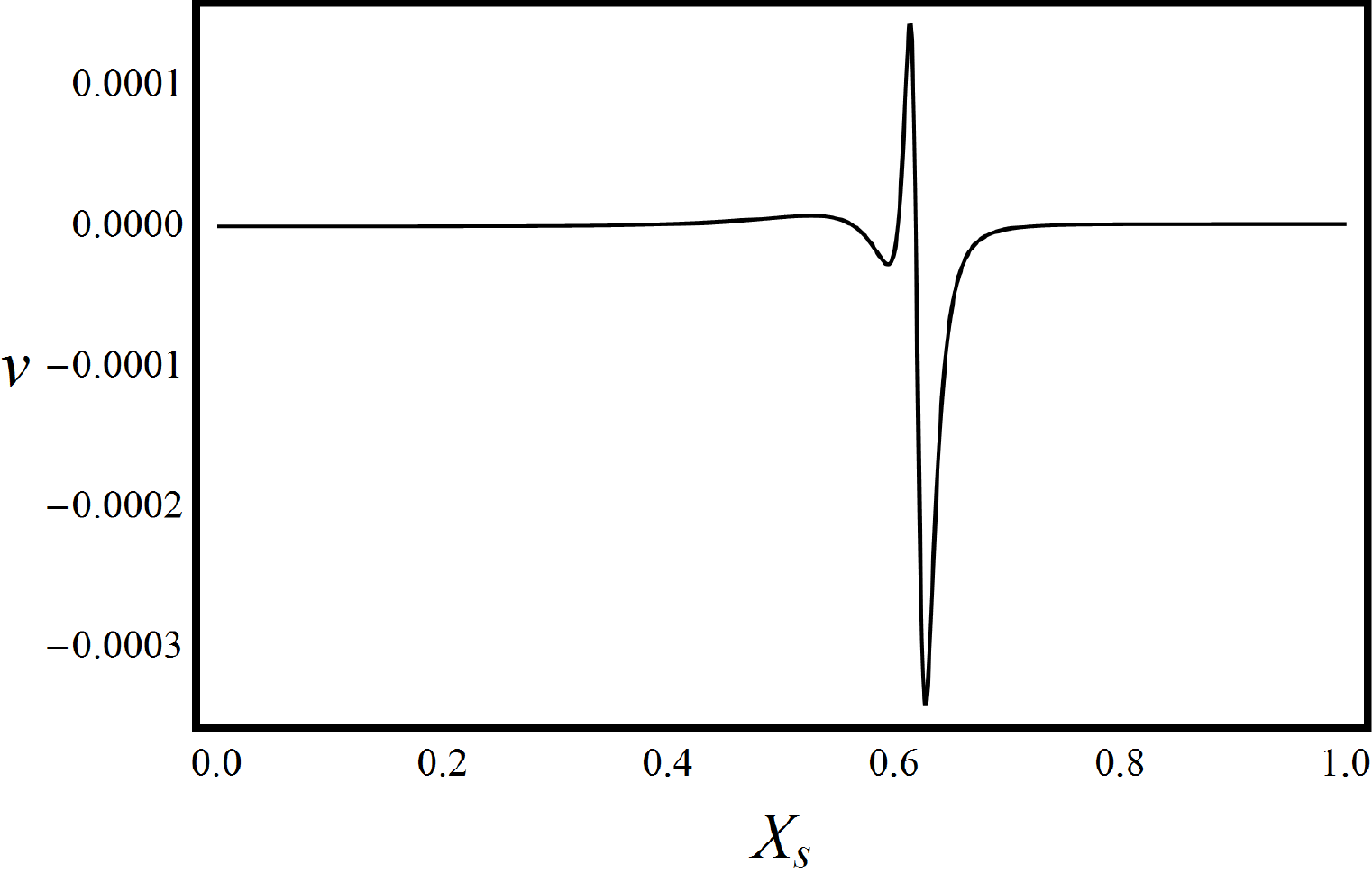}
 \end{minipage}
     \caption{The same as in figure \ref{rettaduepuntivecchio} but with the one side impermeability condition.}
\label{rettaduepuntitappo}
\end{figure}

Figures \ref{rettaduepuntivecchio} and \ref{rettaduepuntitappo} refer
to the evolution of the system starting from spatially linear initial conditions 
for $\varepsilon$ and $m$ connecting the standard (fluid--poor phase) on 
the left to the fluid--rich phase on the right.

The evolution can be divided into two macroscopically different regimes: 
the first one is the formation of the connection profile, while 
the second one is a sort of travelling wave which represents 
the slow propagation of the profile towards the stationary state. 
The transition between the two above--mentioned regimes is associated to 
the showing up of a characteristic velocity profile (see, for instance, 
the three bottom rows in figure~\ref{rettaduepuntivecchio}) 
that we shall call the \emph{late time} velocity profile.
In the following we will analyze separately the two regimes of the 
evolution, comparing the profiles relative to the considered boundary 
conditions. 

Comparing figures \ref{rettaduepuntivecchio} and \ref{rettaduepuntitappo}, 
one can observe that close to the boundary where impermeability is prescribed, 
strain and fluid density variations suffer a jump. This is mainly due to the 
delay of the system in modifying its initial state according with the 
prescribed boundary condition and, in particular, to the fact that no 
flow through the right boundary is allowed. On the other hand, 
when considering the zero chemical potential boundary condition, 
the formation of the profiles do not pass through this intermediate step. 
Consequently the second regime of the evolution process is reached more 
quickly and this is witnessed also by the formation of the characteristic 
late time velocity profile. In both cases the formation of the connection profile
is accompanied by a significant swelling of the solid matrix in that 
part of the domain where a significant fluid admission is observed.

In order to understand the formation of the connection profile, i.e., the first part 
of the evolution, one has to look at the 
first four rows, corresponding to times $t=0.05,\,0.2,\,0.3,\,0.75$, 
in figures~\ref{rettaduepuntivecchio} and \ref{rettaduepuntitappo}.
Close to the left boundary the behavior of the zero chemical potential and 
that of the one--side impermeable problem are very similar.
On the other hand, they are pretty different
(as expected) close to the right boundary. 
Indeed, in both cases a fluid emission occurs through the left boundary point,
as it can be deduced by noting that in the neighbourhood of the left boundary, 
the seepage velocity is negative (see the corresponding third column 
graphs). 

In the zero chemical potential case the profile formation is 
due to two counterbalancing effects: fluid displacement from the central region of 
the sample, where a fluid excess (with respect to the stazionary 
state) is present at the initial times, towards the right and fluid 
supply through the right boundary point. This can be easily deduced by 
looking at the velocity profile which is positive in the central
region and negative close to the right boundary 
(see the first three graphs from the top of the right 
column in figure~\ref{rettaduepuntivecchio}).
On the other hand, in the one--side impermeable case, 
no fluid supply is possible through the right boundary point
(notice that the velocity profile is zero in this case, see 
figure~\ref{rettaduepuntitappo}), so that the profile 
formation is due to the only effect of the central fluid 
displacement. As a result, in this last case the profile formation 
is slower and the profile is formed more to the right with respect 
to the zero chemical potential case. 
Notice, for instance, that the velocity peak at time $t=0.75$ 
is at $0.534$ in the zero chemical potential case and 
at $0.574$ in the one--side impermeable one.

At the end of the formation regime, the velocity profile assumes, 
in both case the peculiar late time shape with two maxima and to minima and 
intensity values significantly lower with respect to the previous time steps. 
This can be observed, for example in the zero chemical potential case, by 
comparing the value of the velocity peak a time $t=0.05$, where its intensity is 
about $5\times 10^{-2}$, and time $t=0.75$
where its intensity is about $1.4\times 10^{-4}$.

As we have already noticed, 
we remark that the profile formation is faster in the zero chemical potential 
boundary conditions than in the other case; indeed, the late time 
velocity profile is neatly formed at time 
$t=0.75$ in the first case, while in the second one it is necessary 
to wait till time $t=4$. 

Once the characteristic late time velocity profile is formed, 
the slow propagation towards the stationary state starts. 
This motion, in this case of linear initial condition and, as we will see, 
in the case of the fluid--rich phase initial condition, 
is characterized by a negative 
peak intensity larger than that of the positive one. 
This property is due to the choice of the initial condition: 
indeed, since the profile motion towards stationarity 
proceeds from the left to the right, an amount of water has to go out 
from the medium, and this is reflected in a bigger negative part 
in the velocity profile.

Finally, we note that the stationary profile is reached 
in the one--side impermeable case faster
than in the zero chemical potential one. It is important to remark 
that this is not a physically relevant effect, but is simply due to the 
choice of the initial condition. Indeed, due to our choice, 
in the one--side impermeable case the profile forms 
very close to the stationary one. 

\subsection{Fluid--poor initial condition}
\label{s:standard}
\begin{figure}[t]
\centering
 \begin{minipage}[b!h!]{3cm}
\centering   
   \includegraphics[width=4cm]{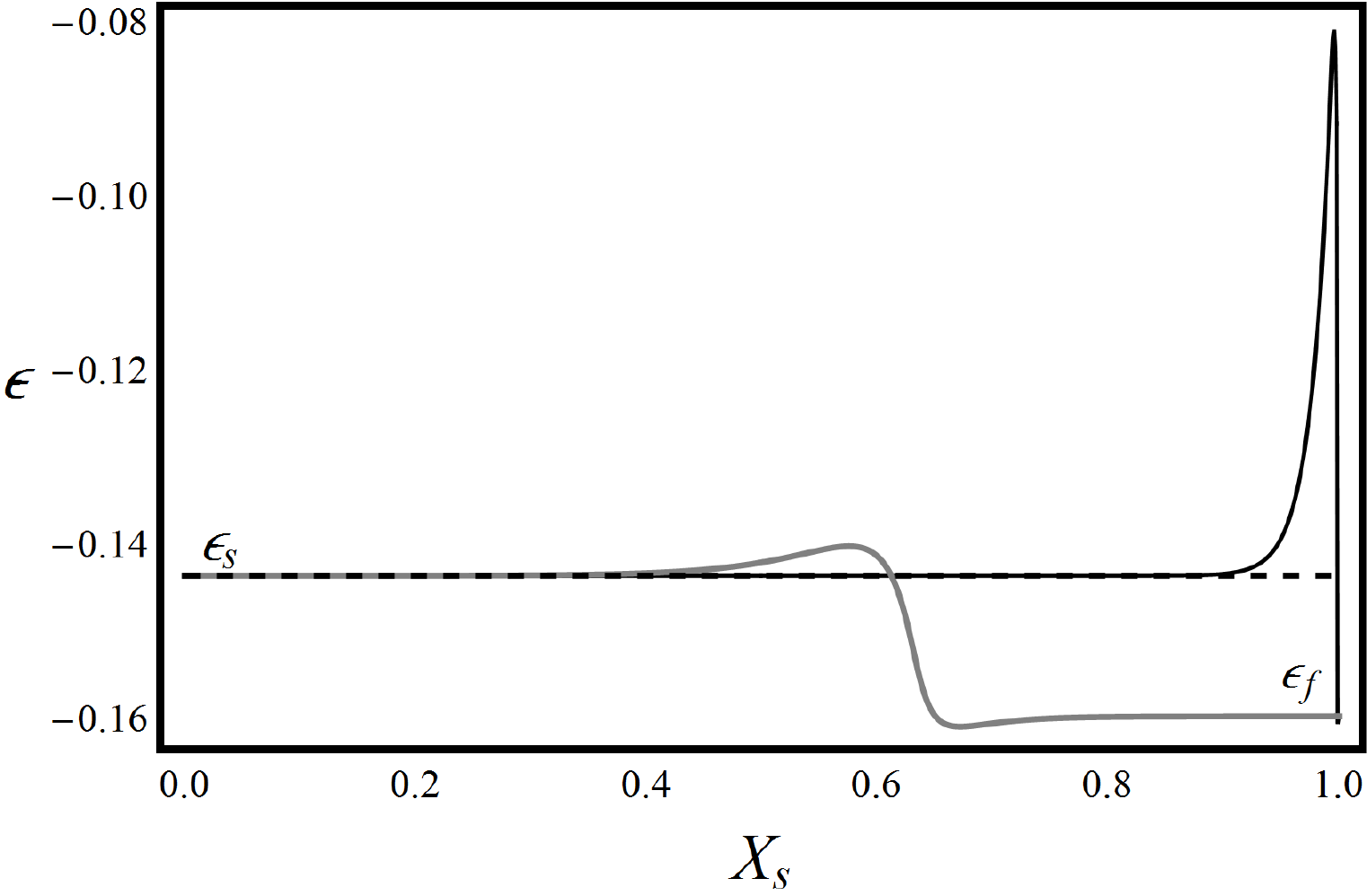}
 \end{minipage}
 \hspace{11mm}  
 \begin{minipage}[b!h!]{3cm}
  \includegraphics[width=4cm]{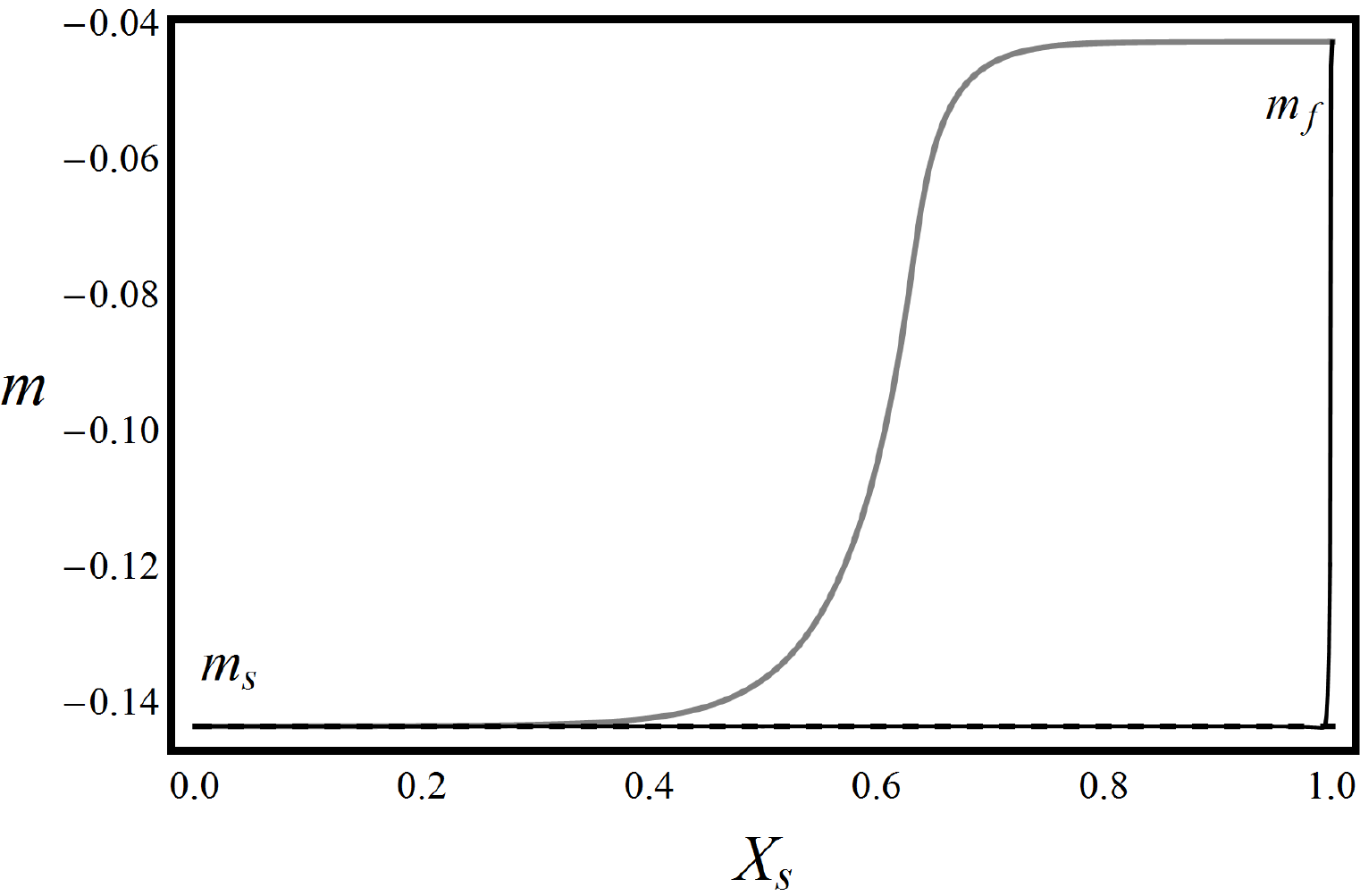}
   \end{minipage}
  \hspace{11mm} 
\begin{minipage}[b!h!]{3cm}
  \centering
   \includegraphics[width=4cm]{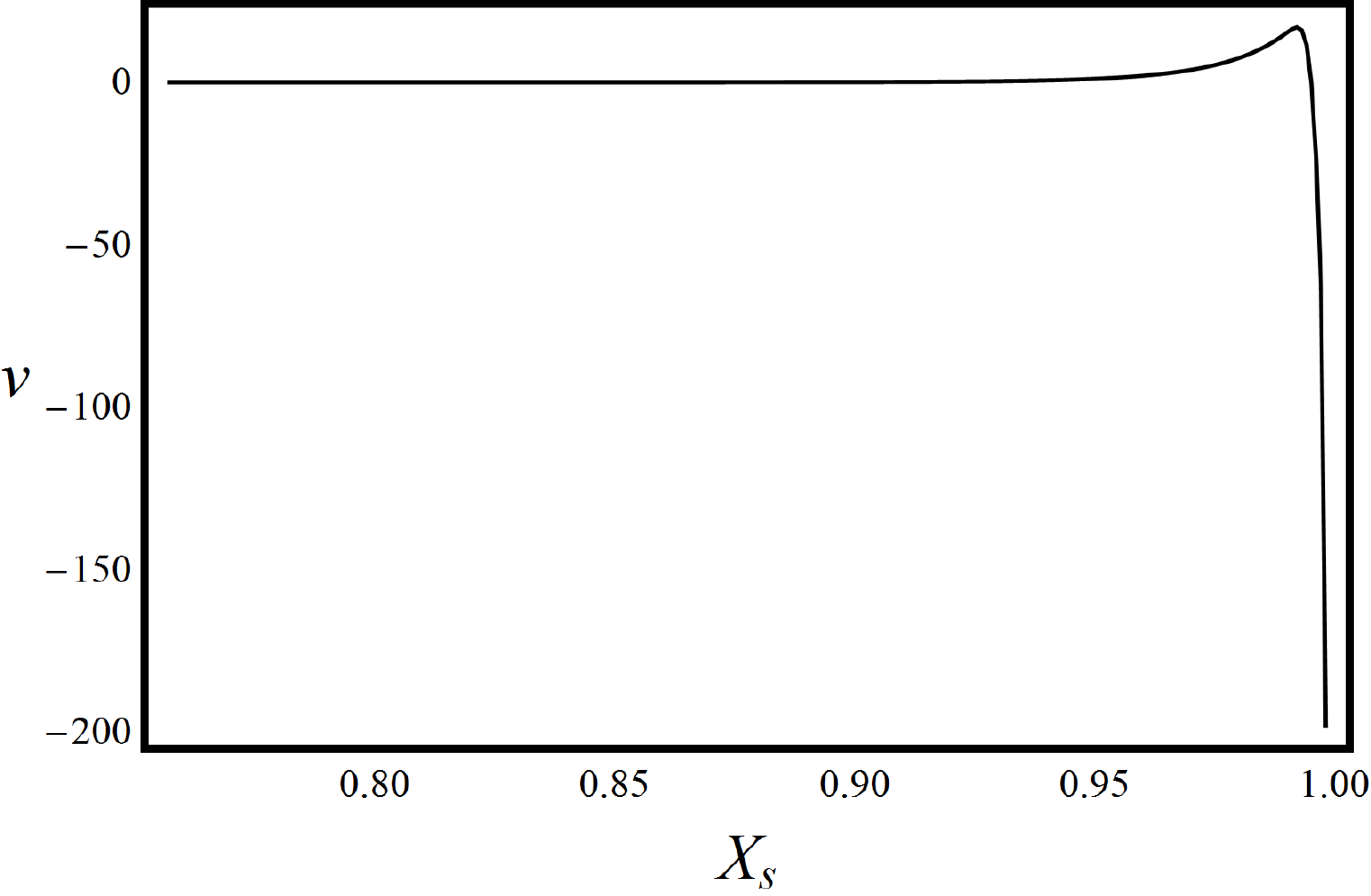}
 \end{minipage}
 \vspace{2mm}
\centering
 \centering \begin{minipage}[b!h!]{3cm}
\centering   
   \includegraphics[width=4cm]{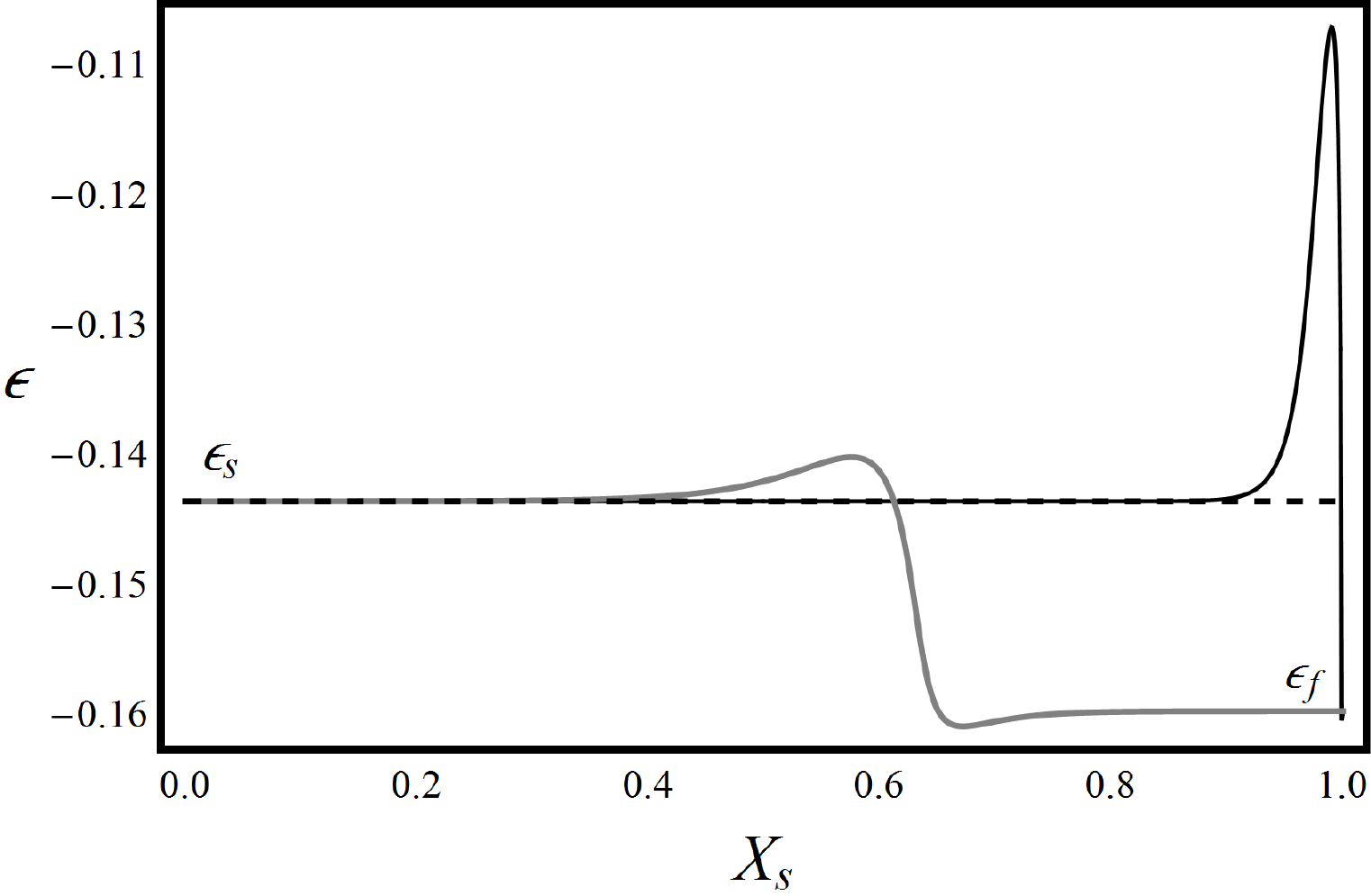}
 \end{minipage}
 \hspace{11mm}  
 \begin{minipage}[b!h!]{3cm}
  \includegraphics[width=4cm]{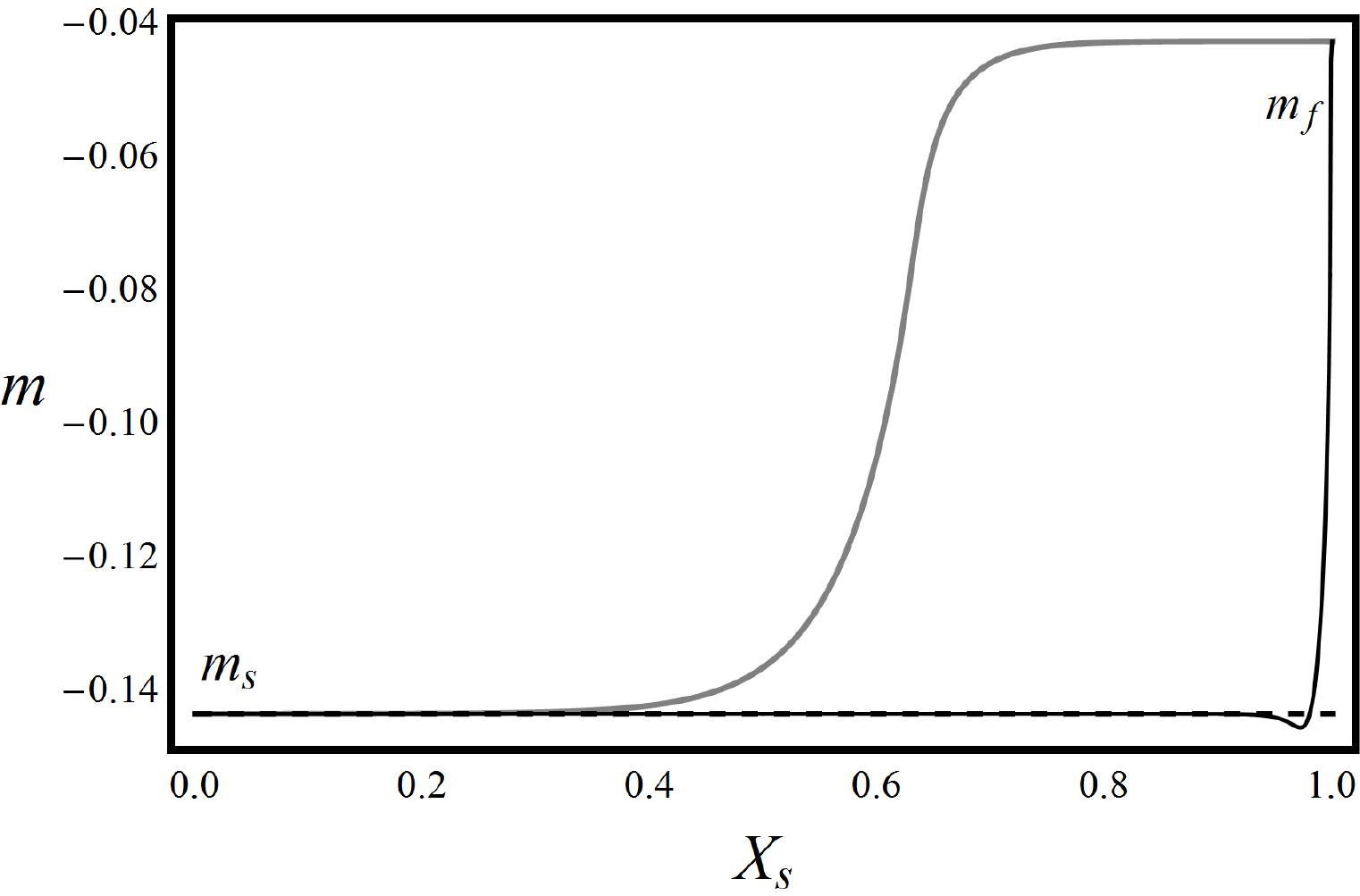}
   \end{minipage}
  \hspace{11mm} 
\begin{minipage}[b!h!]{3cm}
  \centering
   \includegraphics[width=4cm]{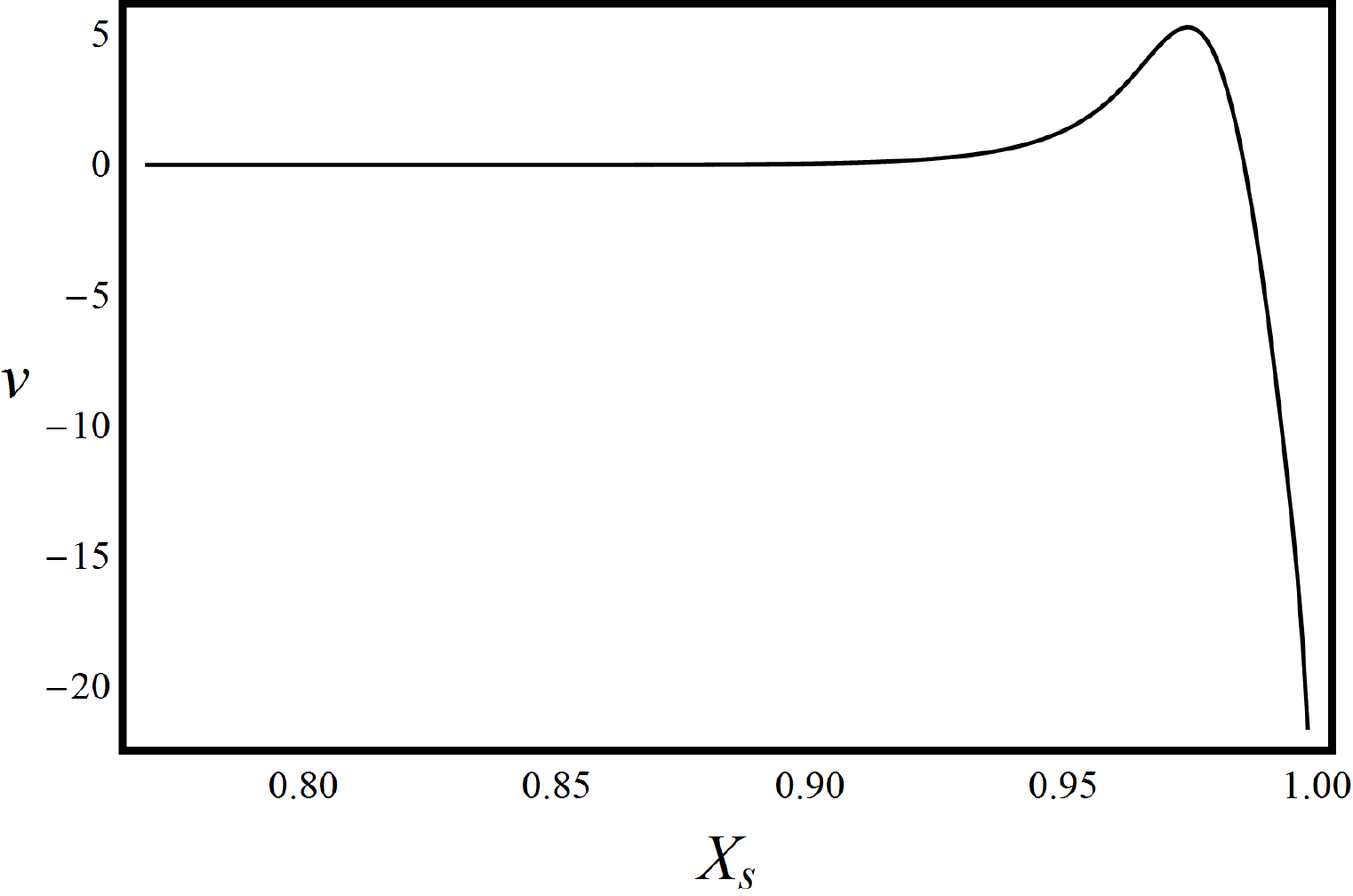} 
   \end{minipage}
 \vspace{2mm}
\centering \begin{minipage}[b!h!]{3cm}
\centering   
   \includegraphics[width=4cm]{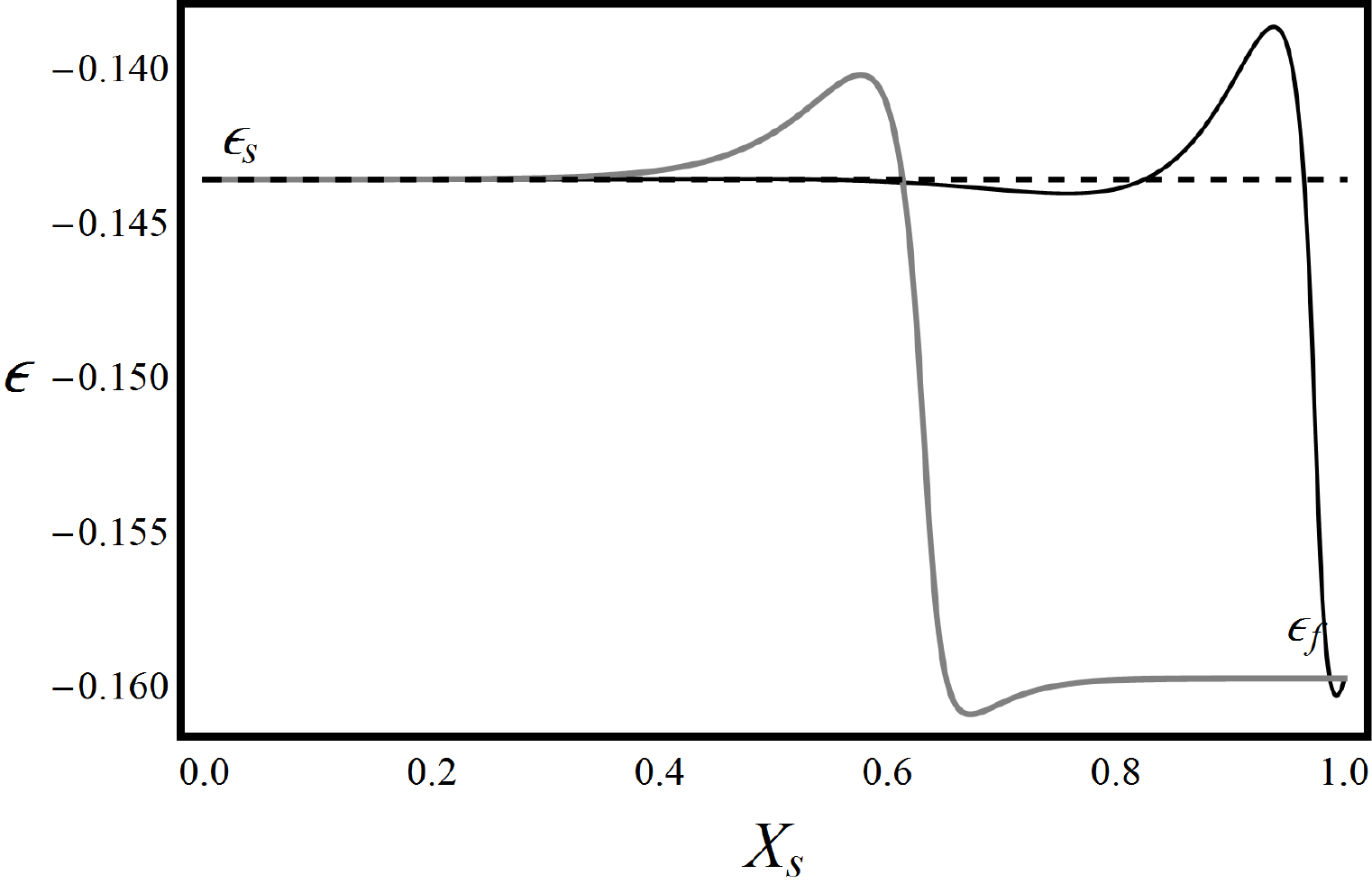}
 \end{minipage}
 \hspace{11mm}  
 \begin{minipage}[b!h!]{3cm}
  \includegraphics[width=4cm]{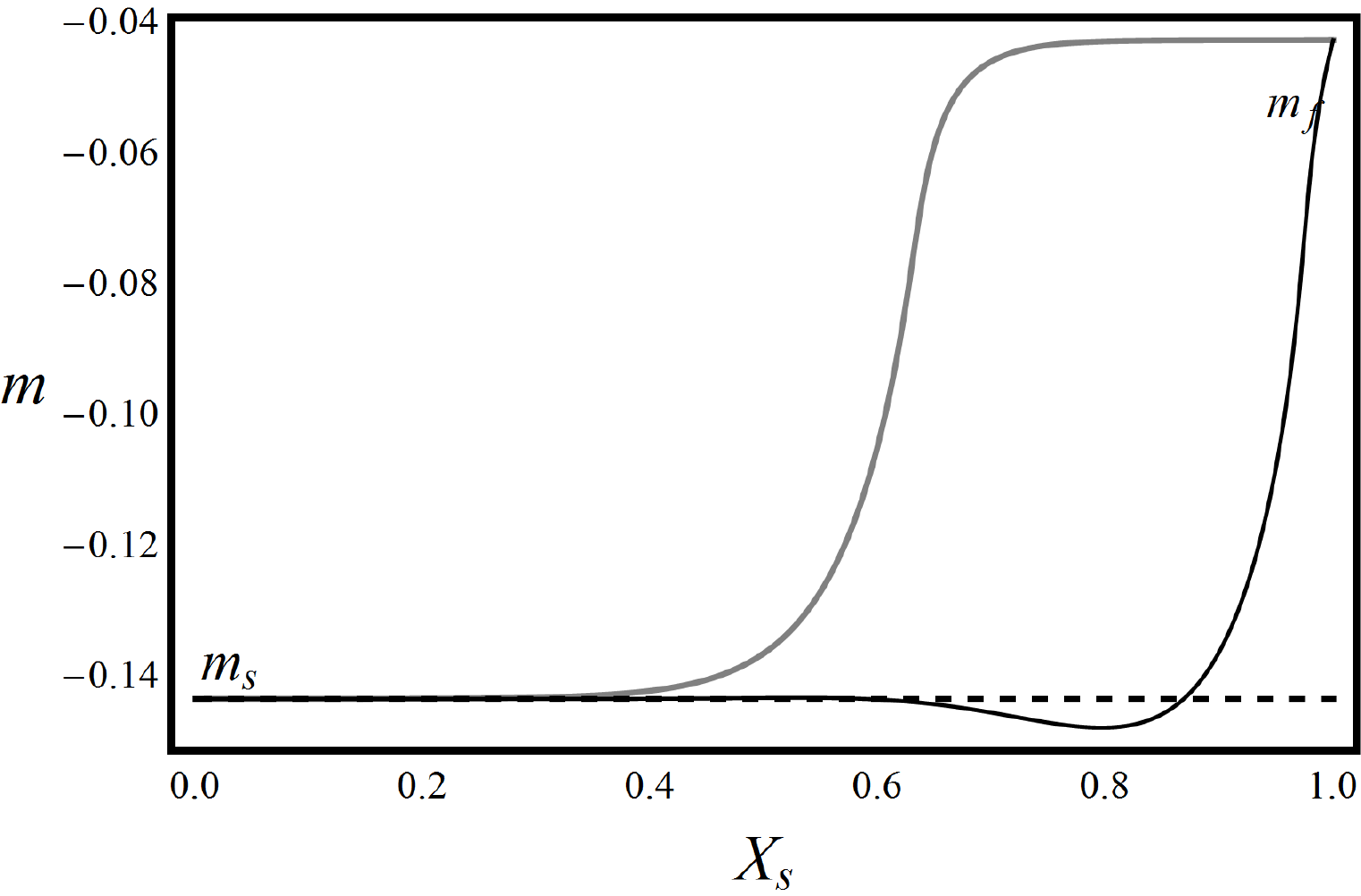}
   \end{minipage}
  \hspace{11mm} 
\begin{minipage}[b!h!]{3cm}
  \centering
   \includegraphics[width=4cm]{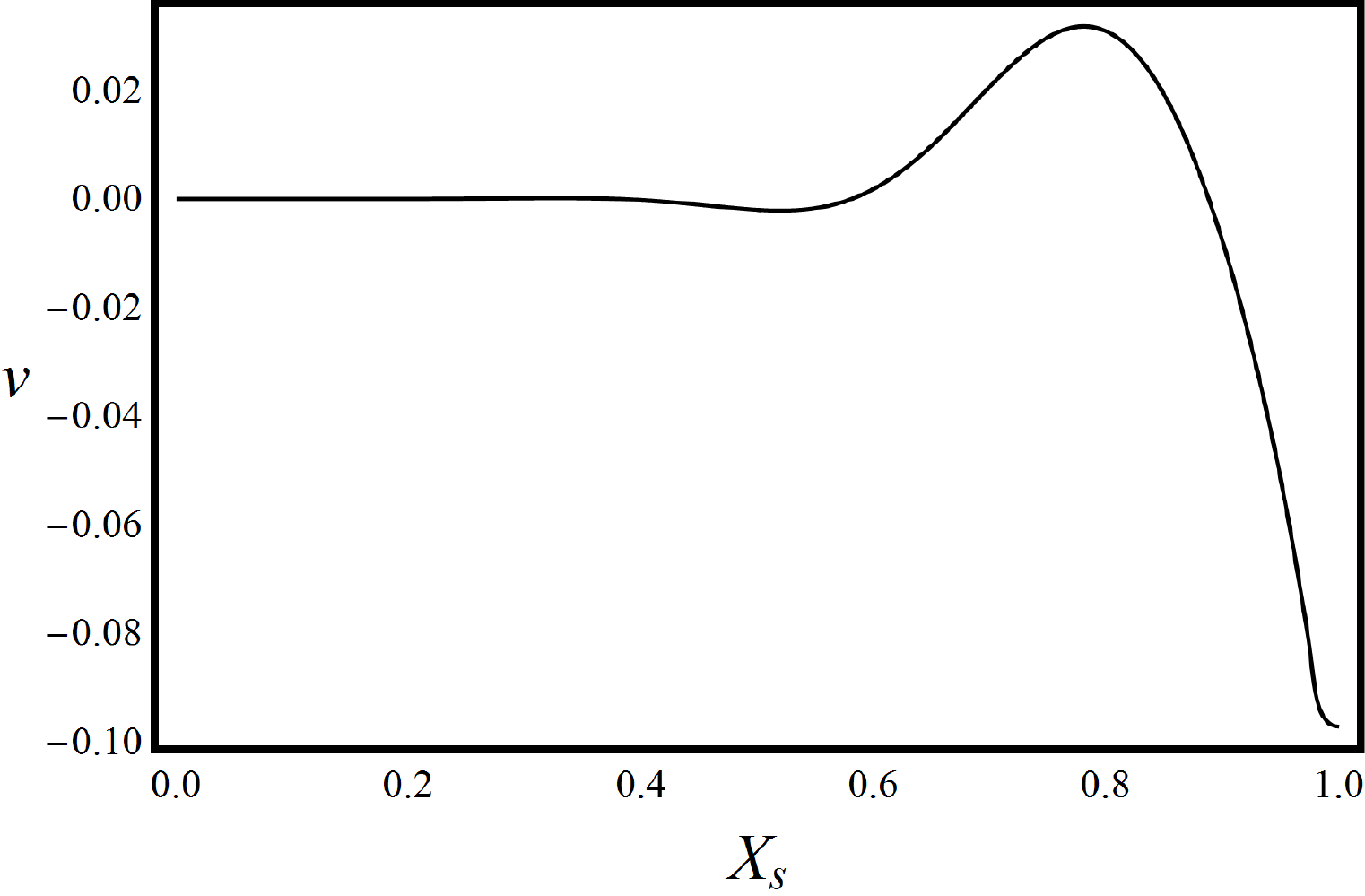}
 \end{minipage}
 \vspace{2mm}
\centering \begin{minipage}[b!h!]{3cm}
\centering   
   \includegraphics[width=4cm]{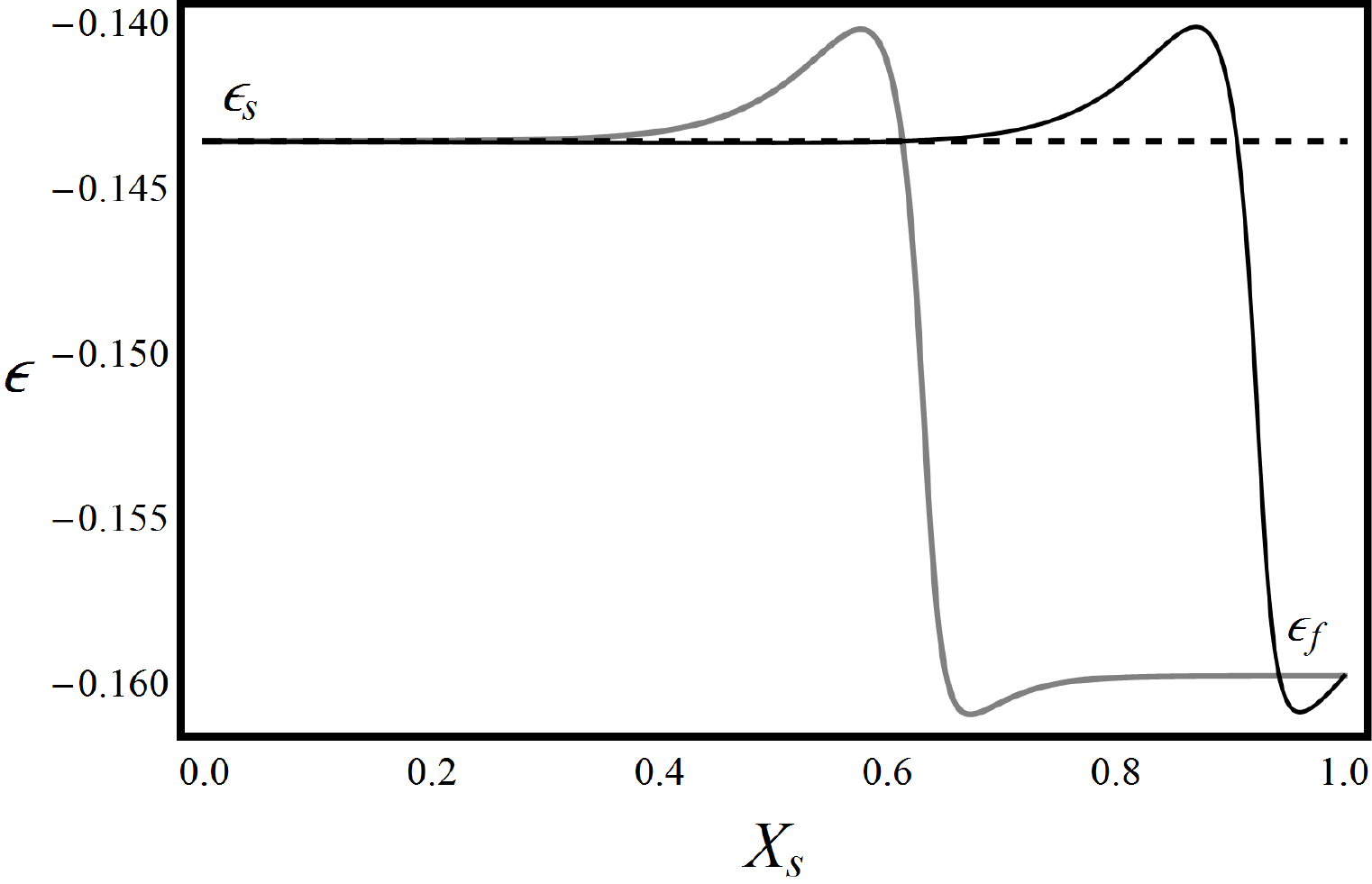}
 \end{minipage}
 \hspace{11mm}  
 \begin{minipage}[b!h!]{3cm}
  \includegraphics[width=4cm]{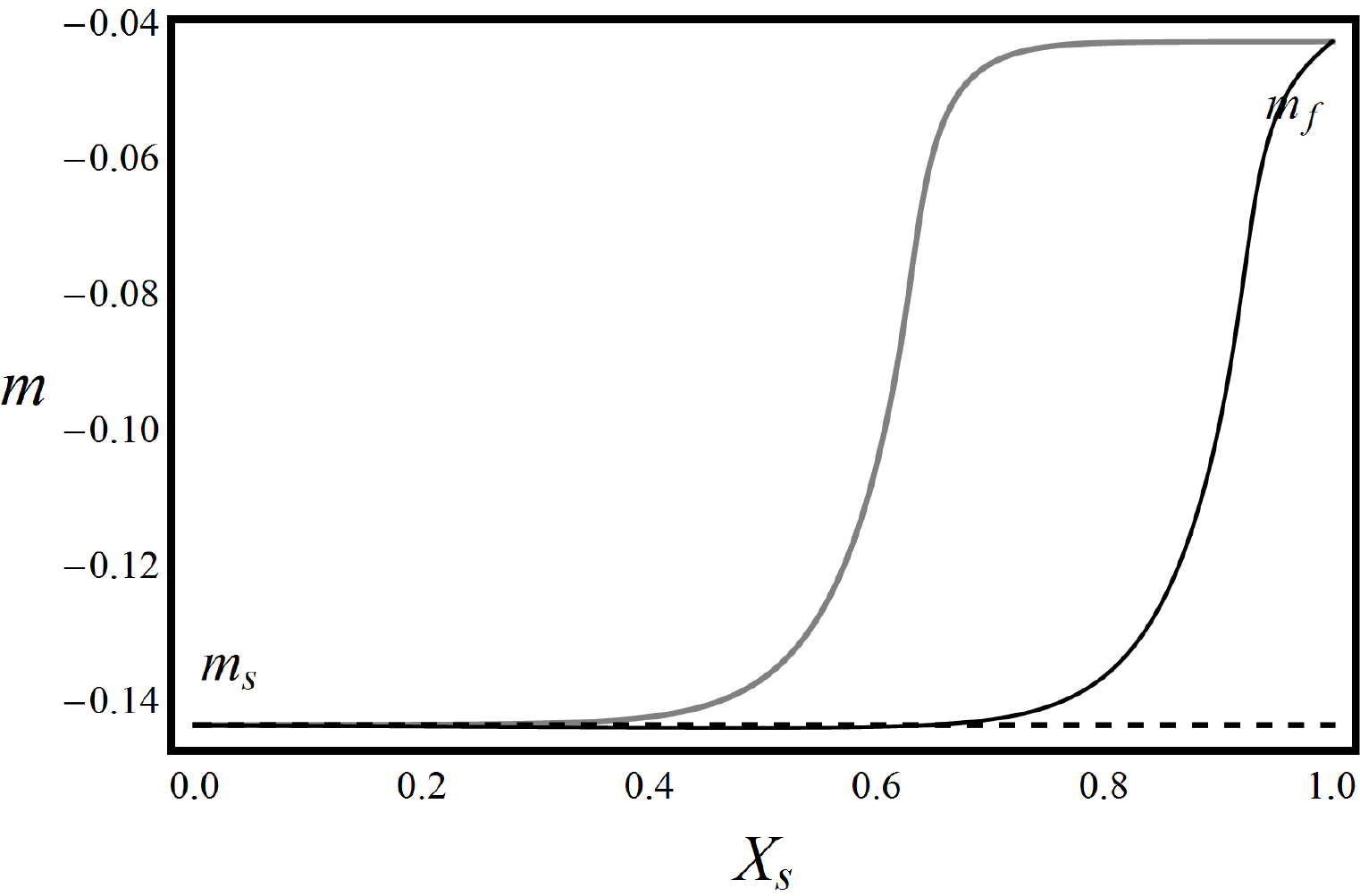}
   \end{minipage}
  \hspace{11mm} 
\begin{minipage}[b!h!]{3cm}
  \centering
   \includegraphics[width=4cm]{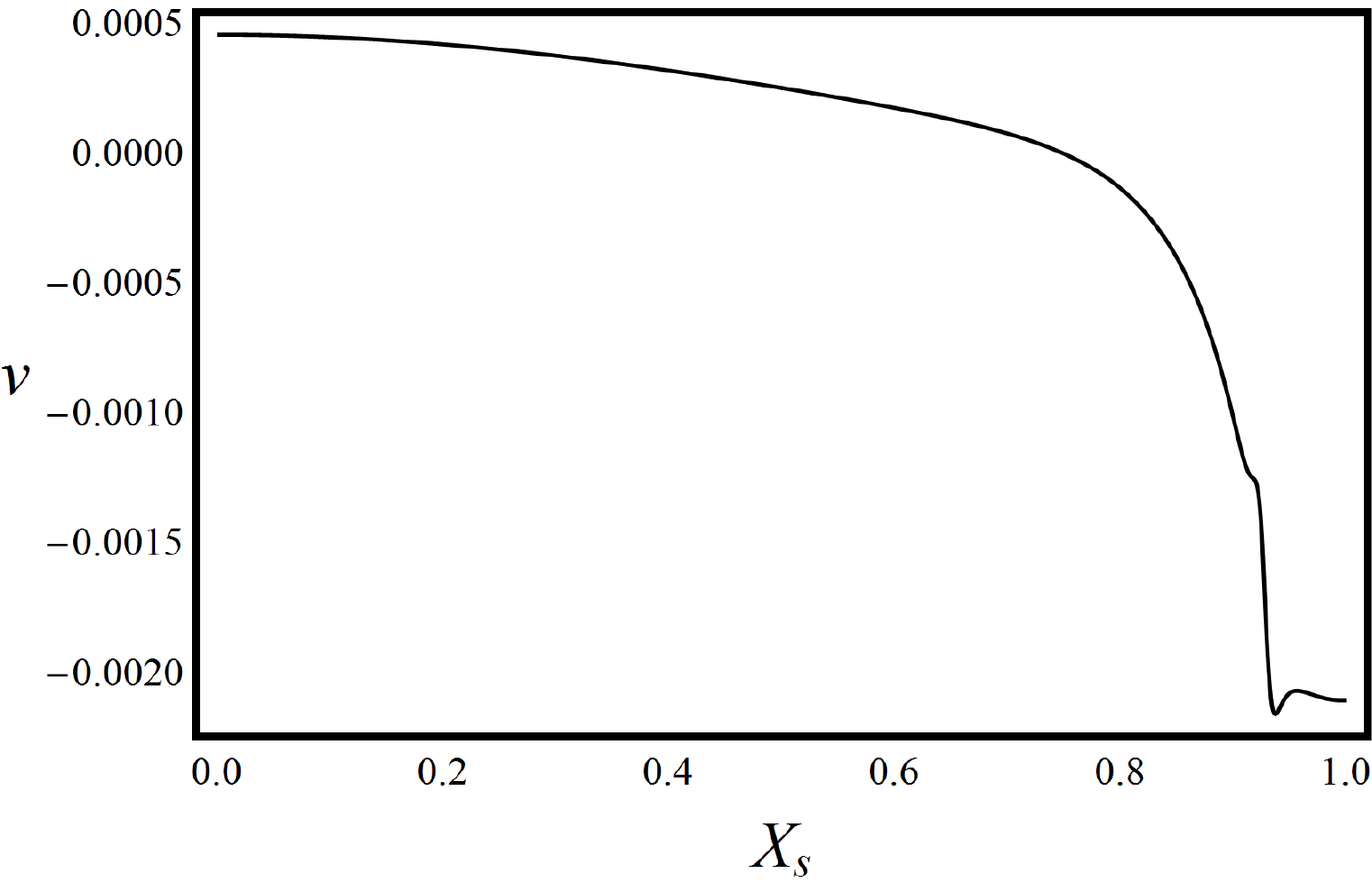}
 \end{minipage}
 \vspace{2mm}
\centering 
\begin{minipage}[b!h!]{3cm}
\centering   
   \includegraphics[width=4cm]{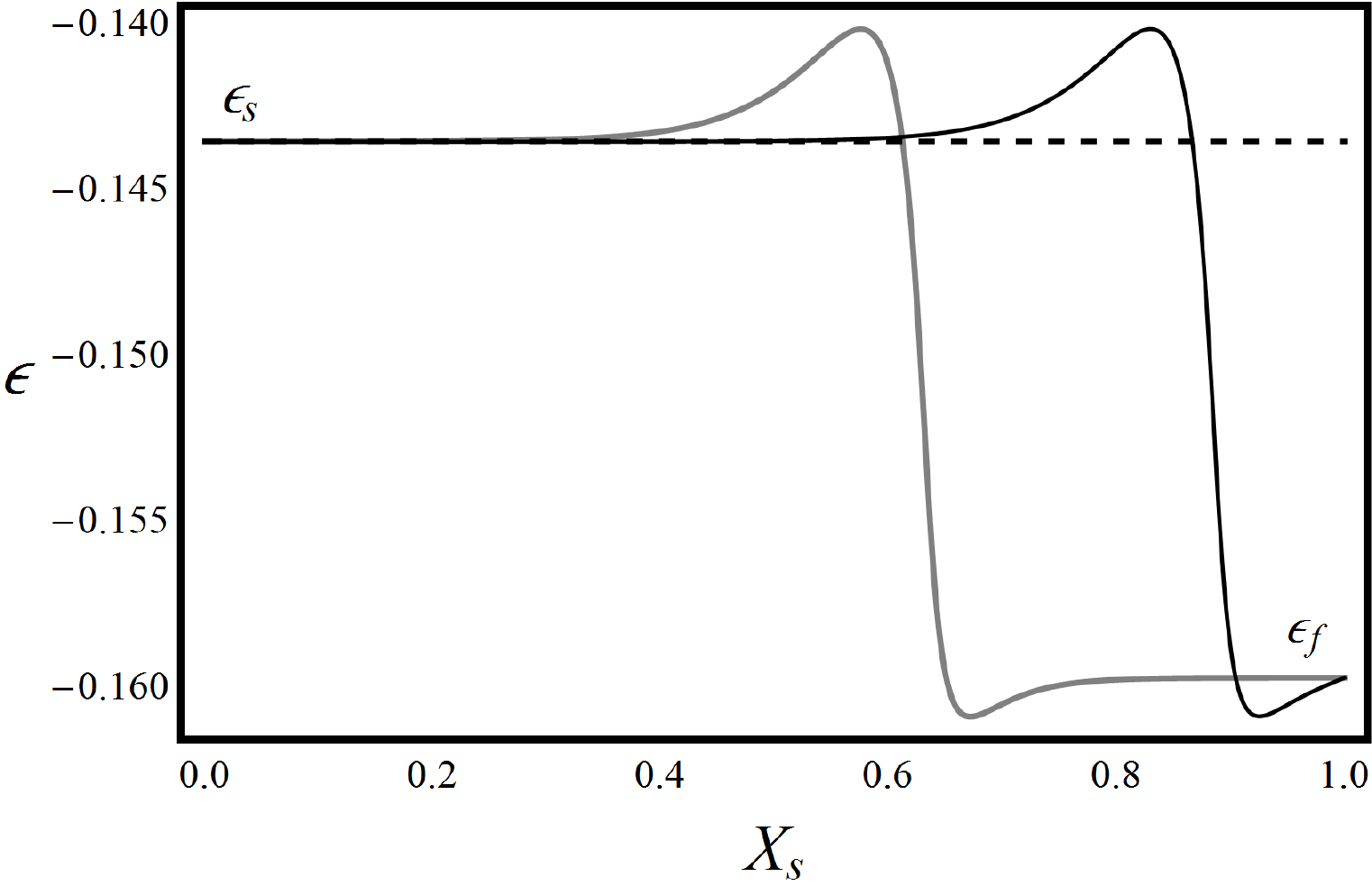}
 \end{minipage}
 \hspace{11mm}  
 \begin{minipage}[b!h!]{3cm}
  \includegraphics[width=4cm]{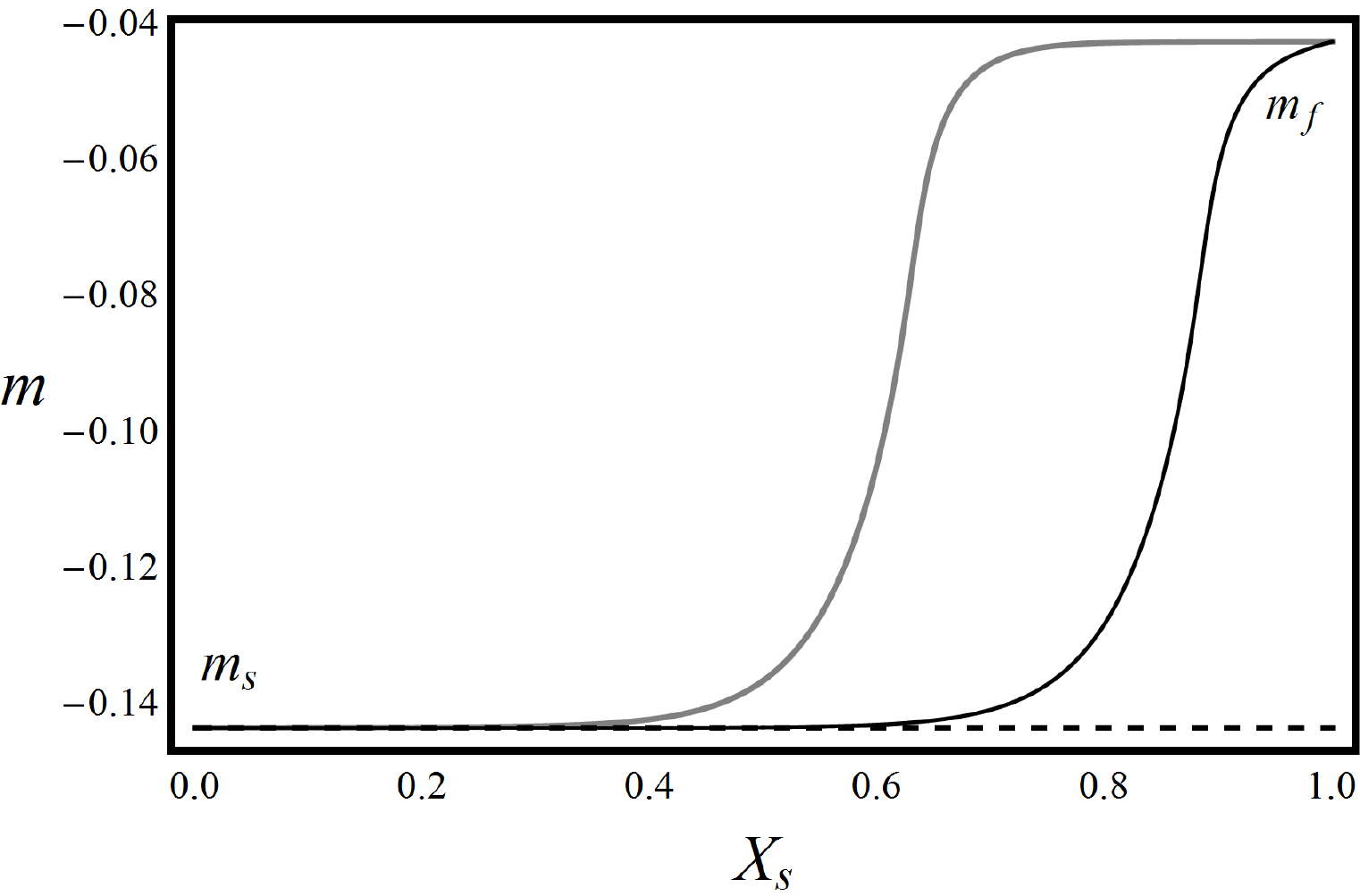}
   \end{minipage}
  \hspace{11mm} 
\begin{minipage}[b!h!]{3cm}
  \centering
   \includegraphics[width=4cm]{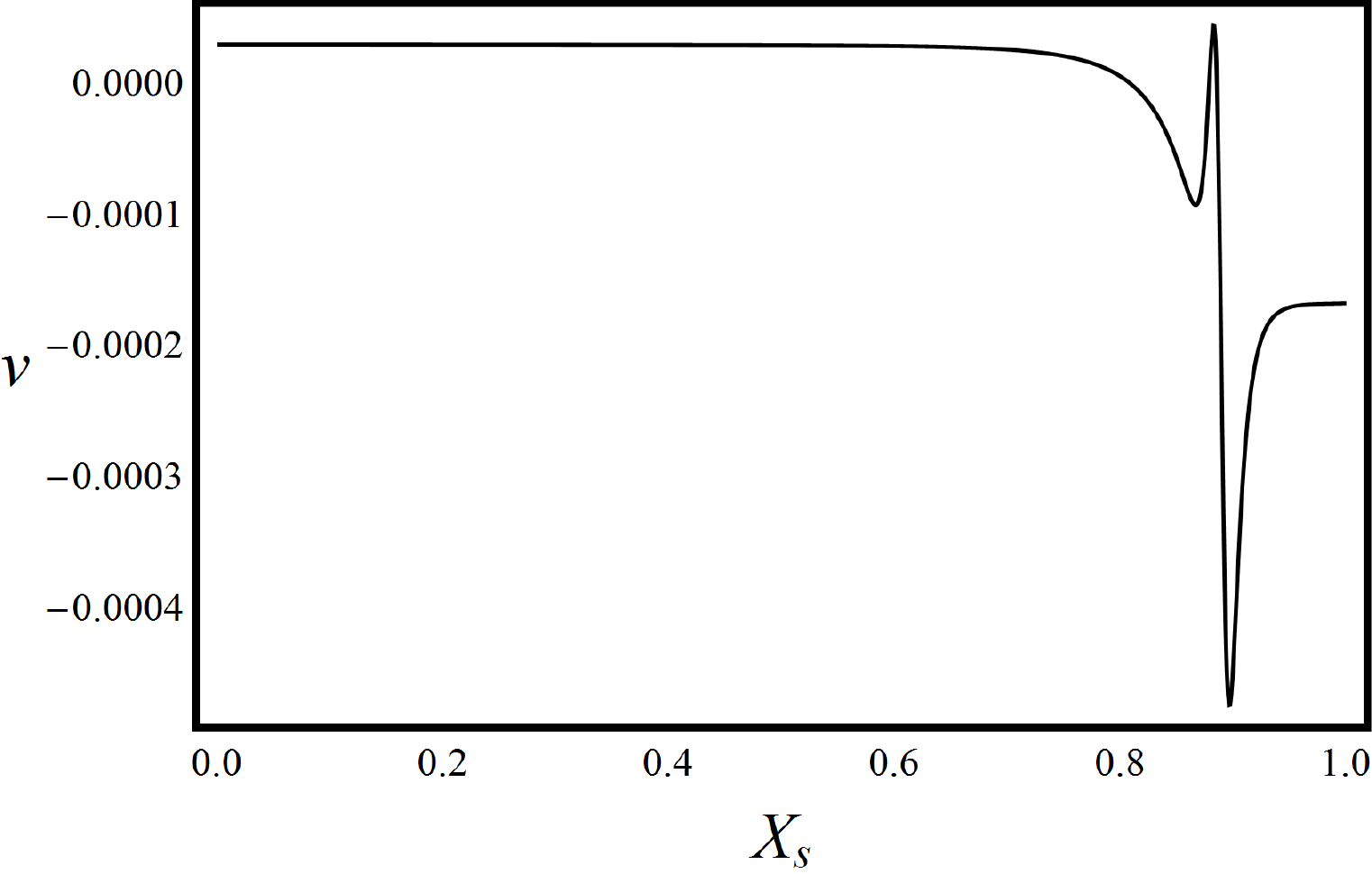}
 \end{minipage}
\vspace{2mm}
\centering 
\begin{minipage}[b!h!]{3cm}
\centering   
   \includegraphics[width=4cm]{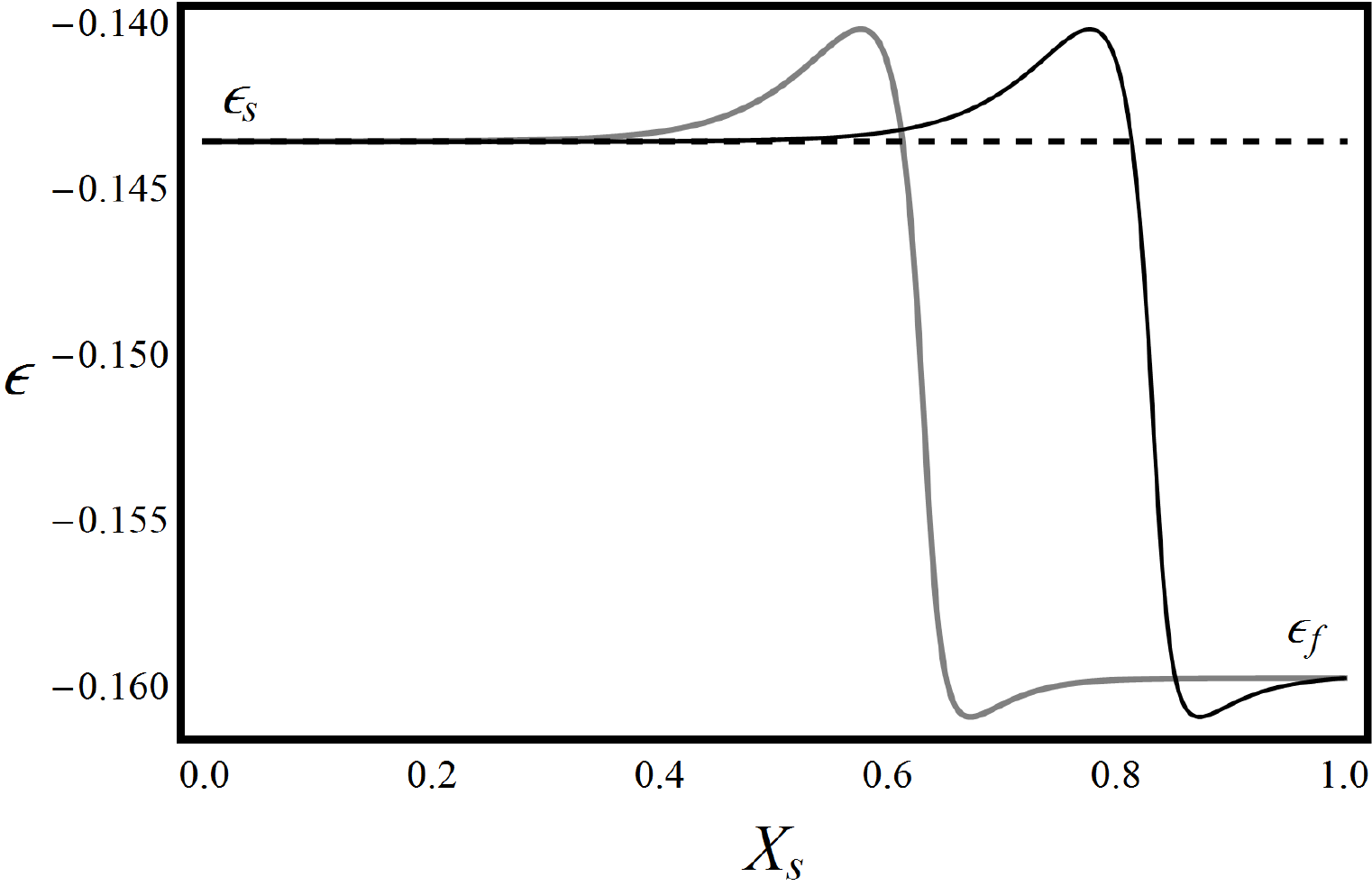}
 \end{minipage}
 \hspace{11mm}  
 \begin{minipage}[b!h!]{3cm}
  \includegraphics[width=4cm]{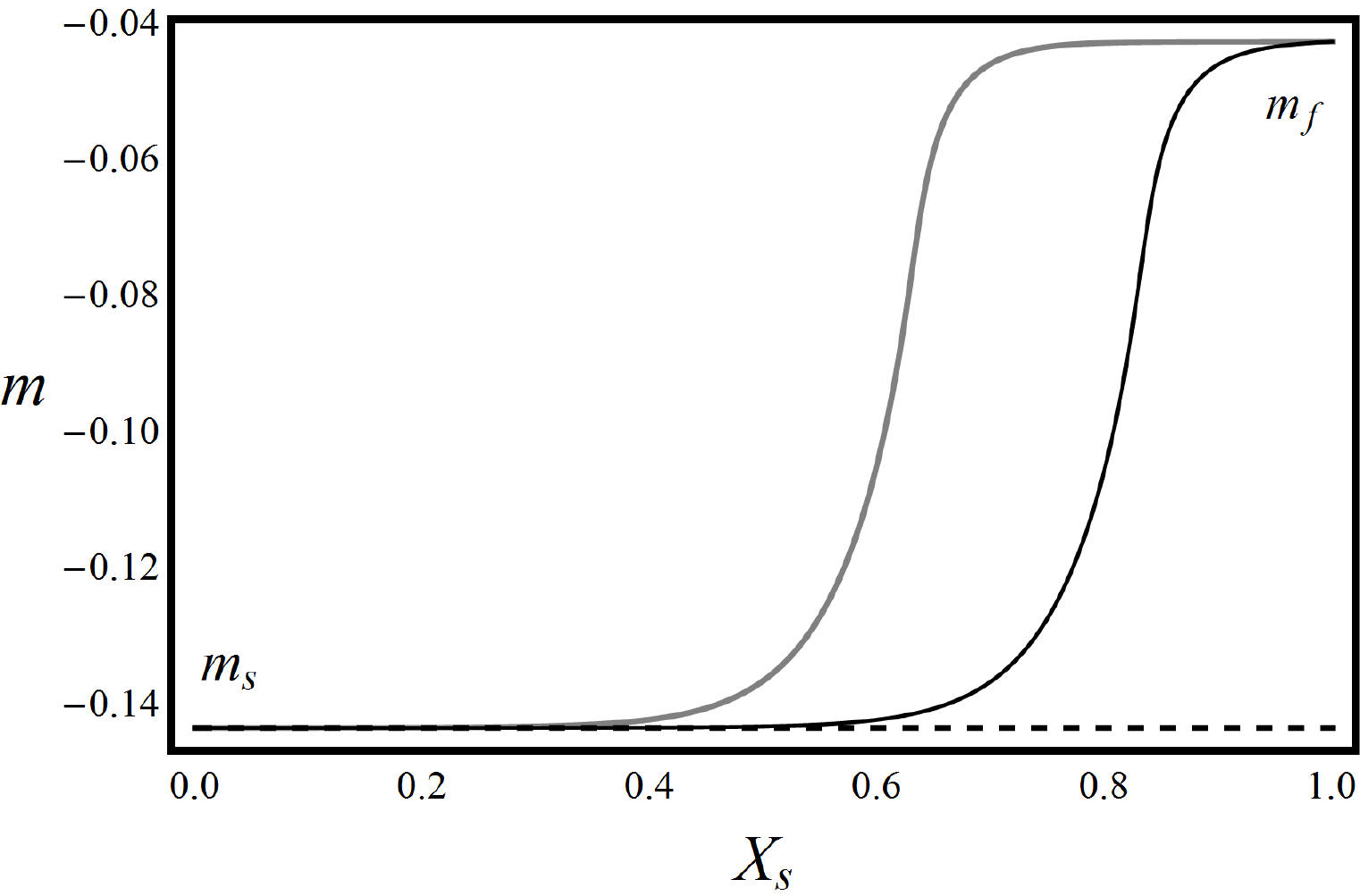}
   \end{minipage}
  \hspace{11mm} 
\begin{minipage}[b!h!]{3cm}
  \centering
   \includegraphics[width=4cm]{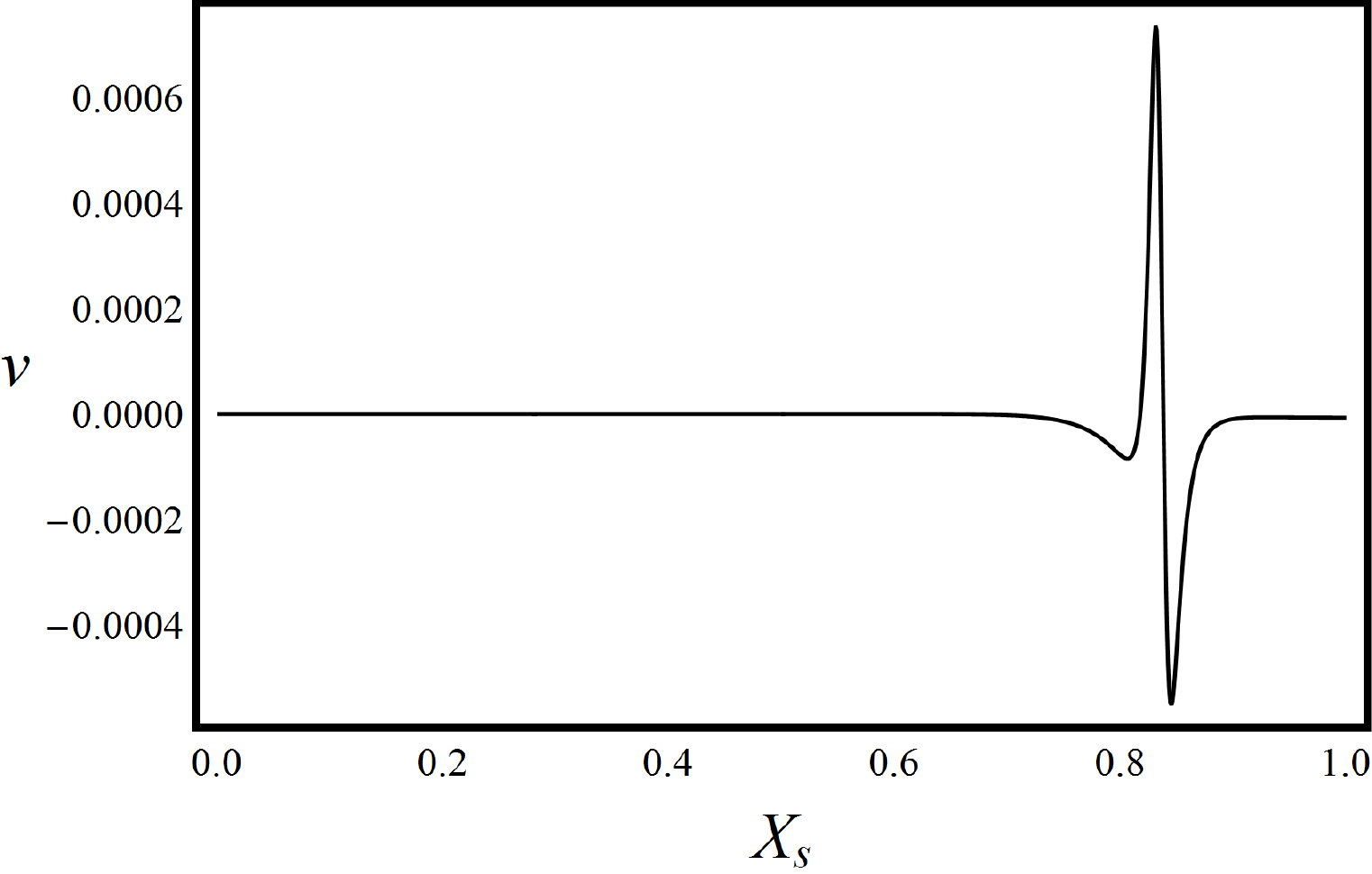}
 \end{minipage}
\caption{Profiles (black lines) $\varepsilon(X_s,t)$, $m(X_s)$ and $v(X_s)$ for the zero chemical potential problem obtained by solving the non-stationary system with a standard phase initial state (dashed lines). We used Dirichelet boundary conditions $m(0)=m_s,\,\varepsilon(0)=\varepsilon_s,\,m(1)=m_f$,  $\varepsilon(1)=\varepsilon_f$ on the finite interval $[0,1]$, at the coexistence pressure for $a=0.5,\,b=1,\,\alpha=100,\,k_1=k_2=k_3=10^{-3}$. Profiles at times $t=2\times 10^{-7},\,t=7.5\times 10^{-6},\,t = 0.0100105,\,t=0.7930105,\,t = 11.10100105,\,t = 1367.10100105$ in lexicographic order. The gray lines represent stationary profiles.}
 \label{vecchiostandard}
\end{figure}

\begin{figure}[t]
\centering
 \begin{minipage}[b!h!]{3cm}
\centering   
   \includegraphics[width=4cm]{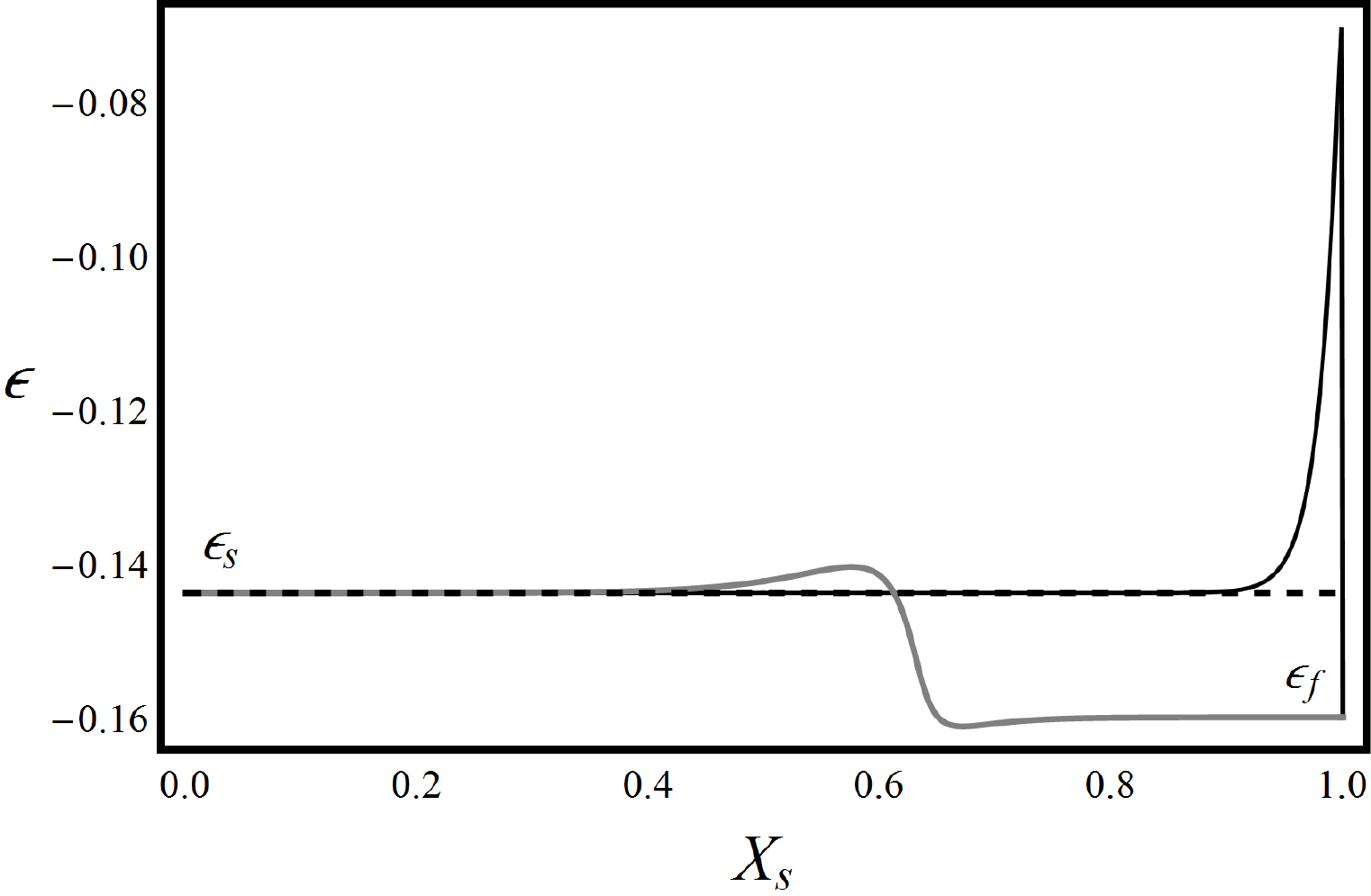}
 \end{minipage}
 \hspace{11mm}  
 \begin{minipage}[b!h!]{3cm}
  \includegraphics[width=4cm]{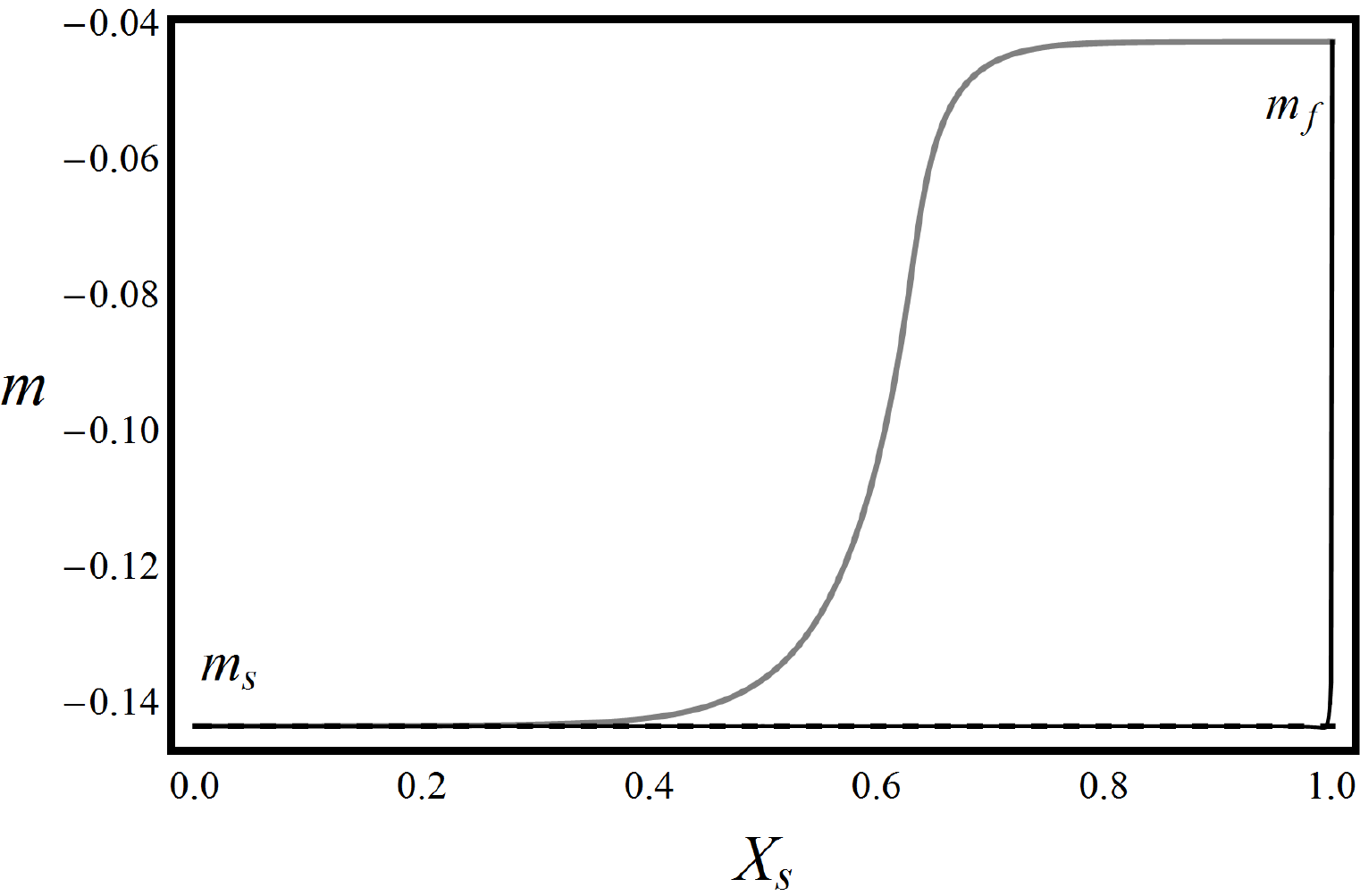}
   \end{minipage}
  \hspace{11mm} 
\begin{minipage}[b!h!]{3cm}
  \centering
   \includegraphics[width=4cm]{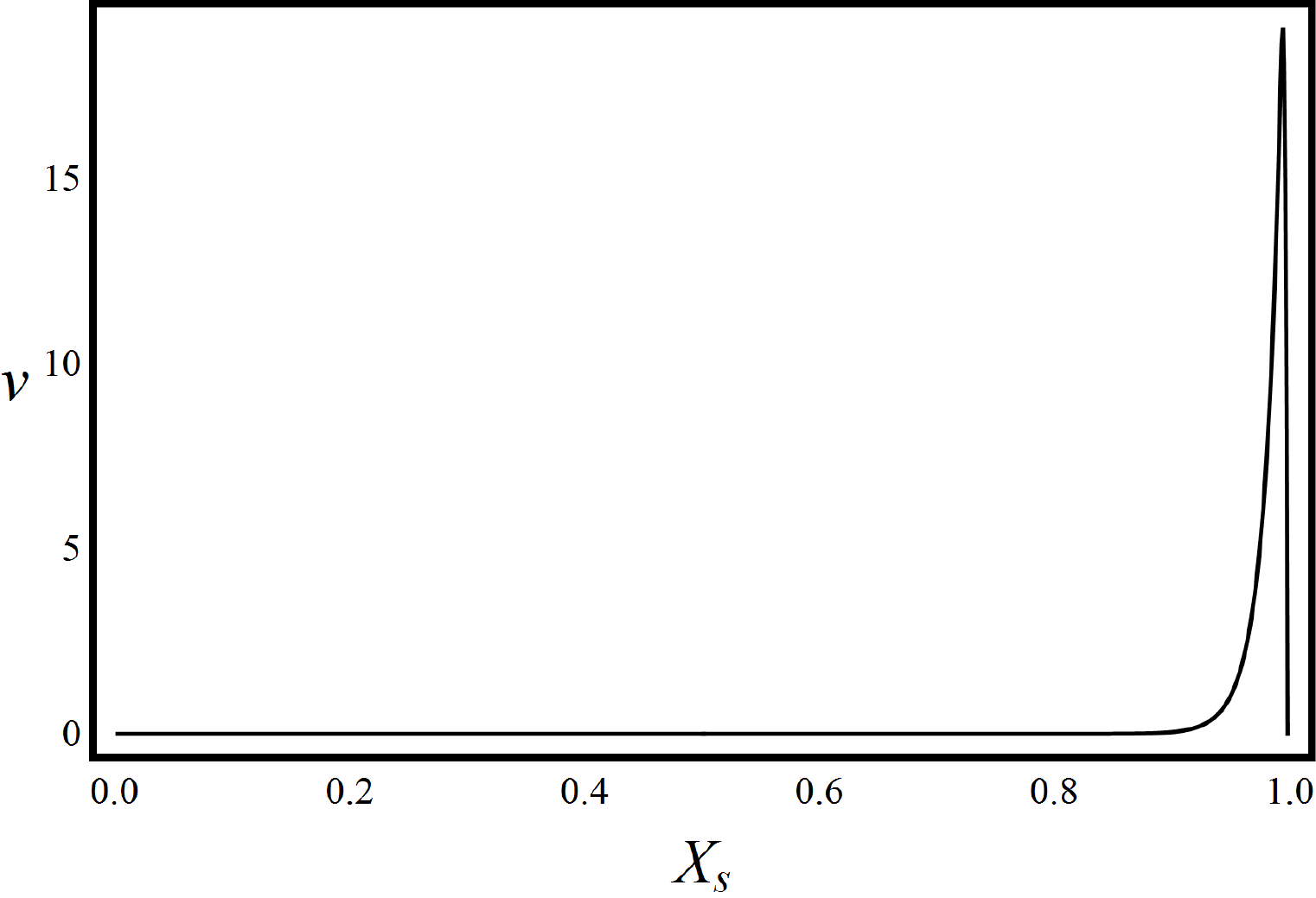}
 \end{minipage}
 \vspace{2mm}
\centering
 \centering \begin{minipage}[b!h!]{3cm}
\centering   
   \includegraphics[width=4cm]{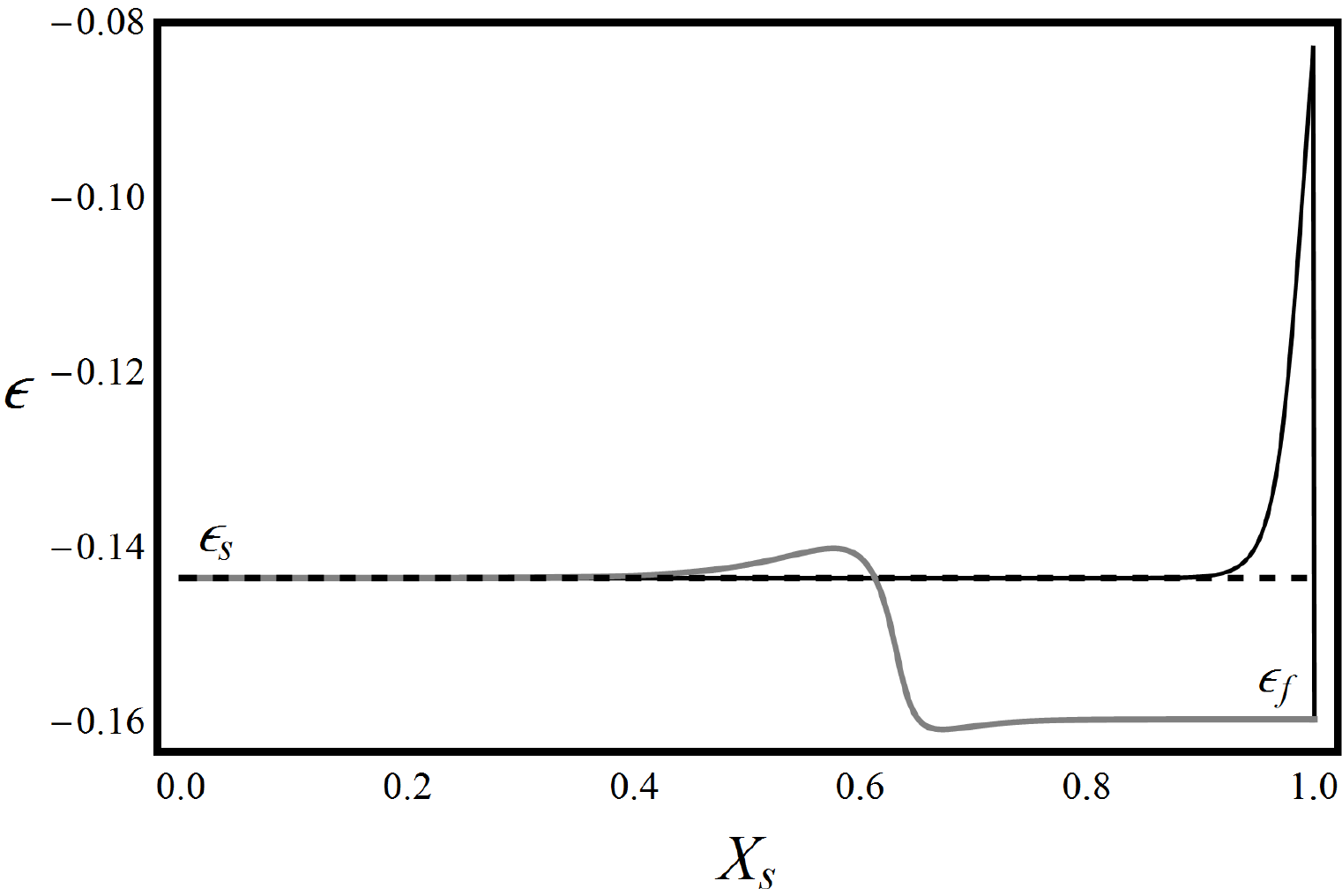}
 \end{minipage}
 \hspace{11mm}  
 \begin{minipage}[b!h!]{3cm}
  \includegraphics[width=4cm]{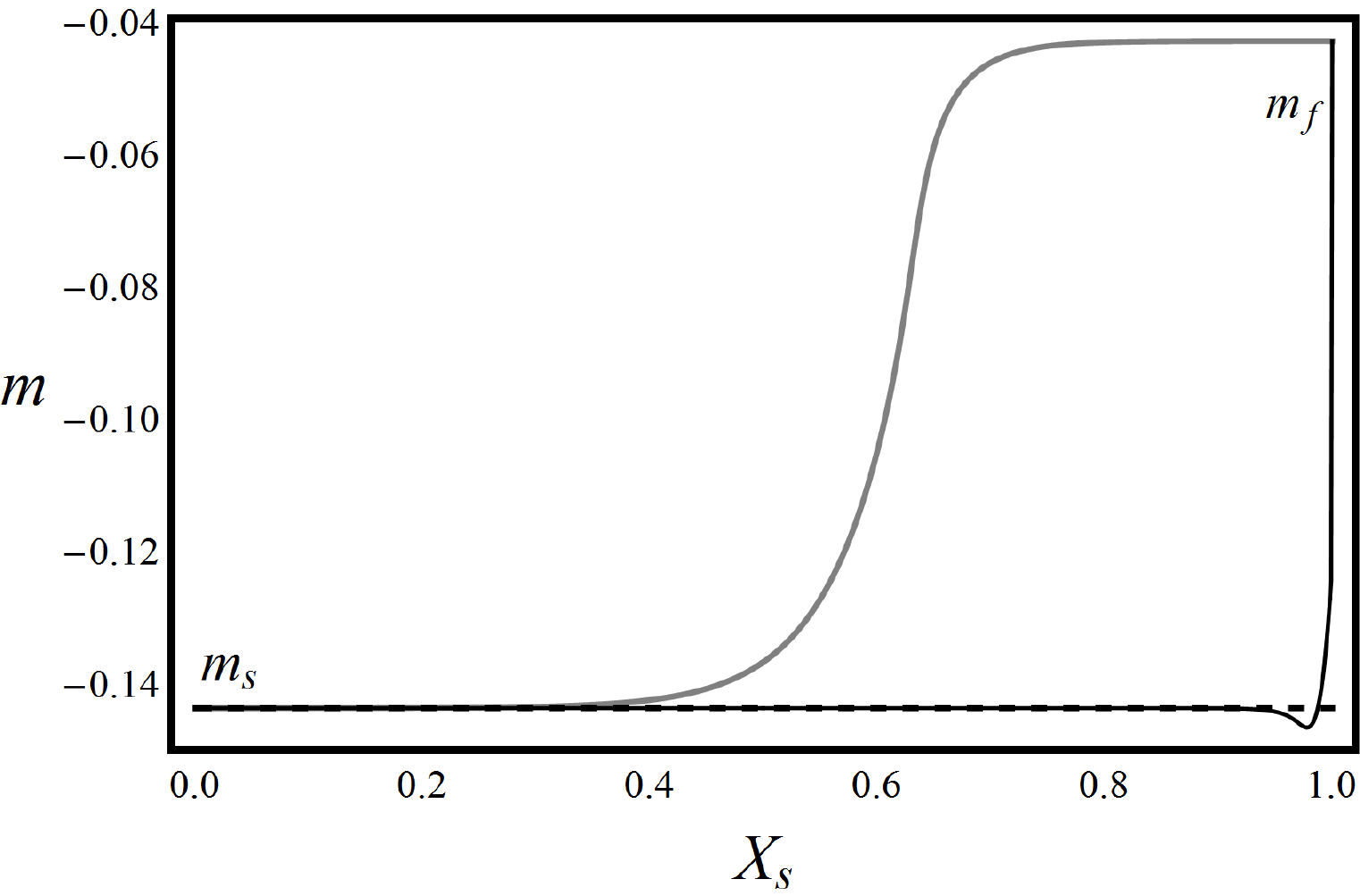}
   \end{minipage}
  \hspace{11mm} 
\begin{minipage}[b!h!]{3cm}
  \centering
   \includegraphics[width=4cm]{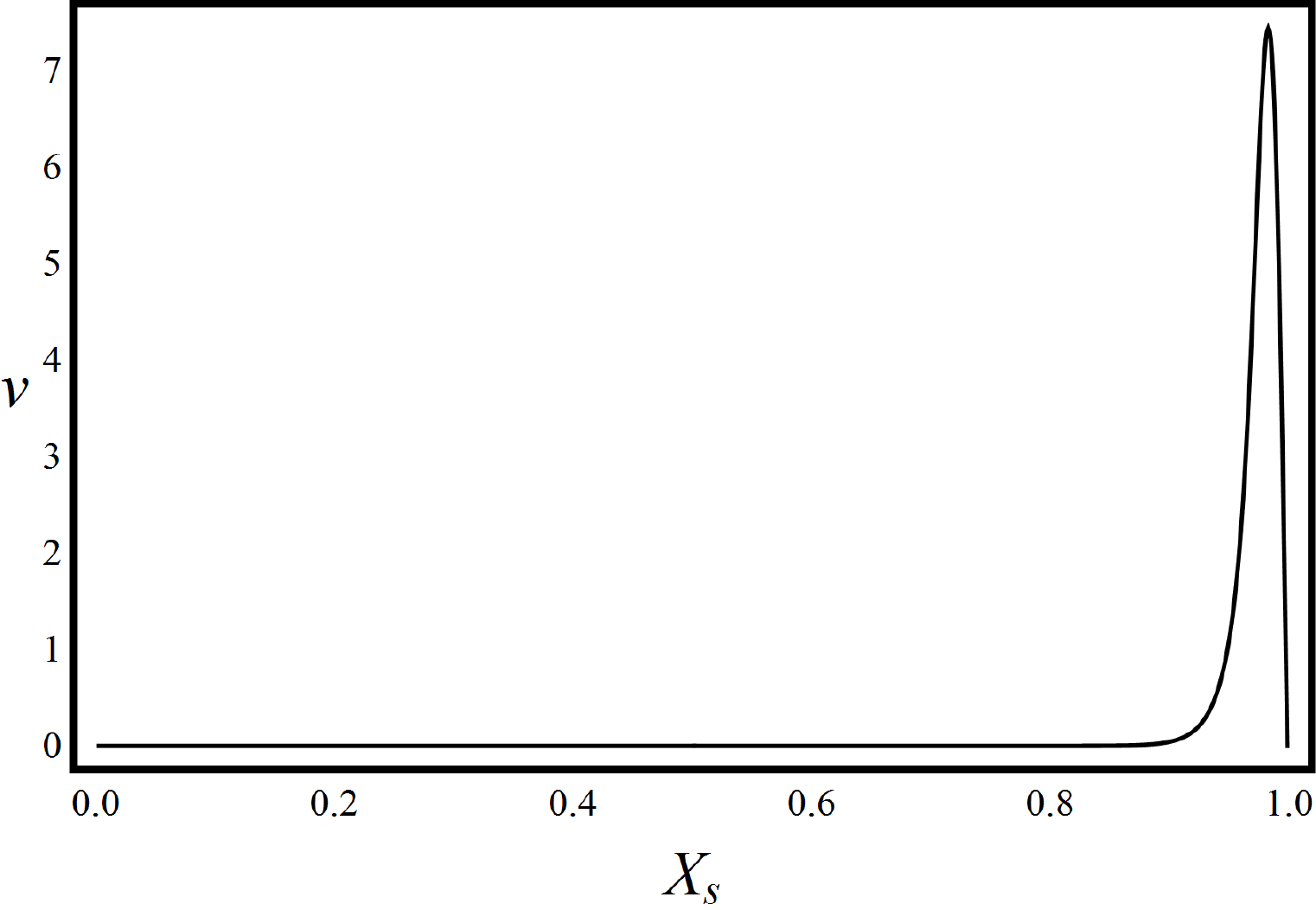} 
   \end{minipage}
 \vspace{2mm}
\centering \begin{minipage}[b!h!]{3cm}
\centering   
   \includegraphics[width=4cm]{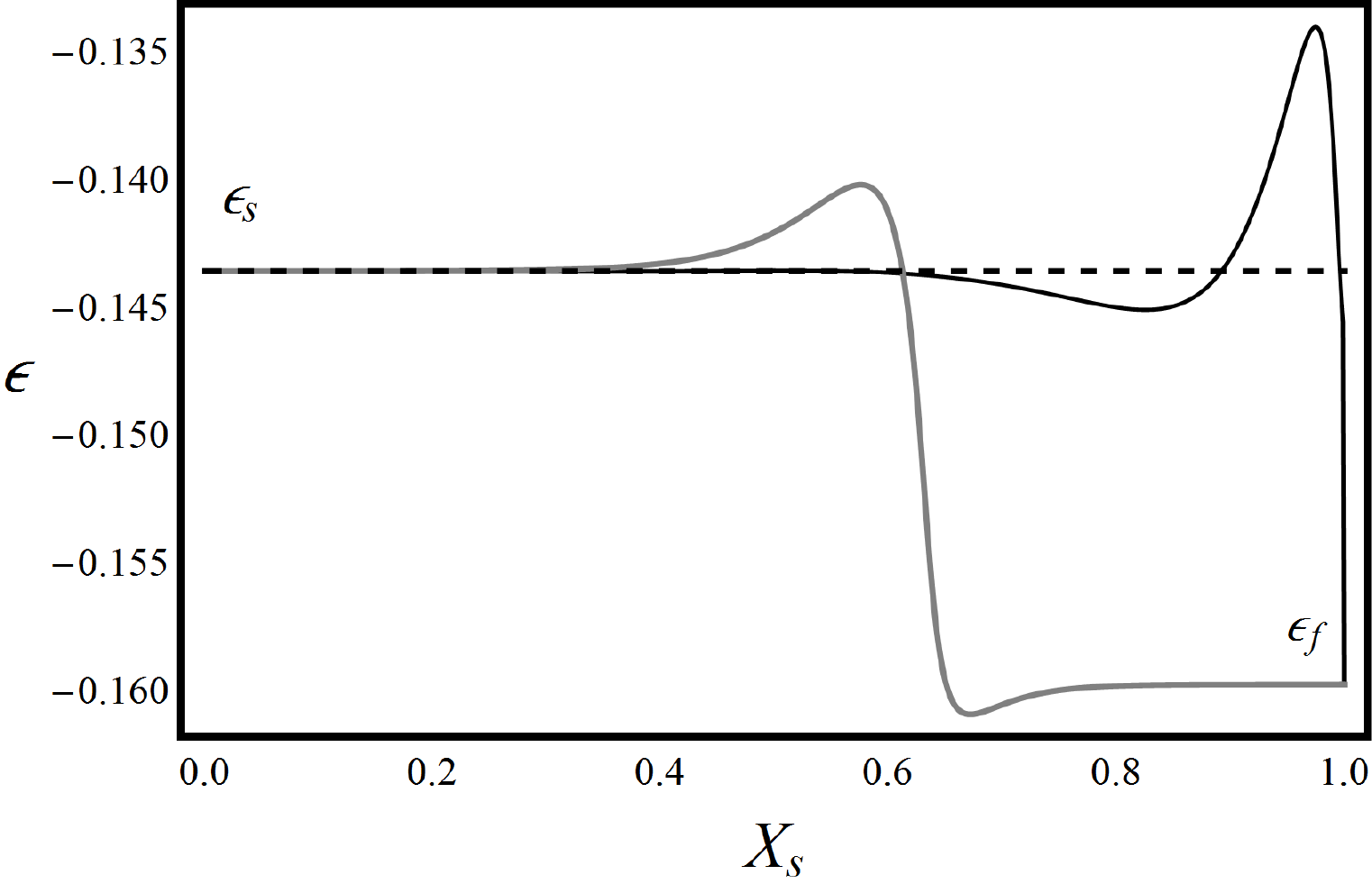}
 \end{minipage}
 \hspace{11mm}  
 \begin{minipage}[b!h!]{3cm}
  \includegraphics[width=4cm]{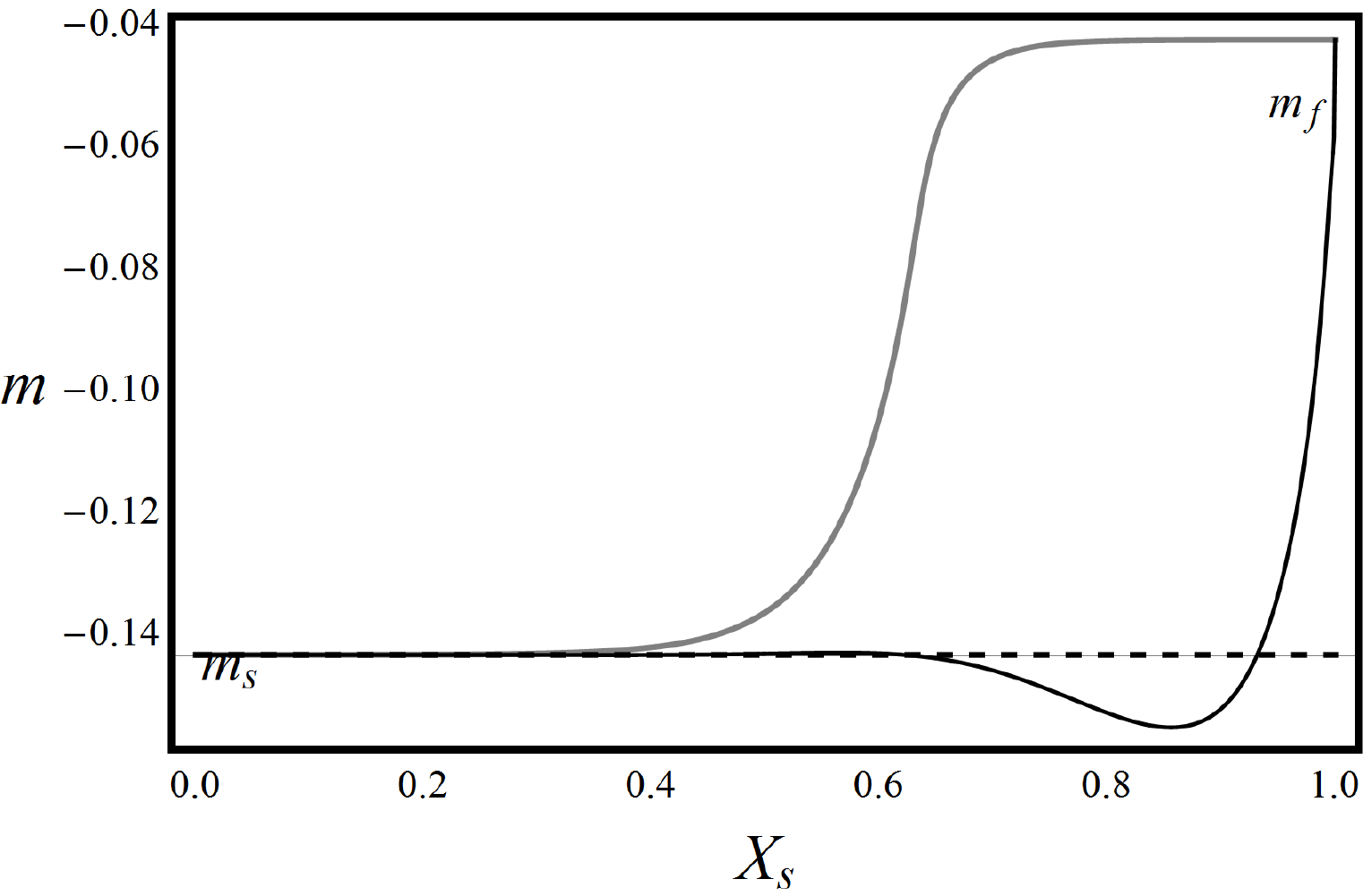}
   \end{minipage}
  \hspace{11mm} 
\begin{minipage}[b!h!]{3cm}
  \centering
   \includegraphics[width=4cm]{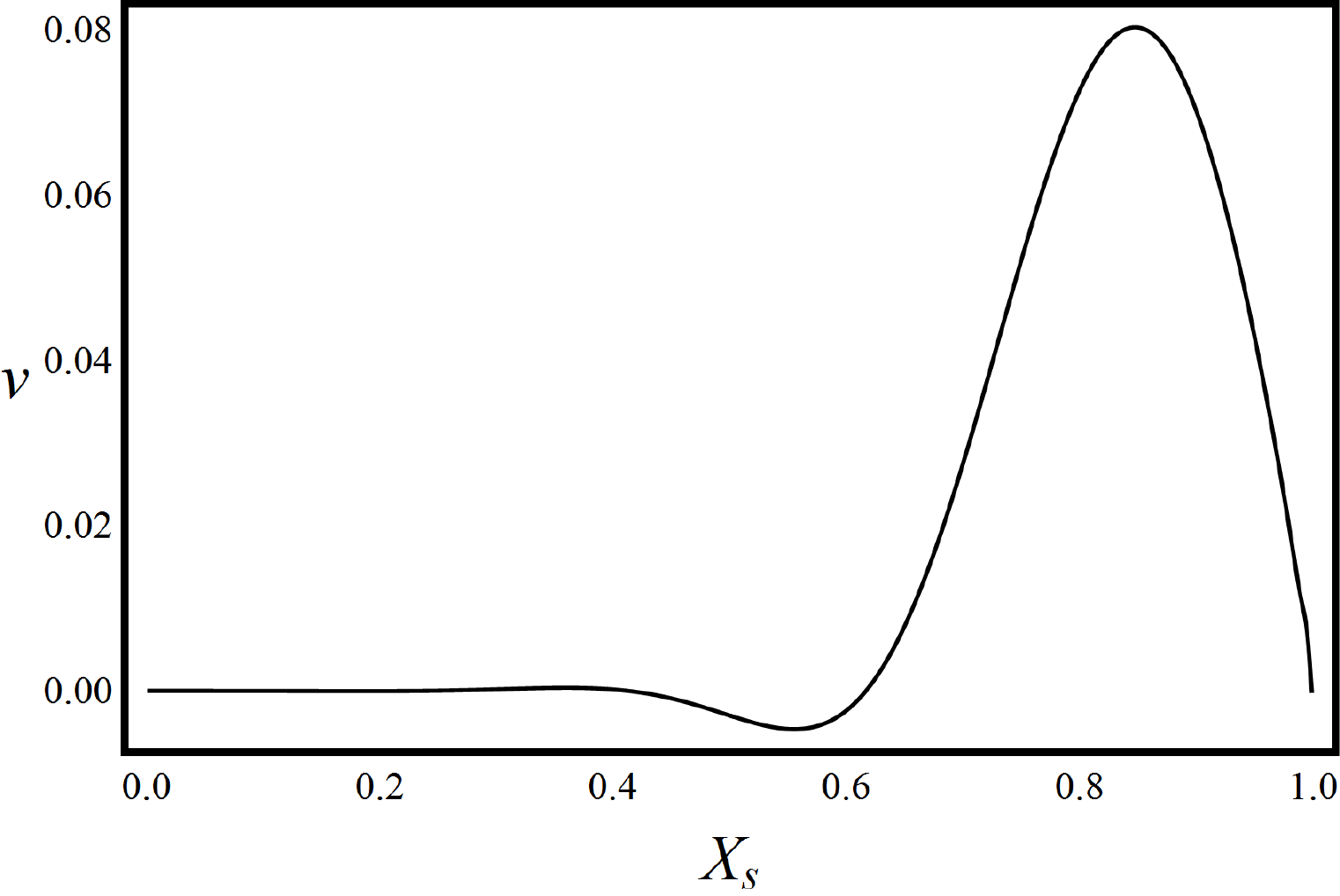}
 \end{minipage}
 \vspace{2mm}
\centering \begin{minipage}[b!h!]{3cm}
\centering   
   \includegraphics[width=4cm]{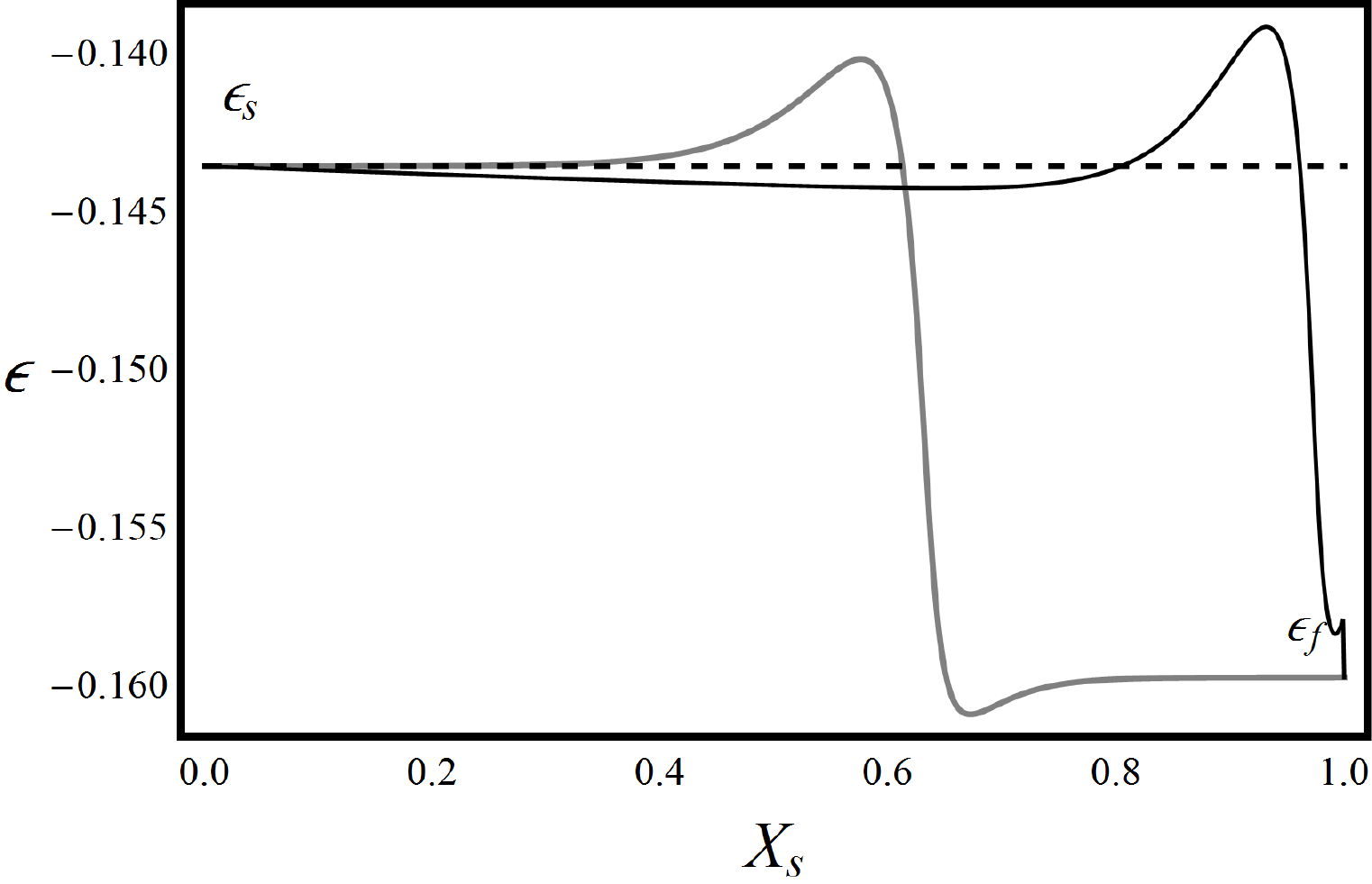}
 \end{minipage}
 \hspace{11mm}  
 \begin{minipage}[b!h!]{3cm}
  \includegraphics[width=4cm]{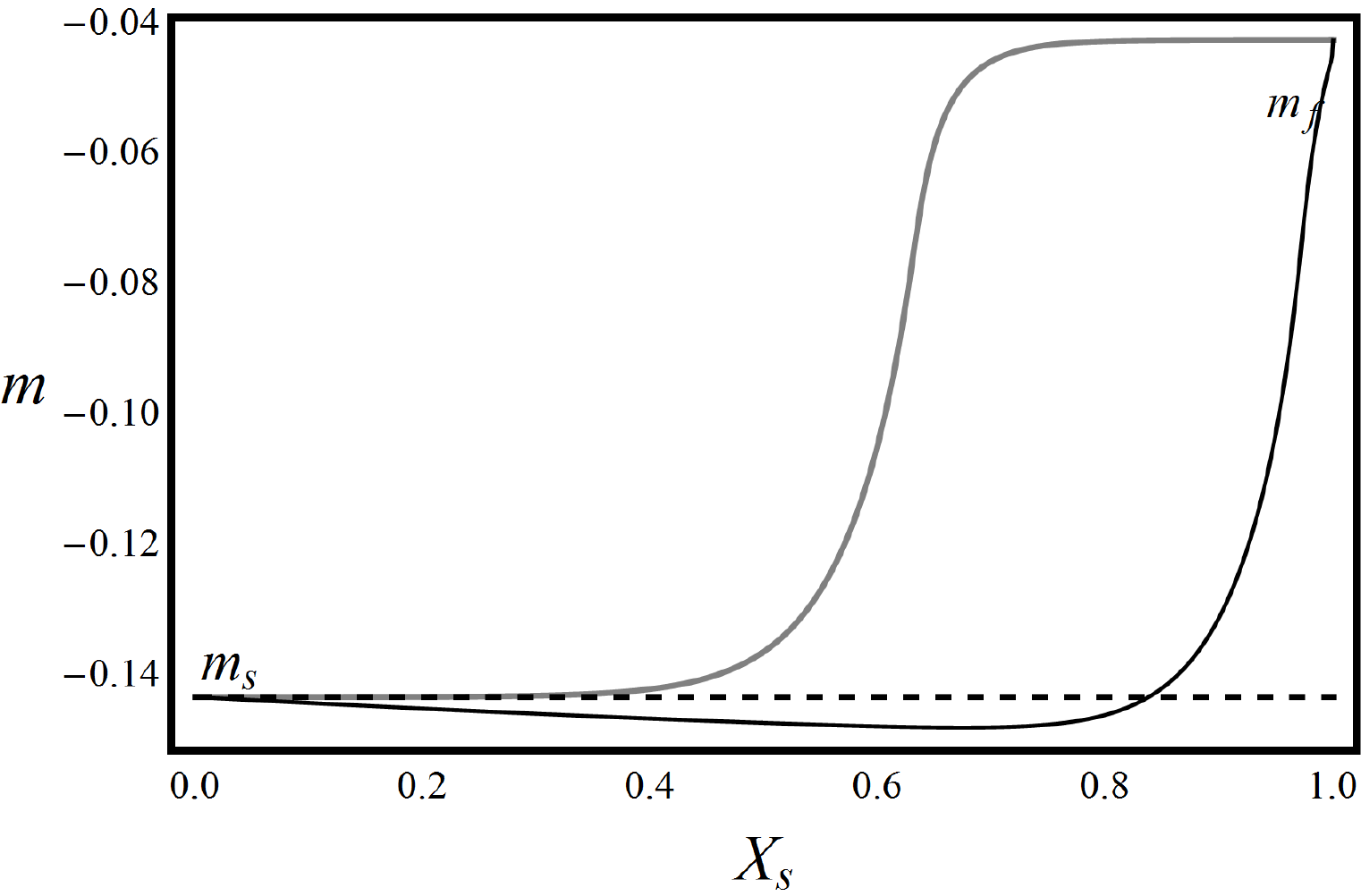}
   \end{minipage}
  \hspace{11mm} 
\begin{minipage}[b!h!]{3cm}
  \centering
   \includegraphics[width=4cm]{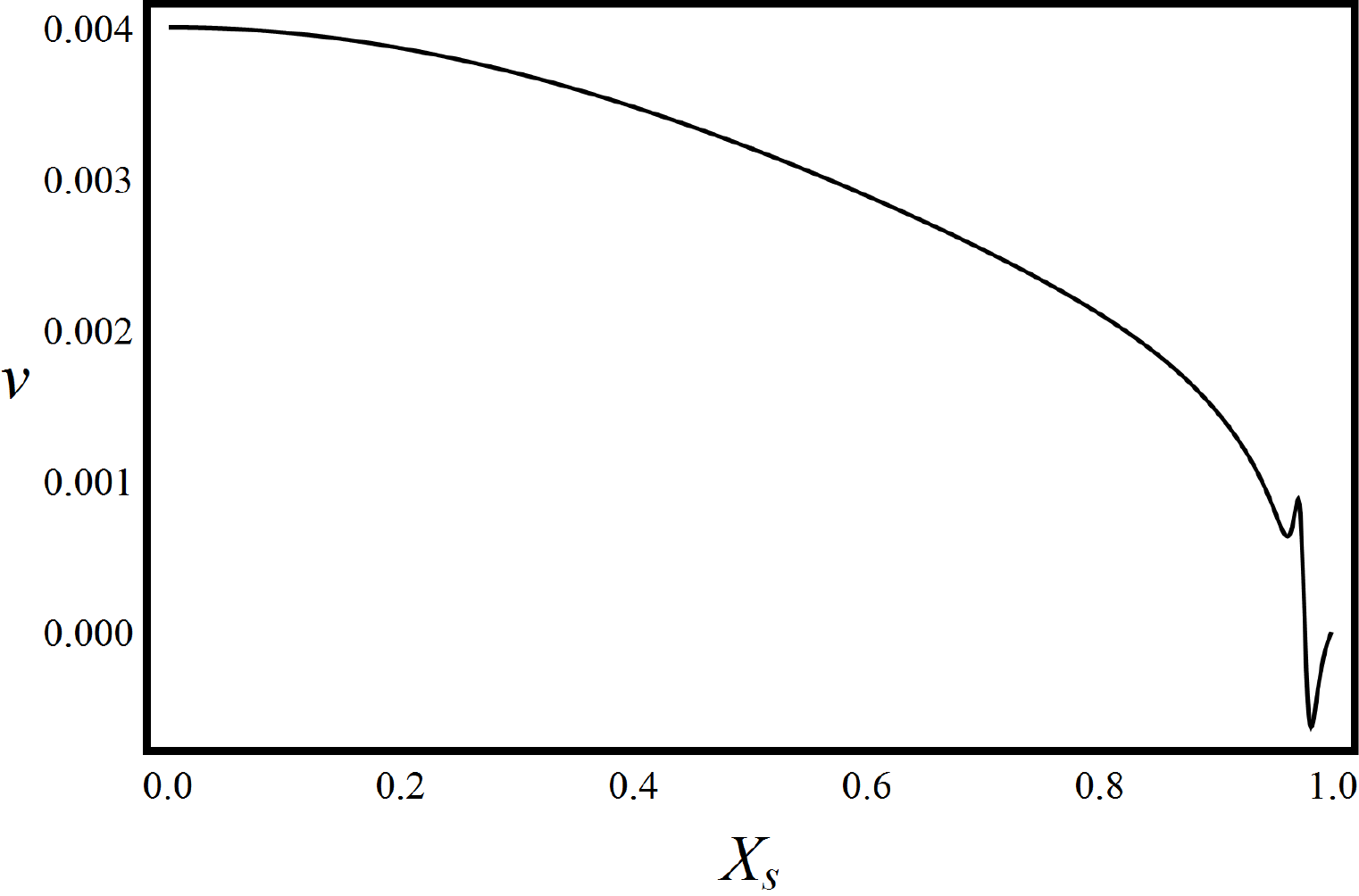}
 \end{minipage}
 \vspace{2mm}
\centering 
\begin{minipage}[b!h!]{3cm}
\centering   
   \includegraphics[width=4cm]{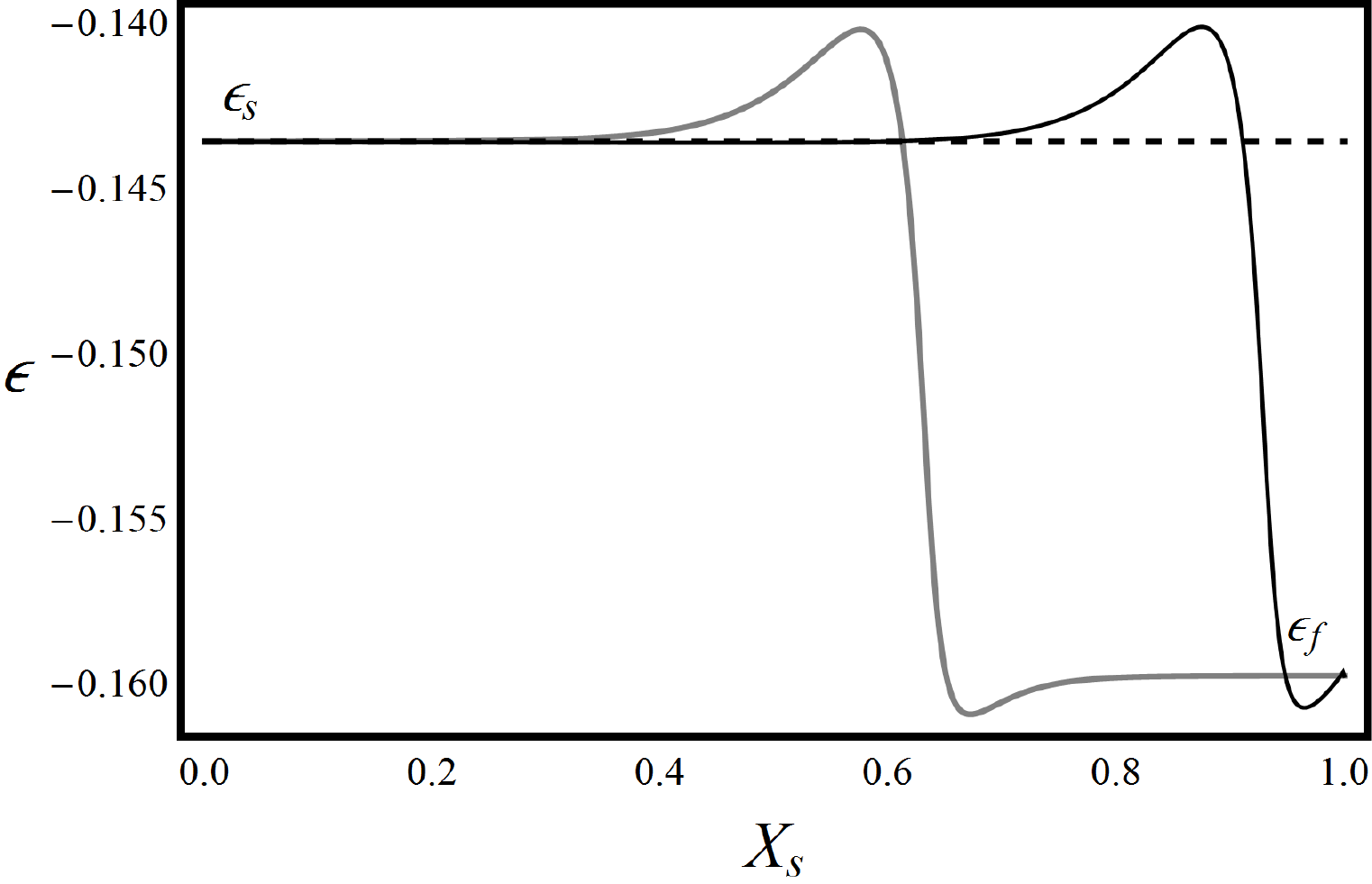}
 \end{minipage}
 \hspace{11mm}  
 \begin{minipage}[b!h!]{3cm}
  \includegraphics[width=4cm]{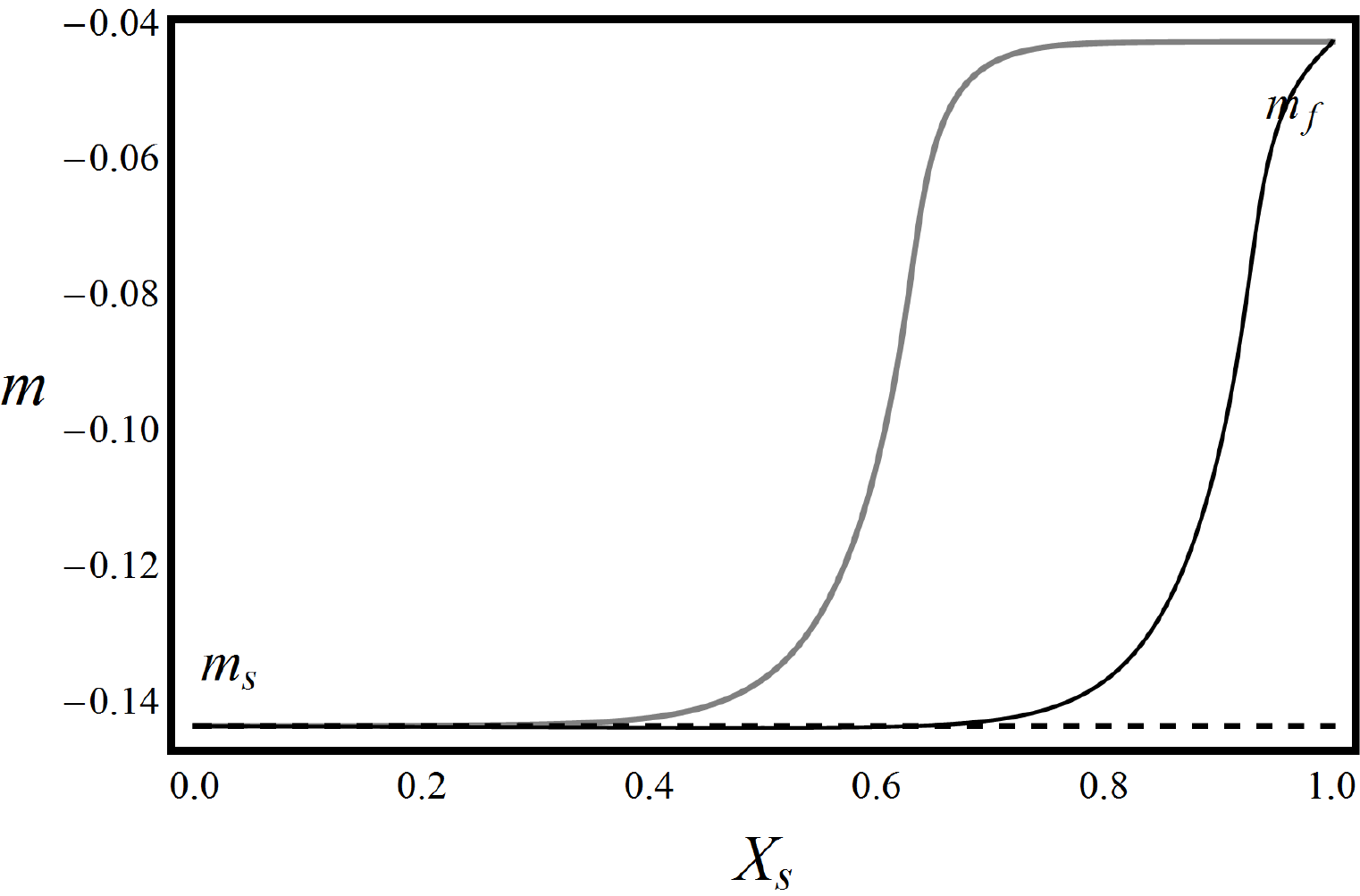}
   \end{minipage}
  \hspace{11mm} 
\begin{minipage}[b!h!]{3cm}
  \centering
   \includegraphics[width=4cm]{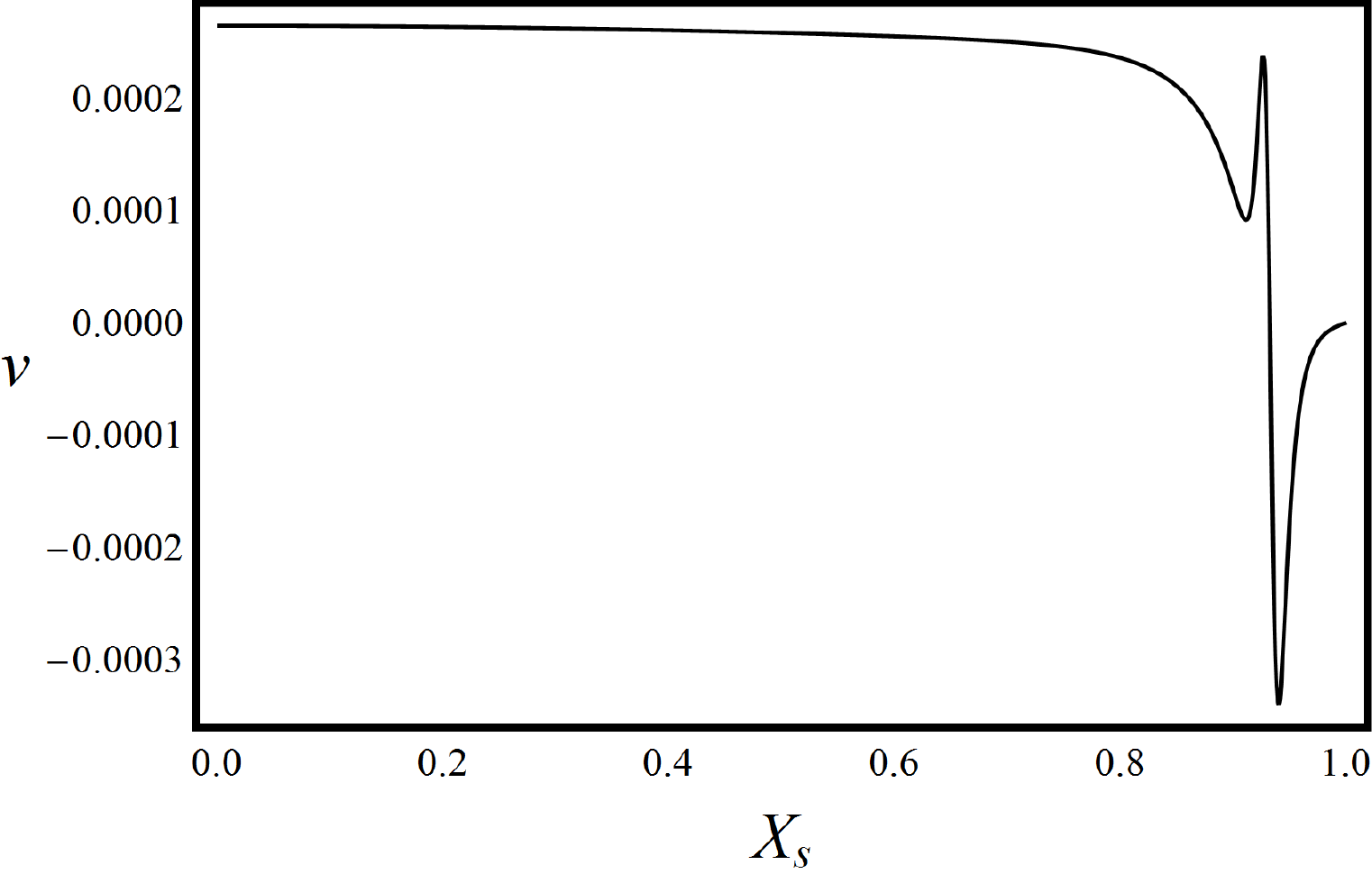}
 \end{minipage}
\vspace{2mm}
\centering 
\begin{minipage}[b!h!]{3cm}
\centering   
   \includegraphics[width=4cm]{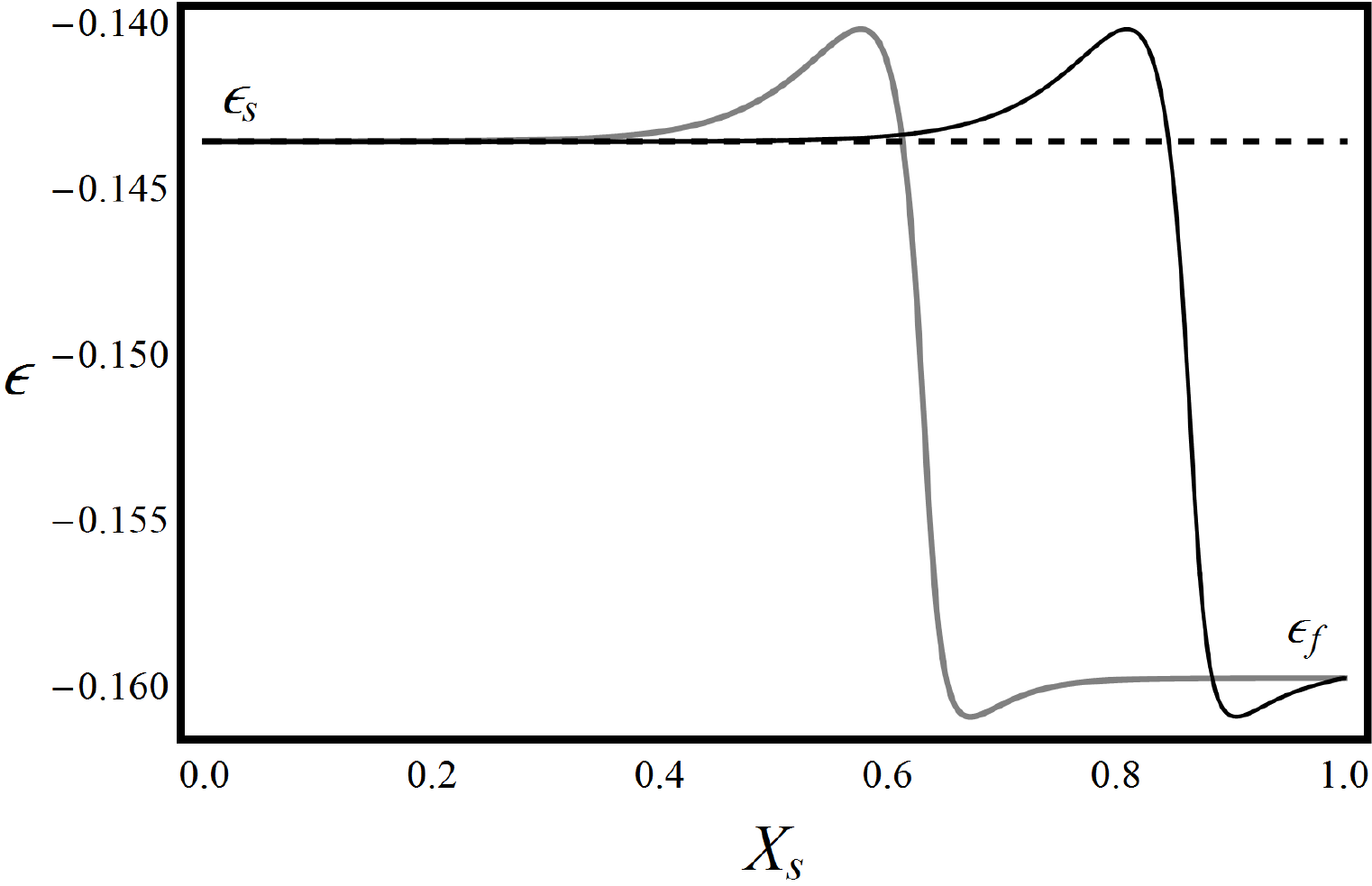}
 \end{minipage}
 \hspace{11mm}  
 \begin{minipage}[b!h!]{3cm}
  \includegraphics[width=4cm]{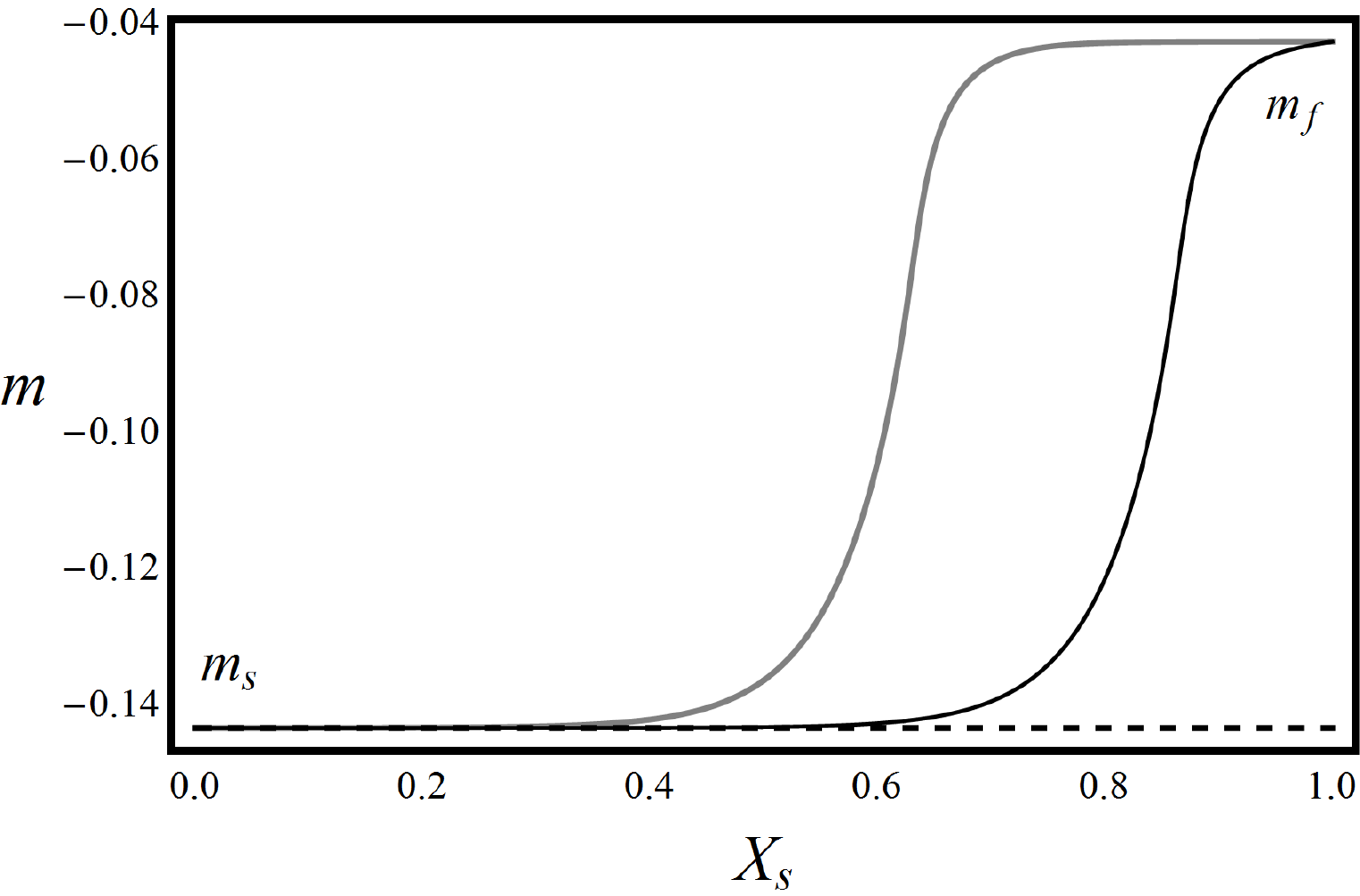}
   \end{minipage}
  \hspace{11mm} 
\begin{minipage}[b!h!]{3cm}
  \centering
   \includegraphics[width=4cm]{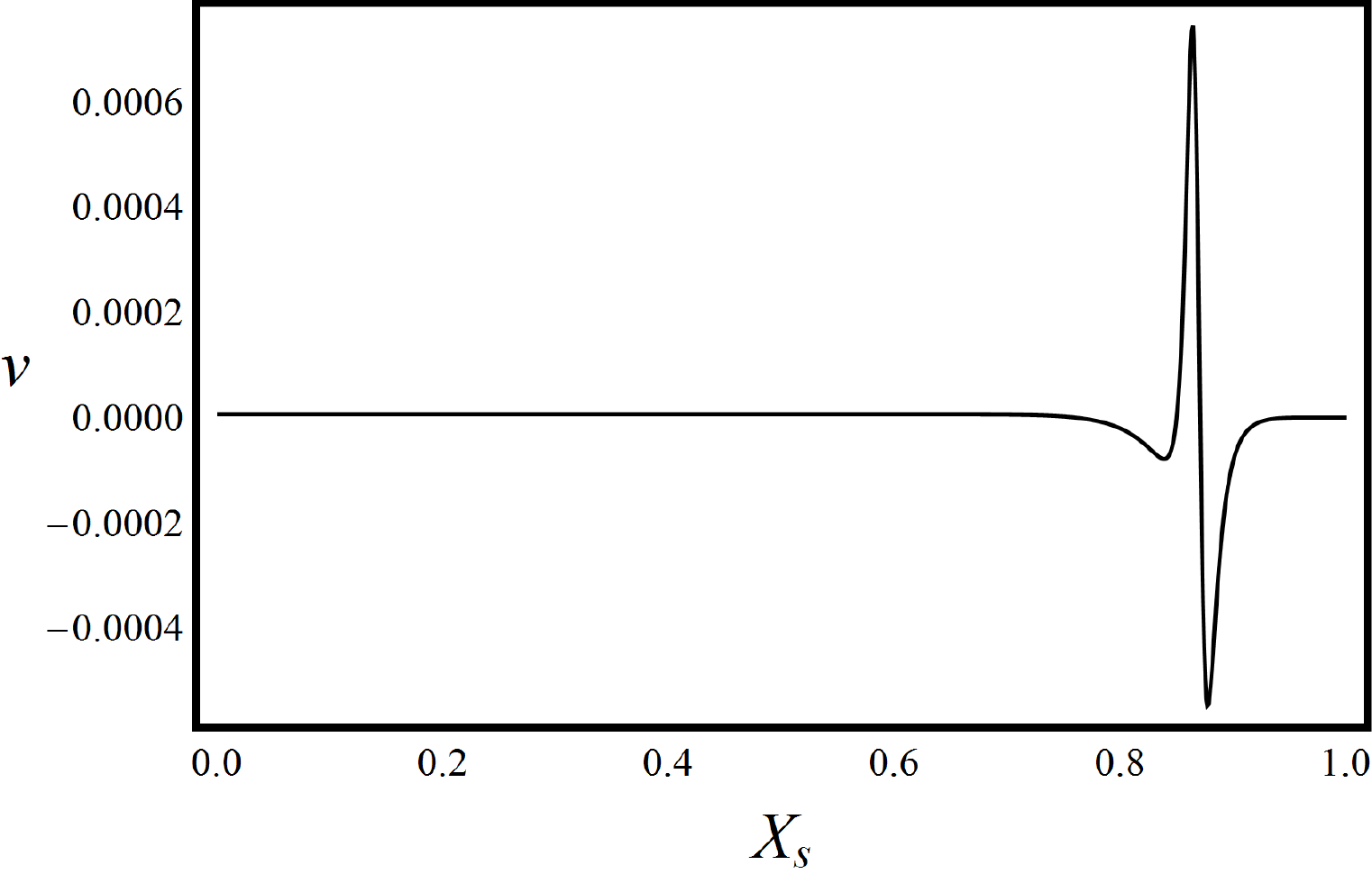}
 \end{minipage}  \caption{The same as in figure \ref{vecchiostandard} but with the one side impermeability condition.}
 \label{tappostandard}
\end{figure}

We discuss now the solution of the problems 
\eqref{num01} e \eqref{num02} with the constant 
fluid--poor profile as initial condition. Results are plotted in 
figures~\ref{vecchiostandard} and \ref{tappostandard}; as before, 
the three columns refer to $\varepsilon$, $m$, and the seepage velocity $v$, 
respectively. 

As in the case of the linear initial condition and, as we will see, in 
the case of the fluid--rich phase initial condition, 
the evolution can be divided into two macroscopically different regimes: 
formation of the connection profile and slow travelling wave regime.
Thus, the same physical 
arguments discussed in the previous case continue to be 
valid even in this case.

In the very early times of the profile formation, 
both for the zero chemical potential and the one--side impermeable 
cases,
the evolution is characterized by a big deviation in the 
$\varepsilon$ profile, from the standard initial value, to large values 
outside the range $(\varepsilon_f,\varepsilon_m)$. 
A similar abrupt behavior is not observed for the fluid 
density profile: the deviation below the $m_s$ value (dashed lines), 
is present, but much smaller.
The corresponding velocity profiles are characterized, in the 
neighbourhood of the right boundary, by large negative values in 
the zero chemical potential case and large positive values 
in the other one. 
This behavior has a rather clear physical interpretation: as a large amount 
of fluid has to be accumulated close to the right boundary, 
in the zero chemical potential case it is retrieved from the 
exterior through the right boundary point, while in the 
other case, due to the impermeability of the right boundary, 
such a fluid has to quickly arrive from the central part of the interior 
of the sample. As in the case of linear initial conditions this fluid accumulation 
implies a swelling of the solid skeleton, whose amplitude is much more
significant with respect to that stemming from linear initial conditions.

It is very interesting to remark that when the density profile 
is almost formed, see the fifth row in figure~\ref{tappostandard} 
corresponding to time $t=11.1$, the seepage velocity attains 
negative values also close to the right boundary. We can intepret 
this fact by imagining that a certain amount of fluid is now present 
in that physical region and it starts to move to the 
right
in order to push the profile towards the stationary position. 

It is again possible to observe that, as in the previous case 
and for the same reason, a sort of discontinuity appears in the 
$\varepsilon$ profile in the one--side impermeable case
(see figure~\ref{tappostandard} at time $t=0.7930105$).

Finally, 
once the profile is formed, again we observe the peculiar late time 
velocity graph. 
In this case, in contrast to the previous one, such a 
profile has a positive part bigger that the negative one.
This is due to the fact that, for the present initial condition, 
it happens that the travelling wave has to propagate in the 
opposite direction.

\subsection{Fluid--rich inital condition}
\label{s:fluid}
\begin{figure}[t]
\centering
 \begin{minipage}[b!h!]{3cm}
 \centering   
   \includegraphics[width=4cm]{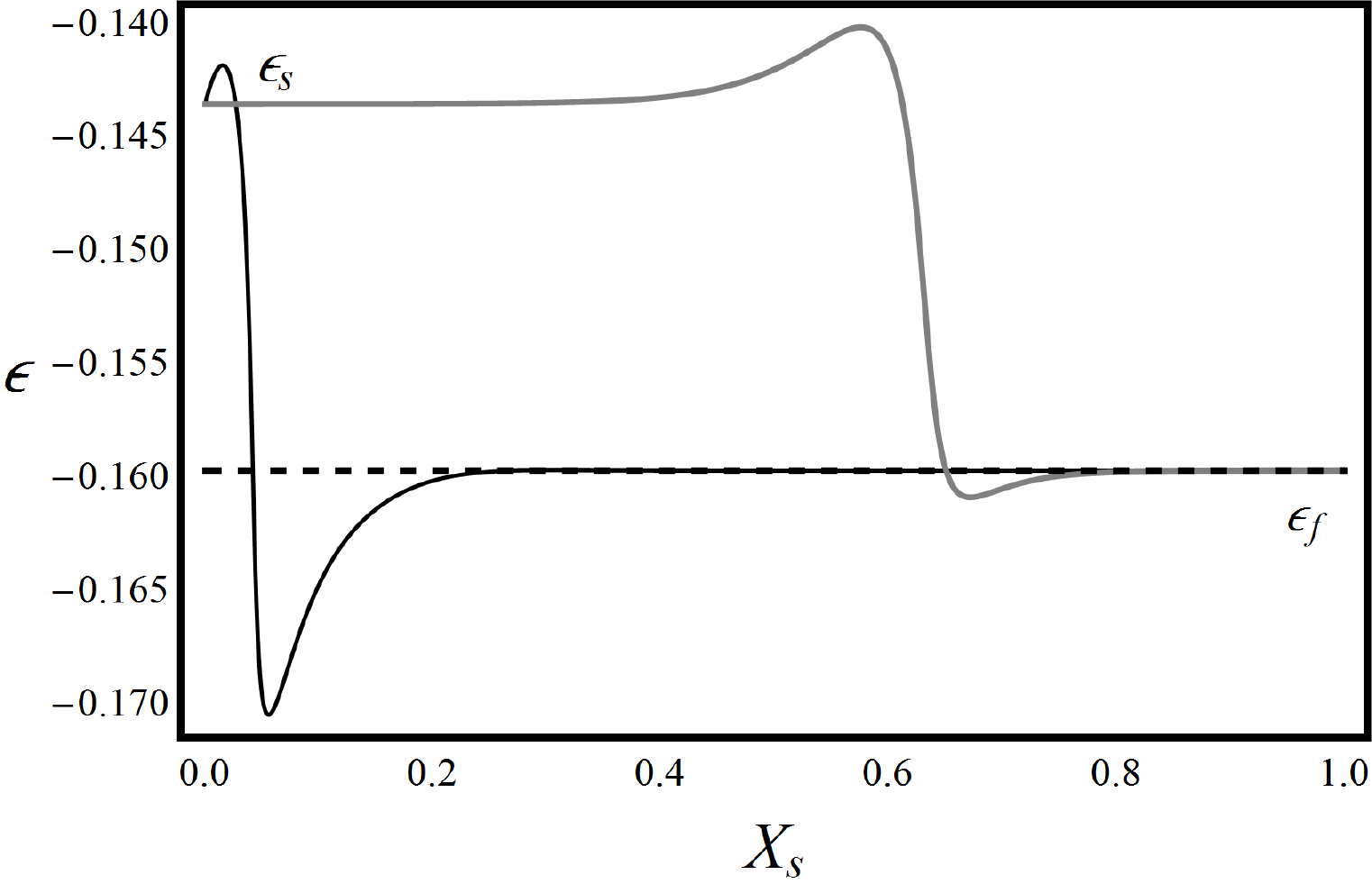}
 \end{minipage}
 \hspace{11mm}  
 \begin{minipage}[b!h!]{3cm}
  \includegraphics[width=4cm]{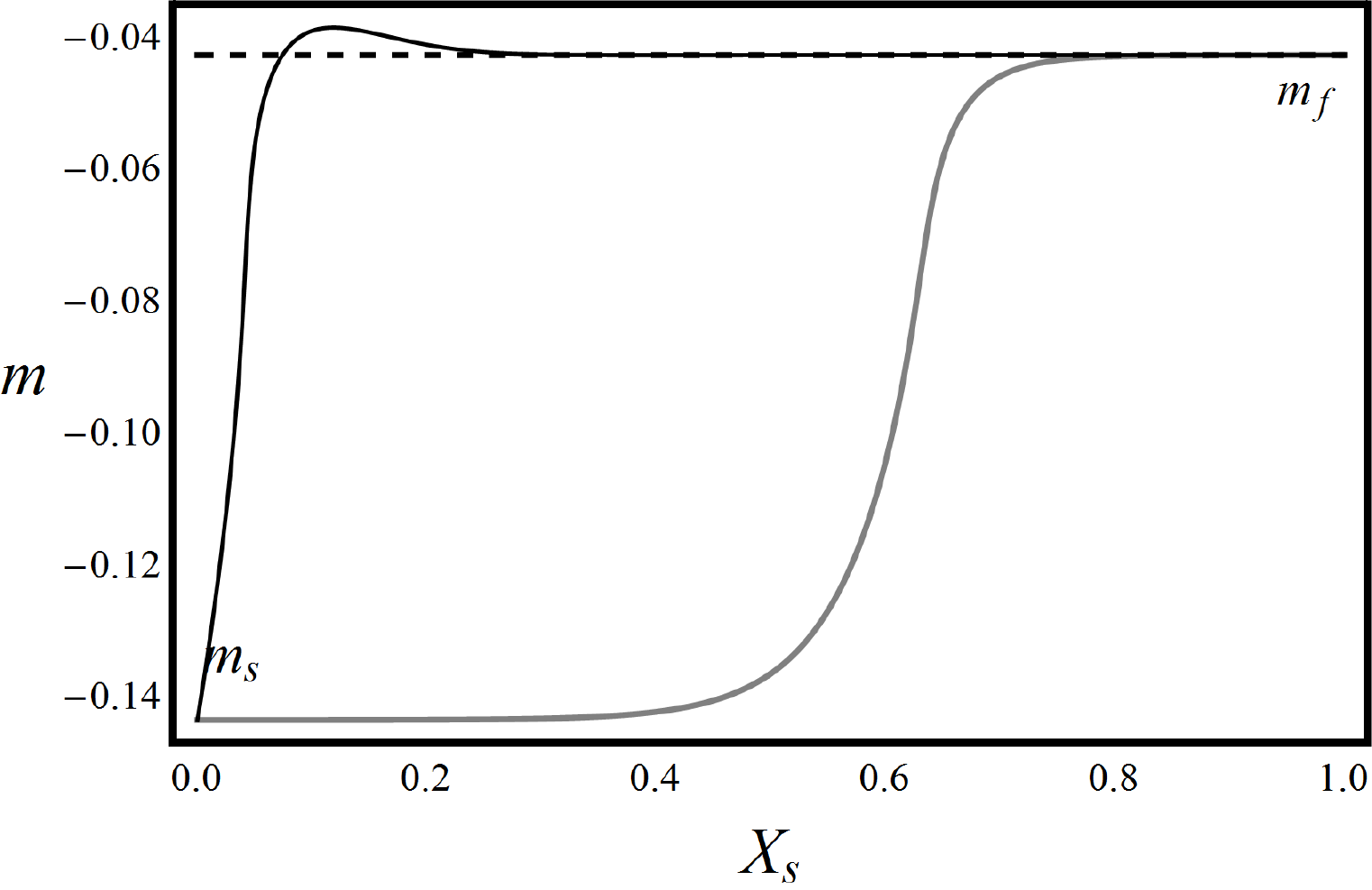}
   \end{minipage}
  \hspace{11mm} 
\begin{minipage}[b!h!]{3cm}
  \centering
   \includegraphics[width=4cm]{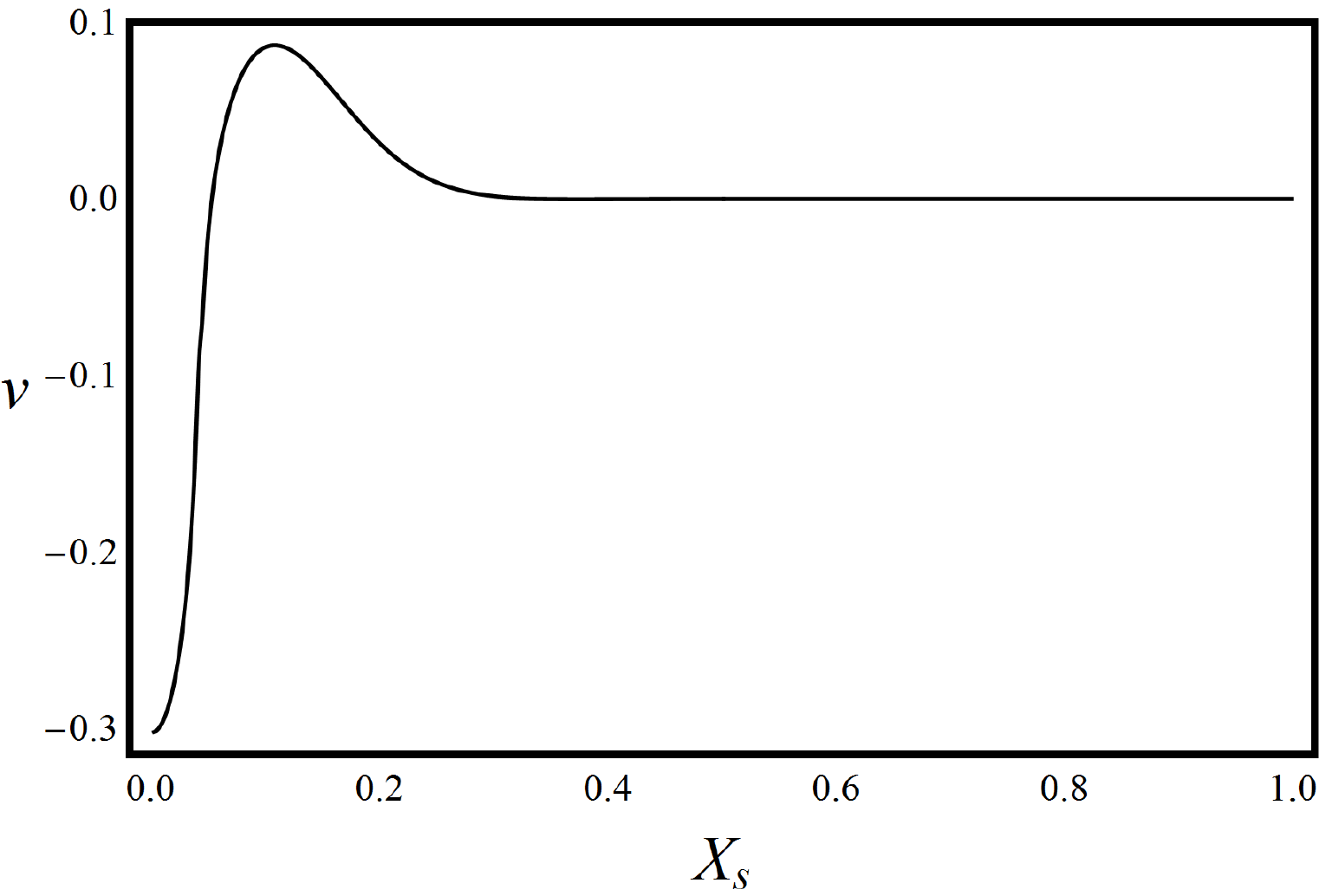}
 \end{minipage}
 \vspace{2mm}
\centering
 \centering \begin{minipage}[b!h!]{3cm}
\centering   
   \includegraphics[width=4cm]{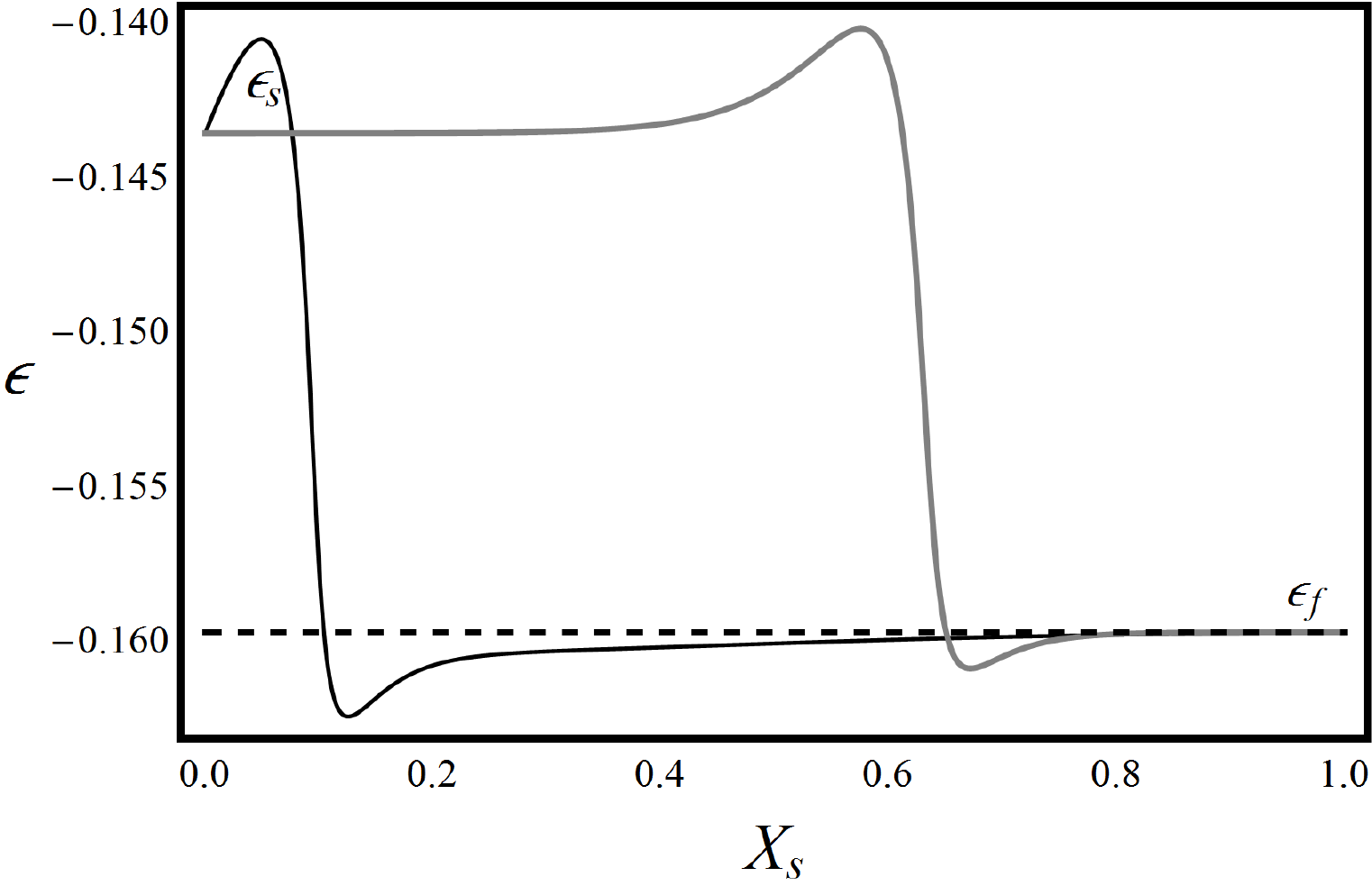}
 \end{minipage}
 \hspace{11mm}  
 \begin{minipage}[b!h!]{3cm}
  \includegraphics[width=4cm]{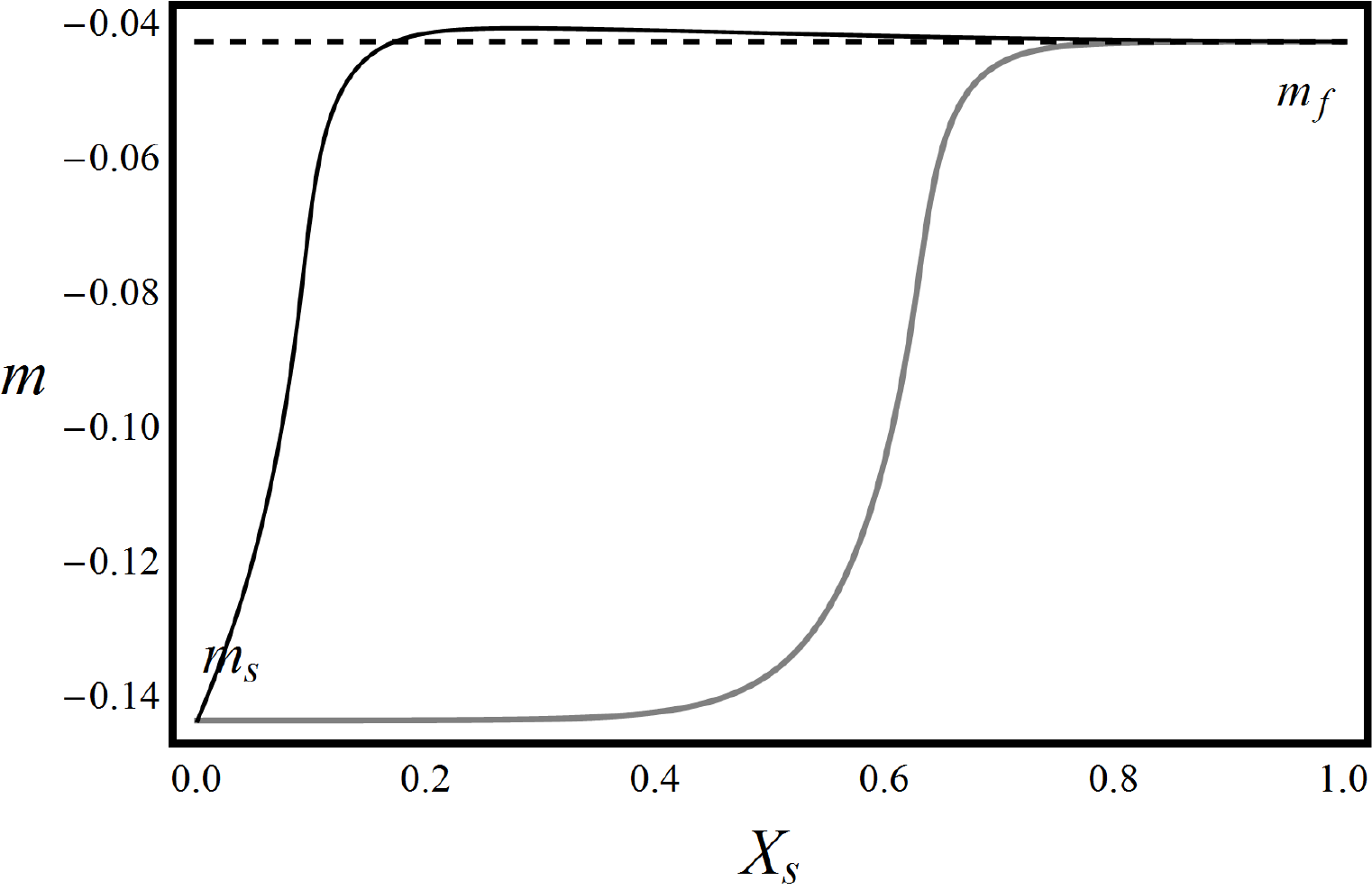}
   \end{minipage}
  \hspace{11mm} 
\begin{minipage}[b!h!]{3cm}
  \centering
   \includegraphics[width=4cm]{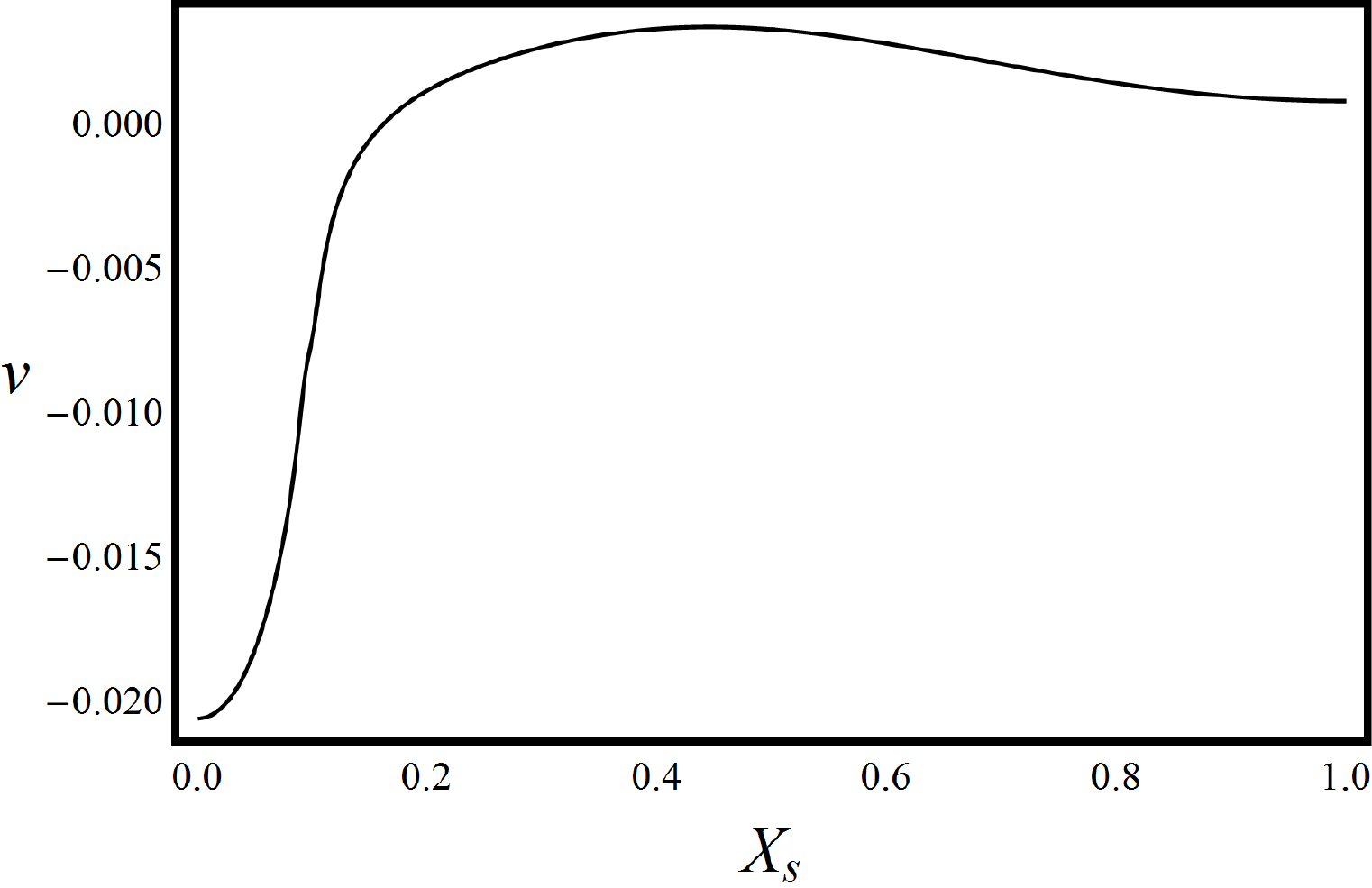}
 \end{minipage}
 \vspace{2mm}
\centering \begin{minipage}[b!h!]{3cm}
\centering   
   \includegraphics[width=4cm]{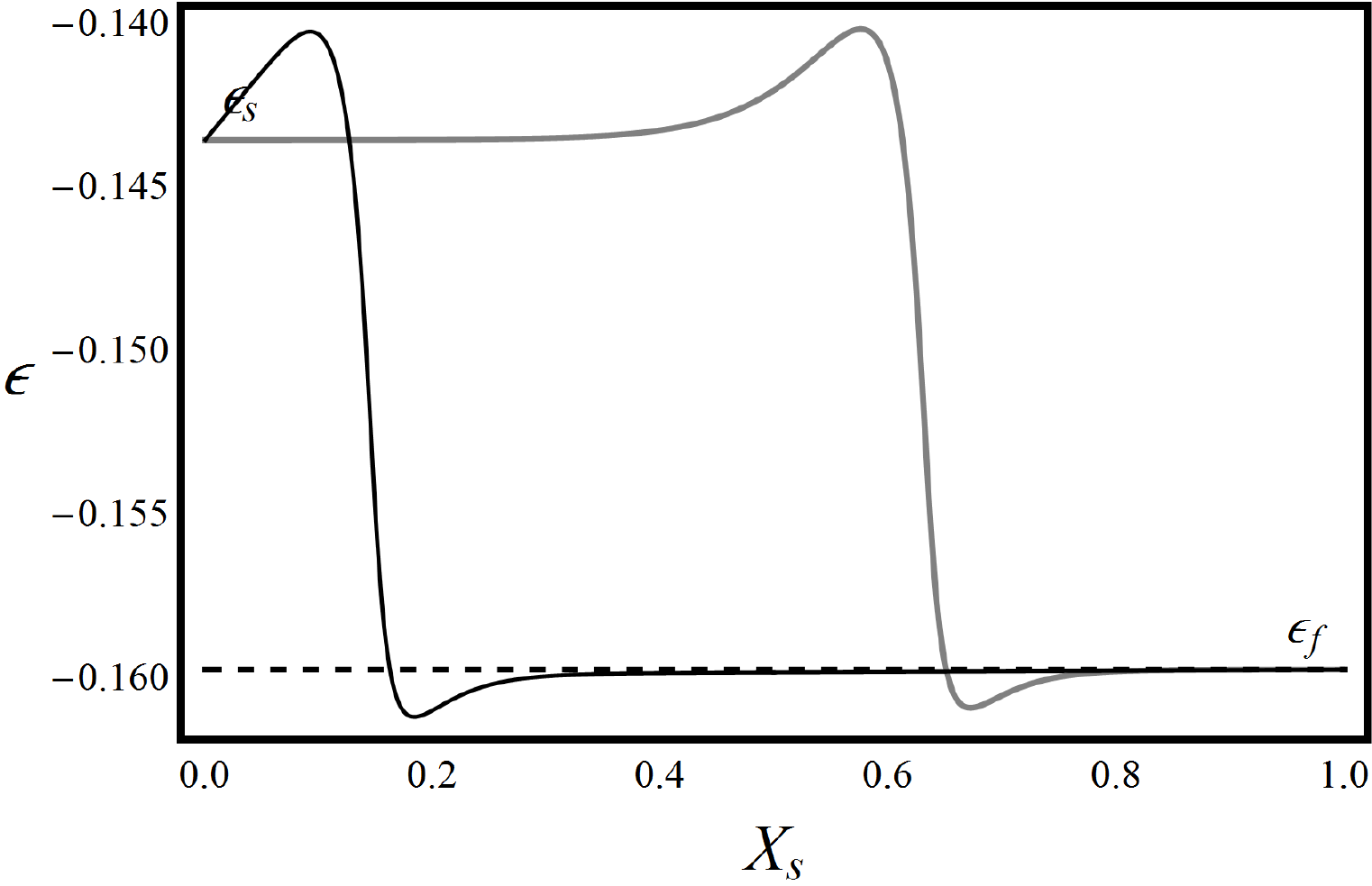}
 \end{minipage}
 \hspace{11mm}  
 \begin{minipage}[b!h!]{3cm}
  \includegraphics[width=4cm]{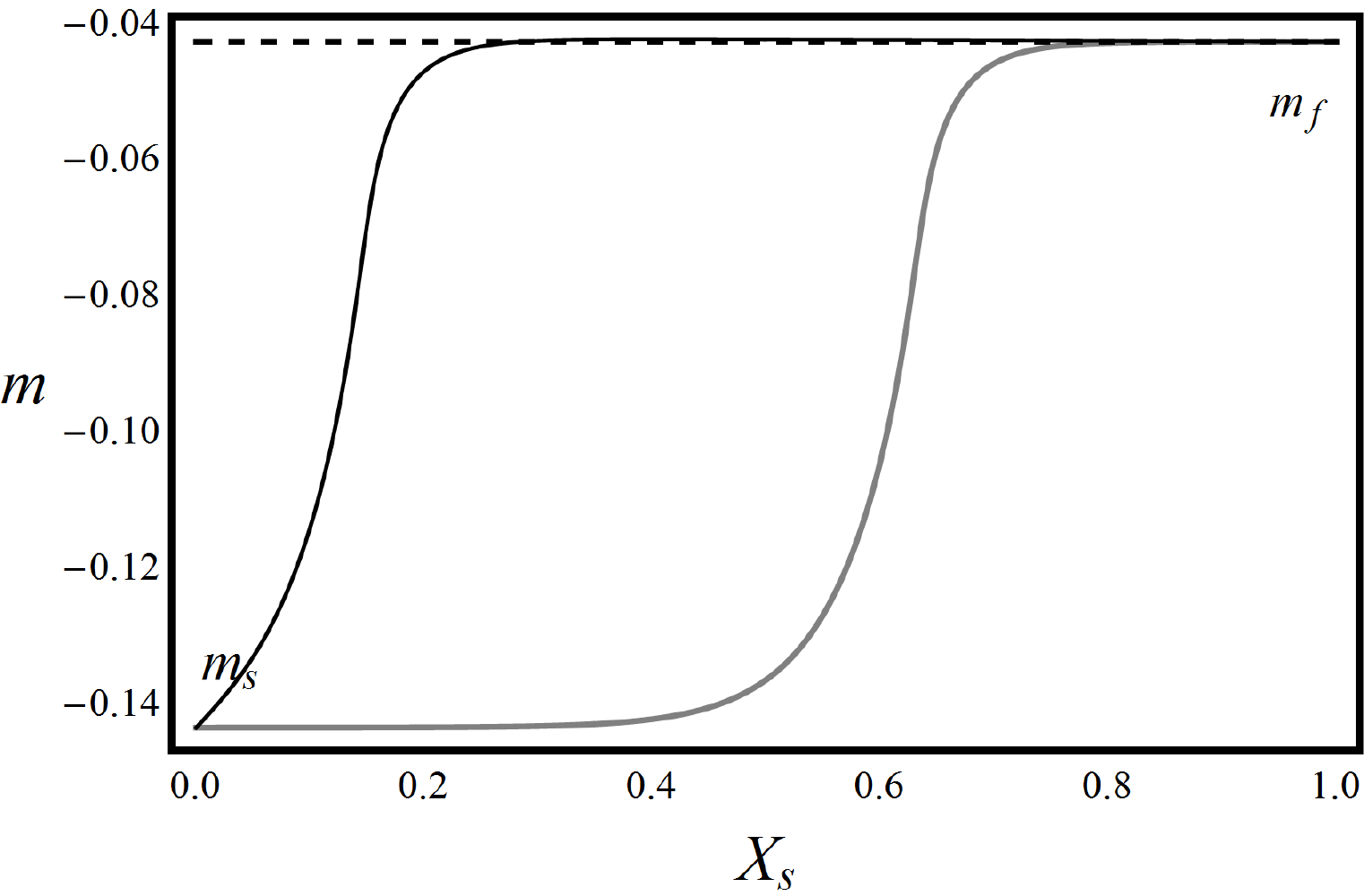}
   \end{minipage}
  \hspace{11mm} 
\begin{minipage}[b!h!]{3cm}
  \centering
   \includegraphics[width=4cm]{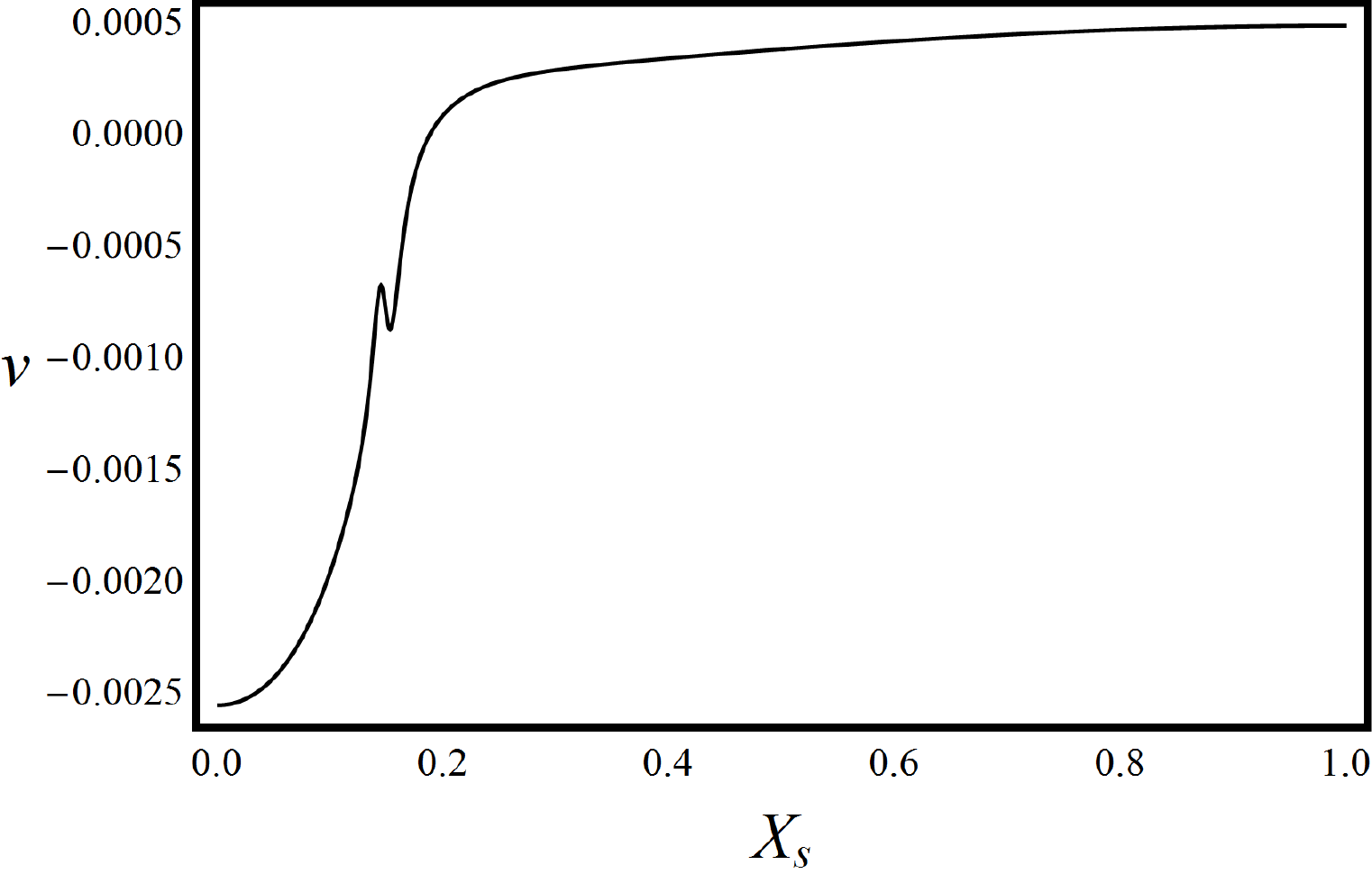}
 \end{minipage}
 \vspace{2mm}
\centering \begin{minipage}[b!h!]{3cm}
\centering   
   \includegraphics[width=4cm]{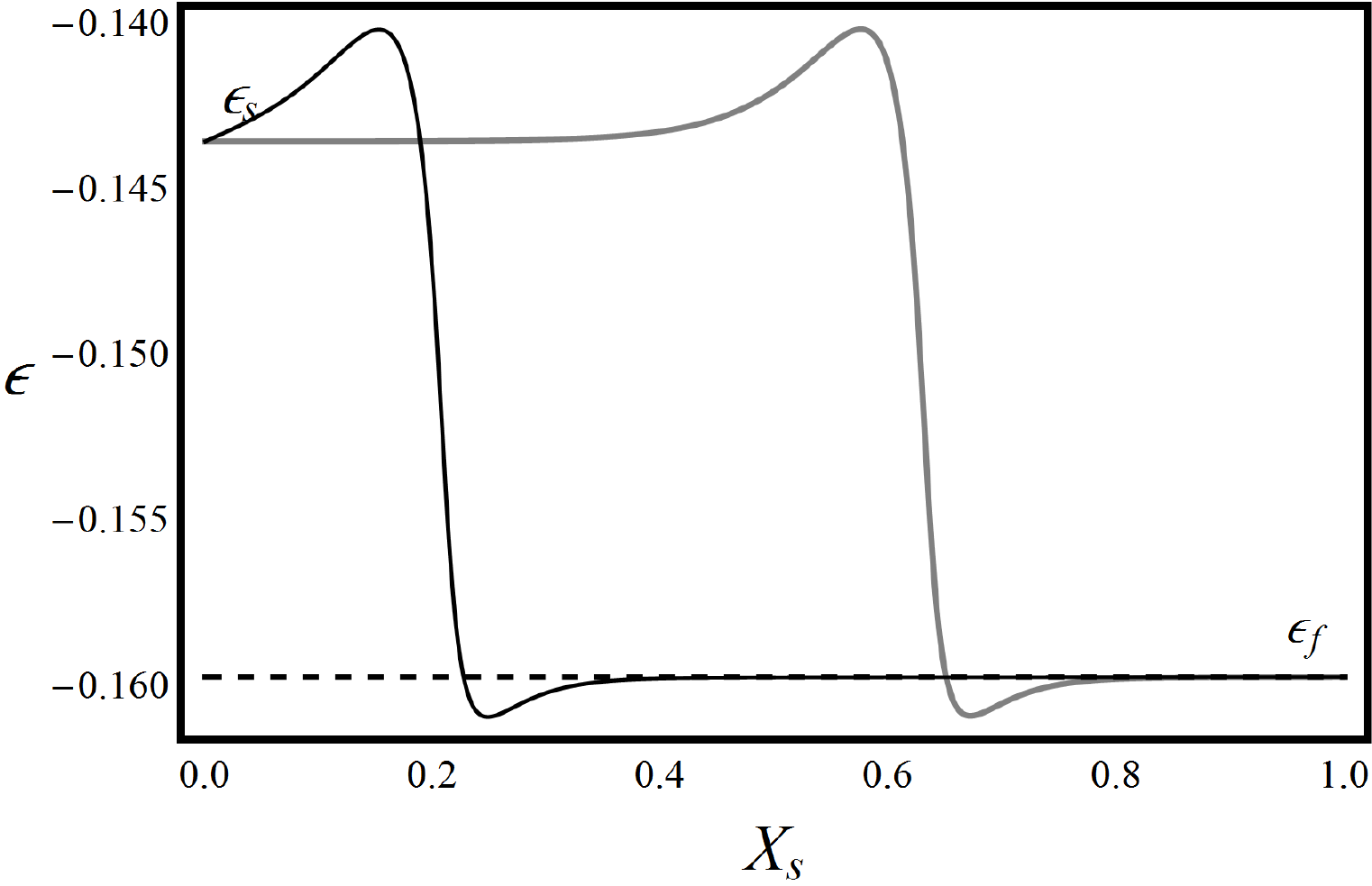}
 \end{minipage}
 \hspace{11mm}  
 \begin{minipage}[b!h!]{3cm}
  \includegraphics[width=4cm]{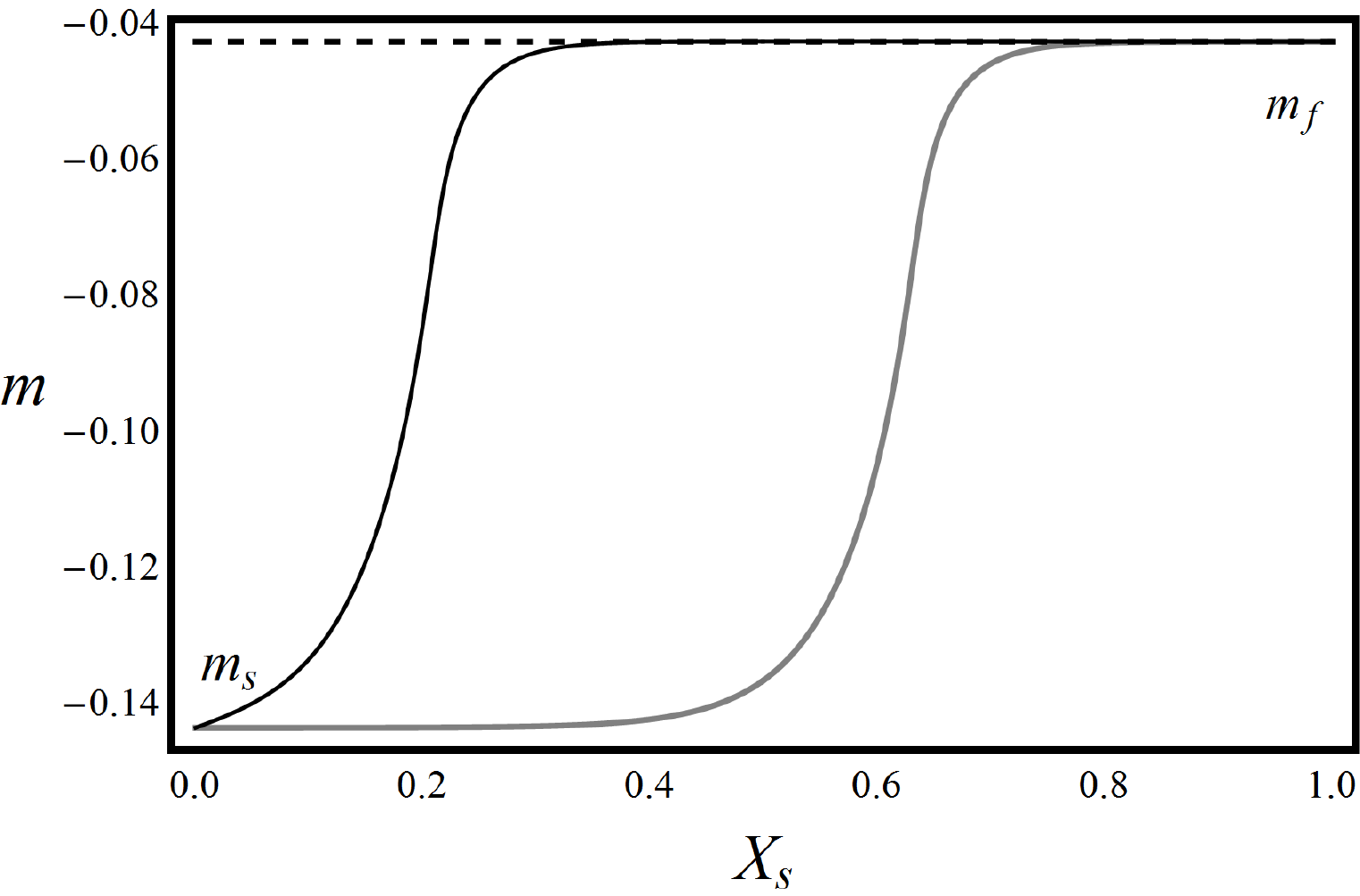}
   \end{minipage}
  \hspace{11mm} 
\begin{minipage}[b!h!]{3cm}
  \centering
   \includegraphics[width=4cm]{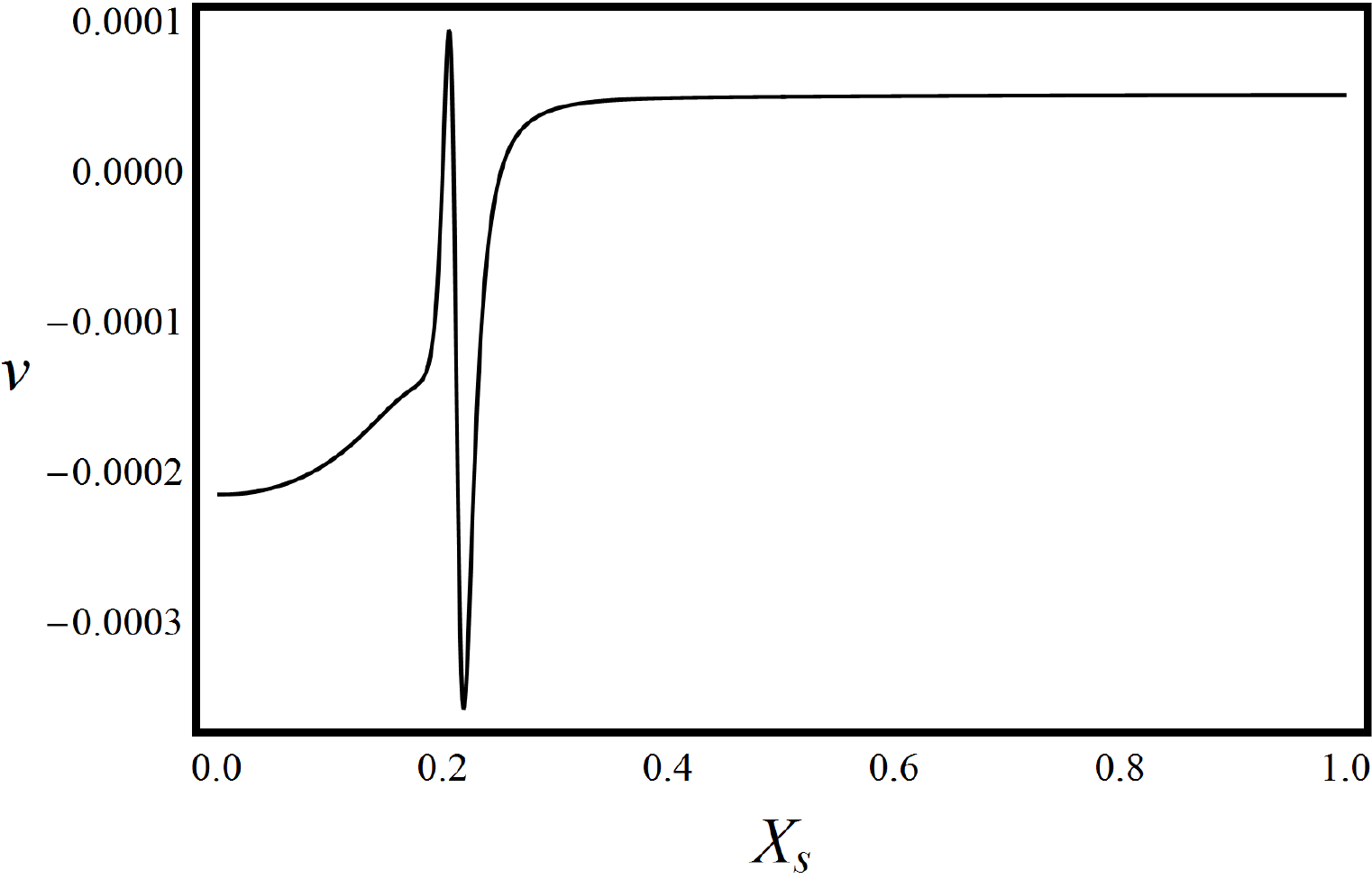}
 \end{minipage}
 \vspace{2mm}
\centering 
\begin{minipage}[b!h!]{3cm}
\centering   
   \includegraphics[width=4cm]{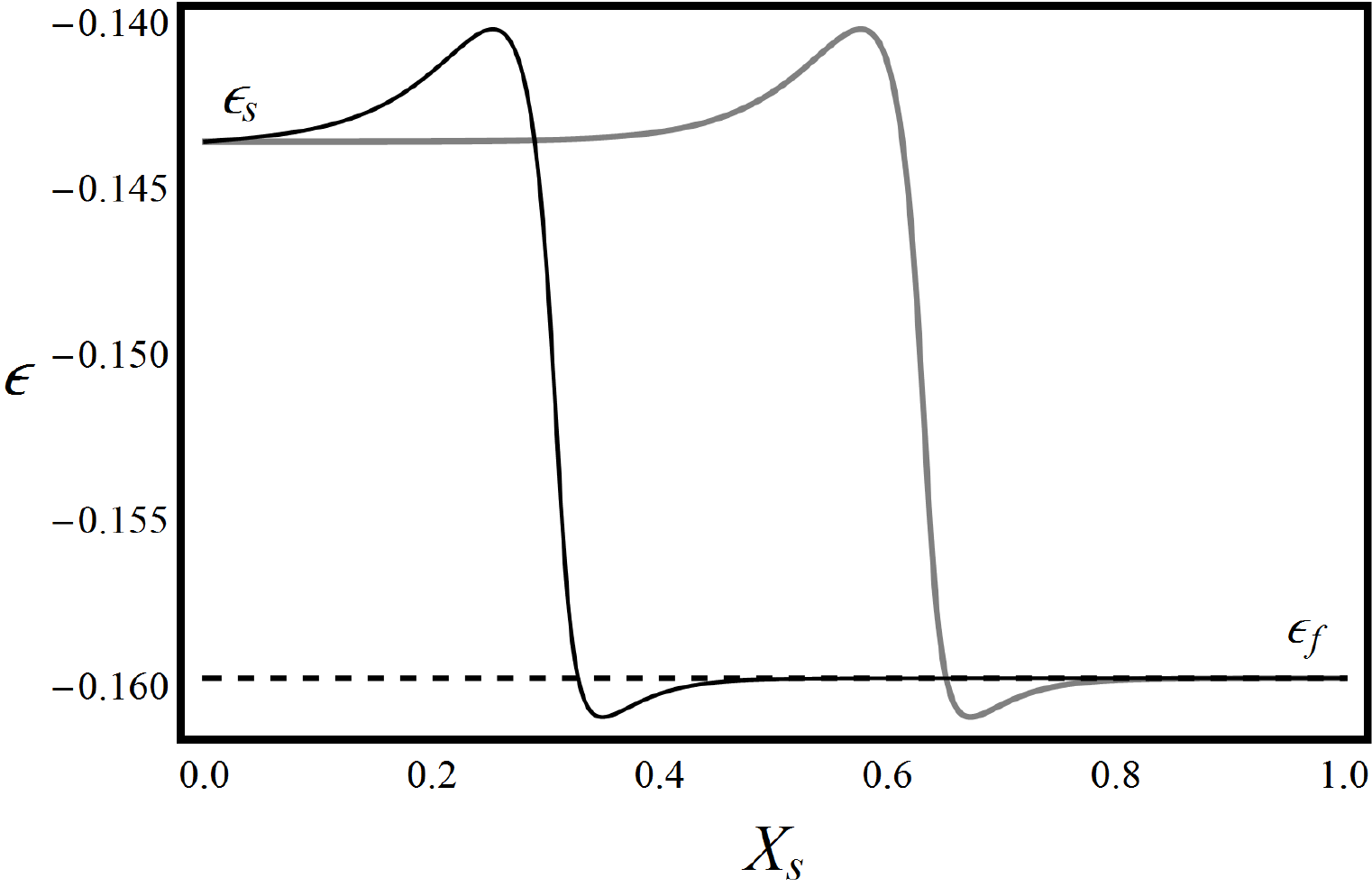}
 \end{minipage}
 \hspace{11mm}  
 \begin{minipage}[b!h!]{3cm}
  \includegraphics[width=4cm]{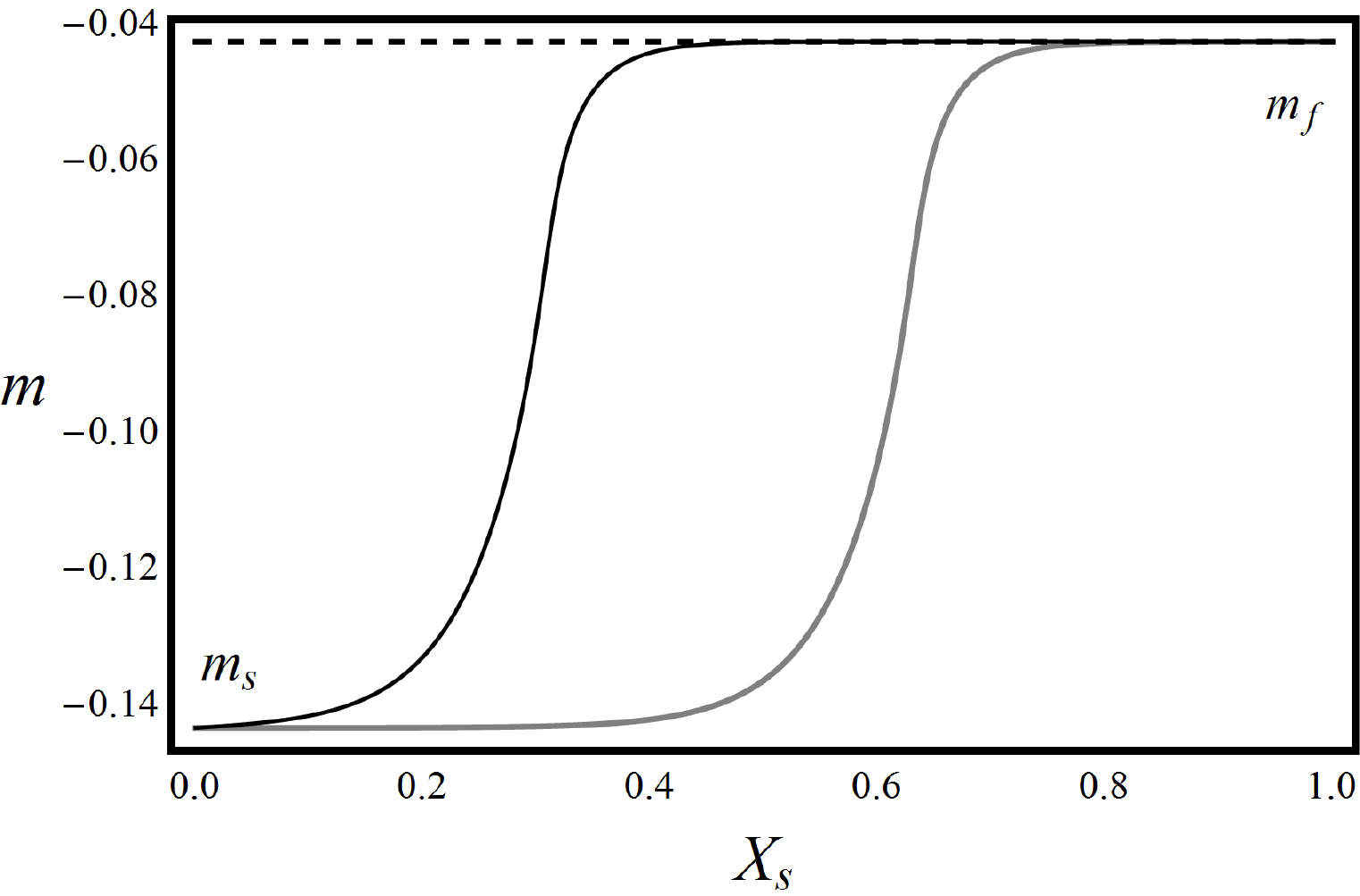}
   \end{minipage}
  \hspace{11mm} 
\begin{minipage}[b!h!]{3cm}
  \centering
   \includegraphics[width=4cm]{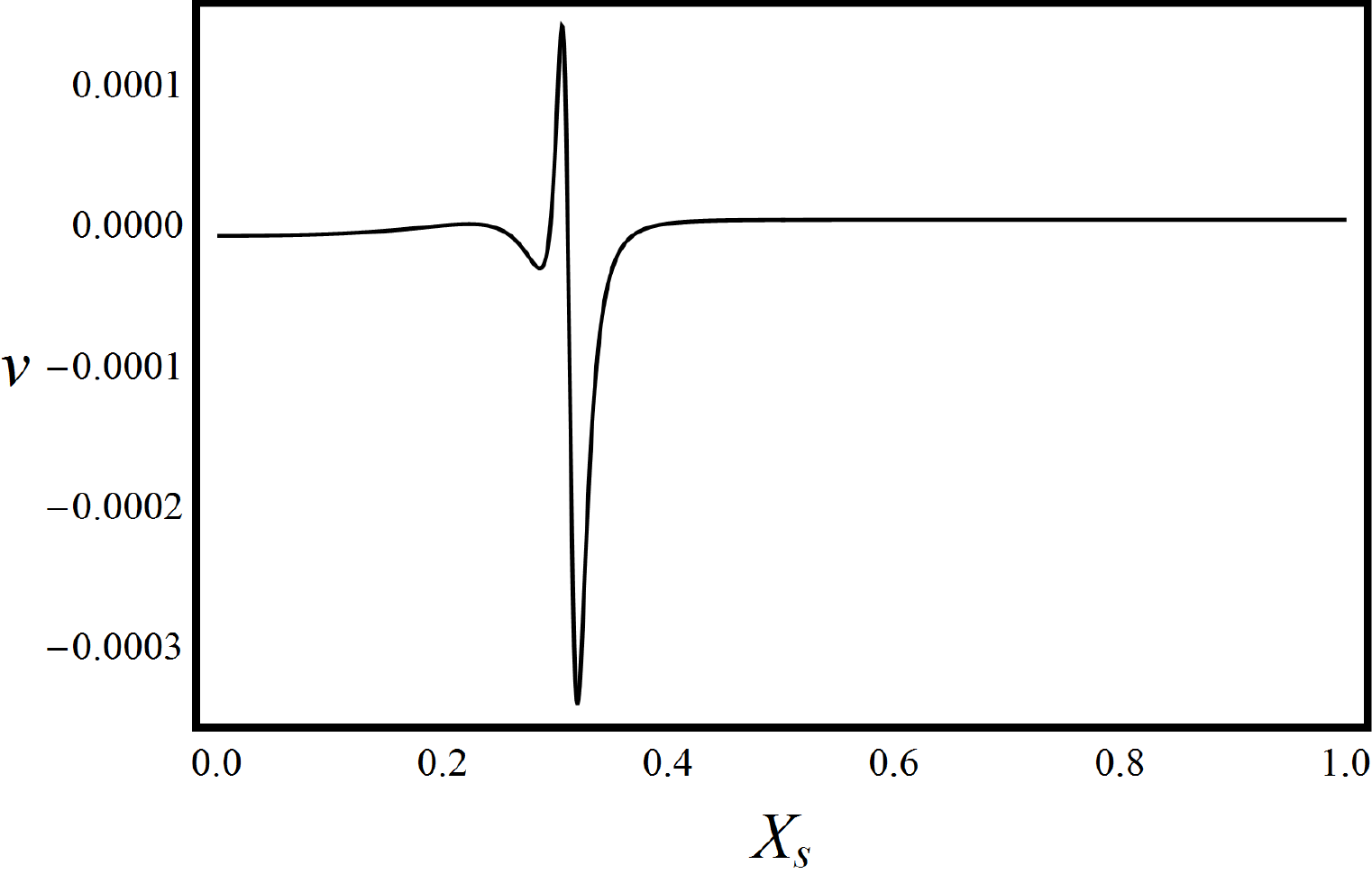}
 \end{minipage}
\vspace{2mm}
\centering 
\begin{minipage}[b!h!]{3cm}
\centering   
   \includegraphics[width=4cm]{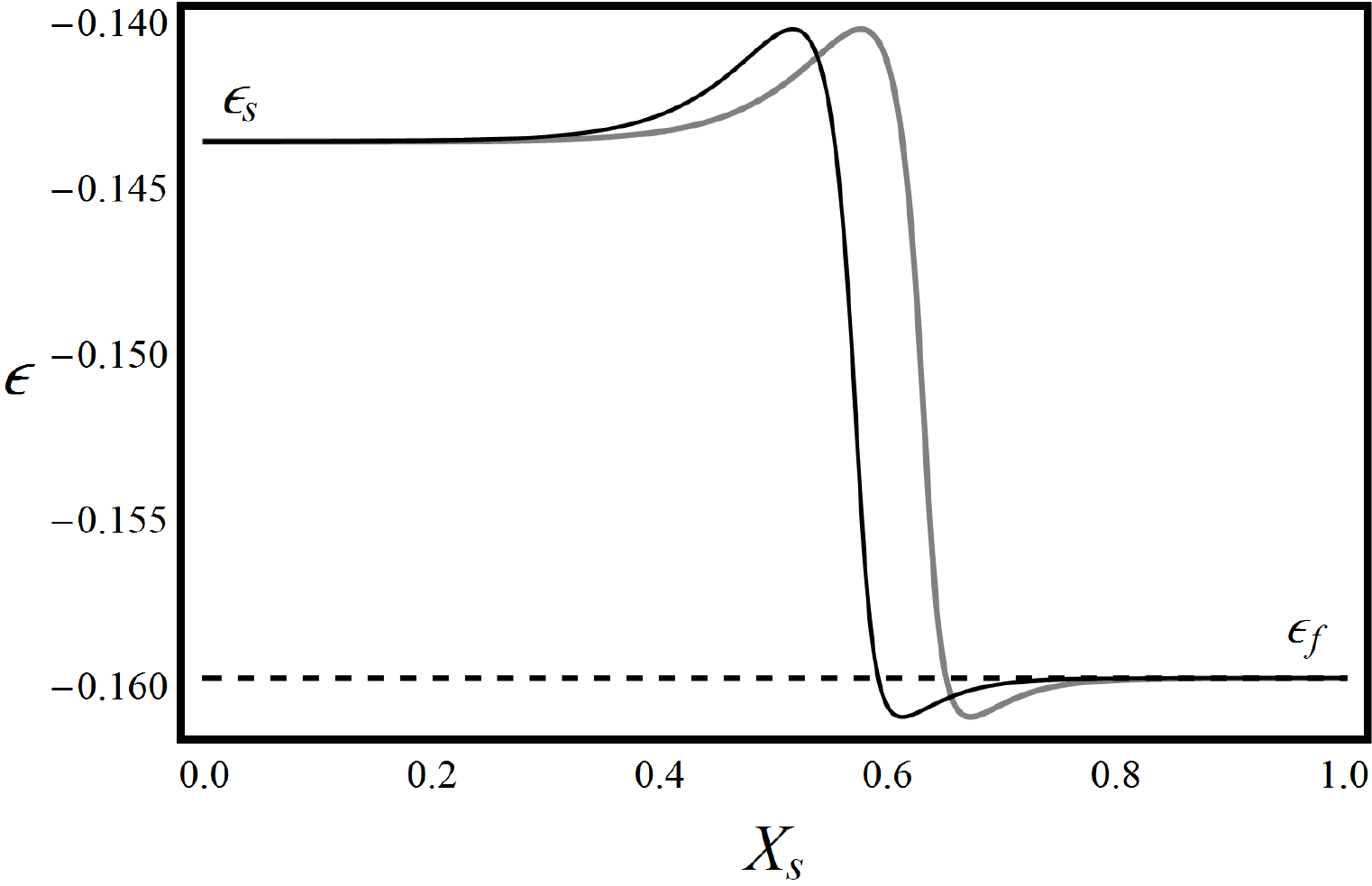}
 \end{minipage}
 \hspace{11mm}  
 \begin{minipage}[b!h!]{3cm}
  \includegraphics[width=4cm]{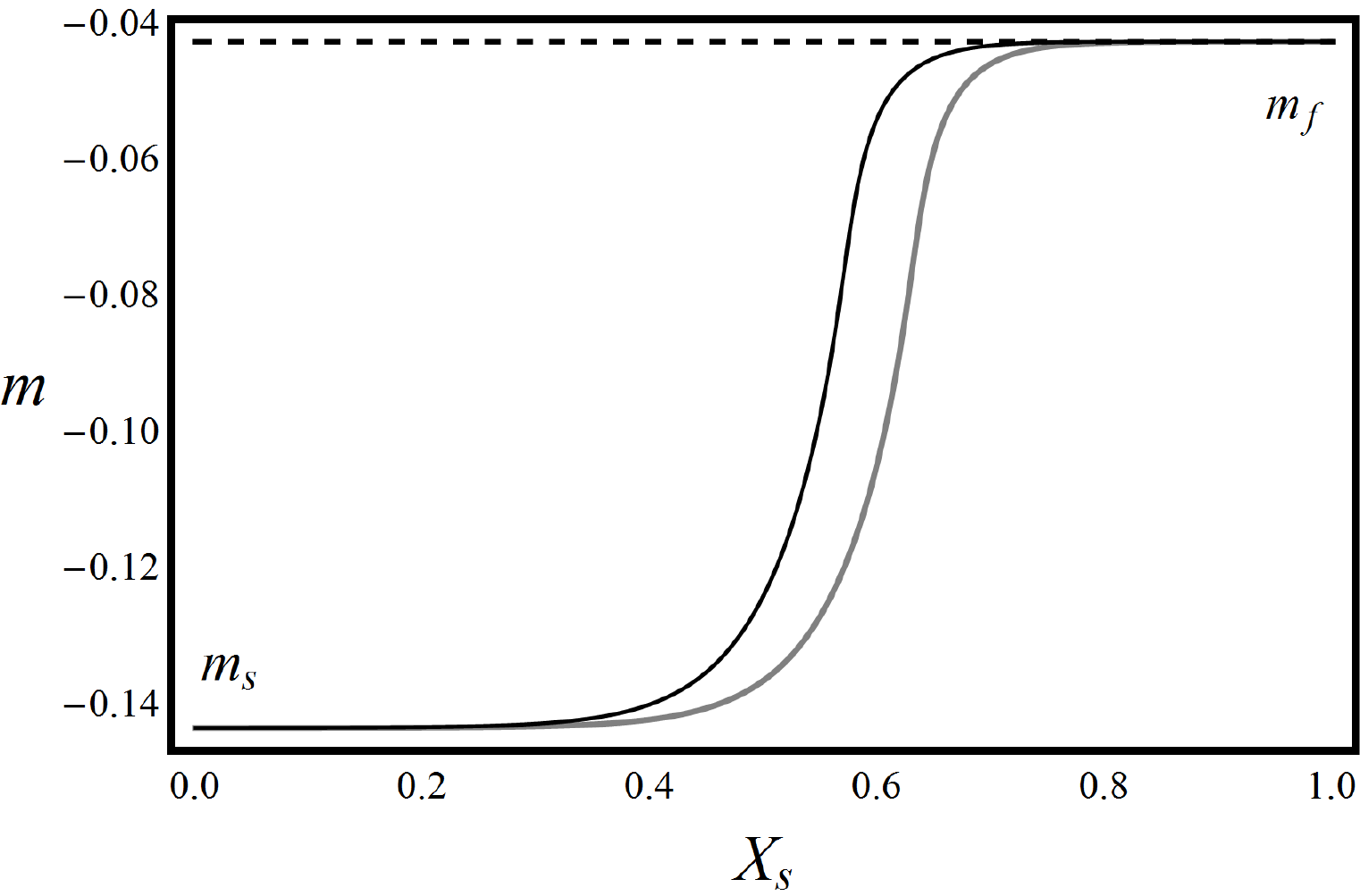}
   \end{minipage}
  \hspace{11mm} 
 \begin{minipage}[b!h!]{3cm}
  \centering
   \includegraphics[width=4cm]{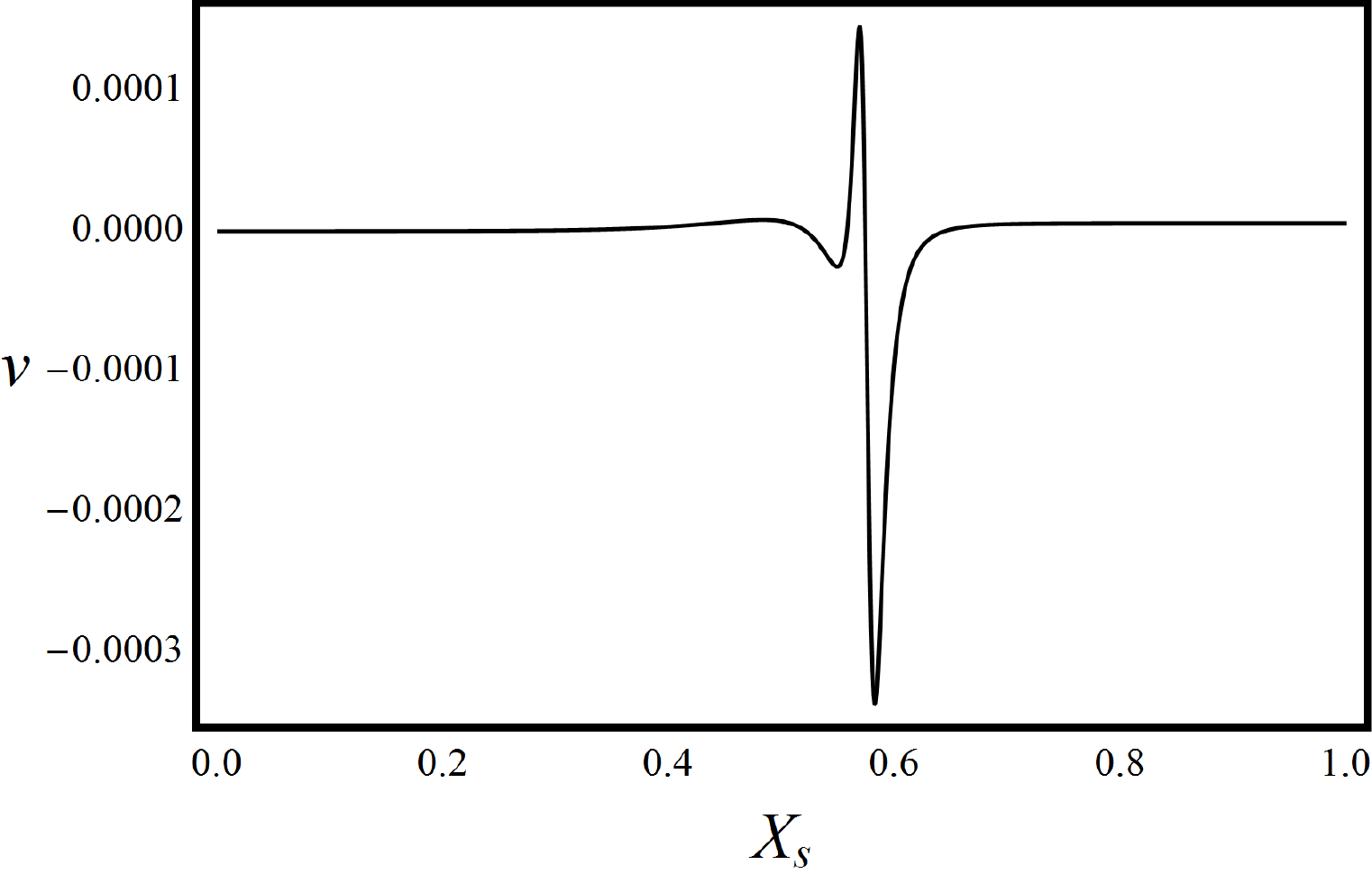}
 \end{minipage}
\caption{Profiles (black lines) $\varepsilon(X_s,t)$, $m(X_s)$ and $v(X_s)$ for the zero chemical potential problem obtained by solving the non-stationary system with a fluid phase initial state (dashed lines). We used Dirichelet boundary conditions $m(0)=m_s,\,\varepsilon(0)=\varepsilon_s,\,m(1)=m_f$, $\varepsilon(1)=\varepsilon_f$ on the finite interval $[0,1]$, at the coexistence pressure for $a=0.5,\,b=1,\,\alpha=100,\,k_1=k_2=k_3=10^{-3}$. Profiles at times $t=0.004,\,t=0.08,\,t=0.8,\,t=8,\,t=83,\,t=333$ in lexicographic order. The gray lines represent stationary profiles.}
 \label{vecchiofluid}
\end{figure}

\begin{figure}[t]
\centering
 \begin{minipage}[b!h!]{3cm}
\centering   
   \includegraphics[width=4cm]{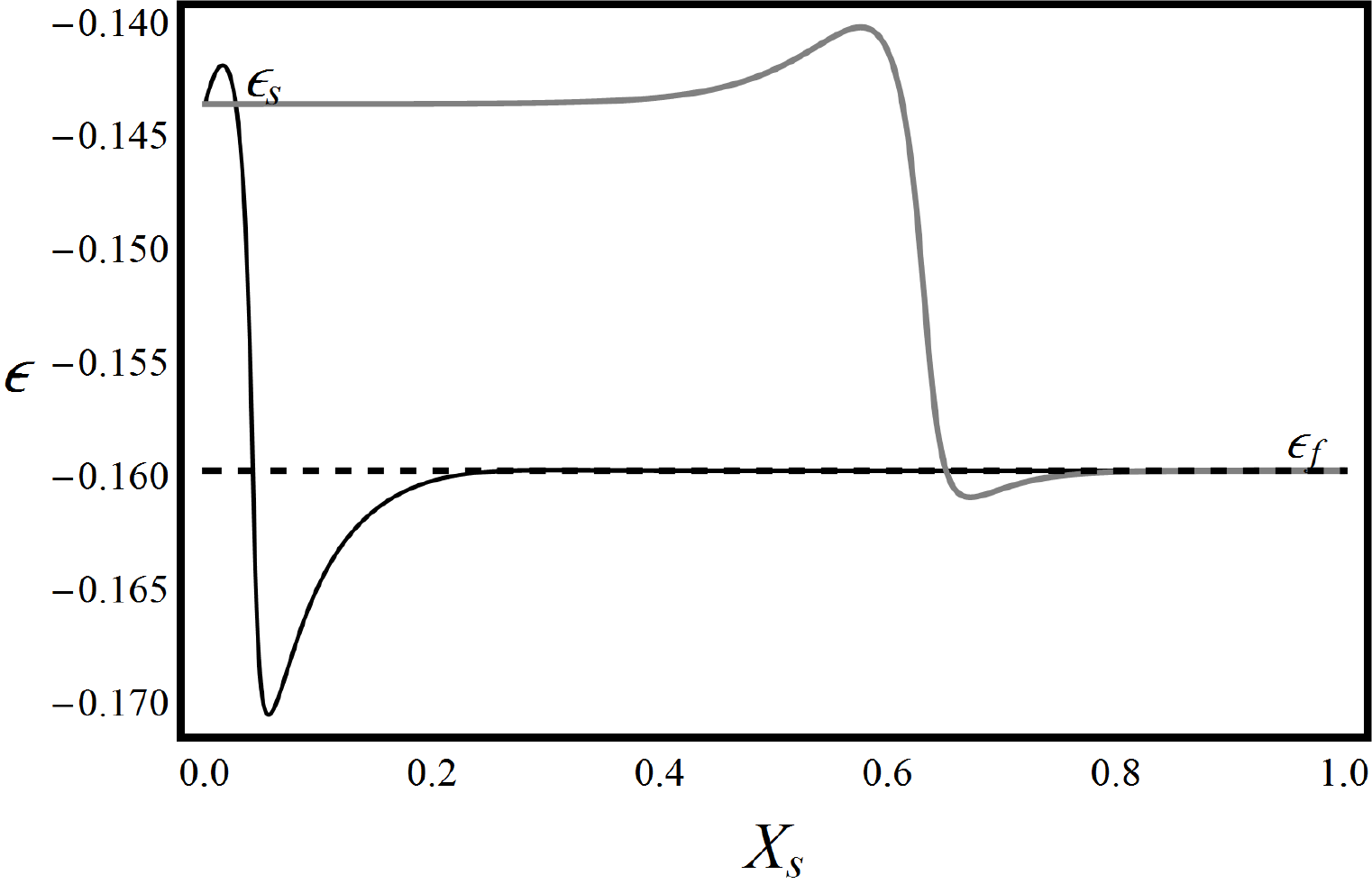}
 \end{minipage}
 \hspace{11mm}  
 \begin{minipage}[b!h!]{3cm}
  \includegraphics[width=4cm]{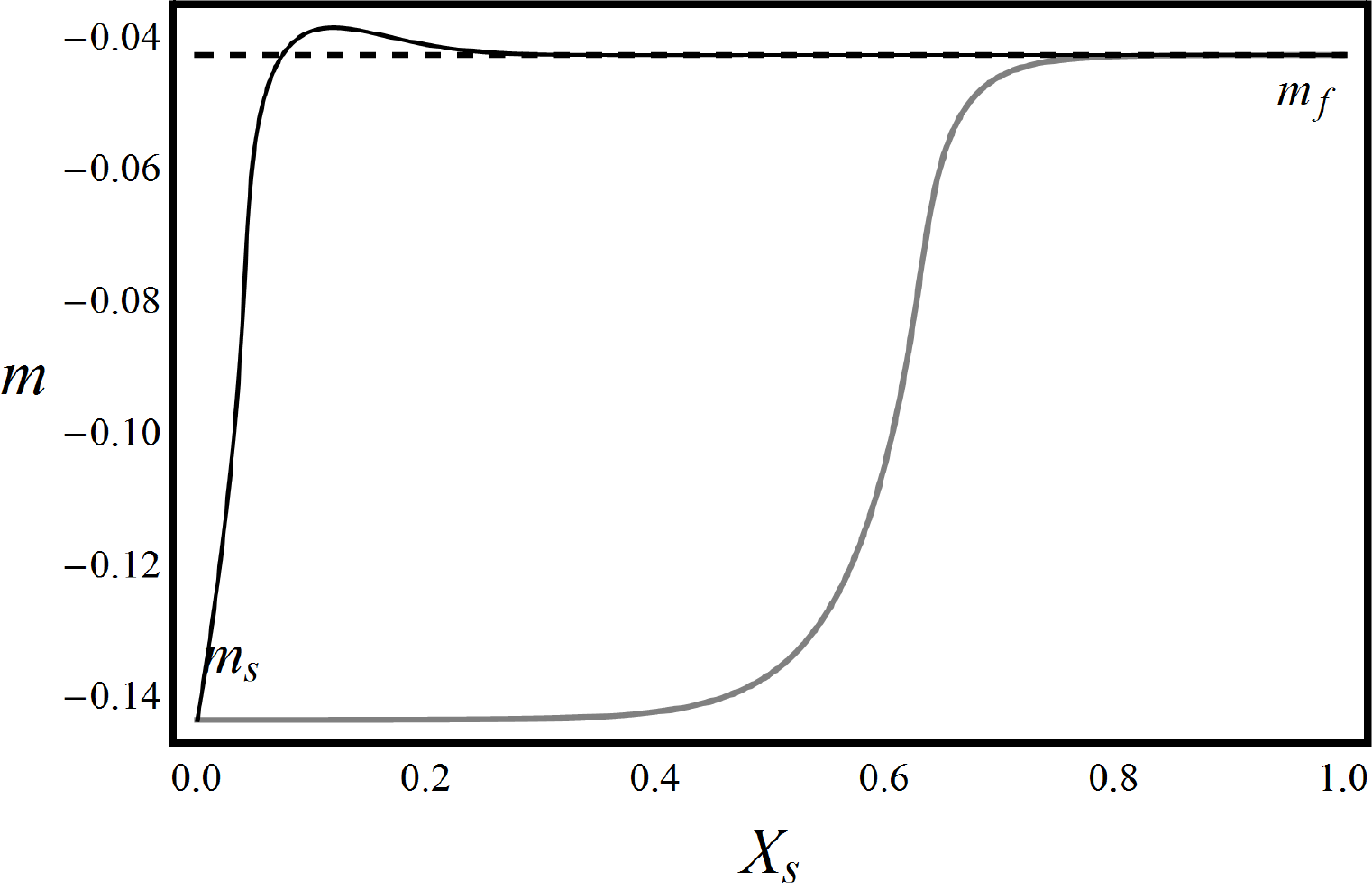}
   \end{minipage}
  \hspace{11mm} 
\begin{minipage}[b!h!]{3cm}
  \centering
   \includegraphics[width=4cm]{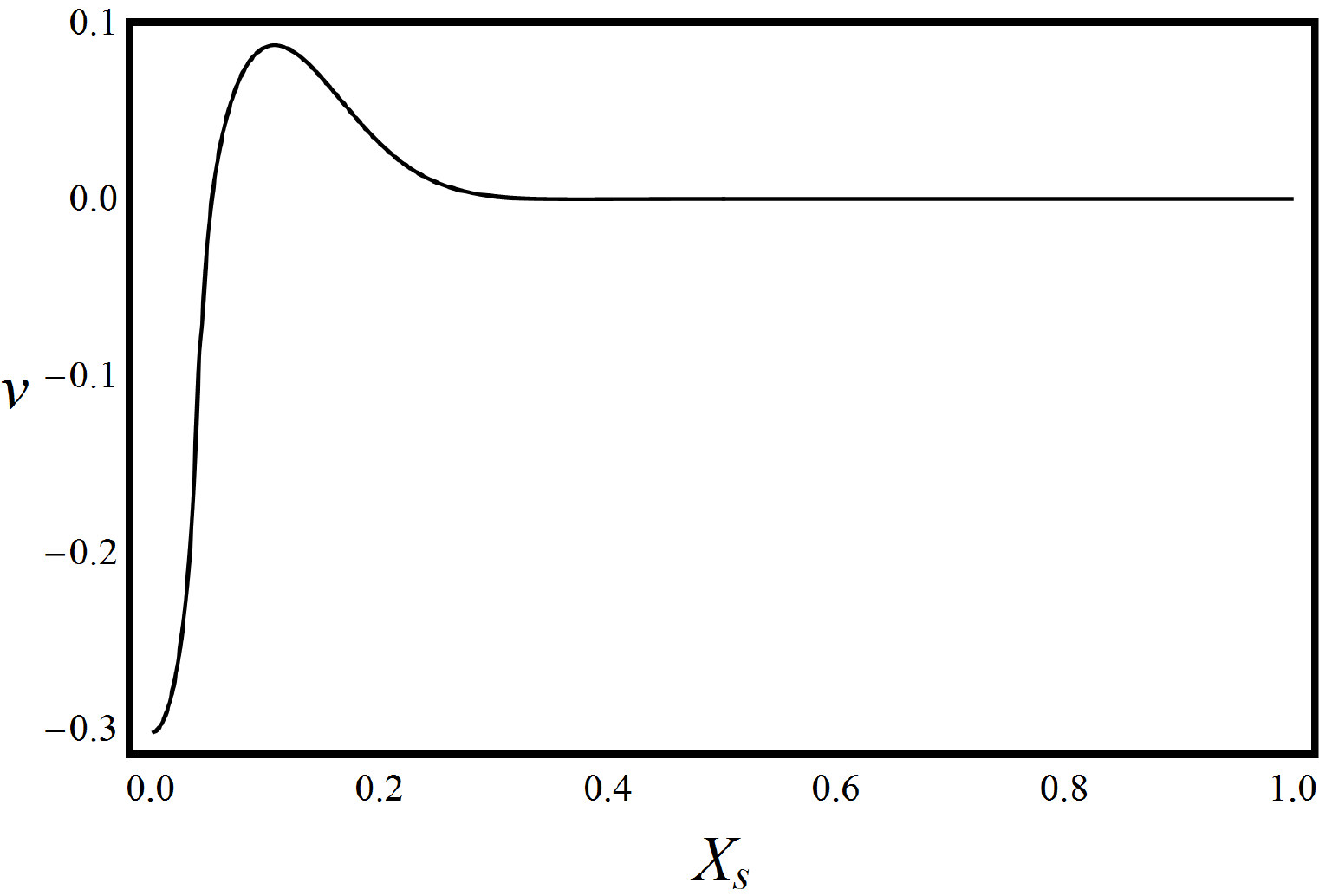}
 \end{minipage}
 \vspace{2mm}
\centering \begin{minipage}[b!h!]{3cm}
\centering   
   \includegraphics[width=4cm]{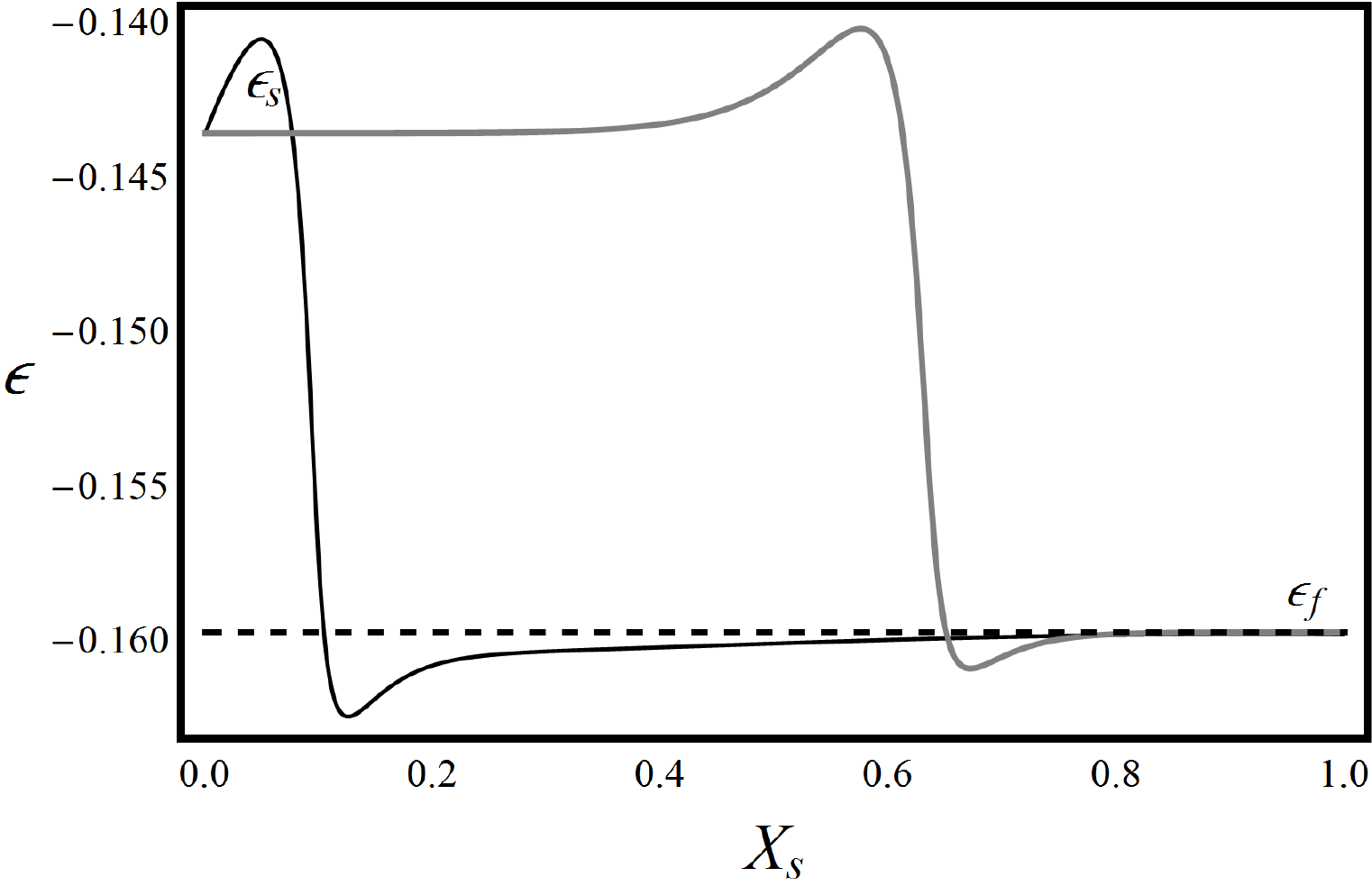}
 \end{minipage}
 \hspace{11mm} 
 \begin{minipage}[b!h!]{3cm}
  \includegraphics[width=4cm]{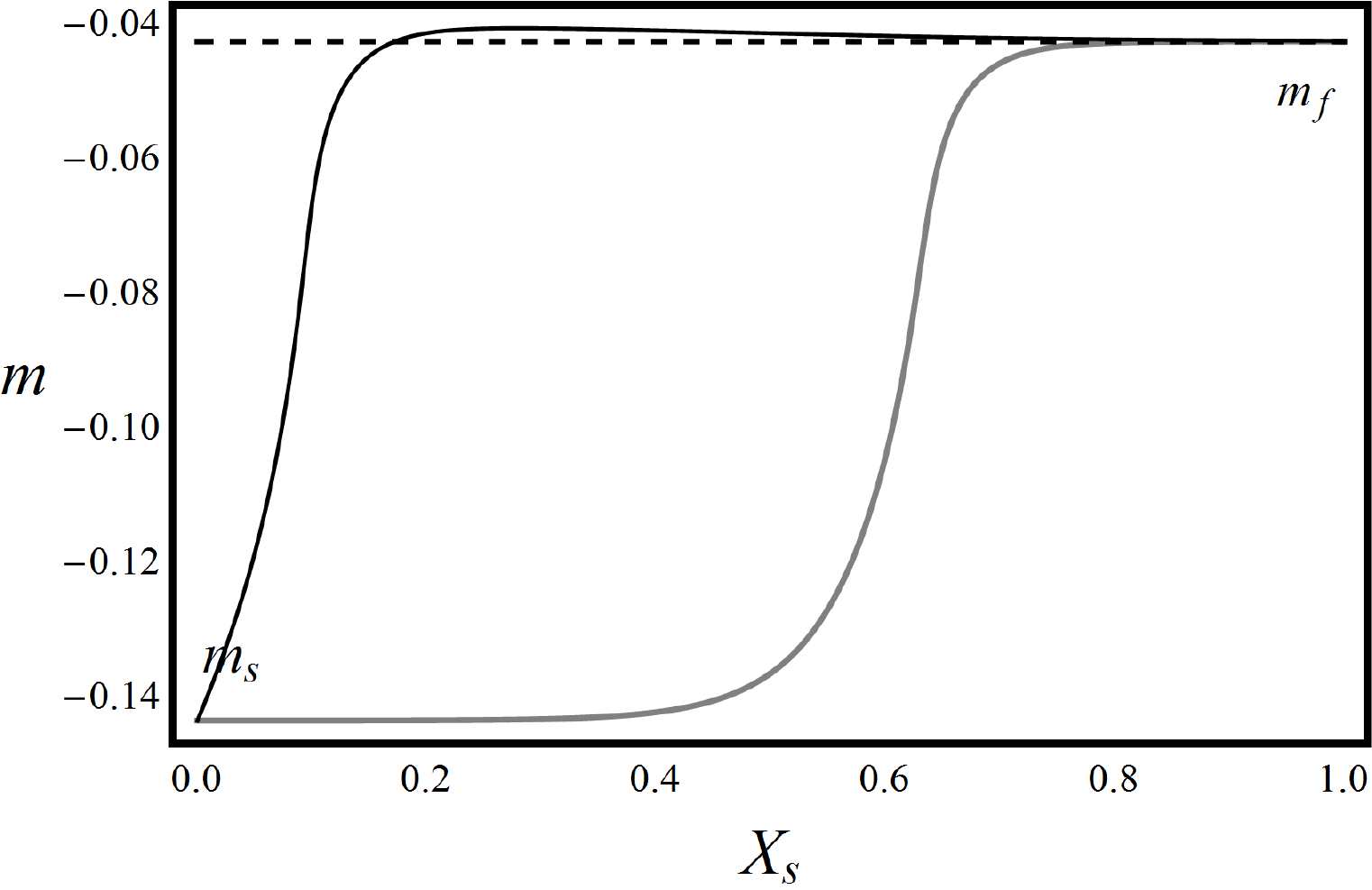}
   \end{minipage}
   \hspace{11mm} 
\begin{minipage}[b!h!]{3cm}
  \centering
   \includegraphics[width=4cm]{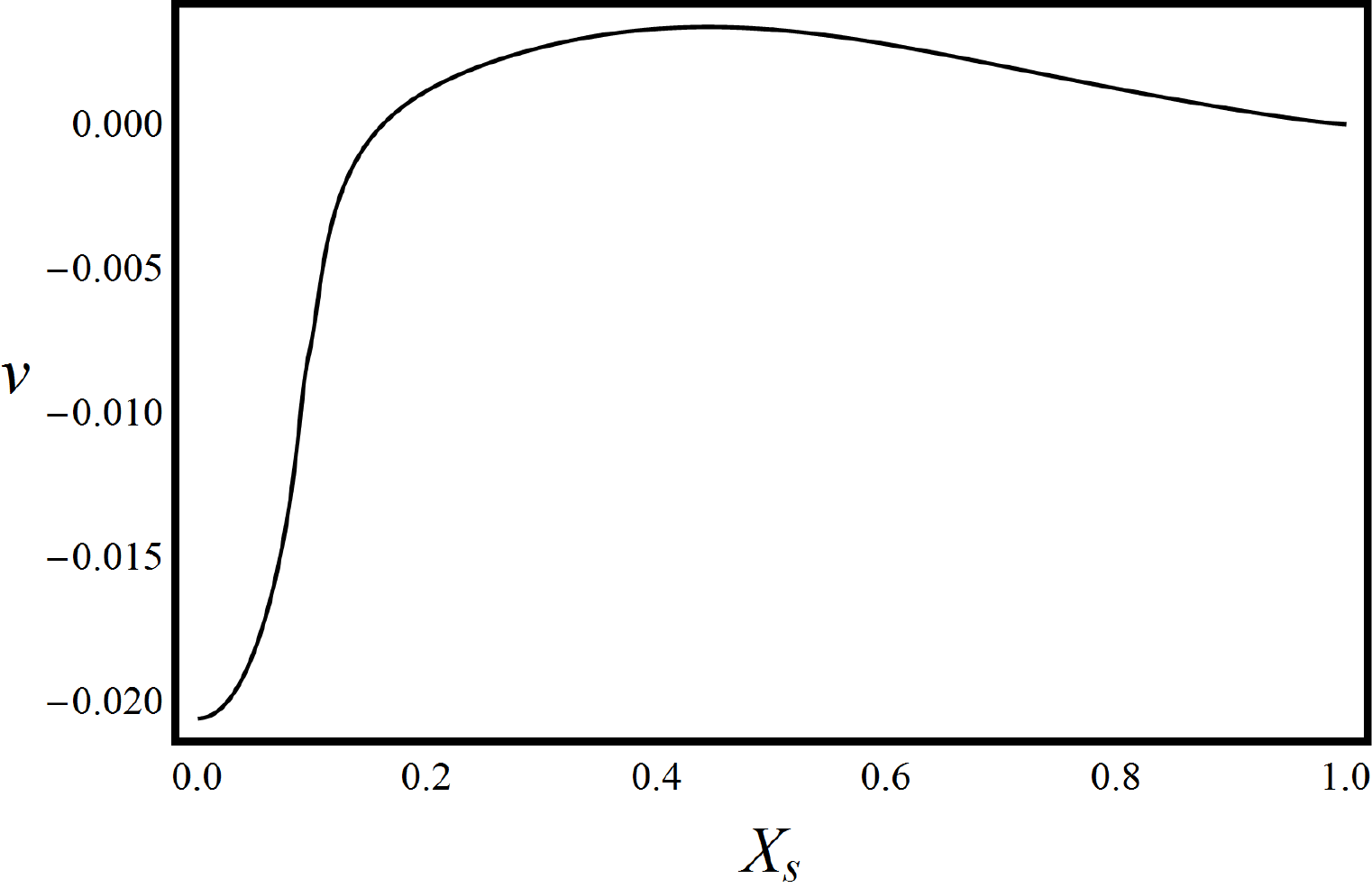}
 \end{minipage}
 \vspace{2mm}
\centering \begin{minipage}[b!h!]{3cm}
\centering   
   \includegraphics[width=4cm]{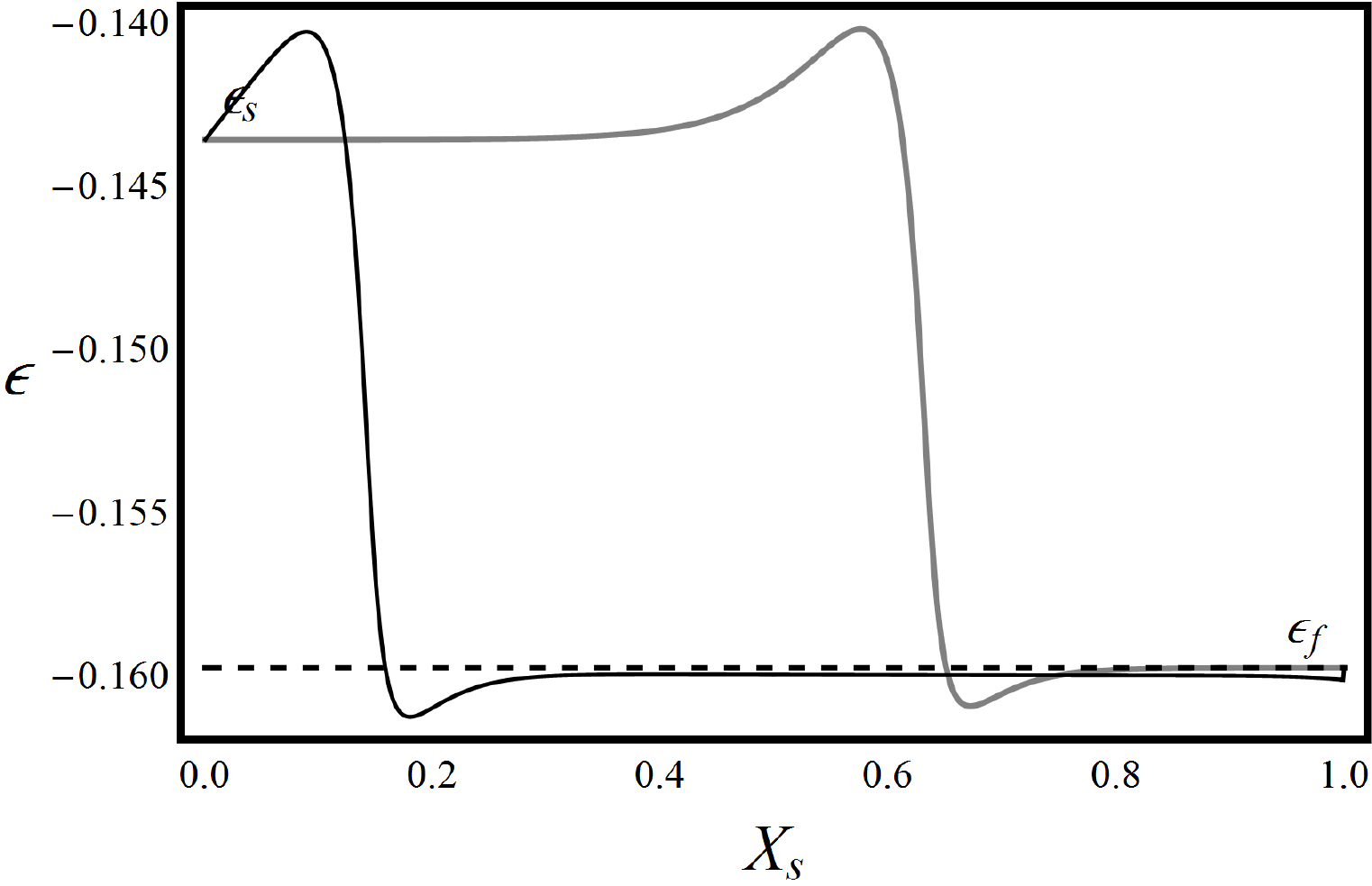}
 \end{minipage}
  \hspace{11mm} 
 \begin{minipage}[b!h!]{3cm}
  \includegraphics[width=4cm]{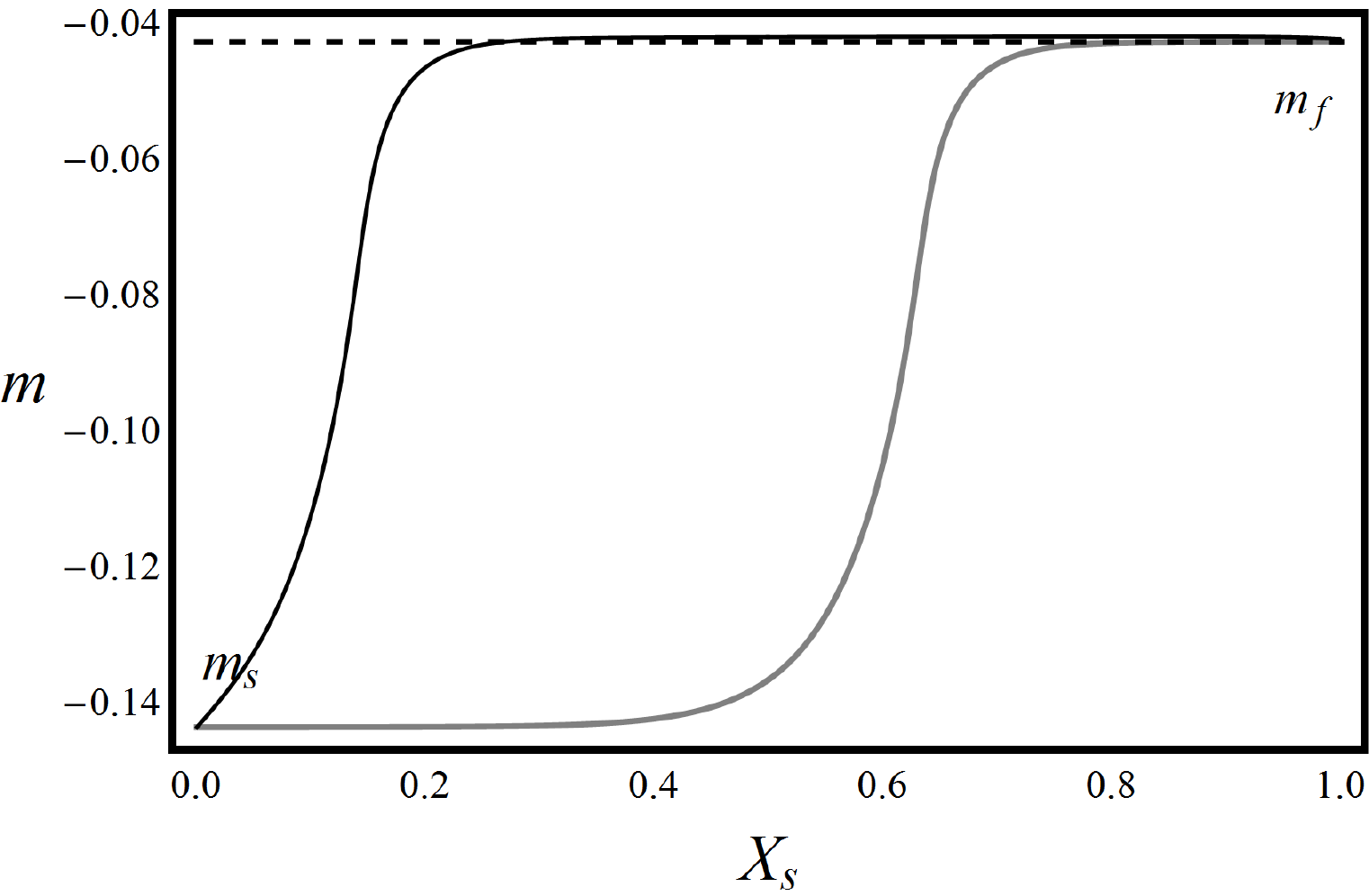}
   \end{minipage}
  \hspace{11mm} 
\begin{minipage}[b!h!]{3cm}
  \centering
   \includegraphics[width=4cm]{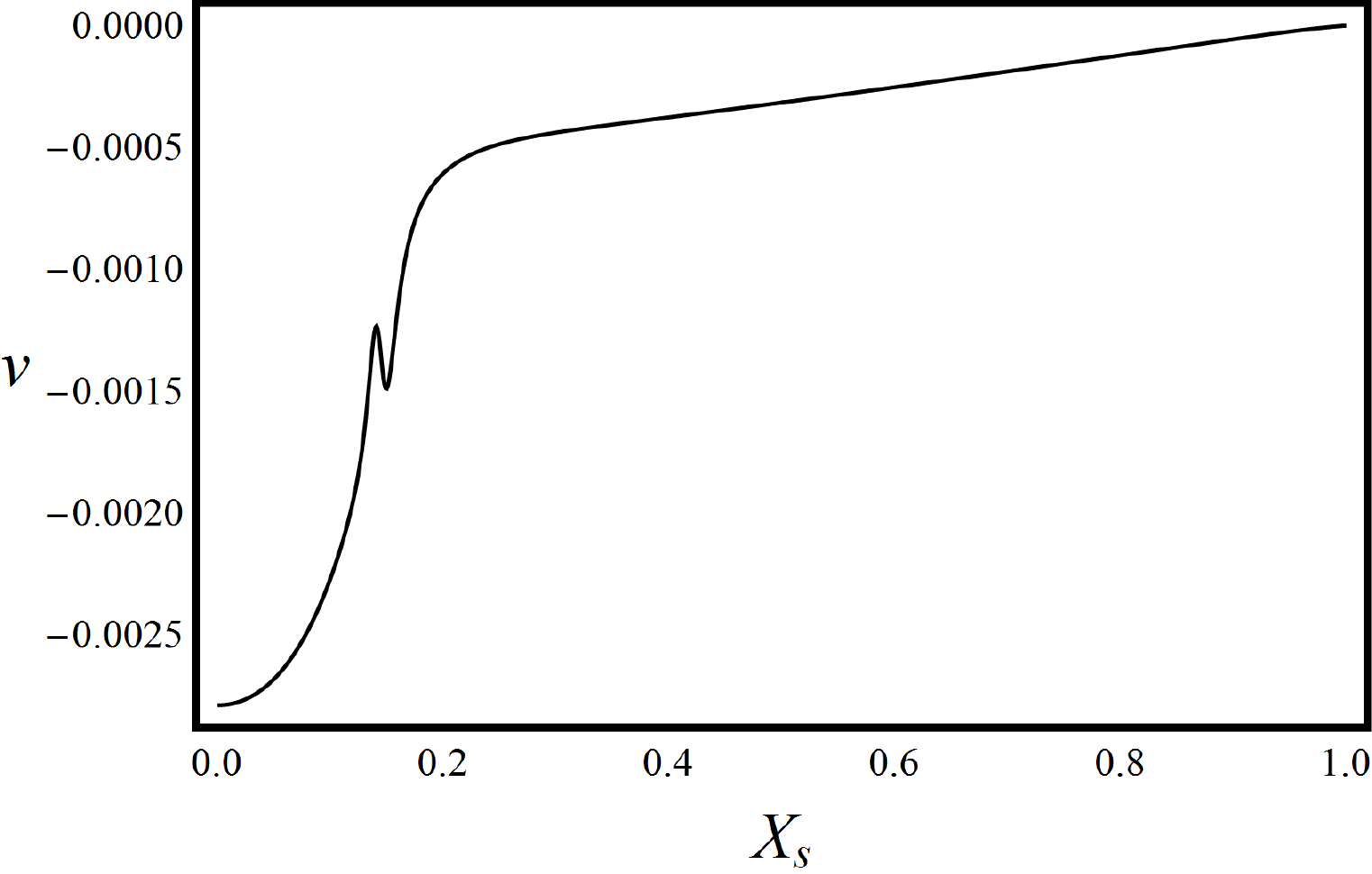}
 \end{minipage}
 \vspace{2mm}
\centering \begin{minipage}[b!h!]{3cm}
\centering   
   \includegraphics[width=4cm]{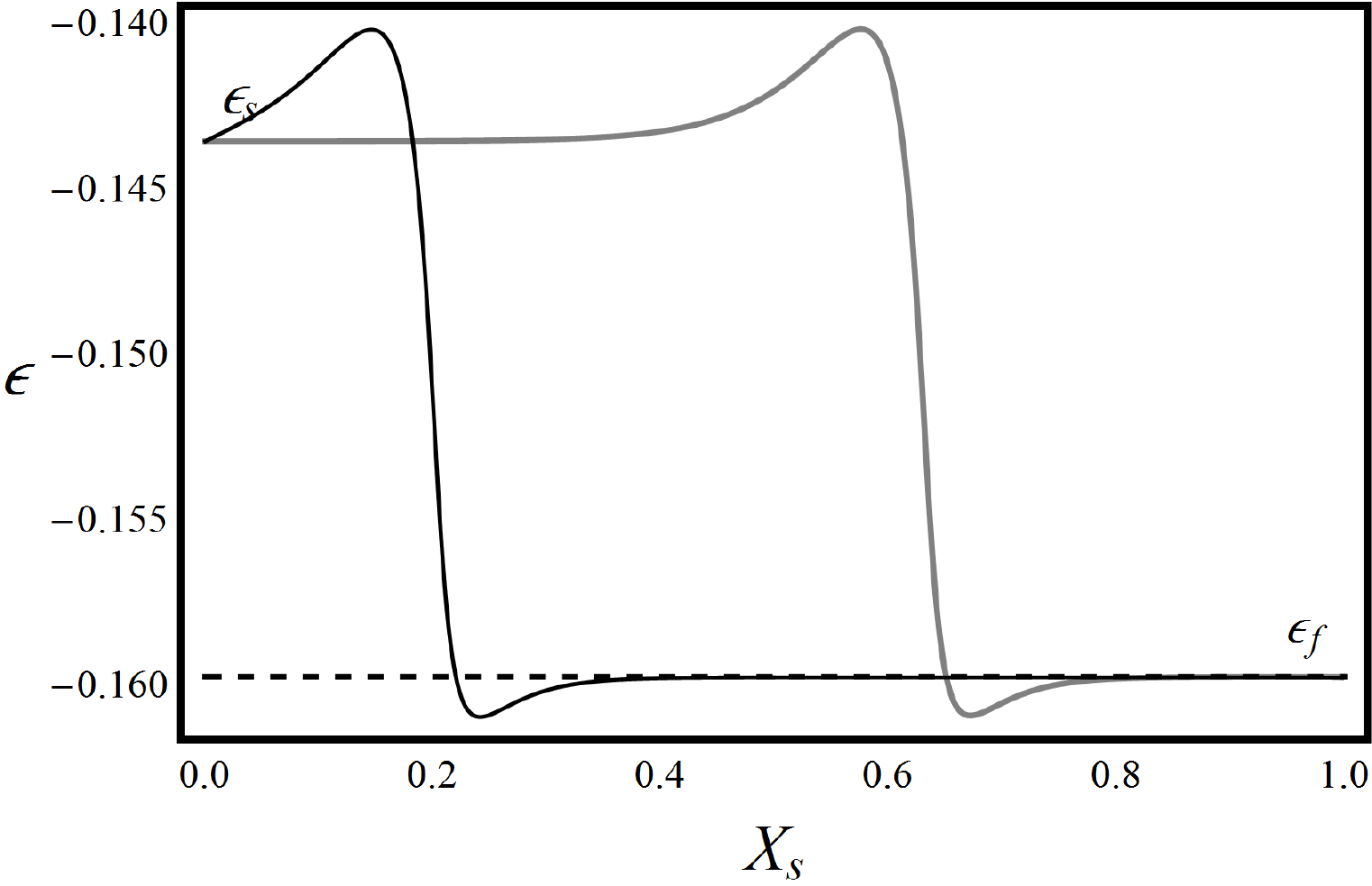}
 \end{minipage}
 \hspace{11mm}  
 \begin{minipage}[b!h!]{3cm}
  \includegraphics[width=4cm]{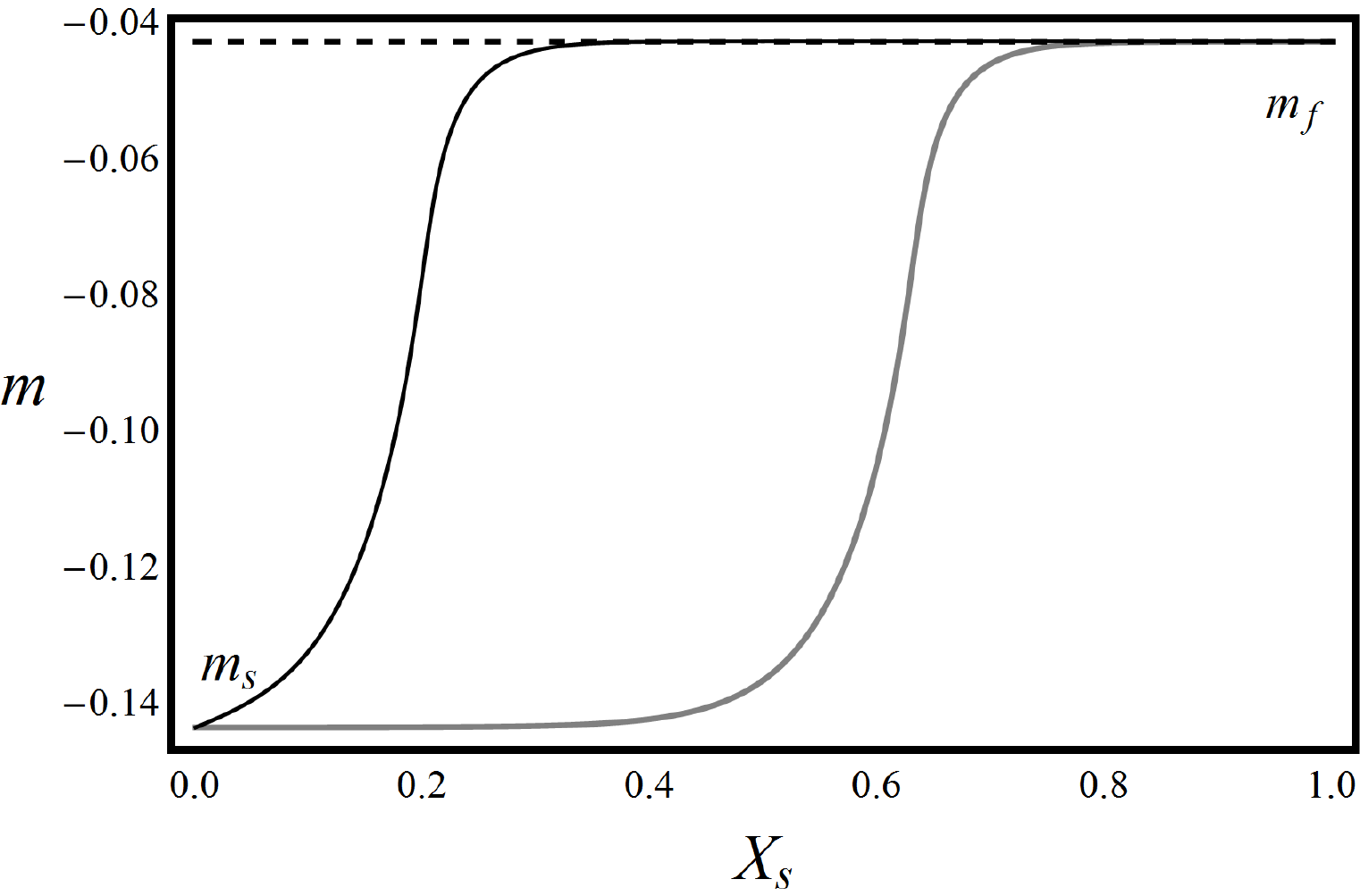}
   \end{minipage}
  \hspace{11mm} 
\begin{minipage}[b!h!]{3cm}
  \centering
   \includegraphics[width=4cm]{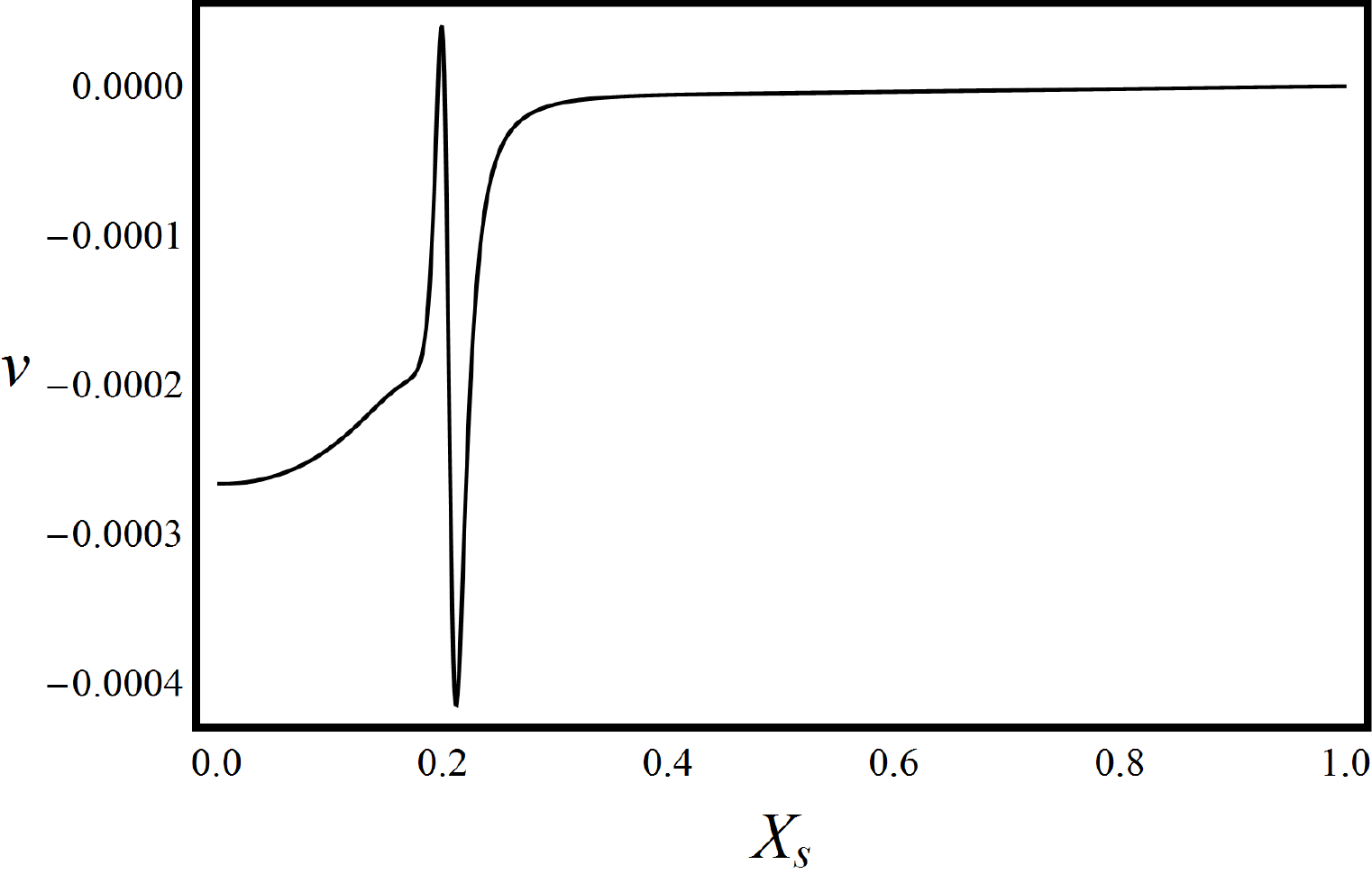}
 \end{minipage}
 \vspace{2mm}
\centering 
\begin{minipage}[b!h!]{3cm}
\centering   
   \includegraphics[width=4cm]{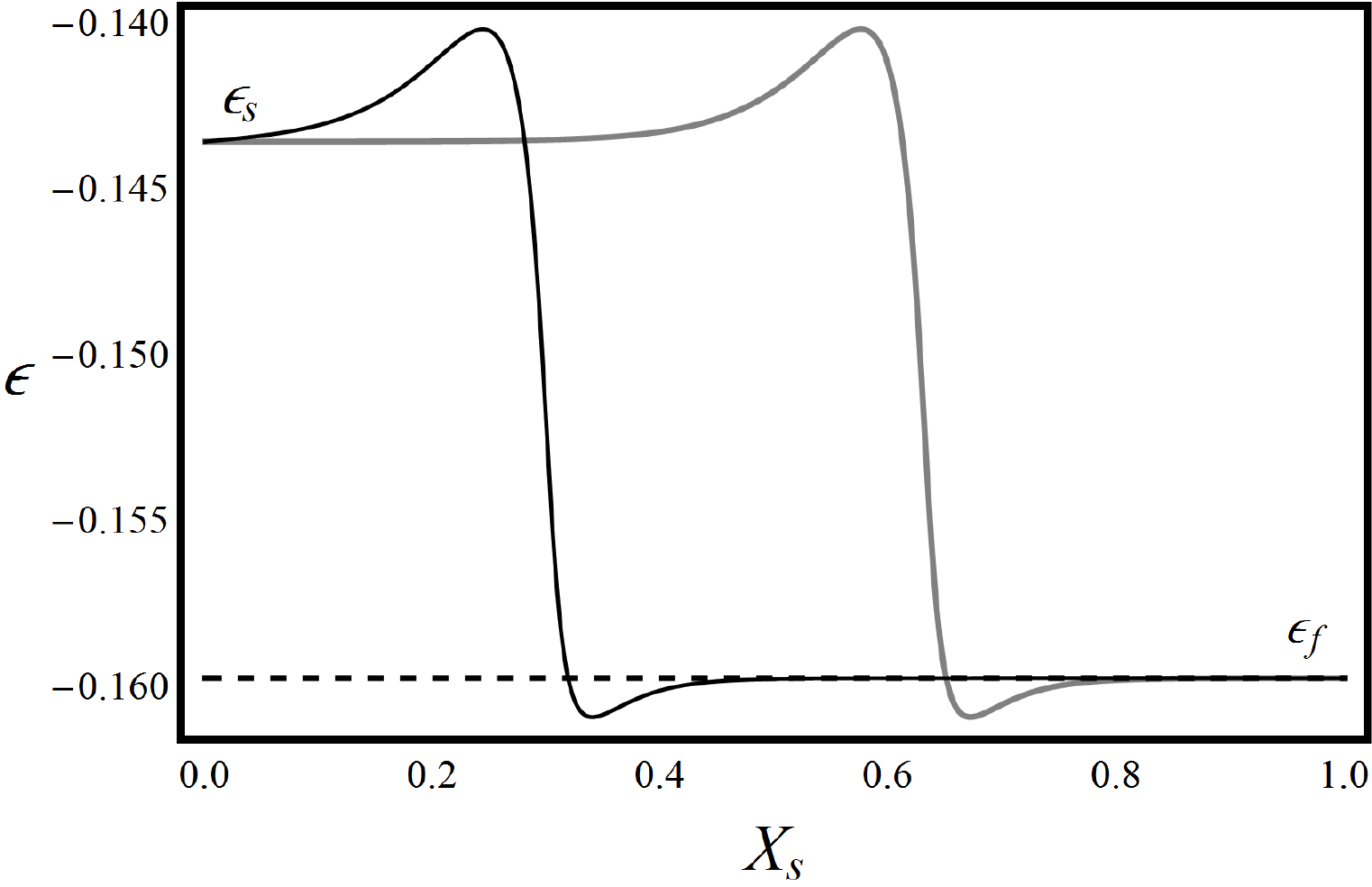}
 \end{minipage}
  \hspace{11mm}   
 \begin{minipage}[b!h!]{3cm}
  \includegraphics[width=4cm]{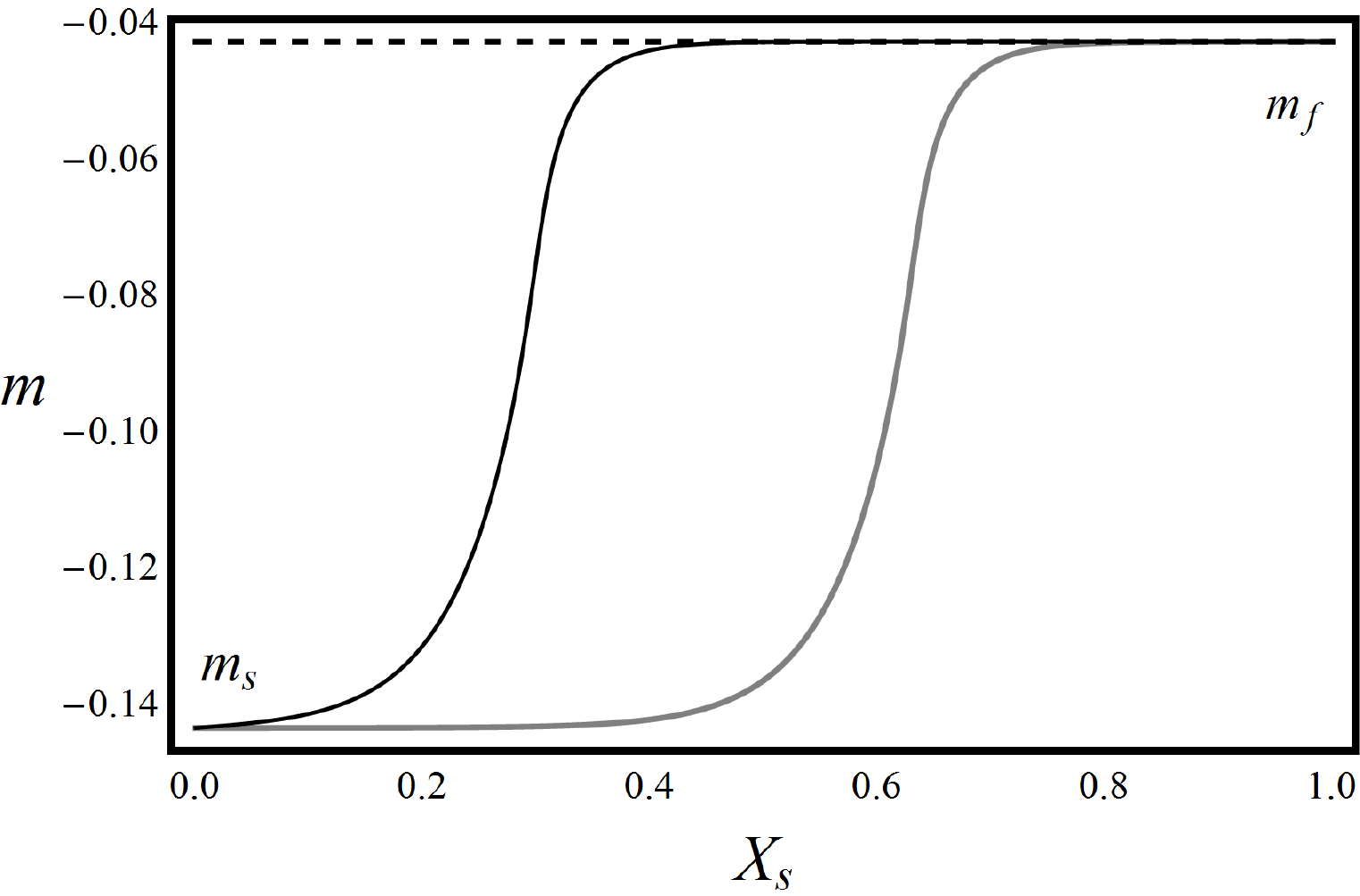}
   \end{minipage}
  \hspace{11mm} 
\begin{minipage}[b!h!]{3cm}
  \centering
   \includegraphics[width=4cm]{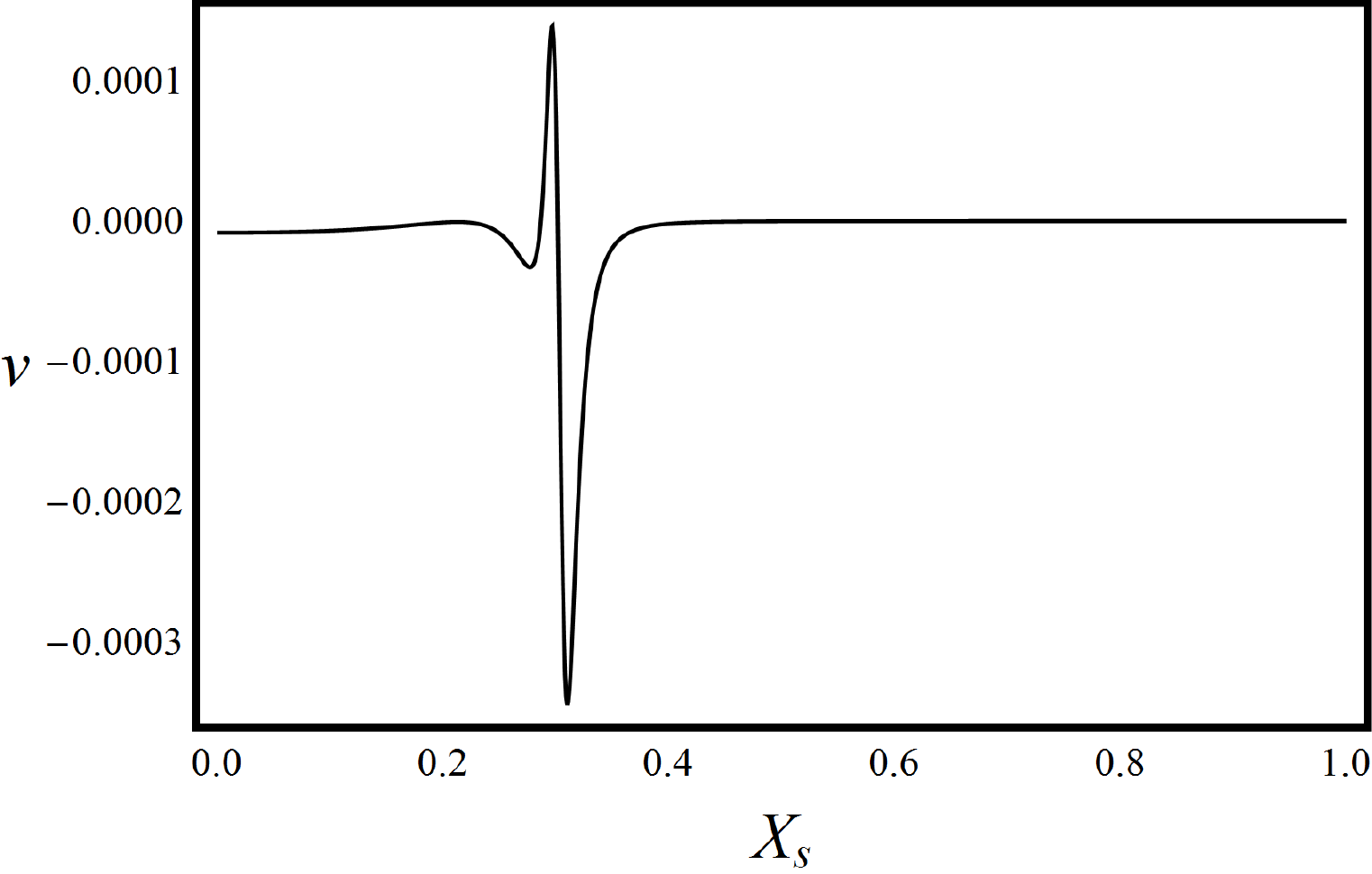}
 \end{minipage}
\vspace{2mm}
\centering 
\begin{minipage}[b!h!]{3cm}
\centering   
   \includegraphics[width=4cm]{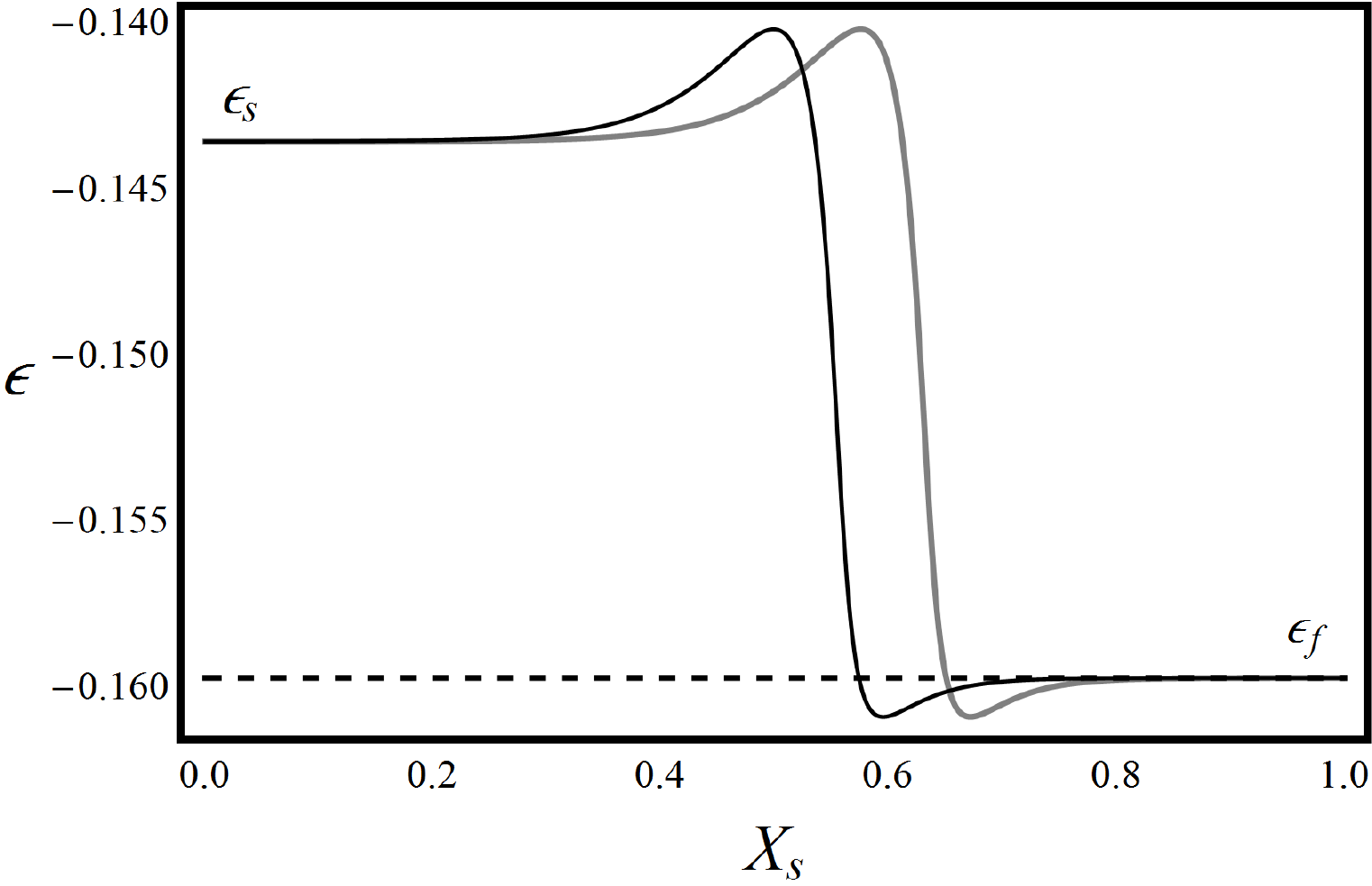}
 \end{minipage}
 \hspace{11mm}   
 \begin{minipage}[b!h!]{3cm}
  \includegraphics[width=4cm]{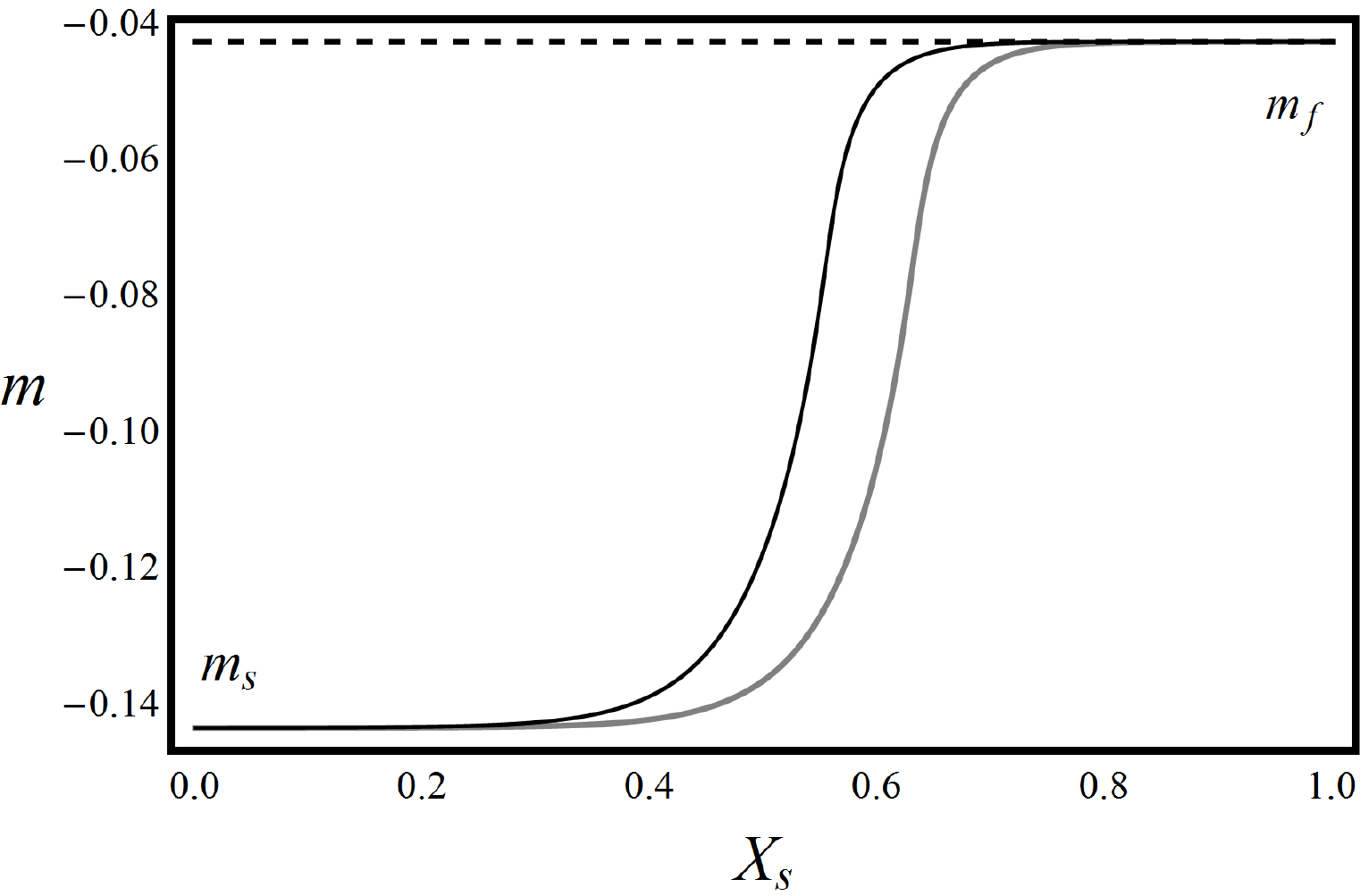}
   \end{minipage}
  \hspace{11mm} 
\begin{minipage}[b!h!]{3cm}
  \centering
   \includegraphics[width=4cm]{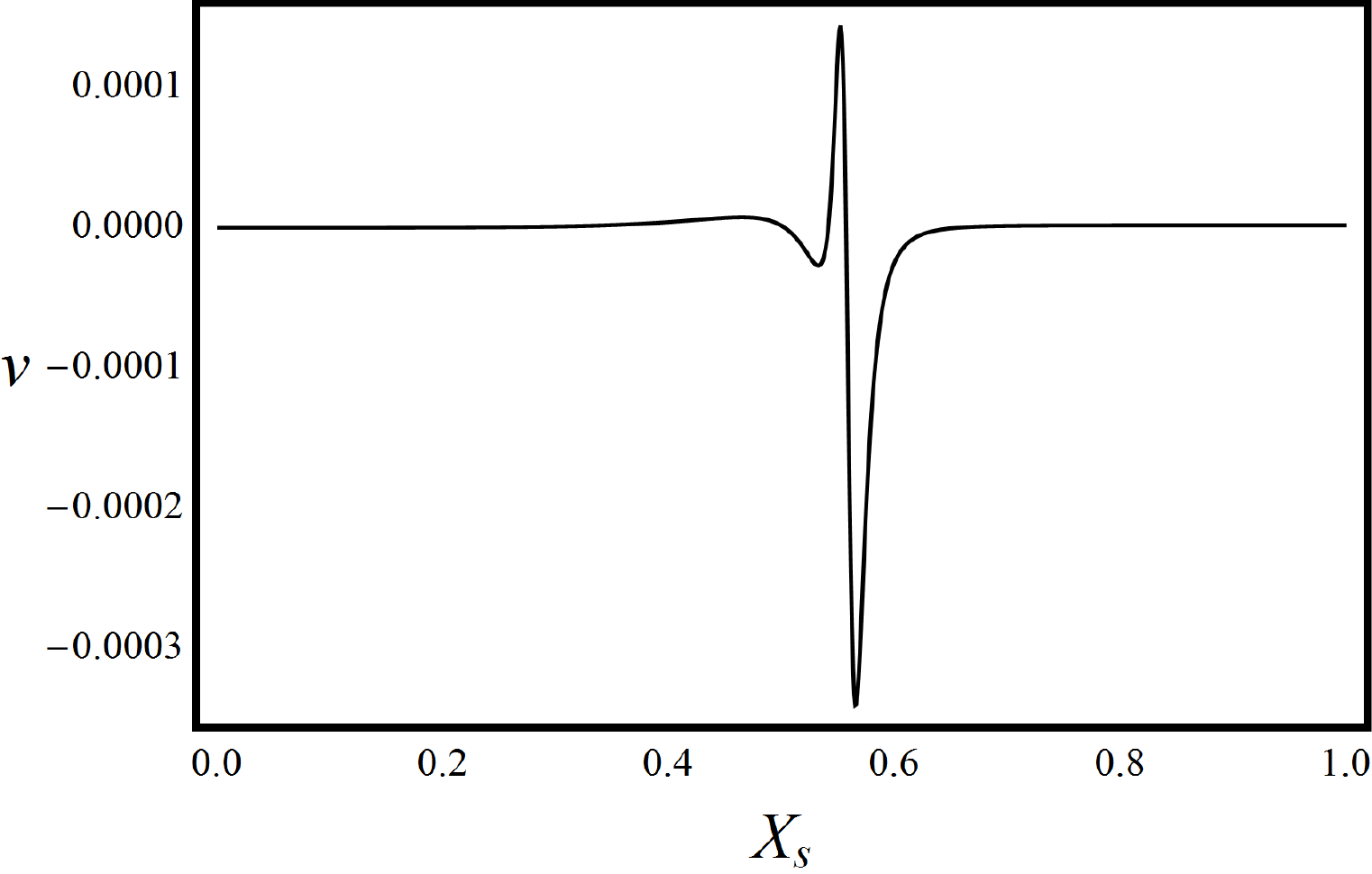}
 \end{minipage}
  \caption{The same as in figure \ref{vecchiofluid} but with the one side impermeability condition.}
\label{tappofluid}
\end{figure}

We discuss now the solution of the problems 
\eqref{num01} e \eqref{num02} with the constant 
fluid--rich profile as initial condition. Results are plotted in 
figures~\ref{vecchiofluid} and \ref{tappofluid}; as before, 
the three columns refer to $\varepsilon$, $m$, and the velocity, 
respectively. 

In this case the system is started with an eccess of fluid throughout 
the sample. The profile matches the Dirichelet boundary condition 
on the right, so that the only phenomenon that has to happen is the 
fluid escape through the left boundary point. For this reason 
the evolutions in the zero chemical potential and the one--side 
impermeable one are very similar. 

With respect to the cases discussed before it is interesting to 
remark that no sort of discontinuity appears in the $\varepsilon$ and $m$ 
profiles close to the right boundary. This is probably due to the fact 
that nothing relevant is happening close to such a boundary point. 

The two evolutions show differences at times $t=0.8$ and $t=8$, see 
figures~\ref{vecchiofluid} and \ref{tappofluid}.
Indeed, the velocity profiles are very similar at early times 
the positive part is almost the same in both cases, 
in the late steps the velocity profile concerning 
the evolution with the zero chemical potential has 
positive values in the middle of the porous medium and in the neighbourhood of the right boundary, 
unlike the velocity profile related to the evolution with the impermeability condition, 
where the velocity profile is all negative in the same region.

Once the profile is formed again the late time velocity graph shows up
and slowly proceeds towards the stationary state moving from the left to the right.
This is consistent with negative peak of the velocity which is greater than the 
positive one.
Finally we can observe that a similar behavior as that relative to the fluid--poor initial condition
can be detected concerning the presence of a kind of jump of the $\varepsilon$ profile outside 
the range $(\varepsilon_f,\varepsilon_m)$, which now concentrates close to the
left boundary, while $m$ has a correspondent overshoot above the $m_f$ value.
In this case however the jump amplitude is smaller than that of the previous one.\\

\section{Conclusions}
The occurrence of skeleton heterogeneous elastic strains and fluid density variations
for a consolidating porous material have been modeled introducing a total potential
energy capable for describing the transition between a fluid--poor and a fluid--rich 
phase inside the porous network. Owing to previous results \cite{CIS2010,CIS2013} an
extended description of the consolidation process has been implemented, within the 
framework of second gradient elasticity, which generalizes that provided by classical Terzaghi's equation.
The capability of the model to accout for phase coexistence allows to capture
the formation of heterogeneous equilibria quite similar to those arising during globally undrained tests when, 
at high level of confinement, the fluid can remain trapped into bands of localized strain.

Further development will be devoted to develop a parametric analysis of the 
consolidation process, with the aim of identifying the ranges of the applied pressure
at which more general heterogeneous equilibria can be attained by the system.

%


\end{document}